\documentclass[%
reprint,
superscriptaddress,
showpacs,preprintnumbers,
 amsmath,amssymb,
 aps,
]{revtex4-1}

\usepackage{slashbox}
\usepackage{graphicx}
\usepackage{dcolumn}
\usepackage{bm}

\usepackage{references}
\usepackage{hyperref} 

\usepackage{equationarray}
\usepackage[normalem]{ulem}
\usepackage[usenames]{xcolor}
\usepackage{booktabs}
\def\calU{{\cal U}}

\newcommand{\be}{\begin{equation}}
\newcommand{\ee}{\end{equation}}
\newcommand{\ba}{\begin{eqnarray}}
\newcommand{\ea}{\end{eqnarray}}
\newcommand{\bd}{\begin{displaymath}}
\newcommand{\ed}{\end{displaymath}}
\newcommand{\mh}{\mathcal{H}}

\usepackage{ulem} 

\begin{document}

\preprint{APS/123-QED}

\title{Thermal properties of hot and dense matter with finite range interactions}

\author{Constantinos Constantinou}
\email{c.constantinou@fz-juelich.de }
\affiliation{Institute for Advanced Simulation, Institut f\"{u}r Kernphysik, and J\"{u}lich Center \\
for Hadron Physics, Forschungszentrum J\"{u}lich, D-52425 J\"{u}lich, Germany}

\author{Brian Muccioli}
\email{bm956810@ohio.edu}
\affiliation{Department of Physics and Astronomy, Ohio University, Athens, OH 45701}

\author{Madappa Prakash}
\email{prakash@ohio.edu}
\affiliation{Department of Physics and Astronomy, Ohio University, Athens, OH 45701}

\author{James M. Lattimer}
\email{james.lattimer@stonybrook.edu}
\affiliation{Department of Physics and Astronomy, Stony Brook University, Stony Brook, NY 11794-3800}

\date{\today}

\begin{abstract}
We explore the thermal properties of hot and dense matter using a model  
that reproduces the empirical properties of isospin symmetric and asymmetric bulk nuclear matter, optical model fits to nucleon-nucleus scattering data, heavy-ion flow data in the energy range 0.5-2 GeV/A, and  the largest well-measured neutron star mass of 2 $\rm{M}_\odot$.  Results of this model which incorporates finite range interactions  
through Yukawa type forces  are contrasted with those of a zero-range Skyrme model that yields nearly identical zero-temperature properties 
at all densities for symmetric and asymmetric nucleonic matter and the maximum neutron star mass, but fails to account for heavy-ion flow data due to the lack of an appropriate momentum dependence in its mean field.  Similarities and differences in the thermal state variables and the specific heats between the two models are highlighted.  
Checks of our exact numerical calculations are performed from formulas derived in the strongly degenerate and non-degenerate limits. 
Our studies of the thermal and adiabatic indices, and the speed of sound  in hot and dense matter for conditions of relevance to core-collapse supernovae, the thermal evolution of neutron stars from their birth and mergers of compact binary stars reveal that substantial variations begin to occur at sub-saturation densities before asymptotic values are reached at supra-nuclear densities.  \\

\noindent Keywords: Hot and dense matter, finite-range potential models, thermal effects.
\end{abstract}

\pacs{21.65.Mn,26.50.+x,51.30.+i,97.60.Bw}
\maketitle


\section{INTRODUCTION}
\label{Sec:Intro}

The modeling of core-collaspe supernovae, neutron stars from their birth to old age, and binary mergers of compact stars requires a detailed knowledge of the equation of state (EOS) of matter at finite temperature.  For use in large-scale computer simulations of these phenomena, the EOS is generally rendered in tabular forms as  functions of the baryon density $n$, temperature $T$, and the electron concentration $Y_e=n_e/n$, where $n_e$ is the net electron density in matter. Examples of such tabulations can be found,  {\it e.g.}, in Refs. \cite{LS,Shen1,Shen2a,Shen2b,Steiner,Hempel,SLM94,Oconnor,sys07}. Entries in such tables include thermodynamic state variables such as the free energy, energy per baryon, pressure, entropy per baryon, specific heats, chemical potentials of the various species and their derivatives with respect to number densities, {\it etc.}, The calculation of a thermodynamically consistent EOS over a wide range of densities ($n$ of $10^{-7}$ to 1-2~${\rm {fm}^{-3}}$) and temperatures up to 100 MeV involves a detailed examination of inhomogeneous phases of matter (with neutron-rich nuclei, pasta-like geometrical configurations, leptons)  at sub-nuclear densities and low enough temperatures, as well as homogeneous phases at supra-nuclear densities with possible  non-nucleonic degrees of freedom (Bose condensates, strangeness-bearing particles, mesons, quarks, and leptons).  

Constraints on the EOS are largely restricted to zero- or low-temperature matter from laboratory experiments involving nuclei  and neutron star observations.  The various sources from which such constraints from experiments and astronomical observations arise are highlighted  below beginning with those at zero-temperature.

Decades of laboratory experiments involving stable and radio-active nuclei have given us a wealth of data about nuclear masses, charge radii, neuron skin thicknesses, 
nucleon effective masses, giant resonances, dipole polarizabilities, {\it etc.}, close to the nuclear equilibrium density of $n_0\simeq 0.16~{\rm fm}^{-3}$ and  isospin 
asymmetry $\alpha=(N-Z)/(N+Z) \simeq (0.35-0.4)$, $N$ and $Z$ being the neutron and proton numbers of nuclei. Through measurements of the collective 
flow of matter, momentum and energy flow,  heavy-ion collisions in the range   $E_{\rm {lab}}/A$ = 0.5-2 GeV have shed light on the EOS of matter up to about $3n_0$. 
Astronomical observations  of neutron stars that have central densities and isospin asymmetries  several times larger than those of laboratory nuclei have  begun to 
compile accurate data on neutron star masses, rotation periods and their time derivatives,  estimates of radii, cooling behaviors that shed light on nucleon superfluidity, 
{\it etc.} Combined with microscopic and model calculations of dense nuclear matter, and the structural aspects and thermal evolution of neutron stars, the EOS of cold 
dense matter is beginning to be pinned down albeit with error bands imposed by laboratory and astronomical data. 

In specific  terms, several properties of nuclei,  extrapolated to 
bulk matter, have yielded values for quantities that characterize the 
key properties of isospin symmetric nuclear matter such as the equilibrium density $n_0$, energy per particle $E_0$ at $n_0$, compression modulus $K_0=9n_0^2(d^2E/dn^2)|_{n{_0}}$, 
and the Landau effective mass $m^*(n_0)$, with small enough one-sigma errors, although smaller errors would be desirable. For isospin asymmetric bulk matter, nuclear data 
have yielded values for the bulk symmetry energy $S_v=S_2(n_0)=(1/2)(\partial E(n,\alpha)/\partial \alpha^2)|_{\alpha=0}$, where now the neutron excess parameter 
$\alpha=(n_n-n_p)/(n=n_n+n_p)$,  $n_n$ and $n_p$ being the neutron and proton densities in bulk matter, respectively.  Percentage wise, the one-sigma error on $S_v$ is larger 
than those of the key parameters that characterize  symmetric nuclear matter. Additionally, constraints on the derivatives of the symmetry energy at nuclear density, 
$L_v=3n_0(dS_2(n)/dn)|_{n{_{0}}}$ and $K_v = 9n_0^2(d^2S_2(n)/dn^2)|_{n{_0}}$ have also emerged, albeit with one-sigma errors that are somewhat large.

Heavy-ion collisions in the energy range $E_{lab}/A = 0.5-2$ GeV have shed light on the EOS  at supra-nuclear densities (up to 
3-4 $n_0$) through studies of matter, momentum, and energy flow of nucleons (for  a clear exposition, see \cite{Bertsch88}).
Collective flow in heavy-ion collisions has been characterized by  (i) the mean transverse momentum per nucleon $\langle p_x \rangle /A$ versus rapidity $y/y_{proj}$ \cite{Danielewicz85}, 
(ii) flow angle from asphericity analysis \cite{Gustafsson84}, (iii) azimuthal distributions \cite{Welke88}, and (iv) radial flow \cite{Siemens79}.  
Detection of neutrons being more difficult than that of protons, flow data gathered to date are largely for protons 
and for collisions of nuclei in which the isospin asymmetry is not large.  
Confrontation of data with theoretical calculations have
generally been performed using Boltzmann-type kinetic equations. One
such equation for the time evolution of the phase space distribution
function $f({\vec r},{\vec p},t)$ of a nucleon that incorporates both
the mean field $U$ and a collision term with Pauli blocking of final
states is (see  Ref. \cite{Bertsch88} for a lucid account):
\def\sst{\scriptscriptstyle}
\begin{eqnarray}
\frac {\partial f}{\partial t} &+& {\vec \nabla}_p U \cdot {\vec \nabla}_r f
- {\vec \nabla}_r U \cdot {\vec \nabla}_p f = \nonumber \\
&-& \frac {1}{(2\pi)^3} \int d^3p_2\,d^3p_{2^{\prime}}\,d\Omega 
\frac {d\sigma_{\sst NN}}{d\Omega} \, v_{12} 
\,\delta^3( {\vec p} + {\vec p}_2 - {\vec p}_{1^{\prime}} 
- {\vec p}_{2^{\prime}} ) \nonumber \\ 
&  \times &
\left[ ff_2 (1-  f_{1^{\prime}}) (1-  f_{2^{\prime}})
- f_{1^{\prime}} f_{2^{\prime}} (1- f) (1- f_2) \right] \,.
\label{BUU}
\end{eqnarray}
In general, the mean field $U$, felt by a nucleon due to the presence of all surrounding nucleons,  depends on both the local density $n$ and the momentum ${\vec p}$ of the nucleon.  
Operationally, $U$ is obtained as the functional derivative of the energy density ${\cal H}$ of matter {\em at zero temperature}: 
$U(n,{\vec p}) \equiv \delta {\cal H}/\delta n$ and serves as an input on the left hand side of  Eq. (\ref{BUU}).
Note that the evolution of the local density $n$ is governed by the off-equilibrium evolution of  $f({\vec r},{\vec p},t)$, which is driven both by spatial and momentum gradients of $U$ 
on the one hand and by hard collisions on the other. 
The collision integral on the right hand side above features the relative velocity $v_{12}$
and the nucleon-nucleon differential cross section ${d\sigma_{\sst NN}}/{d\Omega}$. Thus, Eq. (\ref{BUU}) contains
effects due to both soft interactions (the left hand side) and hard collisions  (the right hand side) albeit at a semiclassical level insofar as phase space distribution functions are evolved in time classically using the Boltzmann-Uehling-Uhlenbeck  or the Vlasov-Uehling-Uhlenbeck   approximation (see Ref. \cite{Bertsch88} for original references, and a clear exposition)  instead of the full quantum evolution of wave functions.

Early theoretical studies that confronted data 
used isospin averaged nucleon-nucleon cross sections and mean
fields of symmetric nuclear matter.  The lesson learned was that
much of the collective behavior observed in experiments stems from
momentum dependent forces at play during the early stages of the
collision \cite{Gale87,Prakash88b,Welke88,Gale90,Danielewicz:00,Danielewicz:02}. The conclusion that emerged from several
studies was that as long as momentum dependent forces are employed in
models that analyze the data, a symmetric matter compression modulus
of $\sim 240\pm 20$ MeV, as suggested by the analysis of the giant monopole
resonance data \cite{Youngblood99,Garg04,Colo04}, fits the heavy-ion data as well
\cite{Danielewicz02}.
 
The prospects of rare-isotope accelerators (RIA's) that can collide
highly neutron-rich nuclei has spurred further work to study a system
of neutrons and protons at high neutron excess
\cite{Das03,Li04,Li04b}. Generalizing Eq.~(\ref{BUU}) to a mixture,
the kinetic equation for neutrons is
\begin{equation}
\frac {\partial f_n}{\partial t} + {\vec \nabla}_p U \cdot {\vec \nabla}_r f_n
- {\vec \nabla}_r U \cdot {\vec \nabla}_p f_n = J_n = \sum_{i=n,p} J_{ni}\,, 
\end{equation}
where $J_n$ describes collisions of a neutron with all other neutrons
and protons.  A similar equation can be written down for protons with
appropriate modifications.  On the left hand side of each coupled
equation the mean field $U\equiv U(n_n,n_p;{\vec p}\,)$ depends
explicitly on the neutron-proton asymmetry. The connection to the
symmetry energy arises from the fact that $U$ is now obtained from a
functional differentiation of the Hamiltonian density ${\cal H}(n_n,n_p)$ of isospin asymmetric matter.
Examples of such mean fields may be found in
Refs.~\cite{Prakash97,Li04,Li04b}.  Isospin asymmetry can be expected to
influence observables such as the neutron-proton differential flow and the ratio of neutron to
proton multiplicity as a function of transverse momentum at mid-rapidity.  Experimental investigations of these signatures await
the development of RIA's at GeV energies.  

On the astrophysical front, precisely measured neutron star masses and radii severely constrain the EOS of isospin asymmetric matter. The recently well-measured neutron star 
masses $1.97\pm 0.04~{\rm M}_\odot$ \cite{Demorest} and $2.01 \pm 0.04~{\rm M}_\odot$ \cite{Antoniadis} have served to eliminate many, but not all,  EOS's in which a substantial 
softening of  the neutron-matter EOS occurred due to the presence of Bose condensates, hyperons, quark matter, {\it etc.}, \cite{LP:11}. A precise measurement of the radius of the same neutron 
star for which a mass has been well measured is yet lacking, but reasonable estimates have been made by analysis of X-ray emission from isolated neutron stars, intermittently quiescent 
neutron stars undergoing accretion from  a companion star, and neutron stars that display type I X-ray bursts from their surfaces \cite{Steiner10,Steiner13,Lat12}. 

A common feature shared by all of the EOS's in Refs.  \cite{LS,Shen1,Shen2a,Shen2b,Steiner,Hempel,SLM94,Oconnor,sys07} is that at zero-temperature they would fail to reproduce heavy-ion flow data. The reason is in both non-relativistic and relativistic mean field-theoretical models, the momentum dependence of the mean-field would lead to a linear dependence on the energy as shown in Refs. \cite{Gale87,Prakash88b,Welke88,Gale90}
and \cite{Jaminon:81,ABBCP} and hence at odds with optical model fits to nucleon-nucleus scattering data \cite{Hama:90,Cooper:93,Danielewicz:00,Danielewicz:02}.  
Potential models with finite-range interactions ({\it e.g.}, \cite{Decharge:80}) solve this problem \cite{Prakash88b}, but with attendant changes in their thermal properties compared to zero-range models as we will show in this work. A similar resolution in the case of field-theoretical models requires further work and is not addressed here.

One of the objectives in this paper is to inquire whether the EOS's that have successfully explained heavy-ion flow data are able to support the largest well-measured neutron star 
mass and also if the radii of 1.4 ${\rm M}_\odot$ stars are in accord with bounds established by analyses of currently available X-ray data. In order to address this issue, we examine models in 
which finite range interactions are employed as they yield a momentum dependence of the mean field $U$ that differs significantly at high momenta from that of zero-range Skyrme-like 
models. Specifically, we  use the MDI model of Das {\em et al}. \cite{dggl03}  in which the momentum dependence was generated through the use of a Yukawa type finite-range interaction to 
match the results of  microscopic calculations and to reproduce optical model fits to nucleon-nucleus scattering data. A revised parametrization of the MDI model 
(labeled MDI(A)) was found necessary to support a 2 solar mass neutron star. We note that Danielewicz {\em et al}., show the EOS for pure neutron matter in Fig. 5 of  Ref. \cite{Danielewicz02}, but the maximum mass of a beta-stable neutron star was not quoted.

Another objective of the present work is to examine the thermal properties of models with finite range interactions and contrast them with those of a zero-range Skyrme model. For this 
purpose, we have chosen the SkO$^\prime$ parametrization \cite{skop} of the zero-range Skyrme model that yields nearly identical zero-temperature properties such as the energy per particle, pressure, {\it etc.}, 
as the MDI(A) model.  However, the single particle potentials $U(n_n,n_p,p)$ differ substantially between these models, the MDI(A) model being constant at  high momenta  in contrast to the quadratic rise of the SkO$^\prime$ model. Consequently, the neutron and proton effective masses, $m_n^*$ and $m_p^*$, also exhibit distinctly different density dependences 
although the isospin splitting of the effective masses is similar in that $m_n^* > m_p^*$ in neutron-rich matter in accordance with Brueckner-Hartree-Fock (BHF) and relativistic BHF 
calculations \cite{Zuo06}. Insofar as the effective masses control the thermal properties, attendant differences in several of the state variables become evident.  
Several of the Skyrme parametrizations exhibit a reversal in behavior of the isospin splitting so that $m_n^*<m_p^*$ in neutron-rich matter (see, for example, Ref. \cite{LiHan} for a compilation, and \cite{chakraborty2015neutron}). 
This has led us to 
establish conditions on the strength and range parameters of the Skyrme and MDI models in which $m_n^*>m_p^*$. 
The numerical results of the models studied here are supplemented with analytical results in the limiting cases of degenerate and non-degenerate matter both as a check of our numerical evaluations and to gain physical insights.  
For use in astrophysical simulations, a detailed analysis of hot and dense matter  containing leptons (electrons and positrons, and in some cases trapped neutrinos) and photons is also performed.
Inhomogeneous phases below sub-nuclear densities and exotic phases of matter at supra-nuclear densities are not considered here as they fall beyond the scope of this work, but will be taken up in a subsequent study.

The organization of this paper is as follows. 
Section  \ref{Sec:Astro} provides an overview of the role of thermal effects in core-collapse supernovae, evolution of proto-neutron stars and binary mergers of compact  stars.  
In Sec. \ref{Sec:models}, we describe the finite-range (and hence momentum dependent) and zero-range models used to study effects of finite 
temperature.  For the extraction of effects purely thermal in origin, the formalism to evaluate the zero temperature state variables (energies, pressures, chemical potentials, {\it etc.}) of 
both these models is also presented in this section.  Section  \ref{Sec:ResultsT0}  presents an analysis of the zero-temperature results based on different parametrizations of the models 
chosen.  Thereafter, differences in the momentum dependence of the single particle potentials between these two models are highlighted.  
Particular emphasis is placed on the density and isospin dependences of the Landau effective masses which mainly control the thermal effects to be discussed in subsequent sections. In 
Sec. \ref{Sec:Teffects}, results of the exact, albeit numerical, calculations of the thermal effects are presented. This section also contains comparisons with analytical 
results in the limiting  cases of degenerate and non degenerate matter. 
Thermal and adiabatic indices, and the speed of sound in hot and dense matter of relevance to hydrodynamical simulations of astrophysical phenomena involving supernovae and compact stars are discussed in Sec. \ref{Sec:Tindices}. 
Our conclusions are in Sec. \ref{Sec:Conclusion}. 
Formulas that are helpful in computing the various state variables of the MDI model at zero temperature are collected in  Appendix \ref{Sec:AppendixA}. 
In Appendix \ref{Sec:AppendixB}, the analytical method by which 
the non degenerate limit of the MDI model is addressed is presented. 
Details concerning the evaluations of the 
specific heats at constant volume and pressure for the MDI and Skyrme models are given in  Appendix \ref{Sec:AppendixC}.

\section{Thermal Effects in Astrophysical Simulations}
\label{Sec:Astro}
The effects of temperature in astrophysical simulations involving
dense matter are most visible in gravitational collapse supernovae,
the evolution of proto-neutron stars, and mergers involving neutron
stars -- either neutron star-neutron star (NS-NS) or black hole-neutron
star (BH-NS) mergers.  

\subsection{Thermal effects in supernovae and proto-neutron stars}

The effects of temperature in supernovae and in
proto-neutron stars remain largely unexplored in detail, but due to
the fact that for the most part matter in such environments is
degenerate, uncertainties in the thermal aspects of the equation of
state (EOS) do not play major roles in the early core-collapse phase.
For example, maximum central densities at bounce are practically
independent of the assumed EOS (see, {\it e.g.}, \cite{Oconnor}). 
The evolution of proto-neutron stars formed following core-collapse will be more
sensitive to thermal effects, as temperatures beyond 50 MeV are reached
in the stellar cores and specific entropies of order 10$k_B$ are reached
in the stellar mantles \cite{LB86, Burrows88}. 
The maximum proto-neutron star mass and the evolution towards black hole
formation will be dependent upon thermal effects.  While the relative
stiffness of the EOS (defined through the incompressibility at
saturation) largely controls the timescale for black hole formation,
thermal effects are important in determining the highest central
densities reached after bounce. Reference \cite{Oconnor} found that the
thermal behavior was more important than incompressibility in this
regard, largely because of thermal pressure support in the hot,
shocked mantles of proto-neutron stars.  In situations in which black
holes do not form and a successful explosion ensues, binding energy is
largely lost due to neutrino emission and heating in the early
evolution of proto-neutron stars, due to neutrino downscattering from
electrons, and the total neutrino energy dwarfs the thermal energy
reservoir.  Nevertheless, the temperature at the neutrinosphere, which
is located in semi-degenerate regions, will be sensitive to thermal
properties of the EOS.  It remains largely unexplored how the resulting
neutrino spectrum, including average energies and emission timescales,
depend on thermal aspects of the EOS.

\subsection{Thermal effects in mergers of binary stars}

Thermal effects are not expected to play a major role in the evolution
of inspiralling compact objects up to the point of merger.  However,
the evolution of the post-merger remnant and some of the mass ejected
could be significantly affected by thermal properties of matter.
Perhaps the most significant recent development is the emergence of a standard
paradigm concerning mergers of neutron stars.  This has been triggered
by the discovery of pulsars with approximately two solar masses, and
strong indications that even larger mass neutron stars exist from a series
of studies of the so-called black widow and redback pulsar systems (for a
review, see \cite{Lat12}).  Most neutron stars in
close binaries have measured gravitational masses in the range 1.3 to
1.5 ${\rm M_\odot}$.  The gravitational mass of the merger remnant will be
less than twice as large, due to binding and due to the ejection of
mass.  It is unlikely that mass ejection will amount to more than a
few hundredths of a solar mass, but binding energies will absorb
larger masses.  The binding energy fraction, the relative difference
between baryon and gravitational masses, can be expressed by a
relatively universal relation, {\em i.e.,} independent of the neutron star
equation of state (EOS), involving only mass and radius \cite{LP:07}:
\begin{equation}
{\rm {BE\over M}}\simeq(0.60\pm0.05){\beta\over1-\beta/2},
\end{equation}
where $\beta=GM/(Rc^2)$.  EOSs capable of supporting $2{\rm M_\odot}$
maximum masses have the property that, for intermediate mass stars,
the radii are nearly independent of the mass.  Furthermore, a
concordance of experimental nuclear physics data, theoretical neutron
matter studies, and astrophysical observations suggests this radius is
about $R=12\pm0.5$ km \cite{Lat12}.  Thus, two equal-mass stars with gravitational
masses of $1.3{\rm M_\odot}$ will have a total baryon mass of $2.87{\rm M_\odot}$.
In a merger event with no mass loss, a remnant of gravitational mass
$2.28{\rm M_\odot}$ would form assuming its radius is also 12 km,
representing an additional mass defect of $0.05{\rm M_\odot}$ relative to
the initial stars.  Should its radius decrease to about 10 km, its
gravitational mass would be approximately $1.9{\rm M_\odot}$, and the
additional mass defect would steepen to $0.43{\rm M_\odot}$.  This
gravitational mass could well be below the cold maximum mass.
Repeating the above estimates for two $1.5 {\rm M_\odot}$ gravitational mass
stars, we find a combined gravitational mass in the range
${\rm 2.16M_\odot}$-${\rm 2.59M_\odot}$, with mass defects larger than for the
previous case.  Therefore, it seems likely that the merged object will
be close to its cold maximum mass.

Studies show that the merged star will be rapidly rotating, and the
rotation may be highly differential \citep{Baiotti08,Rezzolla10,Sekiguchi11}.  
Uniform rotation can increase the maximum mass by a few tenths of percent (see, {\it e.g}., Ref. \cite{CPL:01}), and
differentially rotating objects can support further mass increases.
In all likelihood,  the merged object will be 
metastable.  This possibility is enhanced if the stellar core is
surrounded by a nearly Keplerian disc, a configuration with an even
larger metastable mass limit.  If the merged remnant mass is above the
cold gravitational maximum mass, but less than what rotation is
capable of supporting, it is said to be a ``hypermassive'' 
neutron star, or HMNS.  For a uniformly
rotating star, the maximum equatorial radius increase due to rapid
rotation is about 50\%, with a smalller change in the
polar radius.  The average density of an HMNS in that case would be
less than half of its non-rotating value.  As a result, thermal
effects can be expected to play a much larger role in the stability of
an HMNS than for ordinary neutron stars in which higher degeneracies
exist.  The maximum mass of the HMNS will decrease with time due to
loss of thermal energy from neutrino emission and from loss of angular
momentum due to uniformization of the differential rotation  \cite{Shapiro00,Morrison04}.  Early
calculations, for example those of \citep{Baiotti08,Rezzolla10},
showed that collapse of the metastable HMNS to a black
hole was induced by dissipation of differential rotation and
subsequent loss of angular momentum.  In contrast, \citep{Sekiguchi11} 
argued that thermal
effects are much more important than rotation in determining the
stability and eventual collapse of an HMNS.  In either case, the HMNS
lifetime will crucially depend on the relative difference between the
HMNS mass and the value for the cold maximum mass for the same number
of baryons. The HMNS lifetime, which can range from 10 ms to several seconds
\cite{Sekiguchi11} is potentially measurable through the duration of
short gamma-ray bursts associated with neutron star mergers or the
duration of gravitational wave signals from these mergers.  Such a
measurement therefore has the potential of illuminating the EOS.

Kaplan et al. \cite{Kaplan14} 
have studied the thermal enhancement of the pressure comparing two
tabulated hot EOSs,  LS220 \cite{Lattimer91} 
and Hshen \cite{Shen11}, 
often used in merger simulations.
Generally, thermal enhancements above 3 times
the nuclear saturation density $\rho_s\simeq3\times10^{14}$ g
cm$^{-3}$ are less than about 5\%, but at $\rho_s/3$-$\rho_s/2$ the
pressure is 3 (for Hshen) or 5 (for LS220) times the cold value.  As
we will see, this difference is related to the behavior of the nucleon
effective mass.  In the stellar envelope, for densities from $10^{12}$
to $10^{14}$ g cm$^{-3}$, the thermal enhancements are even larger,
ranging from a factor of 10 (Hshen) to 20 (LS220) for a thermal
profile in which the average temperature in this density range is
about 10 MeV, similar to that found in the merger simulations
from \citep{Sekiguchi11}, who employed the Hshen EOS.  In the
envelope, temperatures are high enough such that nuclear dissociation
is virtually complete, and the thermal pressure differences can once
again be traced to the behavior of the nucleon effective mass.

Thermal pressure alone is capable of increasing the maximum mass more
than 10\% above the cold, catalyzed value for a given EOS \cite{Prakash:97}.  For 
constant temperature $T=50$ MeV profiles studied by \citep{Kaplan14},
the maximum mass of a hot Hshen star was increased by 15\%, while that
of a hot LS220 star was increased by approximately 5\%.  For thermal
profiles similar to the simulations of \citep{Sekiguchi11}, the
increases were more modest: 3.5\% and 1.8\%, respectively.

Hot rotating configurations can show either an increase or decrease in
mass limits relative to cold rotating configurations \cite{Kaplan14}.  When the
central densities are less than about $10^{15}$ g cm$^{-3}$, thermal
effects increase the masses supported at the mass-shedding rotational
limit, and this effect reverses at higher densities.  However, the
rotational frequency and maximum mass at the mass-shedding limit
always decrease due to thermal support: the mass-shedding limit is
very sensitive to the equatorial radius, which increases with thermal
support.  Thermal effects essentially disappear in determining the
maximum masses of extremely differentially rotating configurations
because the outer regions are largely Keplerian and therefore
centrifugally supported.

The analysis by \citep{Kaplan14} concludes that thermal pressure
support plays little role during the bulk of the evolution of an HMNS,
but contributes to the increase of its lifetime by affecting its
initial conditions.  As hot configurations with central densities
less than about $10^{15}$ g cm$^{-3}$ support larger masses than
cold ones with the same central density, those remnants with lower
thermal pressure need to evolve to higher central densities to achieve
metastability.

In the case of BH-NS mergers, although the remnant will always involve
a black hole, simulations indicate the temporary existence of a
remnant disc.  Although the disc is likely differentially rotating and
largely supported by centrifugal effects, and its evolution
controlled by angular momentum dissipation, thermal effects will be
important for dissipation due to various neutrino processes and
emission mechanisms.  

Many early simulations of BHNS mergers employed cold $\Gamma$-law EOSs with
thermal contributions that were, unfortunately, thermodynamically
inconsistent.  For example, it has sometimes been assumed that
\begin{equation}
P=\kappa n^\Gamma+{3k_BTn\over2m}+f(T)aT^4,\qquad\varepsilon=P/(\Gamma-1),
\end{equation}
where $\kappa$ and $\Gamma$ are constants and $f(T)$ is a
temperature-dependent factor reflecting the fraction of relativisitic
particles in the gas, ranging from 1 at low temperatures to 8 at high
temperatures.  However, not only should $f$ also be density-dependent,
but also it can be shown from the above expression for $\varepsilon$ that, even
if it is treated as being density-independent,
\begin{equation}
P=\kappa n^\Gamma+aT\int f(T)T^2dT \,.
\end{equation}
Obviously these two expressions for the pressure are incompatible.
Therefore, interpreting the thermal behaviors of merger calculations
with $\Gamma$-law EOSs is problematic.

Direct comparisons of BHNS merger simulations with tabulated,
temperature-dependent EOSs have yet to be made.  Individual
simulations with tabulated EOSs have been performed in \citep{Duez10} 
with the Hshen EOS and in \citep{Deaton13} 
with the LS220 EOS.  Most
focus has been on the properties of the remnant disc formed from the
disrupted neutron star which survives on timescales ranging from tens
of ms to several s.  Typically, densities in the early evolution of
the disc range from 1-4$\times10^{11}$ g cm$^{-3}$, proton fractions
are around $Y_e\sim0.1$, and specific entropies range from 7-9$k_B$.
As is the case with HMNS evolutions, the thermal properties control
neutrino emissions and the ultimate disc lifetimes.

\section{MODELS WITH FINITE- AND ZERO- RANGE INTERACTIONS}  
\label{Sec:models}
\subsection{Finite range interactions}

We adopt the model of Das {\em et al}. \cite{Das03}  who have generalized the earlier model of Welke {\em et al}. \cite{Welke88} to the case of isospin asymmetric nuclear matter. In this model, exchange contributions arising from  finite range Yukawa interactions between nucleons  give rise to a momentum dependent mean field. BUU calculations performed with such a mean field have been able to account for data from nuclear reactions  induced by neutron rich nuclei \cite{Li04c}.  
Recently, Ref. \cite{xu2015thermal}
has reported results of thermal properties of asymmetric nuclear matter.  
The model's predictions for the structural properties of neutron stars, and thermal effects for conditions of relevance to astrophysical situations have not been investigated so far and are undertaken here. 
Explicitly, the MDI Hamiltonian density is given by \cite{Das03}
\be
\mathcal{H} = \frac{1}{2m}(\tau_n+\tau_p) + V(n_n,n_p,T) \,,
\label{hmdyi}
\ee
where
\ba
n_i = \int d^3p_i~f_i(\vec r_i,\vec p_i)   \label{ni}~~{\rm and}~~
\tau_i = \int d^3p_i~p_i^2f_i(\vec r_i,\vec p_i)
\ea
are the number densities and kinetic energy densities of nucleon species $i=n,p$,  
respectively. The potential energy density $V \equiv V(n_n,n_p,T)$ is expressed as
\ba
V  &=& \frac{A_1}{2n_0}(n_n+n_p)^2 + \frac{A_2}{2n_0}(n_n-n_p)^2  \\
   &+& \frac{B}{\sigma+1}\frac{(n_n+n_p)^{\sigma+1}}{n_0^{\sigma}}
       \left[1-y\frac{(n_n-n_p)^2}{(n_n+n_p)^2}\right]  \nonumber \\
   &+& \frac{C_l}{n_0}\sum_i \int d^3p_i~d^3p_i'
       ~\frac{f_i(\vec r_i,\vec p_i)f_i'(\vec r_i',\vec p_i')}{1+\left(\frac{\vec p_i-\vec p_i'}{\Lambda}\right)^2} \nonumber \\
   &+& \frac{C_u}{n_0}\sum_i \int d^3p_i~d^3p_j
       ~\frac{f_i(\vec r_i,\vec p_i)f_j(\vec r_j,\vec p_j)}{1+\left(\frac{\vec p_i-\vec p_j}{\Lambda}\right)^2}~~;~~ i\ne j. \nonumber
\ea
Above, $n_0\simeq 0.16~{\rm fm}^{-3}$ is the equilibrium density of isospin symmetric matter. 
We will discuss the choice of the strength parameters $A_1,~A_2,~B,~y,~C_l,~C_u$,  the parameter $\sigma$ that captures the density dependence of higher than two-body interactions, and of the finite range parameter $\Lambda$ in subsequent sections. 
For simplicity, the finite range parameter $\Lambda$ is taken to be the same for both like and unlike pairs of nucleons, but the strength parameters, $C_l$ and $C_u$, are allowed to be different.   
Accounting for 2 spin degrees of freedom, the quantities
\be
f_i(\vec r_i,\vec p_i) = \frac{2}{(2\pi\hbar)^3}f_{p_i} 
                      = \frac{2}{(2\pi\hbar)^3}\frac{1}{1+e^{(\epsilon_{p_i}-\mu_i)/T}}
\ee
are the phase-space distributions of nucleons in a heat bath of temperature $T$, having an energy spectrum
\be
\epsilon_{p_i} = p_i^2 \frac{\partial \mathcal{H}}{\partial \tau_i} + \frac{\partial \mathcal{H}}{\partial n_i}
             = \frac{p_i^2}{2m} + U_i(n_n,n_p,p_i)
             \label{spectrum}
\ee
where 
\ba
U_i(n_i,n_j,p_i) &=& \frac{A_1}{n_0}(n_i+n_j) + \frac{A_2}{n_0}(n_i-n_j)  \nonumber \\
    &+& B\left(\frac{n_i+n_j}{n_0}\right)^{\sigma}\left\{1-y\left(\frac{\sigma-1}{\sigma+1}\right)\right.  \nonumber \\
    &\times& \left. \left(\frac{n_i-n_j}{n_i+n_j}\right)^2\left[1+\frac{2}{\sigma-1}\left(\frac{n_i+n_j}{n_i-n_j}\right)\right]\right\}   \nonumber \\ 
    &+& R_i(n_i,n_j,p_i)  \nonumber \\
    &=& {\cal U}_i (n_i,n_j) + R_i(n_i,n_j,p_i)  ~~;~~i\ne j 
\ea
is the single-particle potential (or the mean field). Above, ${\cal U}_i$ represents contributions arising from the densities alone. The momentum dependence is contained in
\ba
R_i(n_i,n_j,p_i) &=& \frac{2C_l}{n_0}\frac{2}{(2\pi\hbar)^3}\int d^3p_i'~\frac{f_{p_i'}}
          {1+\left(\frac{\vec p_i-\vec p_i'}{\Lambda}\right)^2}   \nonumber \\
  &+& \frac{2C_u}{n_0}\frac{2}{(2\pi\hbar)^3}\int d^3p_j~\frac{f_{p_j}}
        {1+\left(\frac{\vec p_i-\vec p_j}{\Lambda}\right)^2} \,. 
\label{Ris}
\ea
At finite temperature, the determination of $R_i(n_i,n_j,p_i)$ requires the knowledge of $R_i(n_i,n_j,p'_i)$ for all values of $p_i'$. 
As in Hartree-Fock theory, a self-consistency condition must be fulfilled;  this is achieved through an iterative procedure as in Ref. \cite{Gale90}. The initial guess is supplied by the zero-temperature $R_i(n_i,n_j,p_i)$ (analytical expressions are given in the next section) which 
may be used in Eq. (\ref{ni}) to obtain a starting chemical potential $\mu_i^{(0)}$ and energy spectrum $\epsilon_i^{(0)}(p_i)$.  These are then used in 
Eqs. (\ref{spectrum}) and (\ref{ni}) to obtain $R_i(n_i,n_j,p_i)$ from 
Eq. (\ref{Ris}). This in turn leads to $\epsilon^{(1)}(p_i)$   which is used in Eqs. (\ref{spectrum}) and (\ref{ni}) to find $\mu_i^{(1)}$.  Upon repetition of the cycle, convergence is achieved in five or less iterations for most cases. 
The ensuing chemical potentials $\mu_i$, for a given density, 
temperature and composition, are then used in the standard statistical mechanics  expression for the entropy density:
\ba
s &=& -\sum_i 2 \int \frac{d^3p_i} {(2\pi\hbar)^3}  ~[f_{p_i}\ln f_{p_i} + (1-f_{p_i})\ln (1-f_{p_i})] \,. \nonumber \\
\label{sden}
\ea
By integrating this expression twice by parts,  $s$ for the MDI model takes the form
\ba 
s &=& \sum_i \frac 1T \left\{ \frac {5\tau_i}{6m} + n_i({\cal U}_i-\mu_i) \right. \nonumber \\
&+& \left.  2 \int \frac{d^3p_i} {(2\pi\hbar)^3}~ f_{p_i} 
\left[ R_i(p_i) + \frac {p_i}{3} \frac {\partial R_i(p_i)}{\partial p_i}  \right] 
\right\} \,. 
\label{sdenmdi}
\ea
The pressure is acquired through the thermodynamic identity
\be
P = -\varepsilon + Ts + \sum_i\mu_i n_i. \label{pres}
\ee
where the energy density $\varepsilon = \mathcal{H}$.
The result in Eq. (\ref{sdenmdi}) enables pressure to be cast in the form \cite{Gale90}
\ba
P  &=&  \frac{A_1}{2n_0}(n_n+n_p)^2+\frac{A_2}{2n_0}(n_n-n_p)^2   \nonumber \\
   &+&    \frac{\sigma B}{\sigma+1}\frac{(n_n+n_p)^{\sigma+1}}{n_0^{\sigma}}  
         \left[1-y\left(\frac{n_n-n_p}{n_n+n_p}\right)^2\right] \nonumber \\
    &+& 2\sum_{i=n,p} \int \frac {d^3p_i}{(2\pi\hbar)^3}~f_{p_i} \left\{ \frac {p_i^2}{3m} + \frac {p_i}{3}  \frac{\partial R_i}{\partial p_i} + \frac {R_i}{2} \right\} .\label{pres2}
\ea
The same expression can also be obtained from Eq. (3.6) of Ref. \cite{Gale90}.

We follow Ref. \cite{LLI} to express the specific heat at constant volume as
\be
C_V = \frac{1}{n}\left.\frac{\partial \varepsilon}{\partial T}\right|_n
\label{cv}
\ee
and the specific heat at constant pressure as \cite{LLI}
\be
C_P = C_V + \frac {T}{n^2} \frac{\left(\left.\frac{\partial P}{\partial T}\right|_n \right)^2 }
               {\left. \frac{\partial P}{\partial n}\right|_T} \,. 
\label{cp}
\ee
By performing a Jacobi transformation to the variables $\mu$ and $T$,  
\ba
C_V &=&\left.\frac{\partial \varepsilon}{\partial T}\right|_{\mu} 
    -\frac{\left.\frac{\partial \varepsilon}{\partial \mu}\right|_{T}
              \left.\frac{\partial n}{\partial T}\right|_{\mu}}
           {\left.\frac{\partial n}{\partial \mu}\right|_{T}}  \\
C_P &=& C_V +\frac{T}{n^2}\frac{\left(
\left.\frac{\partial P}{\partial T}\right|_{\mu} 
    -\frac{\left.\frac{\partial P}{\partial \mu}\right|_{T}
              \left.\frac{\partial n}{\partial T}\right|_{\mu}}
           {\left.\frac{\partial n}{\partial \mu}\right|_{T}}\right)^2}  
     {\frac{\left.\frac{\partial P}{\partial \mu}\right|_{T}}
           {\left.\frac{\partial n}{\partial \mu}\right|_{T}}} \,,
\ea
respectively. 
\medskip

The calculation of $C_V$ and $C_P$ for the MDI model involves some intricacies not encountered in the zero-range Skyrme-like models as $U(n_i,n_p,p)$  in Eq. (\ref{spectrum}) depends only on the nucleon densities and the temperature. 
Consequently, derivatives in the above equations must be evaluated with some care as described in Appendix \ref{Sec:AppendixC}.

\subsection{Zero temperature properties}

The MDI Hamiltonian density can be written as the sum of terms arising from  kinetic sources, $\mathcal{H}_k$,  density-dependent  
interactions, $\mathcal{H}_d$, and  momentum-dependent interactions, $\mathcal{H}_m$:  
\be
\mathcal{H} =\mathcal{H}_k+\mathcal{H}_d+\mathcal{H}_m \,.
\ee
At $T=0$, 
\ba
\mathcal{H}_k &=& \frac{1}{2m}(\tau_n+\tau_p)=\frac{1}{2m}\frac{1}{5\pi^2\hbar^3}(p_{Fn}^5+p_{Fp}^5)  \label{MDYI_H_0T_0}  \\
\mathcal{H}_d &=& \frac{A_1}{2n_0}n^2+\frac{A_2}{2n_0}n^2(1-2x)^2   \nonumber \\
    &+&\frac{B}{\sigma+1}\frac{n^{\sigma+1}}{n_0^{\sigma}}\left[1-y(1-2x)^2\right]  \label{MDYI_H_0T_d}\\
\mathcal{H}_m &=& \frac{C_l}{n_0}(I_{nn}+I_{pp})+\frac{2C_u}{n_0}I_{np}   
\label{MDYI_H_0T}
\ea
with
\ba
p_{Fi} &=& (3\pi^2n_i\hbar^3)^{1/3} \\
I_{ij} &=& \frac{8\pi^2\Lambda^2}{(2\pi\hbar)^6}\left\{p_{Fi}p_{Fj}(p_{Fi}^2+p_{Fj}^2)-\frac{p_{Fi}p_{Fj}\Lambda^2}{3}\right.  \nonumber \\
      &+& \frac{4\Lambda}{3}(p_{Fi}^3-p_{Fj}^3)\arctan\left(\frac{p_{Fi}-p_{Fj}}{\Lambda}\right) \nonumber \\
      &-& \frac{4\Lambda}{3}(p_{Fi}^3+p_{Fj}^3)\arctan\left(\frac{p_{Fi}+p_{Fj}}{\Lambda}\right)   \nonumber \\
      &+& \left[\frac{\Lambda^4}{12}+\frac{(p_{Fi}^2+p_{Fj}^2)\Lambda^2}{2}-\frac{(p_{Fi}^2-p_{Fj}^2)^2}{4}\right] \nonumber \\
      &\times& \left.  \ln\left[\frac{(p_{Fi}+p_{Fj})^2+\Lambda^2}{(p_{Fi}-p_{Fj})^2+\Lambda^2}\right]\right\}\,.
      \label{Iij}
\ea
We note that Eq.~(3.5) in Ref. \cite{dggl03}, which agrees with Eq.~(\ref{Iij}) above for isospin symmetric matter and for pure neutron matter, must be corrected to properly account for properties of bulk matter with intermediate isospin content.

The energy per particle, the pressure, and the chemical potentials 
are obtained from the relations
\ba
E = \frac{\mathcal{H}}{n}\,, \quad 
P = n\frac{\partial \mathcal{H}}{\partial n}-\mathcal{H}\,, \quad {\rm and} \quad
\mu_i = \frac{\partial \mathcal{H}}{\partial n_i}  \label{epmu}
\ea
where $n=n_n+n_p$ is the total baryon number density.

Symmetric nuclear matter properties at the saturation density $n_0 = 0.16$ fm$^{-3}$ 
such as the compression modulus $K_0$, the symmetry energy $S_v$ as well as its 
slope $L_v$ and curvature $K_v$ are obtained from
\ba
K_0 &=& K(n=n_0,x=1/2)= 9n_0\left.\frac{\partial^2 \mathcal{H}}{\partial n^2}\right|_{n=n_0,~x=1/2}  \\
S_v &=& S_2(n_0) ~~;~~ S_2 = \frac{1}{8n}\left.\frac{\partial^2 \mathcal{H}}{\partial x^2}\right|_{x=1/2} \\
L_v &=& L_2(n_0) = 3n_0\left.\frac{dS_2}{dn}\right|_{n=n_0}  \\
K_v &=& K_2(n_0) = 9n_0^2\left.\frac{d^2S_2}{dn^2}\right|_{n=n_0} .
\ea
The proton fraction $x$ is defined as $x=n_p/(n_n+n_p)$.

The $T=0$ single-particle energy spectrum is 
\be
\epsilon_{p_i} = \frac{p_i^2}{2m} + U_i(n_i,n_j,k_i)
\ee
with
\ba
U_i(n_i,n_j,k_i) &=& A_1\frac{(n_i+n_j)}{n_0} + A_2\frac{(n_i-n_j)}{n_0}      \nonumber \\
                &+& B\left(\frac{n_i+n_j}{n_0}\right)^{\sigma}\left\{1-y\left(\frac{\sigma-1}{\sigma+1}\right)\right.  \nonumber \\
                &\times& \left. \left(\frac{n_i-n_j}{n_i+n_j}\right)^2\left[1+\frac{2}{\sigma-1}\left(\frac{n_i+n_j}{n_i-n_j}\right)\right]\right\}  \nonumber \\ 
                &+& \frac{2C_l}{n_0}R_{ii}(n_i,k_i) + \frac{2C_u}{n_0}R_{ij}(n_j,k_i) ~~;~~ i\ne j \nonumber \\
\label{Ui}
\ea
where
\ba
R_{ij}(n_j,p_i) &=& \frac{\Lambda^3}{4\pi^2\hbar^3}\left\{\frac{2p_{Fj}}{\Lambda}\right. \nonumber \\
              &-& 2\left[\arctan\left(\frac{p_i+p_{Fj}}{\Lambda}\right)
                     -\arctan\left(\frac{p_i-p_{Fj}}{\Lambda}\right)\right] \nonumber \\
              &+& \left. \frac{(\Lambda^2+p_{Fj}^2-p_i^2)}{2\Lambda p_i}
                    \ln\left[\frac{(p_i+p_{Fj})^2+\Lambda^2}{(p_i-p_{Fj})^2+\Lambda^2}\right]\right\}
\label{Rij}
\ea
from which the nucleon effective masses can be derived:
\be
m_i^* = p_{Fi}\left(\left.\frac{\partial \epsilon_{p_i}}{\partial p_i}\right|_{p_{Fi}}\right)^{-1}
     = \frac{m}{1+\frac{m}{p_{Fi}}\left.\frac{\partial U_i}{\partial p_i}\right|_{p_{Fi}}}
\label{ms}
\ee
Explicitly, 
\ba
m_i^* &=& p_{Fi}\left(\frac{p_{Fi}}{m}+\frac{2C_l}{n_0}\left.\frac{\partial R_{ii}}{\partial p_i}\right|_{p_{Fi}}
           +\frac{2C_u}{n_0}\left.\frac{\partial R_{ij}}{\partial p_i}\right|_{p_{Fi}}\right)^{-1} \,, \label{msin} \nonumber  \\  \\
\left.\frac{\partial R_{ij}}{\partial p_i}\right|_{p_{Fi}} &=& \frac{\Lambda^2}{2\pi^2\hbar^3}\frac{p_{Fj}}{p_{Fi}}
  \left\{1 -\frac{(\Lambda^2+p_{Fj}^2+p_{Fi}^2)}{4p_{Fi}p_{Fj}}\right. \nonumber  \\
         &\times& \left.  \ln\left[\frac{(p_{Fi}+p_{Fj})^2+\Lambda^2}{(p_{Fi}-p_{Fj})^2+\Lambda^2}\right]\right\}  
\label{msfin}
\ea
For completeness, explicit expressions for $E,~P,~\mu_i,~K,~S_2,~L_2,~K_2$ and $m^*$ as functions of the nucleon 
densities and isospin content are collected in Appendix A.

\subsection{Zero-range Skyrme interactions}

For comparison with the results of the finite range model discussed above, we also consider the often studied  zero-range model   
due to Skyrme \cite{Skyrme59}. In its standard form, the Skyrme Hamiltonian density reads as
\ba
\mathcal{H} &=& \frac{1}{2m_n}\tau_n+\frac{1}{2m_p}\tau_p  \nonumber  \\
            &+& n(\tau_n+\tau_p)\left[\frac{t_1}{4}\left(1+\frac{x_1}{2}\right)  
                                               +\frac{t_2}{4}\left(1+\frac{x_2}{2}\right)\right] \nonumber \\
            &+& (\tau_n n_n+\tau_p n_p)\left[\frac{t_2}{4}\left(\frac{1}{2}+x_2\right)
                                               -\frac{t_1}{4}\left(\frac{1}{2}+x_1\right)\right] \nonumber \\
            &+& \frac{t_o}{2}\left(1+\frac{x_o}{2}\right)n^2
                        -\frac{t_o}{2}\left(\frac{1}{2}+x_o\right)(n_n^2+n_p^2)  \nonumber  \\
            &+& \left[\frac{t_3}{12}\left(1+\frac{x_3}{2}\right)n^2
                        -\frac{t_3}{12}\left(\frac{1}{2}+x_3\right)(n_n^2+n_p^2)\right]n^{\epsilon}. \nonumber \\
\label{hsko}
\ea
Explicit forms of the single-particle potentials for the SkO$^\prime$ Hamiltonian (with $Y_i=n_i/n,~i=n,p$) are
\ba
U_i(n,k) &=& (X_1+Y_iX_2)nk^2      
                   + (X_1+X_2)\tau_i+X_1\tau_j \nonumber \\
                   &+& 2n(X_3+Y_iX_4)   
                   + n^{1+\epsilon}\left\{(2+\epsilon)X_5 \right.   \nonumber  \\
                   &+& \left.[2Y_i+\epsilon({Y_i}^2 +{Y_j}^2)]X_6\right\}; ~~i\ne j  \,,
\label{UnpsSkO'}
\ea
where
\ba
X_1   &=& \frac{1}{4} \left[ t_1 \left(1+\frac{x_1}{2}\right) + t_2 \left(1+\frac{x_2}{2} \right) \right] \nonumber\\
X_2  &=& \frac{1}{4} \left[t_2 \left(\frac{1}{2}+x_2 \right)-t_1\left( \frac{1}{2}+x_1 \right) \right] \nonumber\\
X_3 &=& \frac{t_0}{2} \left(1+\frac{x_0}{2}\right)\,; \quad X_4 = -\frac{t_0}{2} \left(\frac{1}{2}+x_0\right) \nonumber\\
X_5 &=& \frac{t_3}{12} \left(1+\frac{x_3}{2} \right)\,; \quad X_6 = -\frac{t_3}{12} \left(\frac{1}{2}+x_3 \right)\,. 
\ea
From Eq. (\ref{ms}), the density-dependent Landau effective masses are
\ba
\frac {m_i^*}{m} = \left[ 1 + \frac {2m}{\hbar^2} (X_1+Y_iX_2)n   \right]^{-1} \,.
\label{effmSkO'}
\ea

The derivations of the various state variables, nuclear saturation properties and thermal response functions 
proceed as previously described  for MDI. For details of 
evaluating the thermal state variables for Skyrme-like models, we refer the reader to a recent compilation of formulas and numerical methods in Ref. \cite{APRppr}. For numerical values of the various strength parameters above, we choose the SkO$^\prime$ parametrization of Ref. \cite{skop} which will be given in a subsequent section. 

\section{RESULTS FOR ZERO TEMPERATURE}
\label{Sec:ResultsT0}

In this section, we consider the zero temperature properties of the finite-range and zero-range models discussed in the previous section. We begin with the MDI(0) parametrization of the model of Das {\em et al}. \cite{Das03} so that its characteristics extended to neutron-star matter, not considered previously, may be assessed.  

\subsection{MDI models for isospin-asymmetric matter}

Table \ref{pMDI(0)} lists the various parameters employed in the MDI(0) model of Ref. \cite{Das03}. 
In Table \ref{propsMDI(0)} , we list the characteristic properties of this model at the equilibrium density of isospin symmetric nuclear matter. Also included in this table are values of the various physical quantities (the last three rows) accessible to laboratory experiments for small isospin asymmetry.  Note the fairly good agreement with experimental determinations of the various quantities.

\begin{table}[!h]
\begin{ruledtabular} 
\newcolumntype{a}{D{.}{.}{3,13}}
\begin{tabular}{caca}    
Parameter & \multicolumn{1}{c}{Value} & Parameter & \multicolumn{1}{c}{Value} \\
\hline
$A_1 $    & -108.28\phantom{0}\mbox{MeV}       &   $y $        &    0                       \\
$A_2 $    &  -12.30\phantom{0}\mbox{MeV}      &   $C_l $      &    -11.70\phantom{0}\mbox{MeV}   \\        
$B $      &   106.35\phantom{0}\mbox{MeV}     &   $C_u $      &    -103.40\phantom{0}\mbox{MeV}       \\
$\sigma $ &   4/3                          &   $\Lambda $  &   263.04\phantom{0}\mbox{MeV}    \\            
\end{tabular}
\caption[Parameter values for $\mathcal{H}_{MDI(0)}$]{Parameter values for the MDI(0) Hamiltonian density of Das {\em et al}. \cite{Das03}.  
The dimensions are such that the Hamiltonian density is in MeV fm$^{-3}$.}
\label{pMDI(0)}
\end{ruledtabular}
\end{table}
\begin{table}[!h]
\begin{ruledtabular}
\newcolumntype{a}{D{.}{.}{-1}}
\newcolumntype{b}{D{+}{\pm}{-1}}
\begin{tabular}{labc}
 Property  & \multicolumn{1}{c}{Value}  & \multicolumn{1}{c}{Experiment} & Reference\\
  & \multicolumn{1}{c}{[MDI(0)]} & & \\
\hline
 $n_0$ (fm$^{-3}$) & 0.16   & 0.17+0.02                   & \cite{day78,jackson74,myers66,myers96}  \\
 $E_0$ (MeV)      & -16.10  & -16+1                       & \cite{myers66,myers96}                  \\
 $K_0$ (MeV)      & 212.4   & 230+30                      & \cite{Garg04,Colo04}                    \\
                  &         & 240+20                      &\cite{shlomo06}                          \\
 $m_0^*/m$        &  0.67   & 0.8+0.1                     & \cite{bohigas79,krivine80}              \\       \hline              
 $S_v$ (MeV)      & 30.54    & \multicolumn{1}{c}{~~30-35}   & \cite{L,tsang12}                        \\
 $L_v$   (MeV)    & 60.24    & \multicolumn{1}{c}{~~40-70}   & \cite{L,tsang12}                        \\
 $K_v$  (MeV)     & -81.67   & -100+200                    & \cite{APRppr}                                    \\ 
\end{tabular}
\caption[Symmetric nuclear matter saturation properties.]{
Entries in this table are at the equilibrium 
density $n_0$ of symmetric nuclear matter for the MDI(0) model \cite{Das03}.
$E_0$ is the energy per particle, $K_0$ is the compression modulus, $m_0^*/m$ is 
the ratio of the Landau effective mass to mass in vacuum, $S_v$ is the nuclear symmetry 
energy parameter, $L_v$ and $K_v$, are related to the first and second derivatives of the 
symmetry energy, respectively.}
\label{propsMDI(0)}
\end{ruledtabular}
\end{table}

In Table \ref{nspropsMDI(0)}, the structural properties of neutron stars built using the EOS of charge neutral and beta equilibrated matter from the MDI(0) model are summarized. The predicted maximum mass falls slightly short of the largest well-measured mass. The radii of the maximum mass and $1.4~{\rm M}_\odot$ stars are in reasonable agreement with their current estimations from X-ray data. A noteworthy feature is the central baryon chemical potential 
of the maximum mass star, $\mu_c \sim 1.9$ GeV, which is below the limit of $2.1$ GeV set by the maximally compact EOS of a neutron star derived in Ref. \cite{LP:11}.  Note that the density $n_a$, at which the EOS violates causality, that is, the squared speed of sound $c_s^2=dP/d\epsilon >1$  (a common feature of potential models), lies well above the central density of the maximum star as is the case for all models discussed in subsequent sections.

\begin{table}[!h]
\begin{ruledtabular}
\newcolumntype{a}{D{.}{.}{-1}}
\newcolumntype{b}{D{+}{\pm}{-1}}
\begin{tabular}{labc}
 Property  & \multicolumn{1}{c}{Value} & \multicolumn{1}{c}{Observation} & Reference\\
   & \multicolumn{1}{c}{[MDI(0)]} & &  \\
\hline
 ${\rm M}_{max}(M_{\odot})$& 1.884    & 2.01+0.04                   & \cite{Antoniadis}                        \\
 $R_{max}$(km)     & 9.84   & 11.0+ 1.0                   & \cite{slb10}                             \\
 $n_c$(fm$^{-3}$)  &1.3065   & & \\
 $\epsilon_c$(MeV fm$^{-3}$) &1703.3& & \\
 $P_c$(MeV fm$^{-3}$) & 761.22& & \\
 $\mu_c$(MeV)& 1886.3 & & \\
 $n_a$(fm$^{-3}$)& 1.883 & & \\
\hline
 $R_{1.4}$(km)     & 11.77   &  11.5+ 0.7                  & \cite{slb10}                              \\
 $n_c$(fm$^{-3}$)  &0.5782   & & \\
 $\epsilon_c$(MeV fm$^{-3}$) &599.19& & \\
 $P_c$(MeV fm$^{-3}$) & 91.784& & \\
 $\mu_c$(MeV)& 1195.0 & & \\
\end{tabular}
\caption[Neutron star properties.]{
${\rm M}_{max}$ is the maximum neutron star mass predicted by 
the MDI(0) model and $R_{max}$ the radius associated with it. 
Other entries are the central density $n_c$, energy density $\epsilon_c$, pressure $P_c$, and the chemical potential $\mu_c$ for 
both the maximum mass and 1.4 ${\rm M}_{\odot}$ configurations, and the density $n_a$ at which the EOS becomes acausal. 
The value for the radius of a 1.4 ${\rm M}_{\odot}$ neutron star is given by $R_{1.4}$.}
\label{nspropsMDI(0)}
\end{ruledtabular}
\end{table}

\subsection{Revised parameterization of the MDI model}
In this section, we provide a revised set of parameters (see Table \ref{pMDYI_N} ) for the MDI model so that  
the isospin symmetric and asymmetric properties at the nuclear matter equilibrium density are closer to the experimentally derived mean values than given by the MDI(0) parametrization and the neutron star maximum mass also comes close to the recently observed 2 solar mass. We note that this model also allows us to constrain the single particle potential $U(n,p)$ to match optical model fits to data. We have used $U(n_0,p=0)= -74.6$ MeV and $U(n_0,p^2/(2m) = 303.1~{\rm MeV}) = 0$ in determining the constants of the model as the variational Monte Carlo calculations of Wiringa in Ref. \cite{w88} suggest. The asymptotically flat behavior of $U(n_0,p)\simeq 30.6$ MeV with $p$ arises naturally from effects of the finite range interaction in this model.  Before presenting the results of the MDI(A) model, we consider a zero-range model for purposes of comparison. 

\begin{table}[!h]
\begin{ruledtabular} 
\newcolumntype{a}{D{.}{.}{3,13}}
\begin{tabular}{caca}    
Parameter & \multicolumn{1}{c}{Value} & Parameter & \multicolumn{1}{c}{Value} \\
\hline
$A_1 $    & -69.4758\phantom{0}\mbox{MeV}       &   $y $        &     -0.0327929                       \\
$A_2 $    &  -29.2241\phantom{0}\mbox{MeV}      &   $C_l $      &    -23.0576\phantom{0}\mbox{MeV}   \\        
$B $      &   100.084\phantom{00}\mbox{MeV}     &   $C_u $      &    -105.885\phantom{00}\mbox{MeV}       \\
$\sigma $ &   1.36227                          &   $\Lambda $  &   420.864\phantom{00}\mbox{MeV}    \\            
\end{tabular}
\caption[Parameter values for $\mathcal{H}_{MDI}$]{Parameter values for the MDI(A) Hamiltonian density. 
The dimensions are such that the Hamiltonian density is in MeV fm$^{-3}$.}
\label{pMDYI_N}
\end{ruledtabular}
\end{table}

\subsection{Parameters of a prototype Skyrme model }

As an example of a zero-range model, 
we have chosen to work with the Skyrme model of Ref. \cite{skop} known as SkO$^\prime$, the parameters of which are listed 
in Table \ref{pSkO'}.  For an apposite comparison, the parameters of the MDI(A) model were tuned so that its energy per particle vs baryon density closely matches that of the SkO$^\prime$ model for both symmetric matter and pure neutron matter. 
\begin{table}[!h]
\begin{ruledtabular}
\newcolumntype{a}{D{.}{.}{-1}}
\begin{tabular}{caac}
 & \multicolumn{1}{c}{$t_i$} & \multicolumn{1}{c}{$x_i$} & $\epsilon$  \\
\hline
0 & -2099.419\phantom{0}\mbox{MeV~fm}^3  & -0.029503 & 1/4 \\
1 & 301.531\phantom{0}\mbox{MeV~fm}^5       & -1.325732            \\ 
2 & 154.781\phantom{0}\mbox{MeV~fm}^5      & -2.323439             \\
3 & 13526.464\phantom{0}\mbox{MeV~fm}^{3(1+\epsilon)}   & -0.147404      \\
\end{tabular}
\caption[Parameter values for SkO$^\prime$.]{Parameter values for the Skyrme Hamiltonian density SkO$^\prime$  
used in this work. The dimensions are such that the Hamiltonian density is in MeV fm$^{-3}$.}
\label{pSkO'}
\end{ruledtabular} 
\end{table}
%
%

\subsubsection*{Comparison of  MDI(A) and SkO$^\prime$ models}

Attributes of the MDI(A) and SKO$^\prime$ models at their respective equilibrium densities of symmetric nuclear matter
are presented in Table \ref{propsMDYINSkO'}.  The resulting structural aspects of neutrons stars from these two models are 
presented in Table \ref{nspropsMDYINSkO'}. These results indicate the nearly identical nature of the two models at zero temperature. 
As with the MDI(0) model, the central baryon chemical potentials for these models also lie below the value for the maximally compact EOS. A closer examination of the innards of these two models is taken up in the subsequent sections.

\begin{table}[!h]
\begin{ruledtabular}
\newcolumntype{a}{D{.}{.}{-1}}
\newcolumntype{b}{D{+}{\pm}{-1}}
\begin{tabular}{laabc}
 Property  & \multicolumn{1}{c}{Value} & \multicolumn{1}{c}{Value} & \multicolumn{1}{c}{Experiment} & Reference\\
  & \multicolumn{1}{c}{[MDI(A)]} & \multicolumn{1}{c}{[SkO$^\prime$]} & \\
\hline
 $n_0$ (fm$^{-3}$) & 0.160   & 0.160   & 0.17+0.02                   & \cite{day78,jackson74,myers66,myers96}  \\
 $E_0$ (MeV)      & -16.00  & -15.75  & -16+1                       & \cite{myers66,myers96}                  \\
 $K_0$ (MeV)      & 232.0   & 222.3   & 230+30                      & \cite{Garg04,Colo04}                    \\
                  &         &         & 240+20                      &\cite{shlomo06}                          \\
 $m_0^*/m$        &  0.67   &  0.90   & 0.8+0.1                     & \cite{bohigas79,krivine80}              \\  \hline 
 $S_v$ (MeV)      & 30.0    & 31.9    & \multicolumn{1}{c}{~~30-35}   & \cite{L,tsang12}                        \\
 $L_v$   (MeV)    & 65.0    & 68.9    & \multicolumn{1}{c}{~~40-70}   & \cite{L,tsang12}                        \\
 $K_v$  (MeV)     & -72.0   & -78.8   & -100+200                    & \cite{APRppr}                                     \\ 
 \end{tabular}
\caption[Symmetric nuclear matter saturation properties.]{
Same as Table \ref {propsMDI(0)}, but for the  MDI(A) and SkO$^\prime$ models.}
\label{propsMDYINSkO'}
\end{ruledtabular}
\end{table}

\begin{table}[!h]
\begin{ruledtabular}
\newcolumntype{a}{D{.}{.}{-1}}
\newcolumntype{b}{D{+}{\pm}{-1}}
\begin{tabular}{laabc}
 Property  & \multicolumn{1}{c}{Value} & \multicolumn{1}{c}{Value} & \multicolumn{1}{c}{Observation} & Reference\\
  & \multicolumn{1}{c}{[MDI(A)]} & \multicolumn{1}{c}{[SkO$^\prime$]} & \\
\hline
 ${\rm M}_{max}(M_{\odot})$& 1.9725    &  1.9600      & 2.01+0.04                   & \cite{Antoniadis}                        \\
 $R_{max}$(km)     & 10.20   &  10.13      & 11.0+ 1.0                   & \cite{slb10}                             \\
 $n_c$(fm$^{-3}$)  &1.2065   &  1.2233     & & \\
 $\epsilon_c$(MeV fm$^{-3}$) &1573.4&1595.7& & \\
 $P_c$(MeV fm$^{-3}$) & 718.90& 739.42& & \\
 $\mu_c$(MeV)& 1900.0 & 1908.9 & & \\
 $n_a$(fm$^{-3}$)& 1.64 & 1.678 & & \\
\hline
 $R_{1.4}$(km)     & 12.21   &  12.17      &  11.5+ 0.7                  & \cite{slb10}                              \\
 $n_c$(fm$^{-3}$)  &0.5126   &  0.5234     & & \\
 $\epsilon_c$(MeV fm$^{-3}$) &526.36&533.06& & \\
 $P_c$(MeV fm$^{-3}$) & 75.802& 78.421& & \\
 $\mu_c$(MeV)& 1174.7 & 1168.3 & & \\
\end{tabular}
\caption[Neutron star properties.]{
Same as \ref{nspropsMDI(0)}, but for the MDI(A) and SkO$^\prime$ models.}
\label{nspropsMDYINSkO'}
\end{ruledtabular}
\end{table}

We wish to add that neutron star maximum masses in excess of 2 ${\rm M}_\odot$ can also 
be obtained from a reparametrization of the MDI model, but at the expense of losing 
close similarity with the results of the SkO$^\prime$ model. An illustration is provided with  $K_0=260$ MeV, 
$L_v=70$ MeV, and $K_v=-50$ MeV, while keeping the other saturation properties 
the same as for MDI(A). This was achieved through the choice of the constants 
$A_1=-39.0752,~A_2=-27.4916,~B=69.6838,~\sigma=1.56497,~y=-0.105539,~C_l=-26.95,~C_u=-101.993$, 
and $\Lambda=420.864$, the units of these constants being the same as in  Table 
\ref{pMDI(0)}. Results for the structural properties of  a neutron star in beta-equilibrium are: 
${\rm M}_{max}=2.153~{\rm M}_\odot$, $R_{max}=10.58$ km, and $R_{1.4}=12.13$ km.  At the edges 
of the 1-$\sigma$ errors of the empirical saturation properties at the nuclear 
equilibrium density, it is not difficult to raise the maximum mass well above 
2 ${\rm M}_\odot$.

\subsection{Single particle potentials}

In this section, we present results of the single particle potentials for the MDI(A) and
SkO$^\prime$ models from Eqs. (\ref{Ui}) and (\ref{UnpsSkO'}), respectively, and contrast them with  
those from the microscopic calculations of Refs. \cite{w88} and \cite{Zuo14}.

\begin{figure*}[htb]
\centering
\begin{minipage}[b]{0.49\linewidth}
\centering
\includegraphics[width=9.5cm,height=8cm]{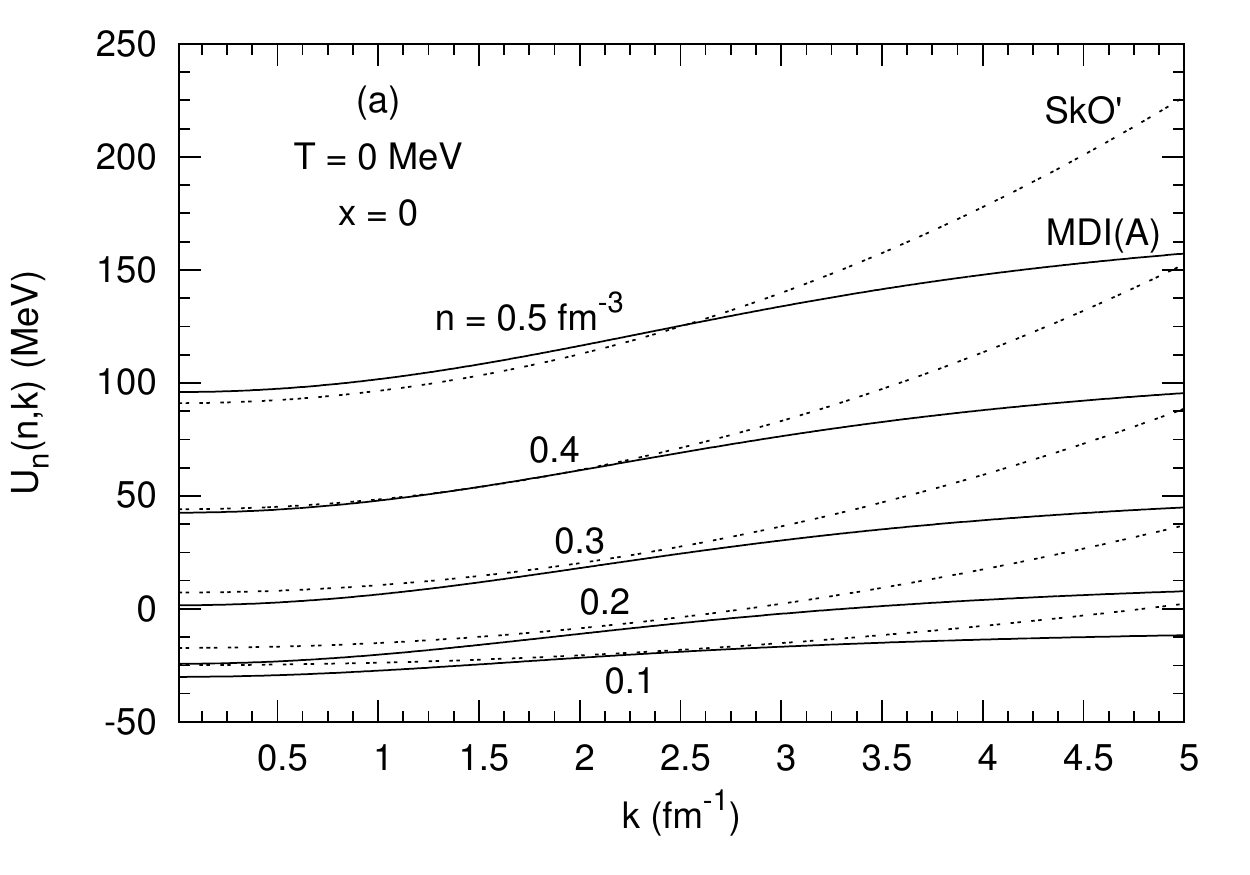}
\end{minipage}
\begin{minipage}[b]{0.49\linewidth}
\centering
\includegraphics[width=9.5cm,height=8cm]{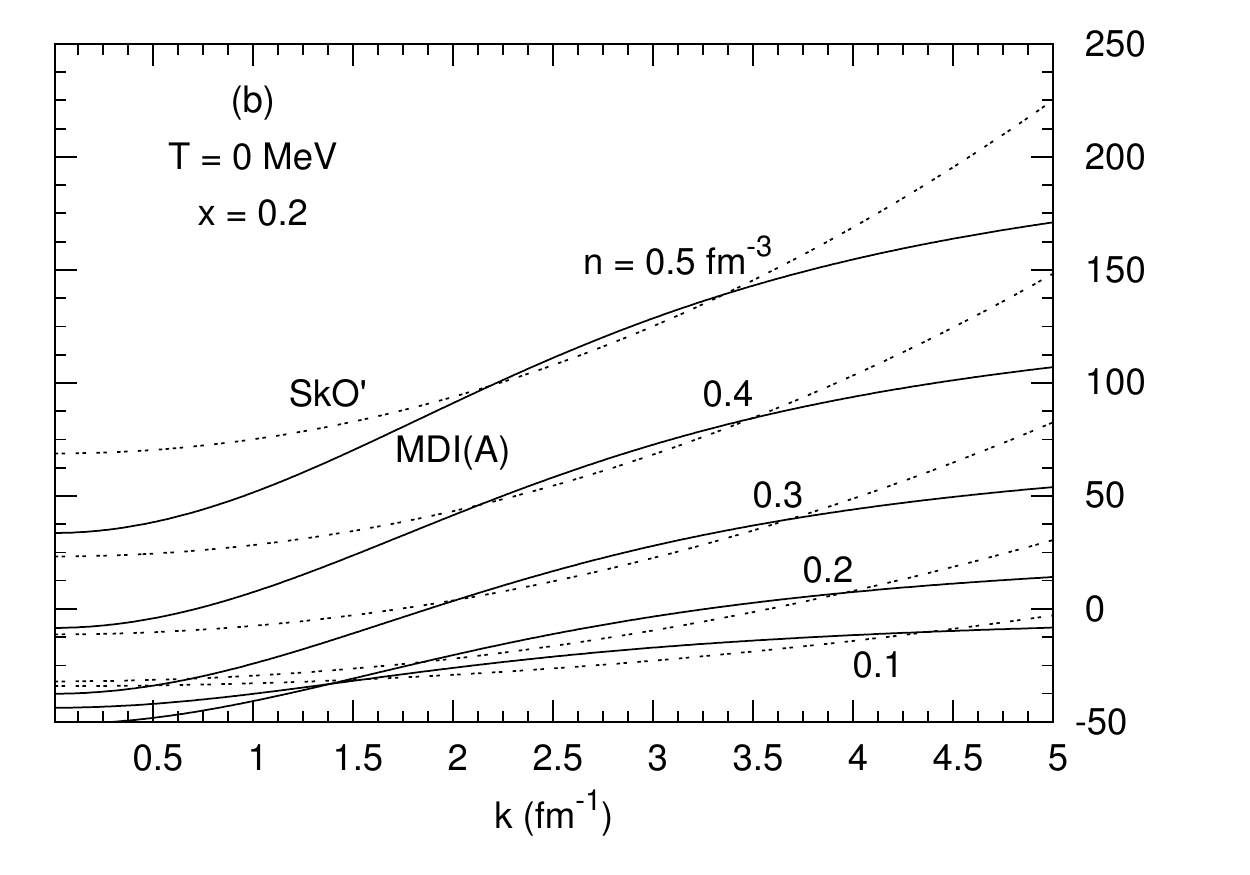}
\end{minipage}
\vskip -0.5cm
\caption{Neutron single particle potentials vs momentum at zero temperature for the baryon densities marked. Results for the MDI(A) model (solid curves) are from from Eqs. (\ref{spectrum}),  (\ref{Ui}), and (\ref{Rij}). Those for the SkO$^\prime$ model (dashed curves) are from   Eq. (\ref{UnpsSkO'}).  (a) Results are for proton fraction $x=0$. (b) Same as (a) but for $x=0.2$. } 
\label{MDYISk_SPP}
\end{figure*}

Figure \ref{MDYISk_SPP} shows the neutron single particle potentials for MDI(A) and
SkO$^\prime$ as functions of momentum for select baryon densities at zero temperature. 
Results shown are for pure neutron matter ($x=0$)
[Fig. 1(a)] and for isospin asymmetric matter with $x=0.2$ [Fig. 1(b)].  
Notice that the results for MDI(A) tend to saturate at large momenta for both proton fractions owing to the logarithmic structure of Eq. (\ref{Rij}).  The SkO$^\prime$ model, however, in common with most Skyrme models, exhibits a quadratic rise with momentum.  
This latter feature is also present in the results of the Akmal, Pandharipande and Ravenhall (APR) model \cite{Akmal98} in which the hamiltonian density of the many-body calculations of Akmal and Pandharipande \cite{Akmal97} is parametrized in Skyrme-like fashion.  For both MDI(A) and SkO$^\prime$ models 
the effect of a finite proton fraction [Fig. 1(b)] is more pronounced at low momenta for which the single particle potential 
becomes more attractive relative to that for pure neutron matter. 

\begin{figure}[!h]
\centering
\makebox[0pt][c]{
\hspace{-1cm}
\begin{minipage}[b]{\linewidth}
\centering
\includegraphics[width=9cm]{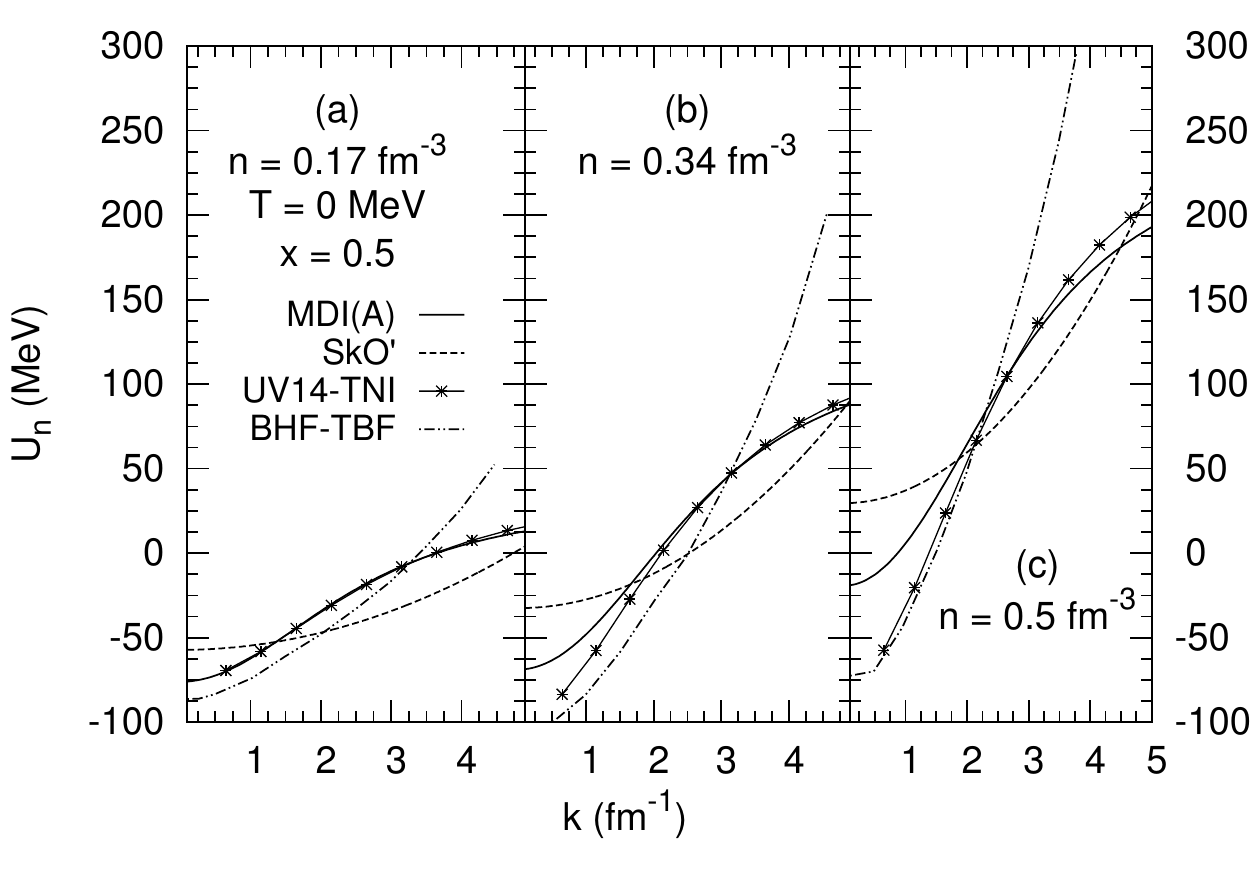}
\end{minipage}
}
\vskip -0.5cm
\caption{Comparison of neutron single particle potentials vs momentum from Eqs. (\ref{Ui}) and (\ref{Rij}) 
with the variational Monte Carlo results of Ref.  
\cite{w88} using the UV14-TNI interaction and Bruekner-Hartree-Fock results of Ref. \cite{Zuo14} with the inclusion of three-body interactions (labeled BHF-TBF). Results shown are for zero temperature symmetric nuclear matter at the 
indicated baryon densities.} 
\label{SPP_Compare}
\end{figure}

In Fig.  \ref{SPP_Compare}, the neutron single particle potentials vs momentum for the MDI(A) model from Eqs. (\ref{Ui}) 
(solid curves) and for the SkO$^\prime$ model from Eq. (\ref{Rij}) (dashed curves) are compared 
with the variational Monte Carlo results (solid curves marked with asterisks) of Ref.  
\cite{w88} using the UV14-TNI interaction and the Bruekner-Hartree-Fock results (dash-dotted curves) of Ref. \cite{Zuo14} with the inclusion of three-body interactions (labeled BHF-TBF).  
Results are for symmetric nuclear matter at about one, two and three times the nuclear matter equilibrium density in 
figures (a), (b), and (c), respectively. 
A curve-to-curve quantitative comparison between the results of models used in this work and those of UV14-TNI and BHF-TBF models is not appropriate because the saturation properties of the latter models differ significantly from those of the former ones. Specifically, for the UV14-TNI model, $E_0=-16.6$ MeV at $n_0=0.157~{\rm fm}^{-3}$, with $K_0=260$ MeV, whereas for the   BHF-TBF model, $E_0=-15.08$ MeV at $n_0=0.198~{\rm fm}^{-3}$, with $K_0=207$ MeV. 
In this work, the single particle potential was designed to mimic closely the behavior of the UV14-TNI model.  
However, the qualitative trends - power-law rise  vs logarithmic rise - at high momenta in other microscopic models are worth noting as discussed below.

At the densities shown, and at high momenta, there is good agreement between the results of the MDI(A) model and those of the UV14-TNI model by design.    As mentioned in the introduction, this saturating behavior at high momenta is in accord with the analysis of optical model fits to nucleon-nucleus scattering data.  The other two models, SkO$^\prime$ and BHF-TBF, rise quadratically with momenta. In the case of the SkO$^\prime$ model, this behavior ensues from the zero-range approximation made for nuclear interactions. In the case of the BHF-TBF calculations, the quadratic rise with momentum stems from the similar behavior of $U(n,k)$ chosen for purposes of convergence during numerical calculations.  Such a behavior of the single particle potential, even with the symmetric nuclear matter compression  moduli around 240 MeV, leads to nucleon collective flows that are larger than those observed in heavy-ion collisions  \cite{Gale87,Prakash88b,Welke88,Gale90}.  We note that Danielewicz reaches  similar conclusions with a different parametrization $U(n,p)$, which also saturates at high momenta (see. Fig. 17 of Ref. \cite{Danielewicz:00}).

\subsection{Isospin dependence of effective masses}
\begin{figure*}[htb]
\centering
\begin{minipage}[b]{0.49\linewidth}
\centering
\includegraphics[width=9.5cm,height=8cm]{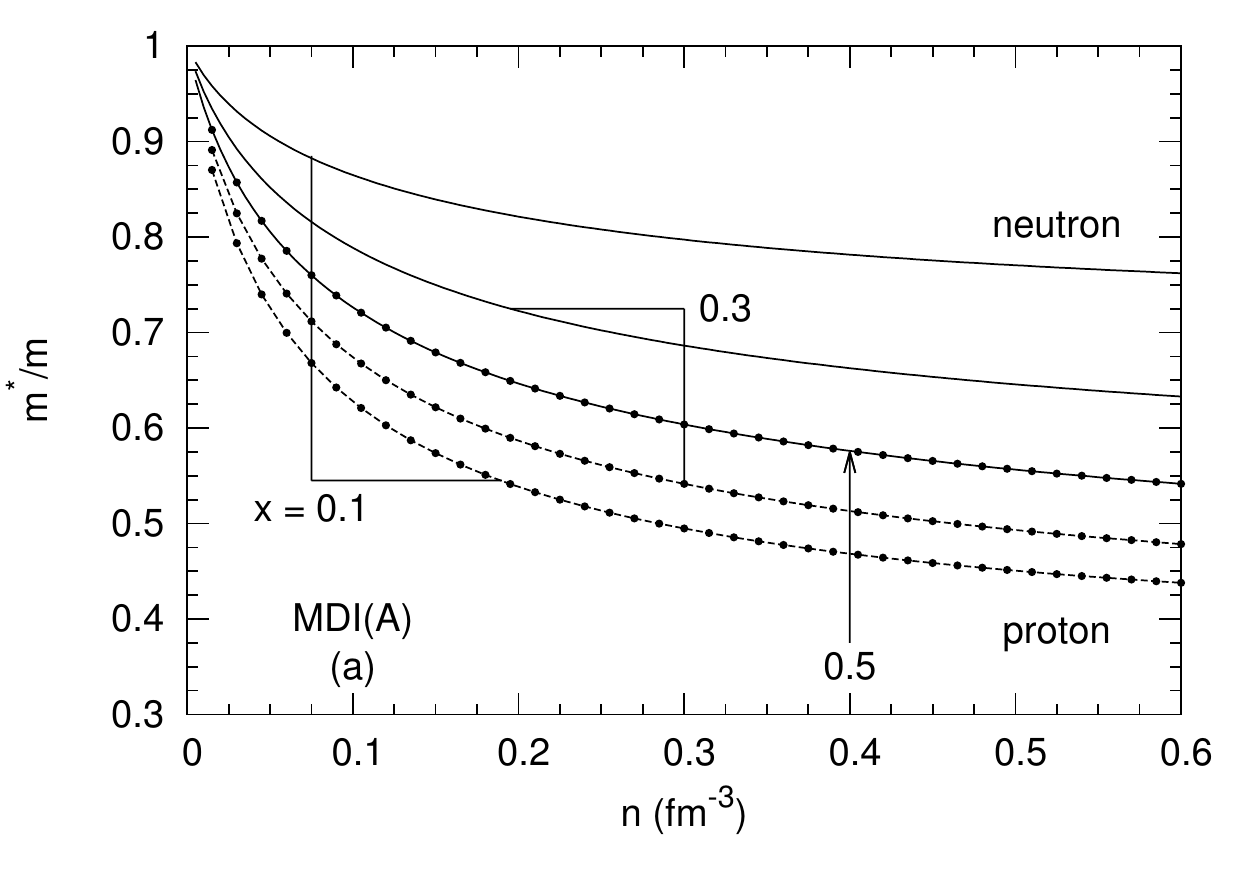}
\end{minipage}
\begin{minipage}[b]{0.49\linewidth}
\centering
\includegraphics[width=9.5cm,height=8cm]{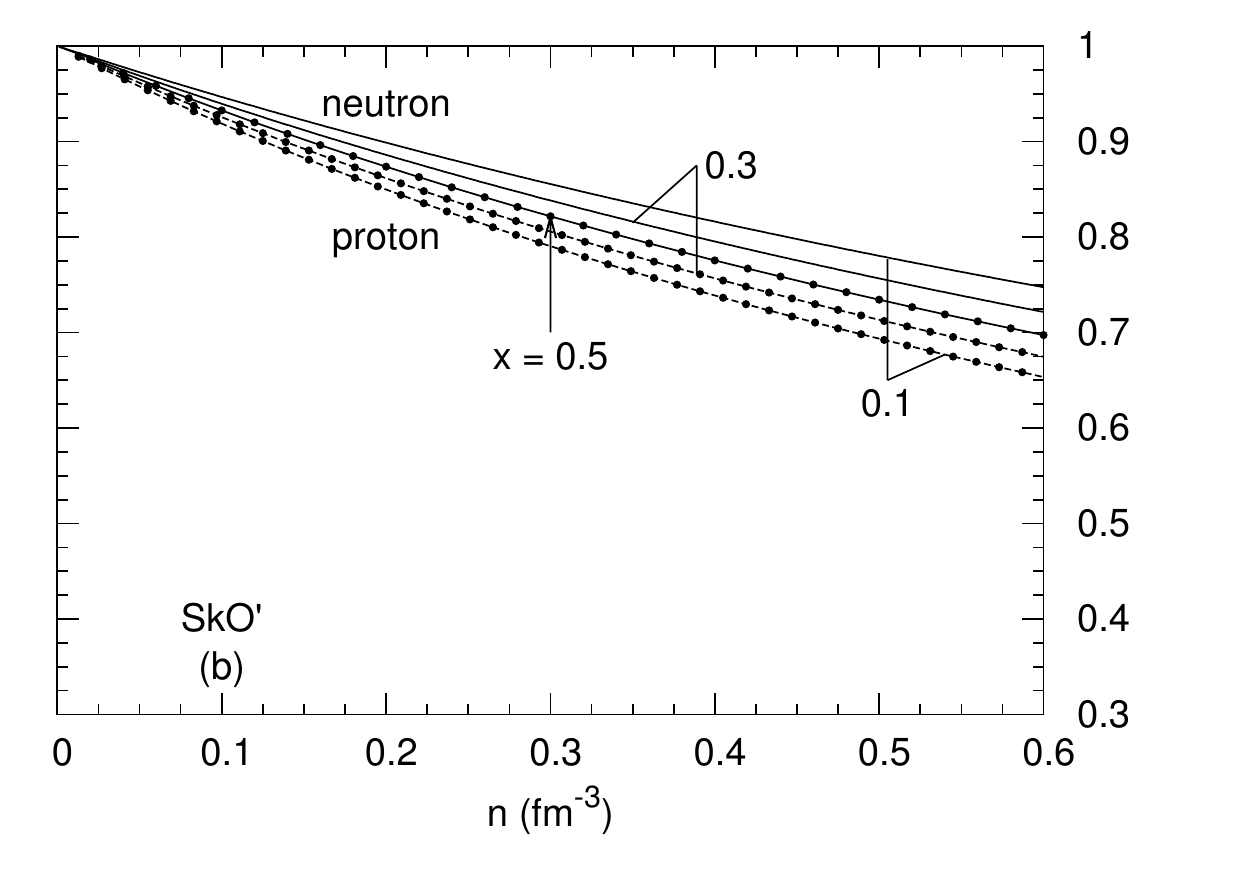}
\end{minipage}
\vskip -0.5cm
\caption{Landau effective masses of the neutron (solid curves)  and proton (dotted curves) scaled with the vacuum nucleon mass  vs baryon density $n$ for the marked values of the proton fraction $x$. (a)  Results are for the MDI(A) model 
from Eqs. (\ref{msin})-(\ref{msfin}). (b) Same as (a), but for the SkO$^\prime$ model from Eq. (\ref{ms}).} 
\label{MDYISk_Ms}
\end{figure*}

The single particle potentials discussed above facilitate the calculation of the Landau effective masses of nucleons according to
Eq. (\ref{ms}).  For the MDI(A) and SkO$^\prime$ models, explicit expressions as functions of density and proton fraction were given in Eqs. (\ref{msfin}) and (\ref{effmSkO'}), respectively.  

In Fig. \ref{MDYISk_Ms}, the neutron and proton effective masses scaled with the vacuum nucleon mass are shown as a function of baryon density for select proton fractions for the MDI(A) and SkO$^\prime$ models in panels (a) and (b),  respectively.
Although these two models yield similar properties for most observables for symmetric nuclear matter at the equilibrium density $n_0$, the effective masses are significantly different - $m^*_0/m= 0.67~(0.9)$ for the MDI(A) (SkO$^\prime$) model (see Table VI).   
The density dependence of $m_{n,p}^*/m$ also differs significantly between the two models - a logarithmic decrease in the 
MDI(A) model vs a $[1+{\rm{(constant)}}\cdot n]^{-1}$ decrease with density in the SkO$^\prime$ model.  Effects of isospin content as it  varies from that of symmetric nuclear matter ($x=0.5$)  toward pure neutron matter ($x\rightarrow 0$) are qualitatively similar, but quantitatively different with MDI(A) producing a significantly larger change compared with SkO$^\prime$. It is worthwhile to note, however, that several parametrizations of Skyrme interactions exist in the literature which yield a larger variation of $m_{n,p}^*/m$ with varying $x$ than is present in the SkO$^\prime$ model, although the logarithmic decline with density of the MDI models would be absent in all of them.  

A noteworthy feature of the results in Figs.  \ref{MDYISk_Ms}(a) and (b) is that $m_n^* > m_p^*$ for all densities as $x$ moves from its symmetric matter value of 0.5 to 0, the value for pure neutron matter.  The cause for this behavior may be traced to the values of strength and range parameters that govern the behavior of effective masses with proton fraction. For example, if we require the condition $m_n^* > m_p^*$  to be satisfied for the MDI models, Eq.  (\ref{msfin}) implies that  to leading order in $n$ and for all $x$
\be
C_l - C_u > 0\,, 
\label{condMDYI1}
\ee
independent of the finite range parameter $\Lambda$.
Next to leading order in $n$ and for all  $x$, the condition becomes
\ba
C_l &\left[ 1 - \frac {20}{3} \left(\frac{p_F}{\Lambda}\right)^2 + 4\left(\frac{n}{n_0} \right) \left(\frac{mC_l}{\Lambda^2}\right)  \right]  \nonumber \\
- C_u &\left[ 1 - 4 \left(\frac{p_F}{\Lambda}\right)^2 + 4\left(\frac{n}{n_0} \right) \left(\frac{mC_u}{\Lambda^2}\right)  \right]  > 0 \,,
\label{condMDYI2}
\ea
where $p_F$ above is the fermi momentum of symmetric nuclear matter. 
As $(20/3) (p_F/\Lambda)^2 > 4 (p_F/\Lambda)^2$, the condition becomes $a~C_l > C_u$ with $a < 1$.
These conditions are met for the MDI(0) (and for all the MDI models in Ref. \cite{Das03}) and MDI(A) models which ensures that
$m_n^* > m_p^*$ for all $x$ in the range $0.5-0$.  

For Skyrme interactions, Eq. (\ref{effmSkO'}) implies that $m_n^* > m_p^*$ in neutron-rich matter as long as
\be
t_1(1+2x_1) > t_2(1+x_2)\,,
\label{condSk}
\ee
which is satisfied by the parameters in Table V. 

When the conditions in Eqs. (\ref{condMDYI1}),  (\ref{condMDYI2}),   and 
(\ref{condSk}), are not satisfied, a reversal in the behavior of neutron and proton effective masses occurs, that is, $m_n^* < m_p^*$ as $x$ moves away from 0.5 toward 0.  Notwithstanding this behavior, the properties of symmetric nuclear matter, symmetry energy attributes, and collective excitations of nuclei have been well described.  
Additionally, the requirement that the EOS of neutron-star matter is able to support stars of 2${\rm M}_\odot$ has also been met. To illustrate this, we 
show in Table VIII parameters of an MDI model and those of a Skyrme model (SLy4 {\cite{sly4})  in Table IX whose predictions at the nuclear  equilibrium density are summarized in Table X. The corresponding results for neutron star structure are given in Table XI. The Landau effective masses for the neutron and proton that exhibit this reversal in behavior as a function of $x$ are shown in Figs.  \ref{MDYISk_Rev_Ms}(a) and (b).  Although we have not done an exhaustive search, we have verified that several parametrizations of Skyrme interactions, GSkI, GSkII, KDE0v1 \cite{gskIandII,kde} to name a few,  exhibit this reversal in behavior of neutron and proton effective masses as a function of $x$.

\begin{table}[htb]
\begin{ruledtabular} 
\newcolumntype{a}{D{.}{.}{3,13}}
\begin{tabular}{caca}    
Parameter & \multicolumn{1}{c}{Value} & Parameter & \multicolumn{1}{c}{Value} \\
\hline
$A_1 $    & 290.942\phantom{0}\mbox{MeV}       &   $y $        &     1.47111                       \\
$A_2 $    & 709.676\phantom{0}\mbox{MeV}      &   $C_l $      &    -525.764\phantom{0}\mbox{MeV}   \\        
$B $      & 151.01\phantom{00}\mbox{MeV}     &   $C_u $      &    -0.0587396\phantom{0}\mbox{MeV}       \\
$\sigma $ &   1.23671                          &   $\Lambda $  &   1315.20\phantom{0}\mbox{MeV}    \\            
\end{tabular}
\caption[Parameter values for $\mathcal{H}_{MDI(B)}$]{Parameter values for the MDI(B) Hamiltonian density. 
The dimensions are such that the Hamiltonian density is in MeV fm$^{-3}$.}
\end{ruledtabular}
\end{table}

\begin{table}[!h]
\begin{ruledtabular}
\newcolumntype{a}{D{.}{.}{-1}}
\begin{tabular}{caac}
i & \multicolumn{1}{c}{$t_i$} & \multicolumn{1}{c}{$x_i$} & $\epsilon$  \\
\hline
0 & -2488.91\phantom{0}\mbox{MeV~fm}^3  & 0.834             & 1/6 \\
1 & 486.82\phantom{0}\mbox{MeV~fm}^5    & -0.344                   \\ 
2 & -546.39\phantom{0}\mbox{MeV~fm}^5   & -1.000                   \\
3 & 13777.0\phantom{00}\mbox{MeV~fm}^{3(1+\epsilon)}   & 1.354          \\
\end{tabular}
\caption[Parameter values for Sly4.]{Parameter values for the Skyrme Hamiltonian density Sly4 \cite{sly4}. 
The dimensions are such that the Hamiltonian density is in MeV fm$^{-3}$.}
\end{ruledtabular} 
\end{table}

\begin{table}[!h]
\begin{ruledtabular}
\newcolumntype{a}{D{.}{.}{-1}}
\newcolumntype{b}{D{+}{\pm}{-1}}
\begin{tabular}{laabc}
 Property  & \multicolumn{1}{c}{Value} & \multicolumn{1}{c}{Value} & \multicolumn{1}{c}{Experiment} & Reference\\
   & \multicolumn{1}{c}{[MDI(B)]} & \multicolumn{1}{c}{[Sly4]} &  &  \\
\hline
 $n_0$ (fm$^{-3}$) & 0.160   & 0.160   & 0.17+0.02                   & \cite{day78,jackson74,myers66,myers96}  \\
 $E_0$ (MeV)      & -16.00  & -15.97  & -16+1                       & \cite{myers66,myers96}                  \\
 $K_0$ (MeV)      & 239.0   & 229.9   & 230+30                      & \cite{Garg04,Colo04}                    \\
                  &         &         & 240+20                      &\cite{shlomo06}                          \\
 $m_0^*/m$        &  0.67   &  0.69   & 0.8+0.1                     & \cite{bohigas79,krivine80}              \\   \hline

 $S_v$ (MeV)      & 30.0    & 32.0    & \multicolumn{1}{c}{~~30-35}   & \cite{L,tsang12}                        \\
 $L_v$   (MeV)    & 52.0    & 46.0    & \multicolumn{1}{c}{~~40-70}   & \cite{L,tsang12}                        \\
 $K_v$  (MeV)     & -81.0   & -119.7   & -100+200                    & \cite{APRppr}                                    \\ 
\end{tabular}
\caption[Symmetric nuclear matter saturation properties.]{
Entries in this table are at the equilibrium 
density $n_0$ of symmetric nuclear matter for the MDI(B) and Sly4 models.
$E_0$ is the energy per particle, $K_0$ is the compression modulus, $m_0^*/m$ is 
the ratio of the Landau effective mass to mass in vacuum, $S_v$ is the nuclear symmetry 
energy parameter, $L_v$ and $K_v$, are related to the first and second derivatives of the 
symmetry energy, respectively. }
\end{ruledtabular}
\end{table}

\begin{table}[!h]
\begin{ruledtabular}
\newcolumntype{a}{D{.}{.}{-1}}
\newcolumntype{b}{D{+}{\pm}{-1}}
\begin{tabular}{laabc}
 Property  & \multicolumn{1}{c}{Value} & \multicolumn{1}{c}{Value} & \multicolumn{1}{c}{Observation} & Reference\\
 & \multicolumn{1}{c}{[MDI(B)]} & \multicolumn{1}{c}{[Sly4]} &  &  \\
\hline
 ${\rm M}_{max}(M_{\odot})$& 2.0295    &  2.037      & 2.01+0.04                   & \cite{Antoniadis}                        \\
 $R_{max}$(km)     & 10.027   & 9.797      & 11.0+ 1.0                   & \cite{slb10}                             \\
 $n_c$(fm$^{-3}$)  &1.1768   &  1.2062     & & \\
 $\epsilon_c$(MeV fm$^{-3}$) &1539.5&1598.8& & \\
 $P_c$(MeV fm$^{-3}$) & 749.91& 856.45& & \\
 $\mu_c$(MeV)& 1945.4 & 2035.54 & & \\
\hline
 $R_{1.4}$(km)     & 11.41  &  11.12      &  11.5+ 0.7                  & \cite{slb10}                              \\
 $n_c$(fm$^{-3}$)  &0.5110   &  0.5413     & & \\
 $\epsilon_c$(MeV fm$^{-3}$) &522.41&555.99& & \\
 $P_c$(MeV fm$^{-3}$) & 77.390& 86.391& & \\
 $\mu_c$(MeV)& 1173.8 & 1186.604 & & \\
\end{tabular}
\caption[Neutron star properties.]{
Same as \ref{nspropsMDI(0)}, but for the MDI(B) and Sly4 models.}
\end{ruledtabular}
\end{table}

\begin{figure*}[htb]
\centering
\begin{minipage}[b]{0.49\linewidth}
\centering
\includegraphics[width=9.5cm,height=8cm]{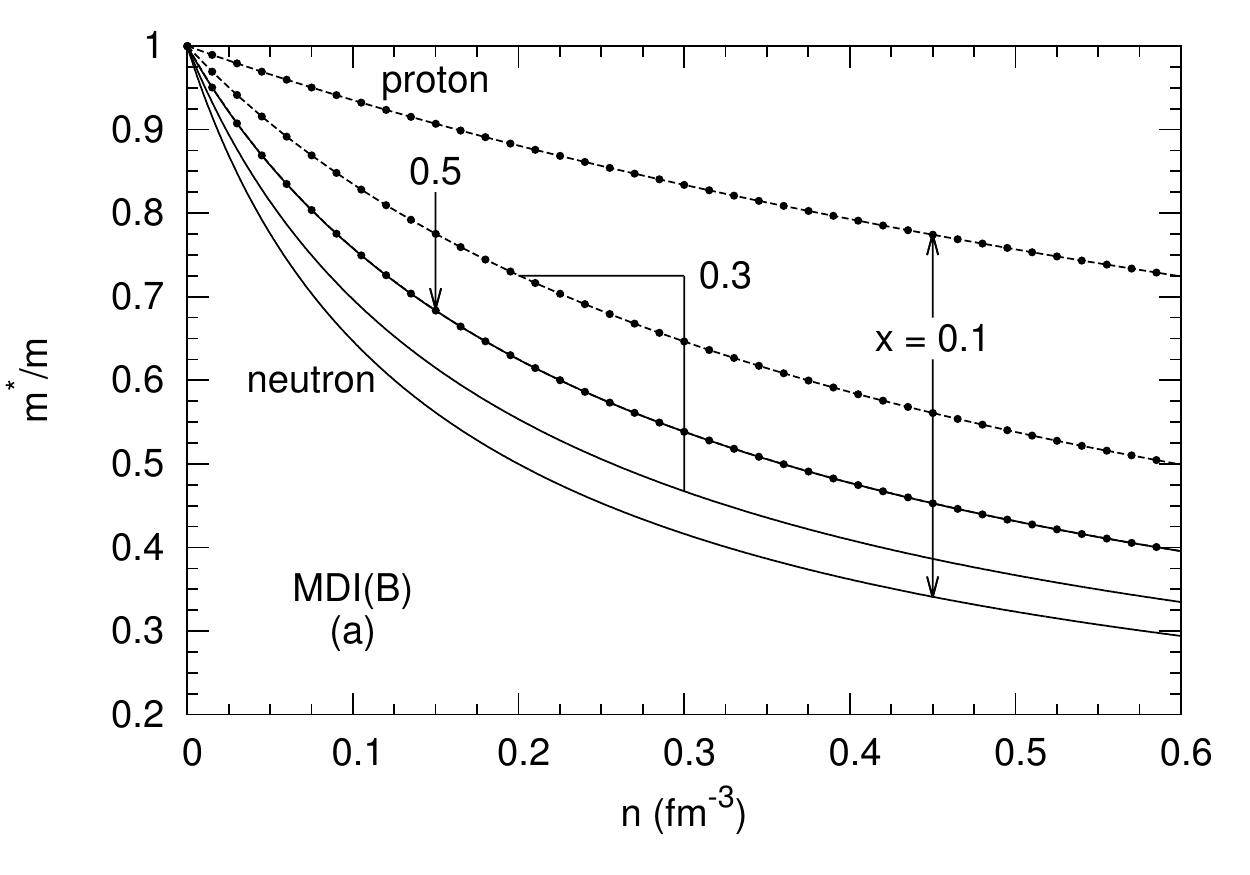}
\end{minipage}
\begin{minipage}[b]{0.49\linewidth}
\centering
\includegraphics[width=9.5cm,height=8cm]{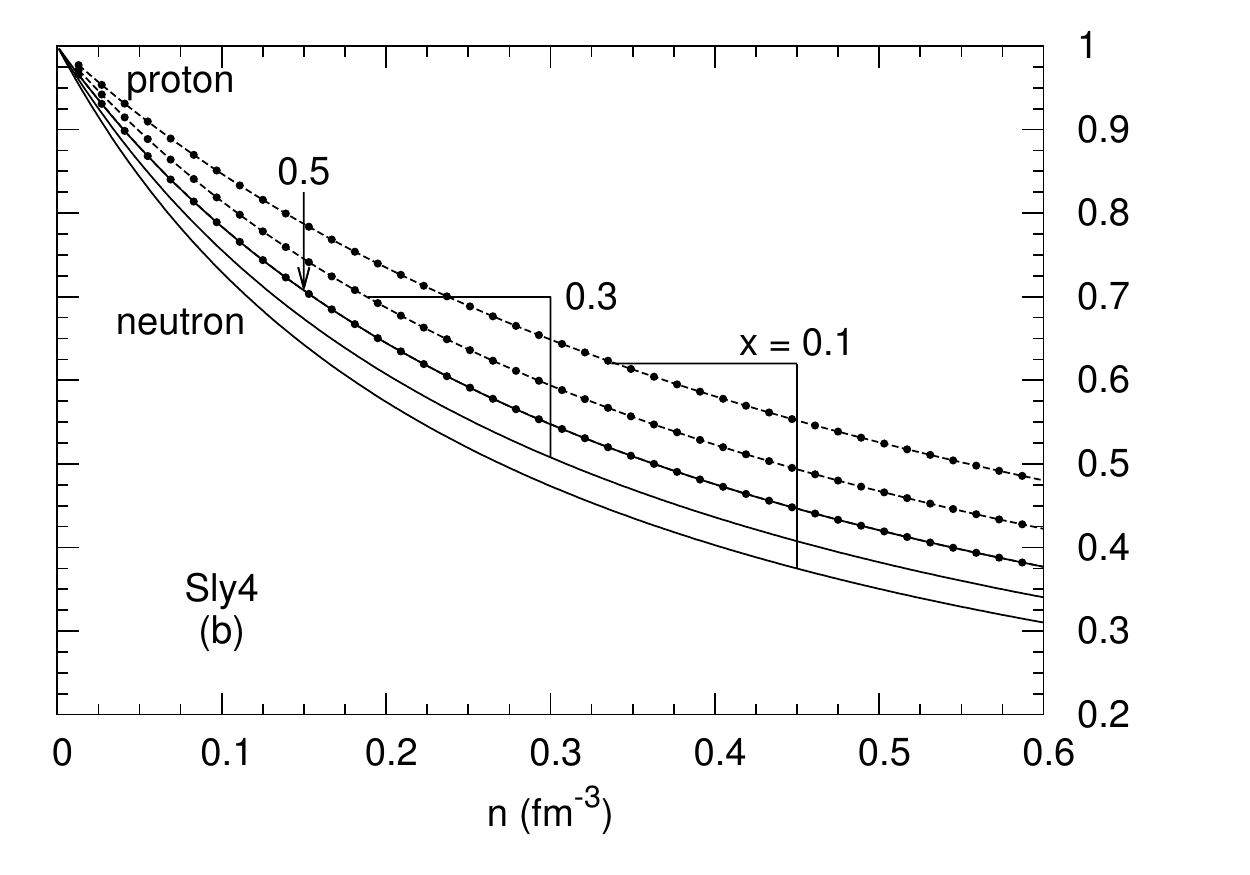}
\end{minipage}
\vskip -0.5cm
\caption{Landau effective masses of the neutron (solid)  and proton (dotted) scaled with the vacuum nucleon mass  
vs baryon density $n$ for the marked values of the proton fraction $x$. (a)  Results are for the MDI(B) model 
from Eqs. (\ref{msin})-(\ref{msfin}). (b) Same as (a), but for the Sly4 model from Eq. (\ref{ms}). Note the reversal in behavior of the effective masses when compared to the results in Figs. \ref{MDYISk_Ms}(a) and (b).} 
\label{MDYISk_Rev_Ms}
\end{figure*}

It is worthwhile noting that in the microscopic Brueckner-Hartree-Fock and Dirac-Brueckner-Hartree-Fock calculations that include 3-body interactions, $m_n^* > m_p^*$ in neutron-rich matter \cite{Ma04,Dalen05,Sammarruca05,Baldo14}. Their 
density dependences, while akin to those of Skyrme-like models with similar splitting exhibit quantitative differences only.  
The isospin splitting of the effective masses in the MDI(A) model is in agreement with the above microscopic models.  

\subsection{Energy, pressure and chemical potentials}

The energy per baryon $E={\cal H}/n$ from the MDI(A) and SkO$^\prime$ models is presented as functions
of baryon density and proton fraction in Fig. \ref{MDYISk_0T_EoA}. Explicit expressions for ${\cal H}$ are  
in Eq. (\ref{MDYI_H_0T}) for the MDI model and in Eq. (\ref{hsko}) for the SkO$^\prime$ model.
For all proton fractions ranging from pure neutron matter to symmetric nuclear matter shown in  Fig. \ref{MDYISk_0T_EoA}(a) 
there is little difference between the energies of the two models, the energy for the SkO$^\prime$ model being slightly larger than 
that of MDI for  pure neutron matter at all densities. 
The inset in Fig. \ref{MDYISk_0T_EoA}(b) shows the energy per baryon of nucleons vs baryon density
for charge neutral matter in beta equilibrium. The two models yield
nearly the same energy at most densities with only small differences around the nuclear matter density.
\begin{figure}[!h]
\centering
\makebox[0pt][c]{
\hspace{0cm}
\begin{minipage}[b]{\linewidth}
\centering
\includegraphics[width=8.5cm]{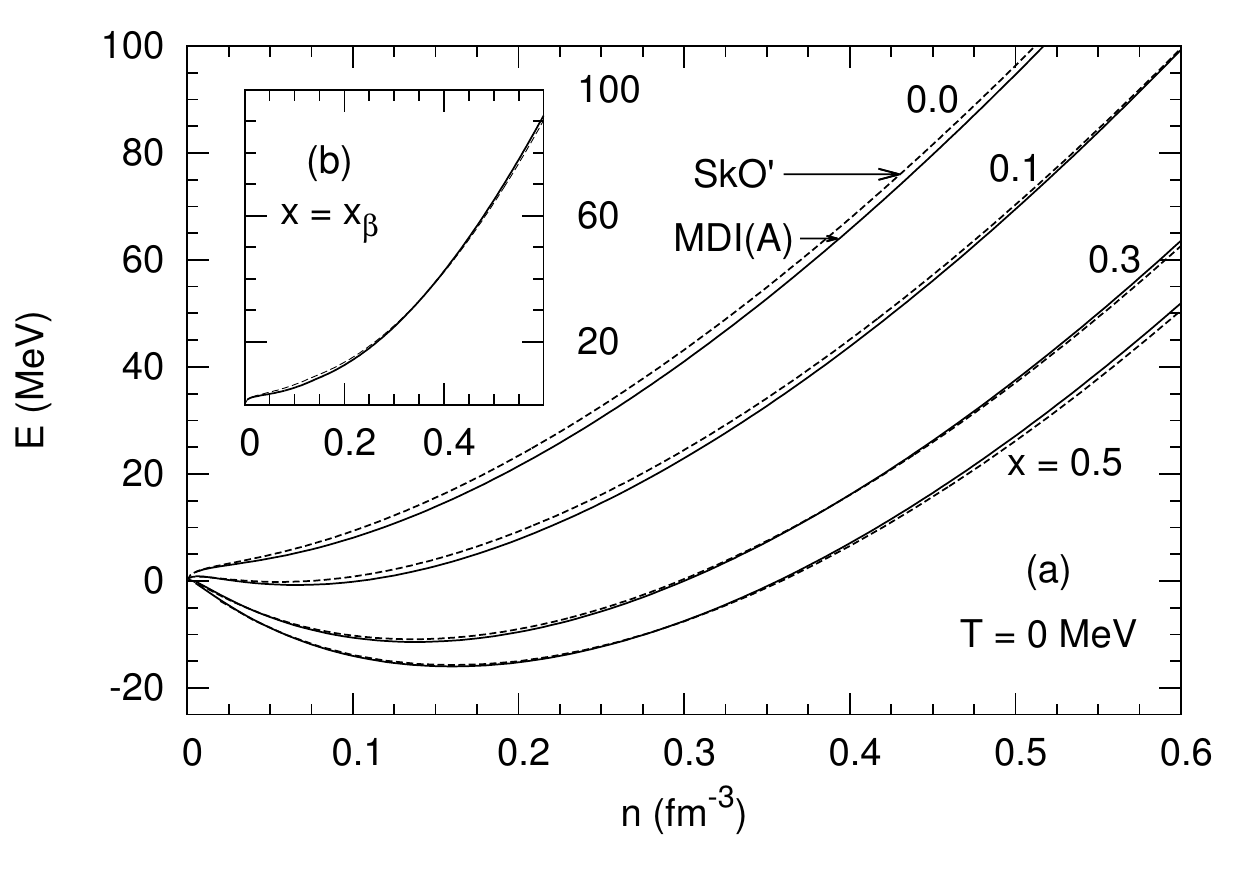}
\end{minipage}
}
\vskip -0.5cm
\caption{(a) Zero temperature energy per particle $E={\cal H}/n$ vs baryon number density $n$ at the 
proton fractions $x$ shown for the MDI(A) model from Eqs. (\ref{MDYI_H_0T_0})-(\ref{MDYI_H_0T}) and the SkO$^\prime$ 
model from Eq. (\ref{hsko}). (b) $E$ vs $n$ for matter with varying 
proton fractions $x_\beta$ determined from charge neutrality and beta-equilibrium. } 
\label{MDYISk_0T_EoA}
\end{figure}

In terms of the neutron excess parameter  $\alpha=(n_n-n_p)/n=1-2x$,  
the PNM to SNM energy difference can be expressed as 
\be
\Delta E = E(n,\alpha) - E(n,0) = \sum_{l=2,4,\dots} S_l(n)\alpha^l \,,  
\ee
where 
\be
S_l = \frac {1}{l!} \left. \frac {\partial^l E(n,\alpha)}{\partial \alpha^l} \right|_{\alpha=0} \,; \quad l = 2,4,\dots
\ee
In Fig. \ref{MDYI_N2_SymE}(a), $\Delta E$ of the MDI(A) model is compared with its symmetry energy $S_2(n)$ which 
provides a reasonable approximation. Contributions of the higher order terms stemming from kinetic  and momentum dependent interaction terms (denoted by $S_{lk}$ and $S_{lm}$, respectively) up to $l=6$ are shown Fig. \ref{MDYI_N2_SymE}(b). The convergence of $\Delta E -\sum_l S_l$ to zero is shown in Fig. \ref{MDYI_N2_SymE}(c).  At the nuclear equilibrium density, $S_2$ falls short of $\Delta E$ by a little over an MeV. The inclusion of higher order terms in $S_l$ drastically  reduces the difference.
Figure \ref{SkOp_SymE_Multi} shows similar results for the SkO$^\prime$ model. Relative to the MDI(A) model, the largest qualitative difference in seen in the density dependence of $S_{4m}$. However, the convergence of   $\Delta E -\sum S_l$ to zero is similar for both models.

\begin{figure}[!h]
\centering
\makebox[0pt][c]{
\hspace{0cm}
\begin{minipage}[b]{\linewidth}
\centering
\includegraphics[width=8.5cm]{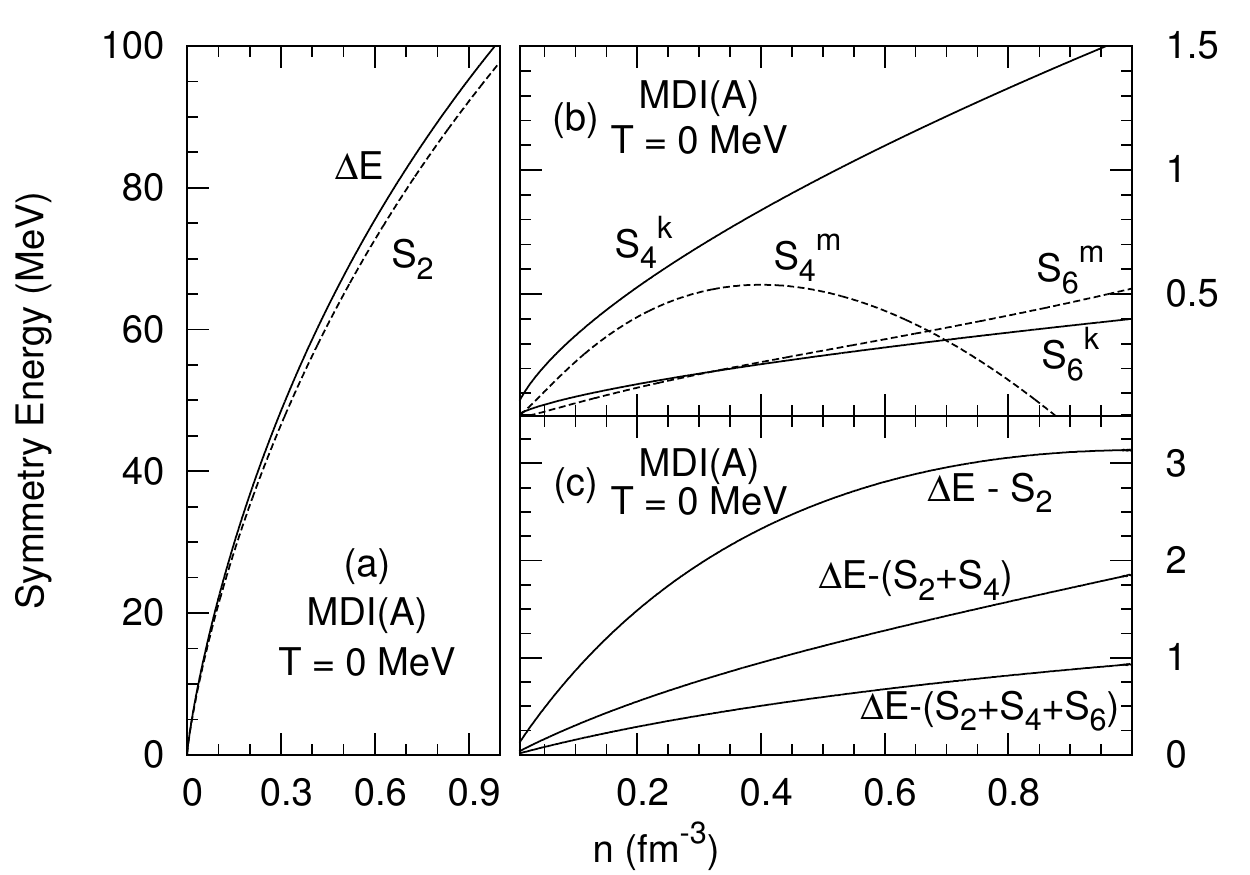}
\end{minipage}
}
\vskip -0.25cm
\caption{(a) The difference $\Delta E = E(n,x=0) - E(n,x=1/2)$ vs $n$ and the symmetry energy $S_2$ vs $n$.  
(b) Higher order contributions of $S_l$ to $\Delta E$. (c) Convergence of $\Delta E -\sum S_l$ up to 6th order terms in $S_l$.} 
\label{MDYI_N2_SymE}
\end{figure}

\begin{figure}[!h]
\centering
\makebox[0pt][c]{
\hspace{0cm}
\begin{minipage}[b]{\linewidth}
\centering
\includegraphics[width=8.5cm]{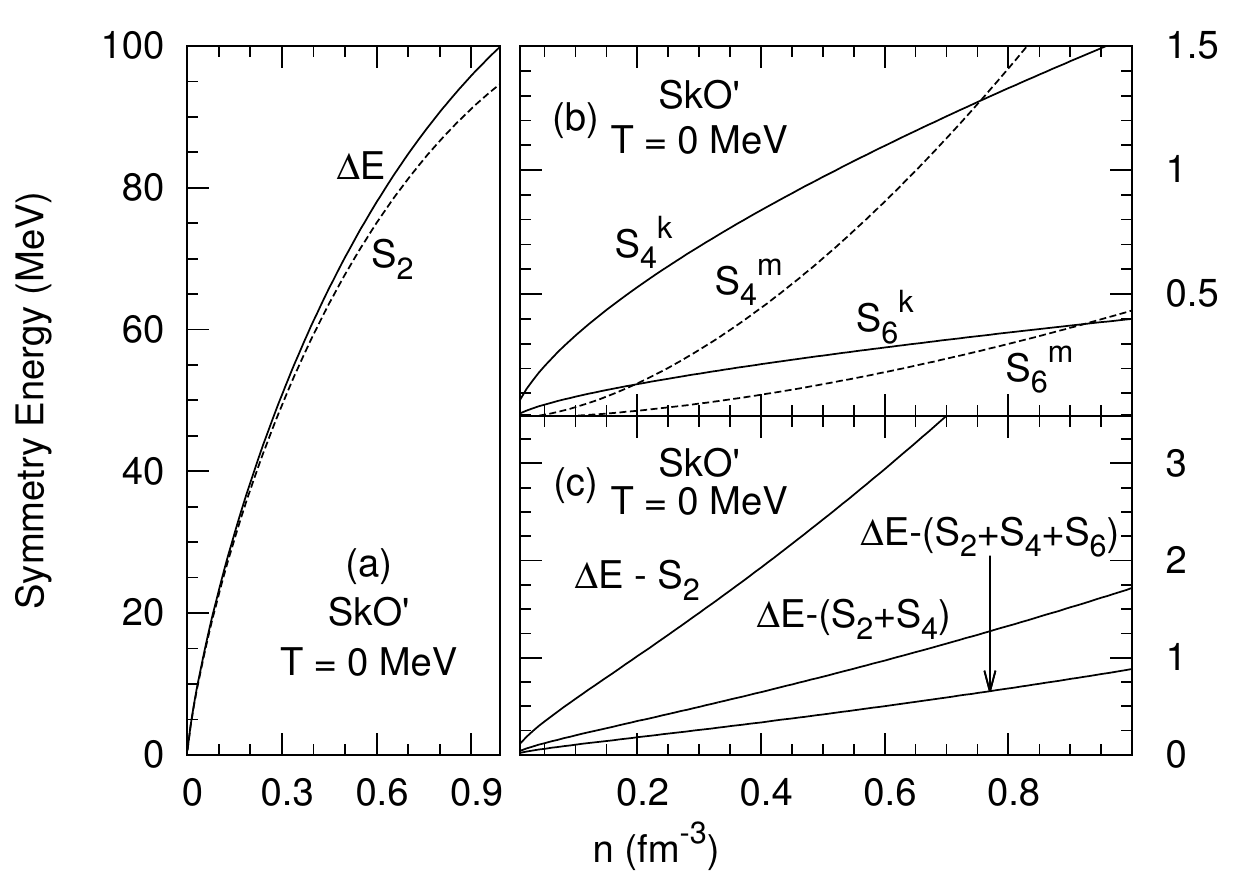}
\end{minipage}
}
\vskip -0.25cm
\caption{ Same as Fig. \ref{MDYI_N2_SymE}, but for the SkO$^\prime$ model. }
\label{SkOp_SymE_Multi}
\end{figure}

In Fig. \ref{MDYISk_0T_P}(a), we show the pressures exerted by nucleons resulting from the MDI(A) and SkO$^\prime$ 
models as  functions of density for symmetric nuclear and pure neutron matter. 
The results in Fig. \ref{MDYISk_0T_P}(b) correspond to charge neutral matter  in beta equilibrium. 
Both models have nearly identical pressures regardless of the proton fraction with  small differences  at the highest densities shown. 

\begin{figure}[!h]
\centering
\makebox[0pt][c]{
\hspace{0cm}
\begin{minipage}[b]{\linewidth}
\centering
\includegraphics[width=8.5cm]{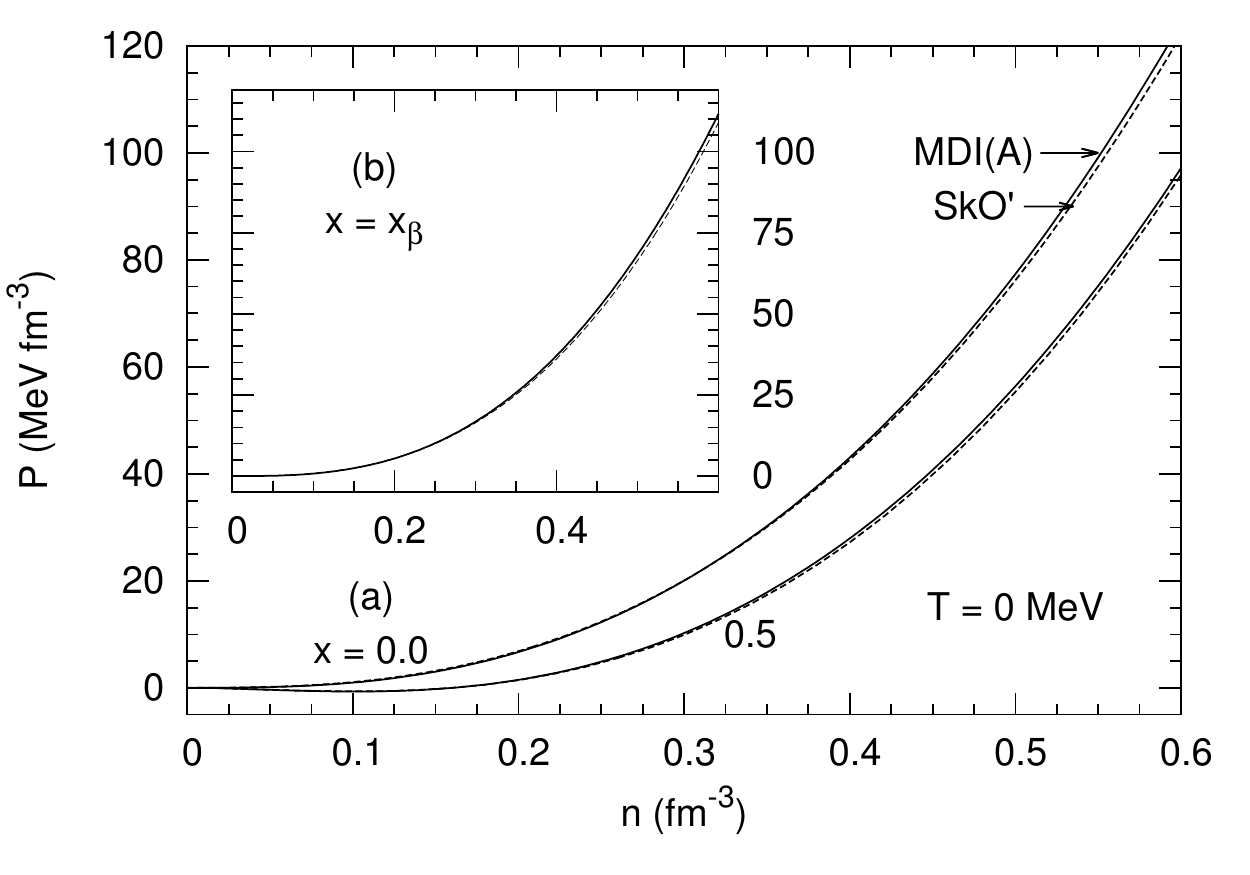}
\end{minipage}
}
\vskip -0.5cm
\caption{(a) Zero temperature pressure $P$ vs baryon density $n$ at the proton fractions shown for the MDI(A) model 
from Eqs. (\ref{Pin})-(\ref{Pfin}) and the SkO$^\prime$ model from Eq. (\ref{epmu}). (b) $P$ vs $n$ for matter with varying 
proton fractions $x_\beta$ determined from charge neutrality and beta-equilibrium. } 
\label{MDYISk_0T_P}
\end{figure}
The density dependence of the neutron and proton chemical potentials  from the MDI(A) and SkO$^\prime$ models are 
presented in Figs. \ref{MDYISk_0T_muNPs}(a) and (b), respectively. The MDI(A) 
chemical potential was calculated using Eqs. (\ref{muin})-(\ref{mufin}). For all proton
fractions presented ($x = 0.1, 0.3~{\rm and}~0.5$), the neutron chemical potentials of the two
models agree at all densities. Small differences between the two models are observed in the 
proton chemical potentials with the largest difference occuring at low proton fractions and
large baryon densities.
Figure \ref{MDYISk_0T_muNPs}(c)  shows the difference between
the neutron and proton chemical potentials ($\hat \mu=\mu_n-\mu_p$) vs density for the MDI(A) and 
SkO$^\prime$ models, respectively. As there is little difference between the neutron chemical
potentials of the two models, the small differences between the $\hat \mu$'s 
stem from differences in the proton chemical potentials. The 
differences in $\hat \mu$ occur mainly for low proton fractions and high densities.  
\begin{figure*}[htb]
\centering
\makebox[0pt][c]{
\hspace{-2cm}
\begin{minipage}[b]{0.33\linewidth}
\centering
\includegraphics[width=7.7cm,height=5.5cm]{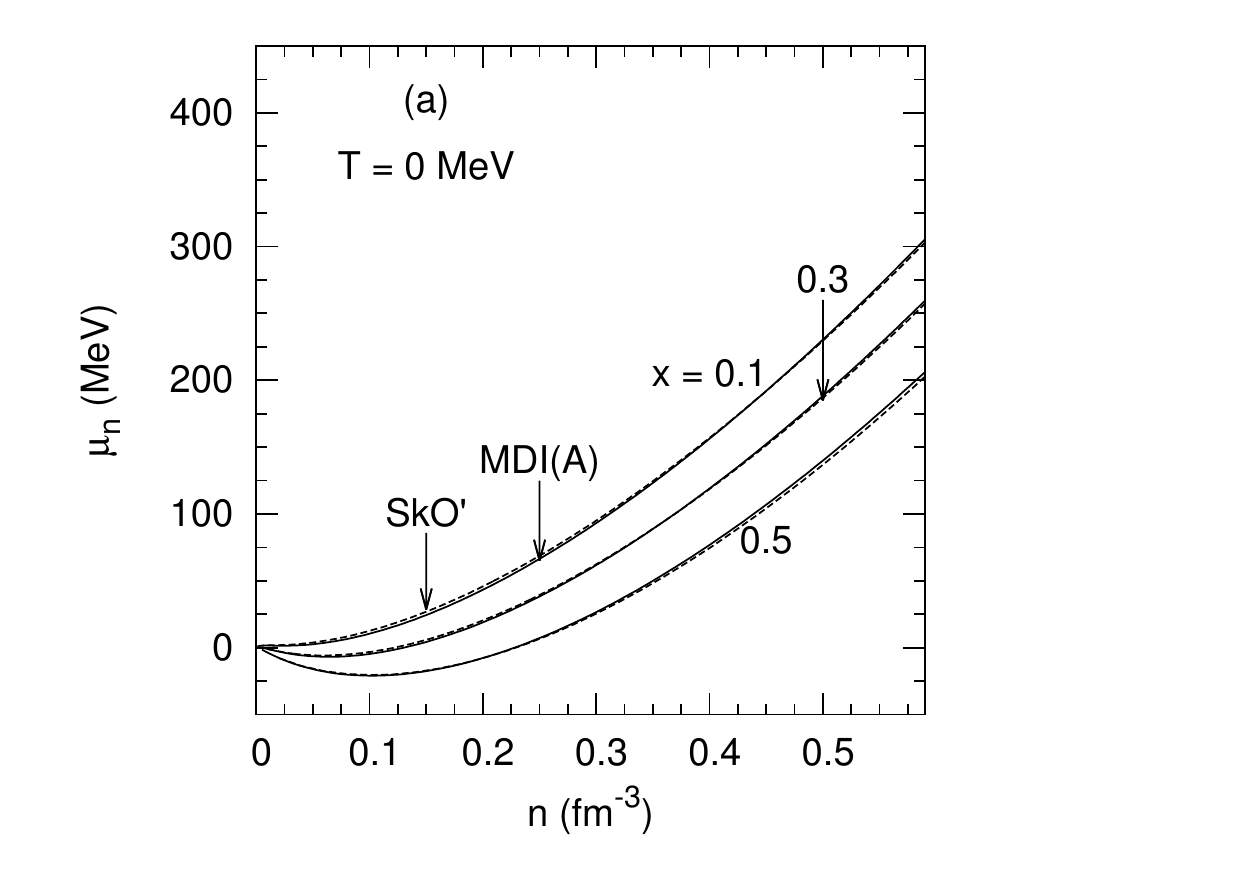}
\end{minipage}
\hspace{-1.25cm}
\begin{minipage}[b]{0.33\linewidth}
\centering
\includegraphics[width=7.7cm,height=5.5cm]{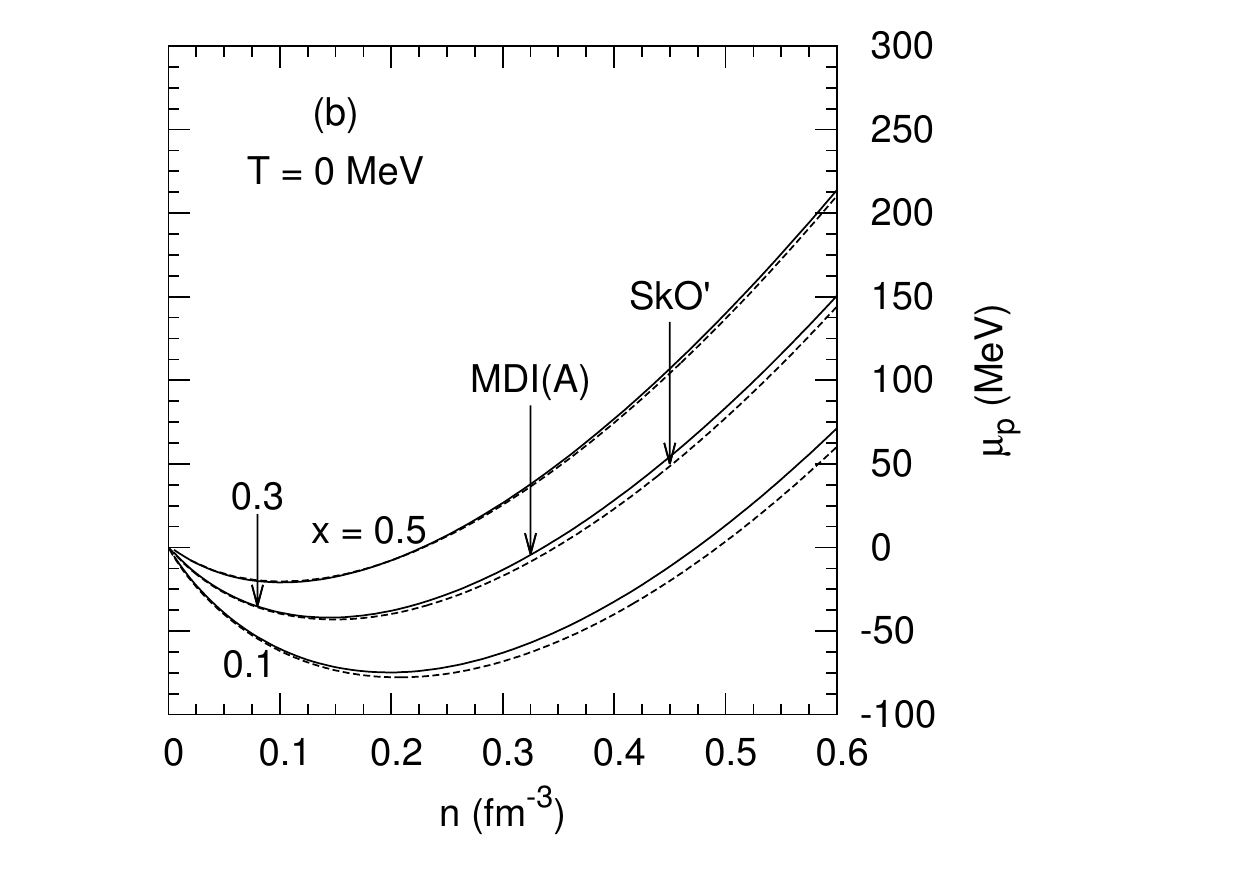}
\end{minipage}
\hspace{0.25cm}
\begin{minipage}[b]{0.32\linewidth}
\centering
\includegraphics[width=7cm,height=5.5cm]{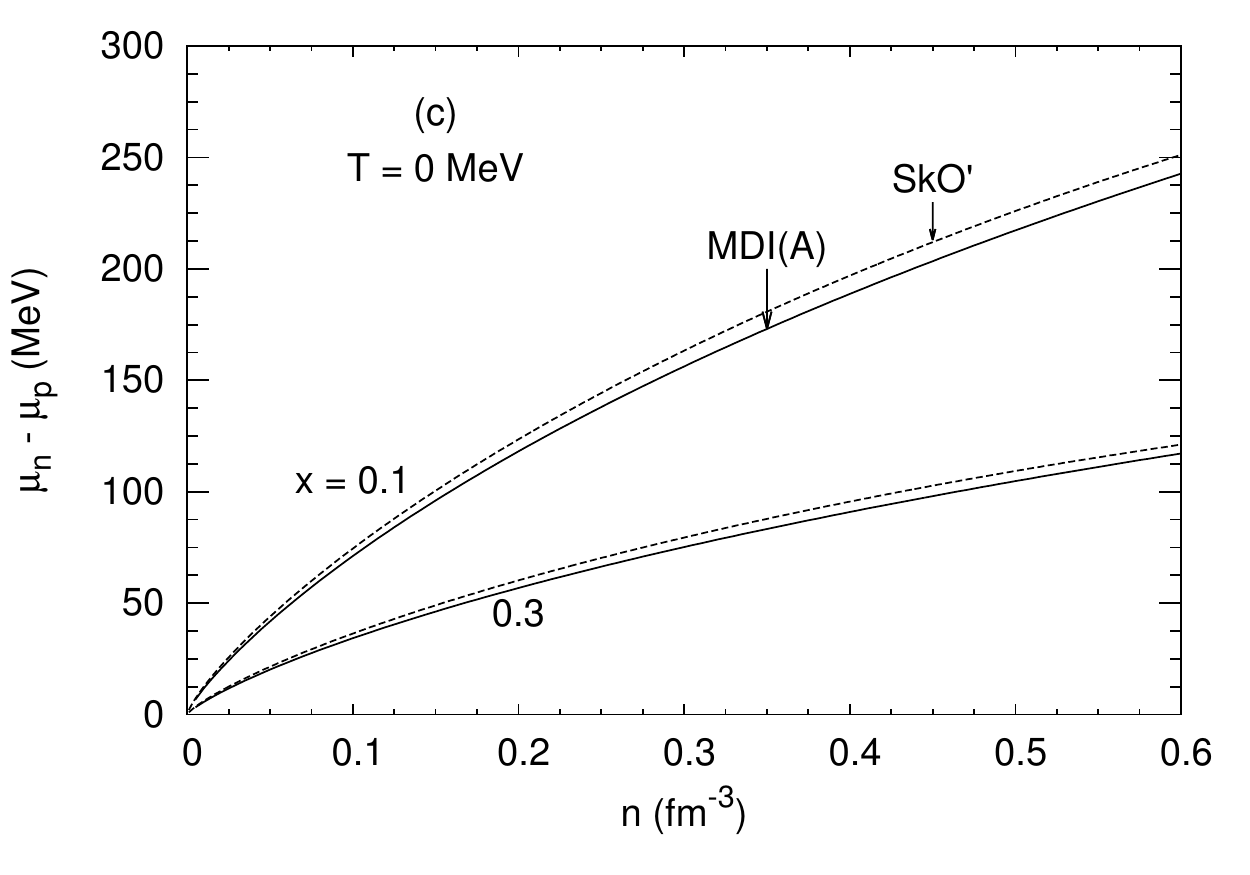}
\end{minipage}
}
\vskip -0.5cm
\caption{[(a) and (b)] The neutron and proton  chemical potentials vs baryon density $n$ for the MDI(A)  
[Eqs. (\ref{muin})- (\ref{mufin})] and the
SkO$^\prime$  [Eq. (\ref{epmu})] models for values of $x$ shown. (c) $\hat \mu = \mu_n - \mu_p$ vs $n$.}
\label{MDYISk_0T_muNPs}
\end{figure*}

\subsection{Properties of cold-catalyzed neutron stars}

\begin{figure}[!h]
\centering
\makebox[0pt][c]{
\hspace{0cm}
\begin{minipage}[b]{\linewidth}
\centering
\includegraphics[width=9.5cm]{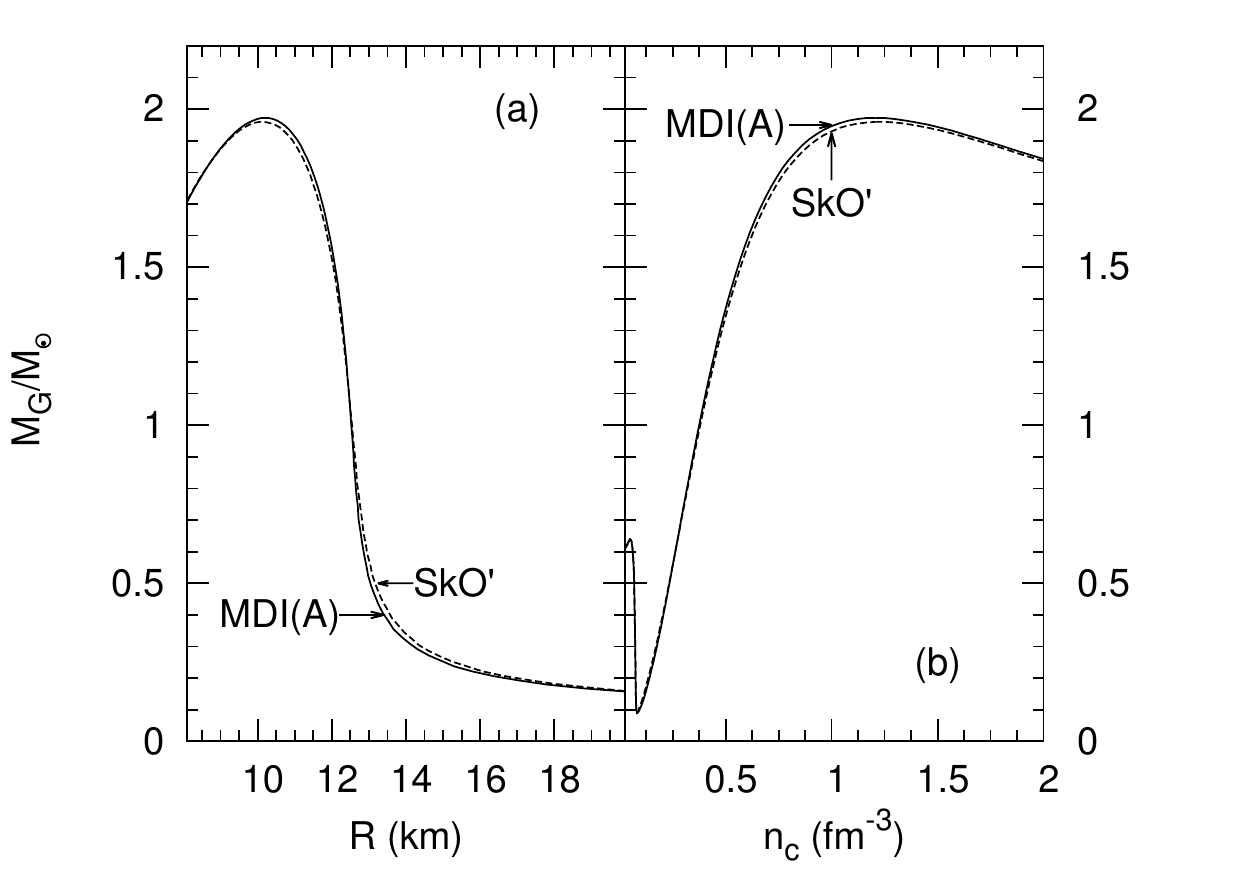}
\end{minipage}
}
\vskip -0.25cm
\caption{(a) Mass vs radius and (b) mass vs central density of cold-catalyzed neutron stars for the MDI(A) and SkO$^\prime$ models. } 
\label{MDYIuN2SkOp_Beq_MR_Mnc}
\end{figure}

In Figs. \ref{MDYIuN2SkOp_Beq_MR_Mnc}(a) and (b), we show the gravitational mass vs radius (${M_G}~{\rm vs}~R)$ and mass vs central density (${M_G}~{\rm vs}~n_c)$ of cold-catalyzed neutron stars for the MDI(A) and SkO$^{\prime}$ models. Not unexpectedly, the results are very nearly the same as the $P~{\rm vs}~n$ relations in neutron-star matter are nearly identical for the two models [see Fig. \ref{MDYISk_0T_P}(b)]. The close to vertical rise of mass with radius  is due to the similar behaviors of the symmetry energies in the two models (characterized by similar values of $L$ at $n_0$).  

The central densities of stars with different masses are of relevance in the long-term cooling of  neutron stars \cite{LPPH}.
For up to a million years after their birth, neutron stars cool primarily through neutrino emission. 
As pointed out in Ref. \cite{LPPH}, the density dependence of $S_2(n)$, more precisely $\Delta E(n)$, determines the threshold densities for the rapid direct Urca processes $n \leftrightarrow p+e^-+\bar \nu_e$   and $n \leftrightarrow p+\mu^-+\bar \nu_\mu$ to occur.  In charge neutral ($n_{e^-} + n_{\mu^-} = n_p$) and beta-equilibrated ($\hat\mu=\mu_{e^-} = \mu_{\mu^-}$) neutron-star matter, the threshold densities for these direct Urca processes are determined when the triangular inequalities
\be
p_{F_i} + p_{F_j} \geq p_{F_k} \,,
\ee
where $i,j$ and $k$ are $p,~e^-~(\mu^-)$, and $n$, and cyclic permutations of them, are satisfied (neutrinos at the relevant temperatures contributing little to the momentum balance conditions).  For the direct Urca process involving electrons (and muons), the threshold density is $0.67(0.93)~{\rm fm}^{-3}$ for the MDI(A) model. Thus, stars with masses $M > 1.7\rm{M_\odot}$ will undergo rapid cooling from neutrino emission from the process involving electrons while those with $M> 1.9\rm{M_\odot}$ will receive an equal and additional contribution to neutrino emissivity from the process with muons. 
The corresponding numbers for the SkO$^\prime$ model are $0.59(0.92)~{\rm fm}^{-3}$ and $1.5(1.9)\rm{M_\odot}$, respectively.

\section{THERMAL EFFECTS}
\label{Sec:Teffects}
Finite-temperature effects can be studied by isolating 
the thermal part of the various functions of interest  defined as the 
difference between the finite-$T$ and $T=0$ expressions for a given 
quantity $X$:
\be
X_{th} = X(n_i,n_j,T)-X(n_i,n_j,0) \,.
\ee
For Skyrme-like models, we refer the reader to Sec. V of Ref. \cite{APRppr} where explicit expressions and numerical notes are provided for evaluations of the thermal state variables and response functions.  Thus, our discussion below will focus on evaluations of the thermal state variables for the MDI models only.  

Suppressing the symbols denoting dependences on the baryon density $n$ and proton fraction $x$ for brevity, the thermal energy is given by 
\be
E_{th} = E(T) - E(0) \,,
\label{eth}
\ee
where $E(T)={\cal H(T)}/n$ is calculated using Eqs. (\ref{hmdyi})-(\ref{Ris}) in Sec. II-A, and  $E(0)={{\cal H}(0)}/n$ is obtained from 
Eqs. (\ref{MDYI_H_0T})-(\ref{Iij}) in Sec. II-B.

Likewise, the thermal pressure is 
\be
P_{th} = P(T) - P(0)\,,
\label{ppth}
\ee
where $P(T)$ and $P(0)$ are calculated using Eq. (\ref{pres2}) in Sec. II-A, and in Appendix \ref{Sec:AppendixA_P}, respectively.
Details regarding the numerical evaluation of the chemical potentials at finite $T$ are described in Sec II-A in connection with Eq. (\ref{Ris}) and the 
discussion thereafter. Explicit expressions for evaluating the chemical potentials at $T=0$ are collected in Appendix \ref{Sec:AppendixA_Mu}. Utilizing 
these results, the thermal chemical potentials are given by
\be
\mu_{th} = \mu(T) -\mu(0)\,.
\label{muth}
\ee
The entropy per baryon $S=s/n$ issues from Eq. (\ref{sden}) of Sec. II-a, the specific heats at constant volume and pressure, $C_V$ and $C_P$, 
from Eqs. (\ref{cv}) and (\ref{cp}), respectively, of the same section.

The question arises as to what plays the role of the degeneracy parameter for a general momentum dependent interaction.
The exposition below is for a 1-component system. Generalization to the 2-component case is straightforward. 
For the MDI model,  the single particle spectrum in Eq. (\ref{spectrum}) can be written as  
\be
\epsilon_p = \frac {p^2}{2m} + U(n,p) = \frac {p^2}{2m} + R(n,p;T) + \calU (n) \,,
\ee
where the explicit $p$-dependence arising from finite-range interactions is contained in $R(n,p;T)$ and $\calU (n)$ is a density-dependent, but $p$-independent term from interactions that are local in space. For a given $n$ at fixed $T$, the chemical potential $\mu$ is determined from Eq. (\ref{ni}) using an iterative procedure as outlined earlier. 
At finite $T$, the term $R$ acquires a $T$-dependence (see, {\it e.g.}, Fig. \ref{MDYISk_SPP}) which has been explicitly indicated. We can write
\ba
\epsilon_p-\mu = \frac {p^2}{2m} &+& R(n,p;T) - [\mu - \calU (n)]  \nonumber \\
&=& \frac {p^2}{2m} +  R(n,p;T) - R(n,p=0;T)\nonumber \\ 
&-& [\mu - \{\calU (n) + R(n,p=0;T)\} ] \,,
\ea
where in the last step we have isolated the $p$-independent, but $n$- and $T$-dependent part  $R(n,p=0;T)$ and grouped it with $\calU (n)$. 
Note that  $R(n,p=0;T)$ provides an $n$- and $T$- dependent pedestal for the momentum dependence of $R(n,p;T)$.
As a result,
\be
\frac {\epsilon_p - \mu}{T } = \frac 1T \left\{ \frac {p^2}{2m} +   R(n,p;T) - R(n,p=0;T) \right\} -\eta
\ee
with
\be
\eta  = \frac {1}{T}  [\mu - \{\calU (n) + R(n,p=0;T)\} ] \,,
\label{eta1}
\ee
which serves as the degeneracy parameter.  The term $R(n,p;T)$ when combined with $p^2/(2m)$ generates an effective mass $m^*$.  (For free gases, $R(n,p;T=0)=0$ and $\calU (n)=0$ so that $\eta = \mu/T$.) 

For Skyrme interactions, $R(n,p;T) = \beta\times np^2/(2m)$, where $\beta$ is an interaction strength-dependent constant.  Combined with the $p^2/(2m)$ term, the $p$-dependent part of $\epsilon_p$  can be written as  $p^2/(2m^*)$  with  $m^*=m(1+\beta n)^{-1}$ being a density-dependent, but $T$-independent, effective mass. Thus, 
\be
\eta  = \frac {\mu - \calU (n)} {T}  \qquad {\rm for~Skyrme~interactions} \,.
\label{etas}
\ee

For a general $R(n,p)$, more complicated dependencies of $m^*$ with $n$ ensue including a $T$-dependence as for the MDI models with Yukawa-type interactions.  

Utilizing  
\be
\mu(n,T=0) = E_F = \frac {p_F^2}{2m} + R(n,p_F;T=0) + \calU (n)
\ee
(as follows from Hugenholtz-Van Hove's theorem) we also have
\ba
\eta &=& \frac {1}{T} \left\{ \mu(n,T=0) + \mu_{th} (n,T) -  [\calU (n) + R(n,p=0;T) ] \right\} \nonumber \\
&=& \frac 1T \left\{  \frac {p_F^2}{2m} + [R(n,p_F;T=0) - R(n,p=0;T] 
+ \mu_{th} (n,T) \right\} \,, \nonumber \\
\label{eta2}
\ea
where the thermal contribution to the chemical potential, $\mu_{th}$, is as defined in Eq. (\ref{muth}).
Above, $p_F$ is the Fermi momentum at $T=0$ determined using $p_F=\hbar (3\pi^2n)^{1/3}$. Recall that the number of particles is fixed be it at $T=0$ or $T \neq 0$. Consequently, $p_F$ serves as   a useful reference momentum with respect to which effects of finite $T$ (for which states with $p<p_F$ and $p>p_F$  up to $\infty$ are all occupied with finite probabilities) can be gauged. For example,
\ba
\eta &=&  \frac 1T \left\{  \frac {p_F^2}{2m} (1+ \beta n) + \mu_{th} (n,T) \right\} \nonumber \\
&=& \frac 1 T  \left\{  \frac {p_F^2}{2m^*}  + \mu_{th} (n,T) \right\}
\label{eta3}
\ea
for Skyrme interactions. For MDI models, an equivalent expression can be derived but it features additional density dependencies and agrees with Eq. (\ref{eta3}) only when the $p$-dependence is truncated at the quadratic term. 

The sign of $\eta$ depends on the magnitudes of the various terms in Eq.~(\ref{eta2}). 
The first two terms are positive, and 
generally, $\mu_{th} <0$ for all degeneracies (see Fig. \ref{MDYISk_Muth}).  In the degenerate limit, the sum of the positive terms dominates over $\mu_{th}$ rendering $\eta$ positive, whereas in the non-degenerate limit the magnitude of $\mu_{th}$ exceeds the summed positive terms making $\eta$ negative.   

The contribution of $R(n,p=0;T)$ to Eq. (\ref{eta1}) is significant as an analysis for the MDI model used here illustrates. 
For like particles,
\be
R(n,p=0;T) = \frac {2C}{n_0}~2~\int \frac {d^3p^\prime}{(2\pi\hbar)^3 }~f_{p^\prime} \frac {1}{1+ (\frac {p^\prime}{\Lambda})^2 }\,,
\ee
where $C=C_l$ for pure neutron matter (PNM) and $C=C_l+C_u$ for symmetric nuclear matter (SNM).
Although temperature corrections are significant  at low momenta (see Fig. \ref{MDYISk_SPP}), it is useful to study the $T=0$ values for orientation. In this case (with $f_{p^\prime}=1$ and $y=p^\prime/\Lambda$), 
\be
R(n,p=0;T=0) = \frac {2C}{n_0}~ \left(\frac {\Lambda}{\hbar} \right)^3 \frac {1}{\pi^2}~\int_0^{y_F}  dy~\frac {y^2}{1+ y^2 }\,,
\ee
where $y_f=p_F/\Lambda$. 
The integral above is easily performed with the result
\ba
R(n,p=0;T=0) = \frac {2C}{n_0}~ \left(\frac {\Lambda}{\hbar} \right)^3 \frac {1}{\pi^2}~
\left[\frac {p_f}{\Lambda} - \tan^{-1} \frac{p_F}{\Lambda}\right] \nonumber \\
= 6C\left(\frac{n}{n_0}\right) \left[ 1 - \frac 15 \left(\frac {p_F}{\Lambda}\right)^2 
+ \frac 17 \left(\frac {p_F}{\Lambda}\right)^4 - \cdots \right] \nonumber \\
\ea
The results in Fig. \ref{MDYISk_SPP} are in agreement with the above result in the degenerate limit. 
Corrections from temperature effects would be of ${\cal O}((T/E_F)^2)$. 

An analytical result for the non-degenerate limit can be obtained from Eq. (B12):
\be
R(n,p=0;T) = \frac {C}{g} \left(\frac {n}{n_0}\right)~ \frac {\Lambda^2}{p_{0R}} 
\left[ \frac 1p  ~\ln \frac {\Lambda^2+(p+p_{0R})^2} {\Lambda^2 + (p-p_{0R})^2} \right]_{p\rightarrow 0} \,,
\ee
where $g=2(4)$ for PNM (SNM) and  $p_{0R} = (2mT)^{1/2}$. From a Taylor expansion of the term in brackets, 
\ba
R(n,p=0;T) &=& \frac {C}{g}~ \left(\frac {n}{n_0}\right)~ \frac {\Lambda^2}{p_{0R}} ~\frac {4p_{0R}}{\Lambda^2 + p^2_{0R}} \nonumber \\
&=& \frac {4C}{g}~ \left(\frac {n}{n_0}\right)~ \frac {\Lambda^2}{\Lambda^2 + 2mT} \,.
\ea
The non-degenerate results in Fig. \ref{MDYISk_SPP} are in accord with  expectations from the above result. 

In what follows, we compare the various state variables 
and the specific heats of the MDI(A) and SkO$^\prime$ models. Results of these calculations 
are presented in Figs. \ref{MDYISk_SPP} through \ref{MDYISk_Cp} for pure neutron matter
and symmetric nuclear matter at $T=20$ and 50 MeV, respectively.

\subsection{Results of numerical calculations}

\begin{figure*}[htb]
\centering
\begin{minipage}[b]{0.49\linewidth}
\centering
\includegraphics[width=9.5cm]{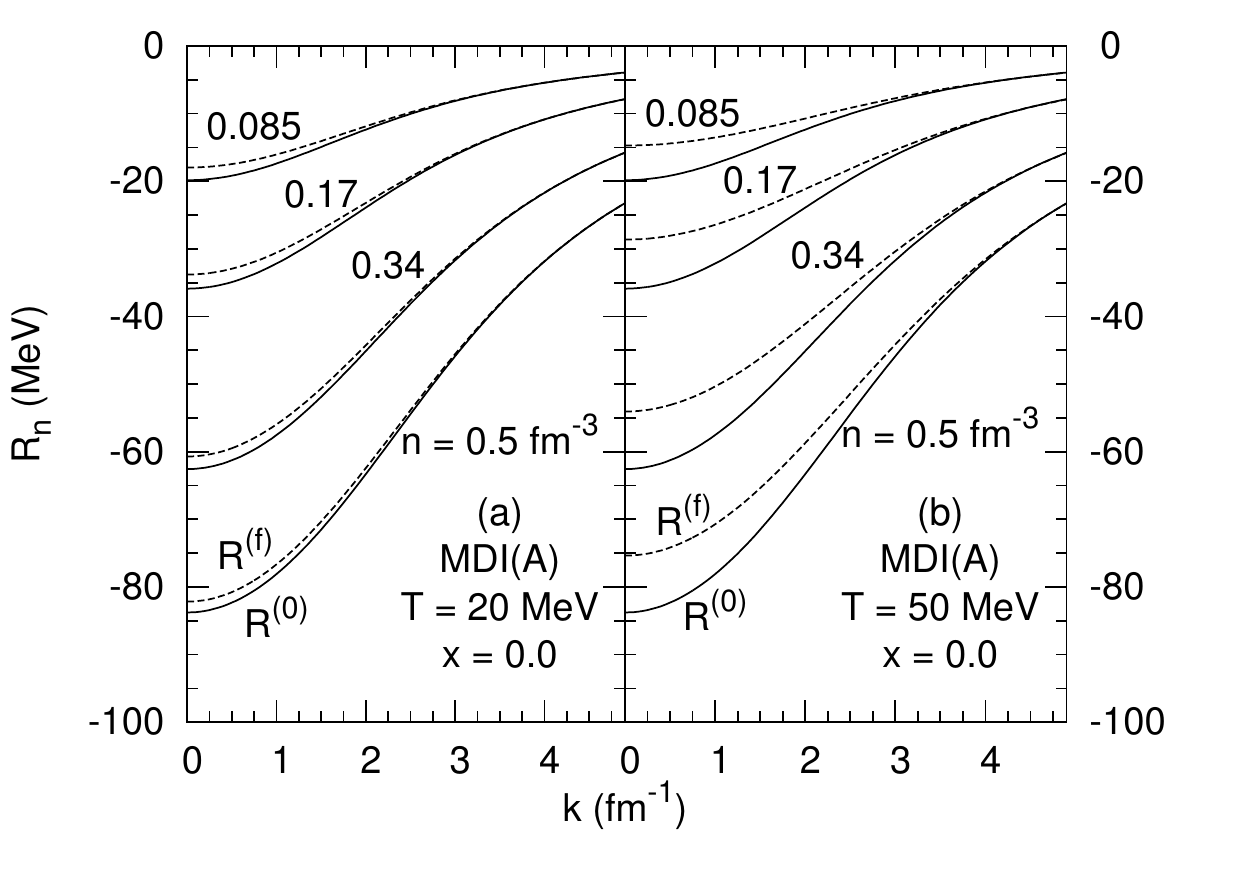}
\end{minipage}
\begin{minipage}[b]{0.49\linewidth}
\centering
\includegraphics[width=9.5cm]{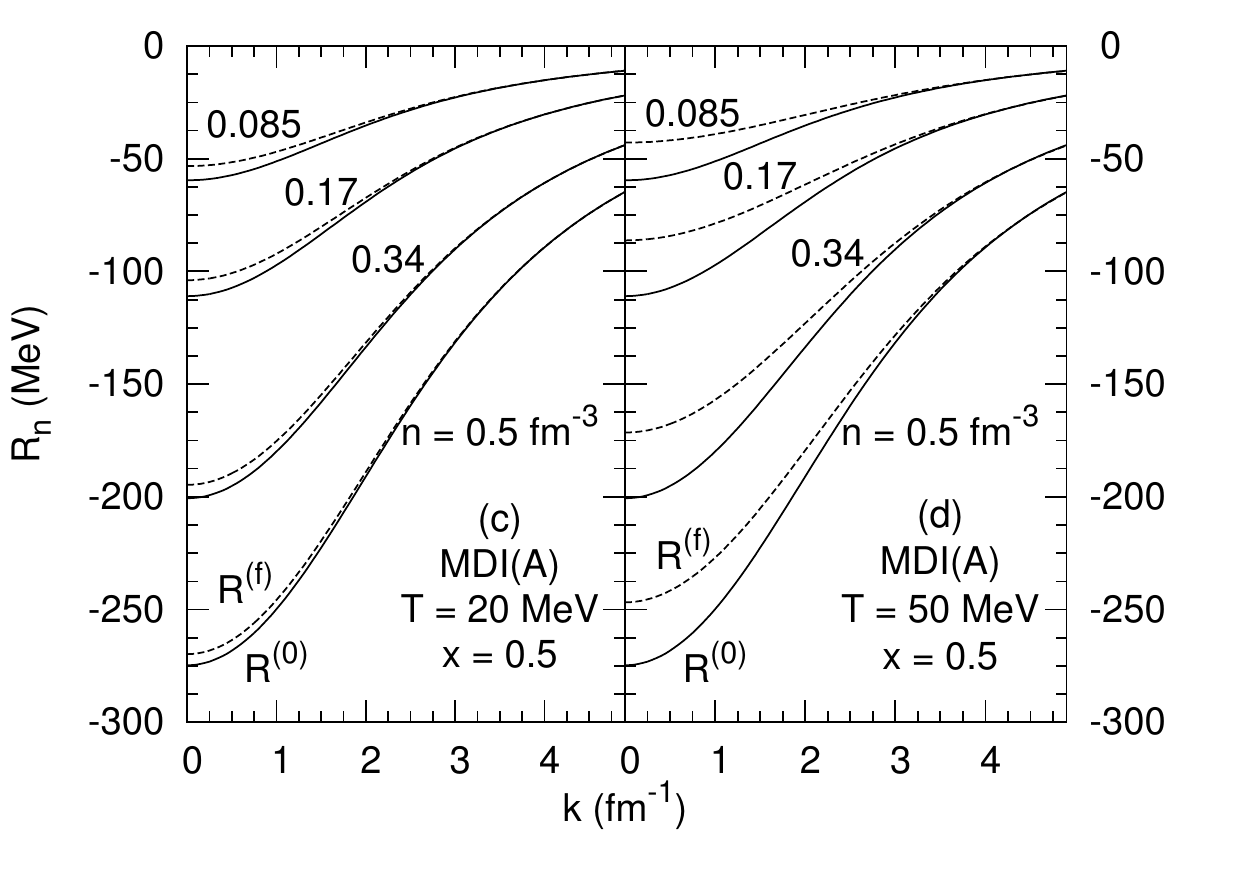}
\end{minipage}
\vskip -0.5cm
\caption{Momentum dependent part of the neutron single particle potential for the baryon 
densities, temperatures, and proton fractions indicated in each panel. The zeroth approximation $R^{(0)}$
is the zero temperature $R_n$ given by the momentum dependent terms in Eq. 
(\ref{Ui}),  whereas the final result $R^{(f)}$ is found via an iterative procedure using Eq. (\ref{Ris}).} 
\label{MDYISk_SPP}
\end{figure*}

For given $n,~x,$ and $T$, the chemical potentials $\mu_n$ and $\mu_p$ for the MDI(A) model can be calculated using the iterative procedure described earlier in connection with Eq.~(\ref{Ris}).   
This entails a self-consistent determination of $R_n(p)$ and $R_p(p)$,  
results for which are shown in Fig. \ref {MDYISk_SPP} for pure neutron matter and isospin symmetric nuclear matter at various densities, and temperatures of $T=20$ and 50 MeV, respectively.  In few iterations, convergence is reached from the starting guess $R^{(0)}$   at $T=0$ to the final result $R^{(f)}$ shown in panels (a) through (d) of this figure. As expected, substantial corrections to the zero-temperature result at low to intermediate momenta are evident as the temperature increases.

The thermal chemical potentials of the neutron  $\mu_{n,th}$ for the two models are  shown in Fig. \ref{MDYISk_Muth}.
Both models predict that the $\mu_{n,th}$'s of symmetric matter and pure neutron matter cross with 
the latter having a larger value at low densities. The density at which the crossing occurs is 
smaller in the case of MDI(A) than that of SkO$^\prime$. Larger temperatures move the 
crossing to larger densities. For pure neutron matter the two models predict nearly the same result
at all densities and temperatures. 
The saturating behavior in both models is a consequence of progressively increasing degeneracy of the fermions and its onset occurs at higher densities for the higher temperature.

\begin{figure}[!htb]
\includegraphics[width=9.2cm]{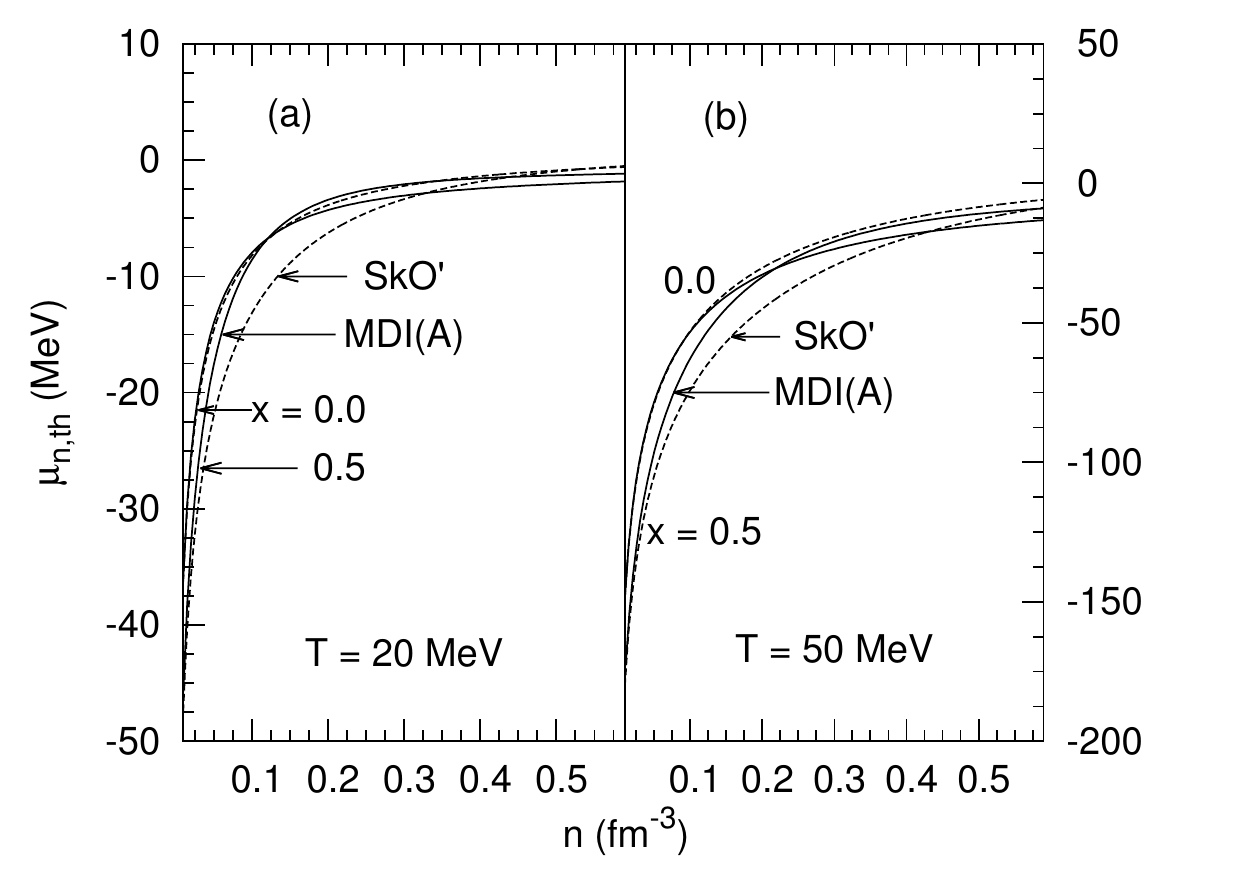}
\caption{Thermal chemical potentials vs $n$ 
at the indicated proton fractions $x$ and temperatures $T$ for the MDI(A) and SkO$^\prime$ models from Eq. (\ref{ni}). }
\label{MDYISk_Muth}
\end{figure}
\begin{figure*}[htb]
\centering
\begin{minipage}[b]{0.49\linewidth}
\centering
\includegraphics[width=9.5cm]{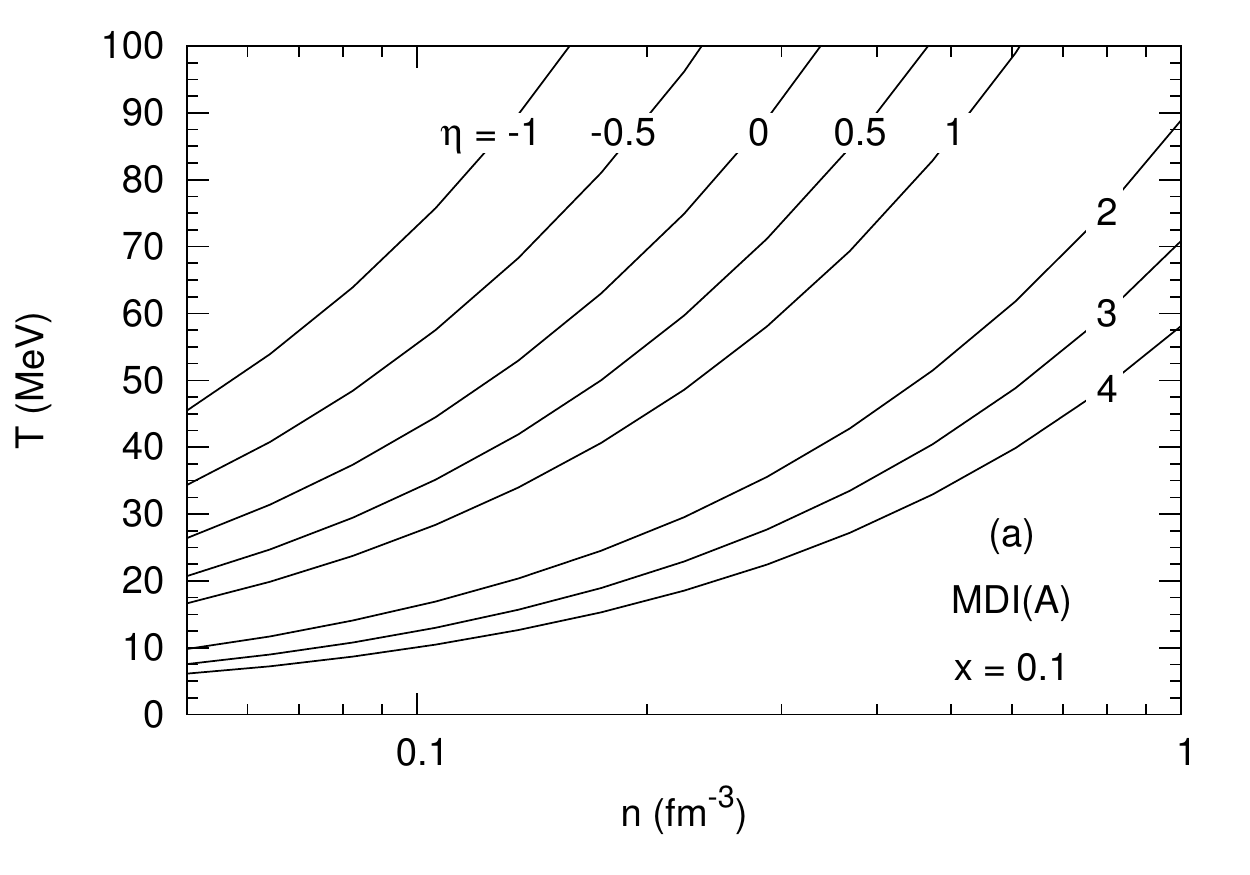}
\end{minipage}
\hspace{0.01cm}
\begin{minipage}[b]{0.49\linewidth}
\centering
\includegraphics[width=9.5cm]{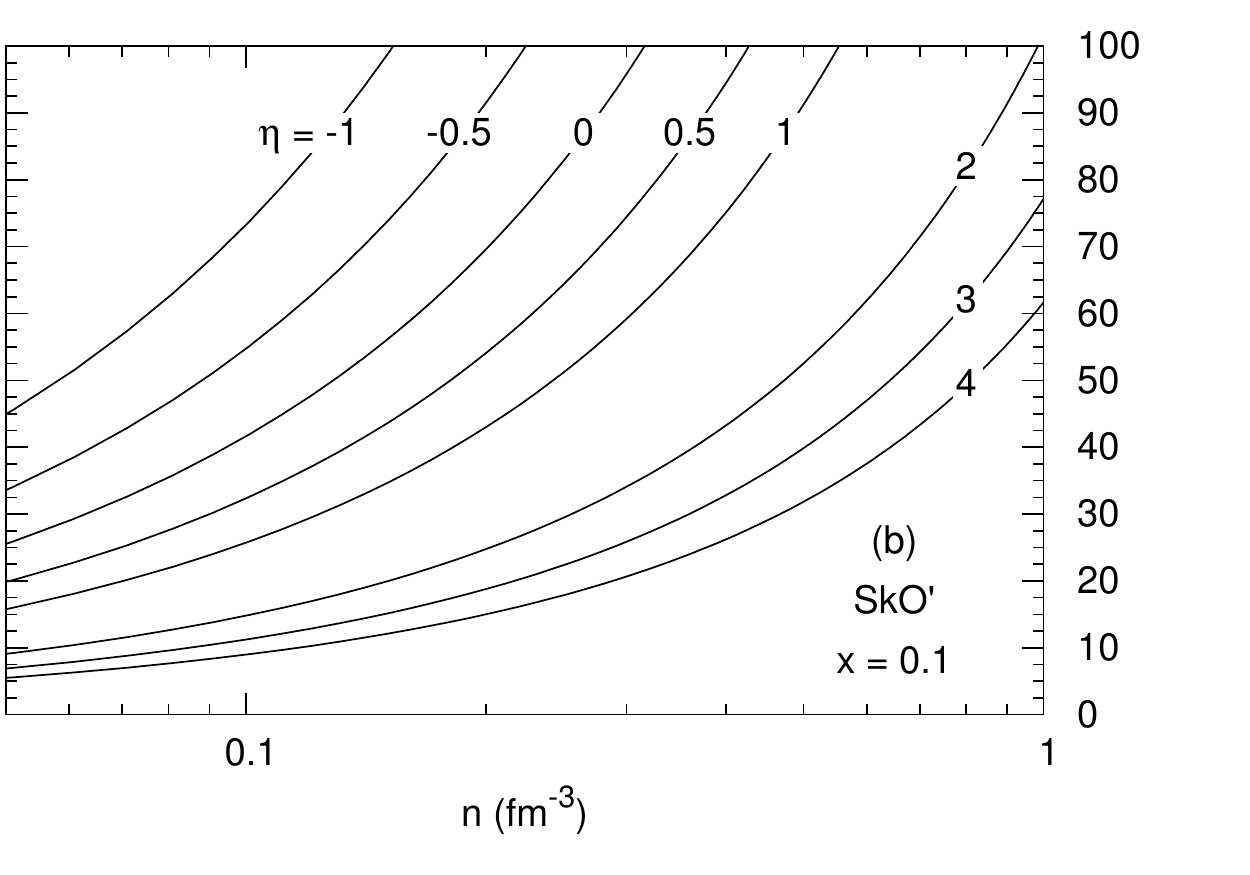}
\end{minipage}
\vskip -0.5cm
\caption{Contours of the degeneracy parameter $\eta$ in the $n$-$T$ plane. Results for the MDI(A) model are from Eq. (\ref{eta1}) and those for the ${\rm SkO^\prime}$ model are from Eq. (\ref{etas}). } 
\label{MDYISk_EtaContours}
\end{figure*}

Contours of the degeneracy parameter $\eta$ are shown in Fig. \ref{MDYISk_EtaContours}. While the qualitative trends are similar for the MDI(A) and SkO$^\prime$ models, quantitative differences exist in the degenerate regime  for $\eta>1$. 
The origin of these differences can be traced to the different behaviors of the effective masses in the two models.

The thermal energy per baryon from the MDI(A) and SkO$^\prime$ models are shown as functions of 
baryon density in Figs. \ref{MDYISk_Eth}(a) and (b).  At $T=20$ MeV, the MDI(A) model has somewhat lower values than those of the SkO$^\prime$ model beyond nuclear densities for symmetric nuclear matter ($x=0.5$). However, results of the two models agree to well beyond the nuclear density for pure neutron matter ($x=0$), significant differences occurring only for densities beyond those shown in the figure. An opposite trend is observed at $T=50$ MeV for which the two models differ slightly around nuclear densities  for $x=0.5$, whereas they yield similar results for $x=0$ at subnuclear densities. We attribute these behaviors to the significantly different behaviors of the effective masses (both their magnitudes and density dependences) in these two models (see Fig. \ref{MDYISk_Ms}) as our analysis in the subsequent section, where analytical results in the limiting cases of degenerate and non degenerate matter are compared with the exact results, shows.  
In order to further appreciate the extent to which thermal energies differ quantitatively between different models, it is instructive to compare these results with those in Fig. 10 of Ref. \cite{APRppr} where predictions for the APR and the Skyrme-Ska models 
were recently reported.

\begin{figure}[!htb]
\includegraphics[width=9.2cm]{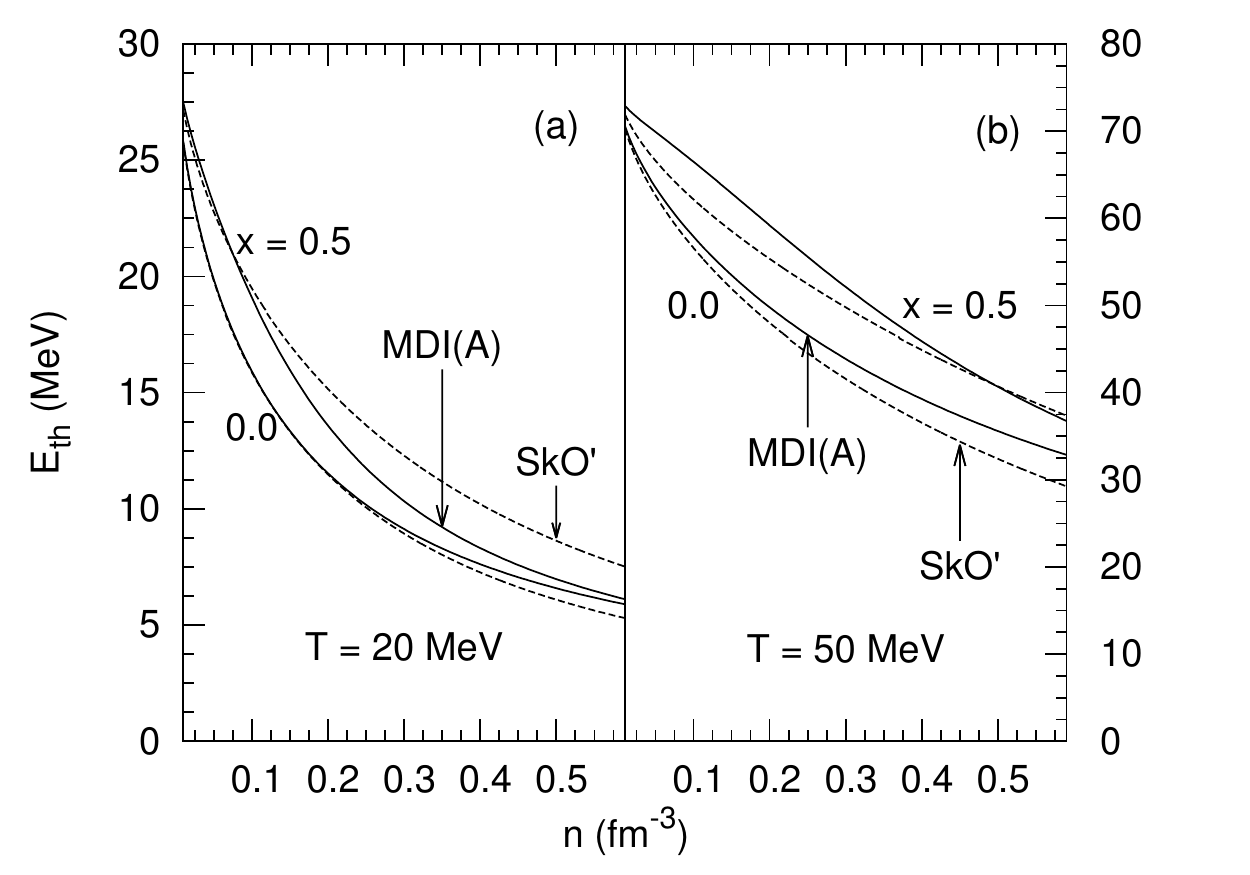}
\caption{Thermal energy per particle vs $n$ at the indicated proton fractions $x$ and temperatures $T$ for the 
MDI(A) and SkO$^\prime$ models from Eqs. (\ref{hmdyi}) and (\ref{hsko}). }
\label{MDYISk_Eth}
\end{figure}

The thermal free energy per baryon is shown in Fig. \ref{MDYISk_Fth} as a function of baryon density
for the MDI(A) and SkO$^\prime$ models. Results for the two models  are indistinguishable at low densities 
$\left(n<0.01~ {\rm fm}^{-3}\right)$. This low density agreement between the two models improves with 
increasing temperature and with lower proton fractions. 
For  $n>0.01~ {\rm fm}^{-3}$, quantitative differences between the two models are due to the different trends with density of nucleon effective masses. These differences become increasingly small for all proton fractions as the limit of extreme degeneracy is approached at very high densities.
\begin{figure}[!htb]
\includegraphics[width=9.2cm]{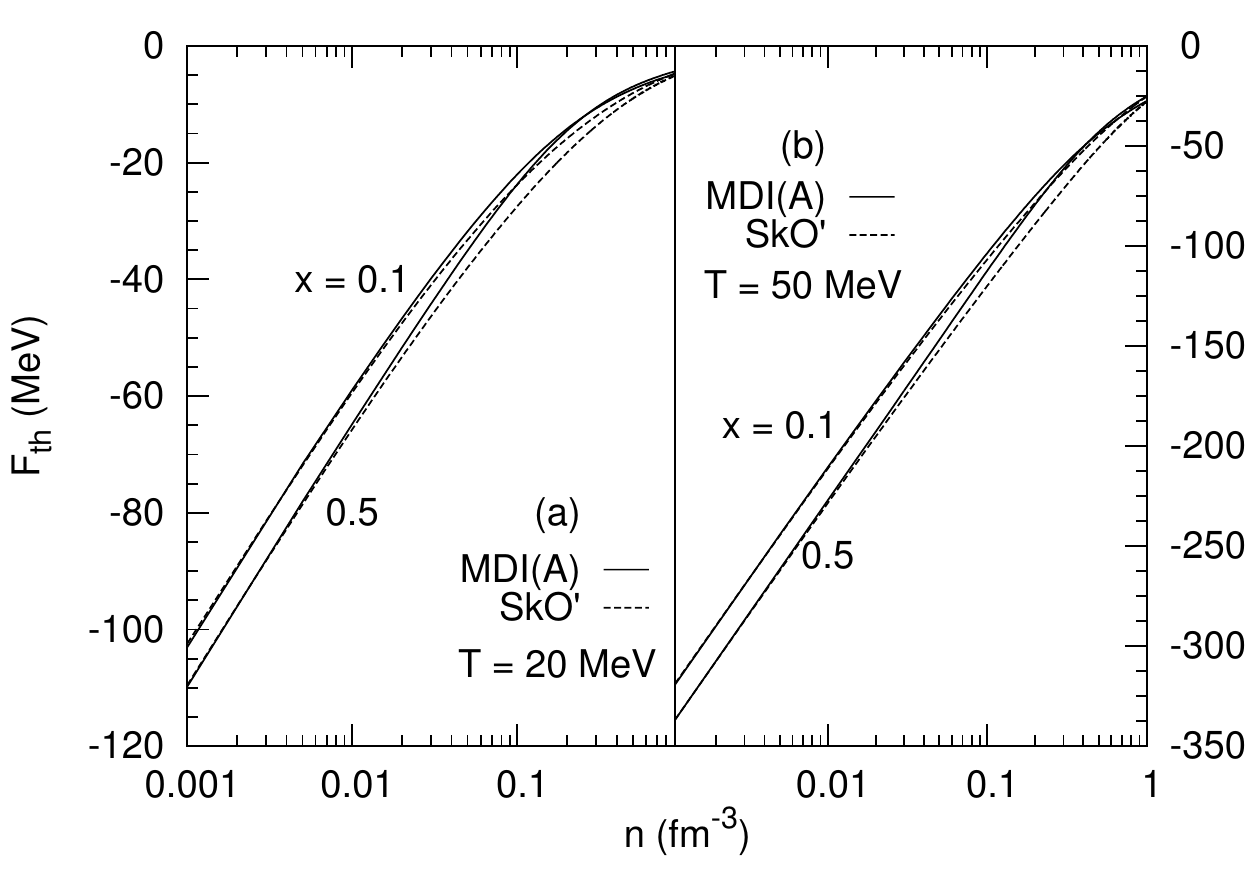}
\caption{Thermal free energy for the MDI(A) and 
SkO$^\prime$ models at the indicated temperatures and proton fractions. Results are from 
Eqs. (\ref{hmdyi}), (\ref{hsko}) and (\ref{sden}).}
\label{MDYISk_Fth}
\end{figure}

In Fig. \ref{MDYISk_Pth}, we present the thermal pressures vs density.  For both temperatures and proton fractions shown, the two models display similar traits in that at around nuclear and subnuclear densities they predict similar values, but begin to differ substantially at supra-nuclear densities.  With increasing density, the thermal pressure of the MDI(A) model is smaller than that of the SkO$^\prime$ model chiefly due to its smaller effective mass and relatively flat variation with density.  

\begin{figure}[!htb]
\includegraphics[width=9.2cm]{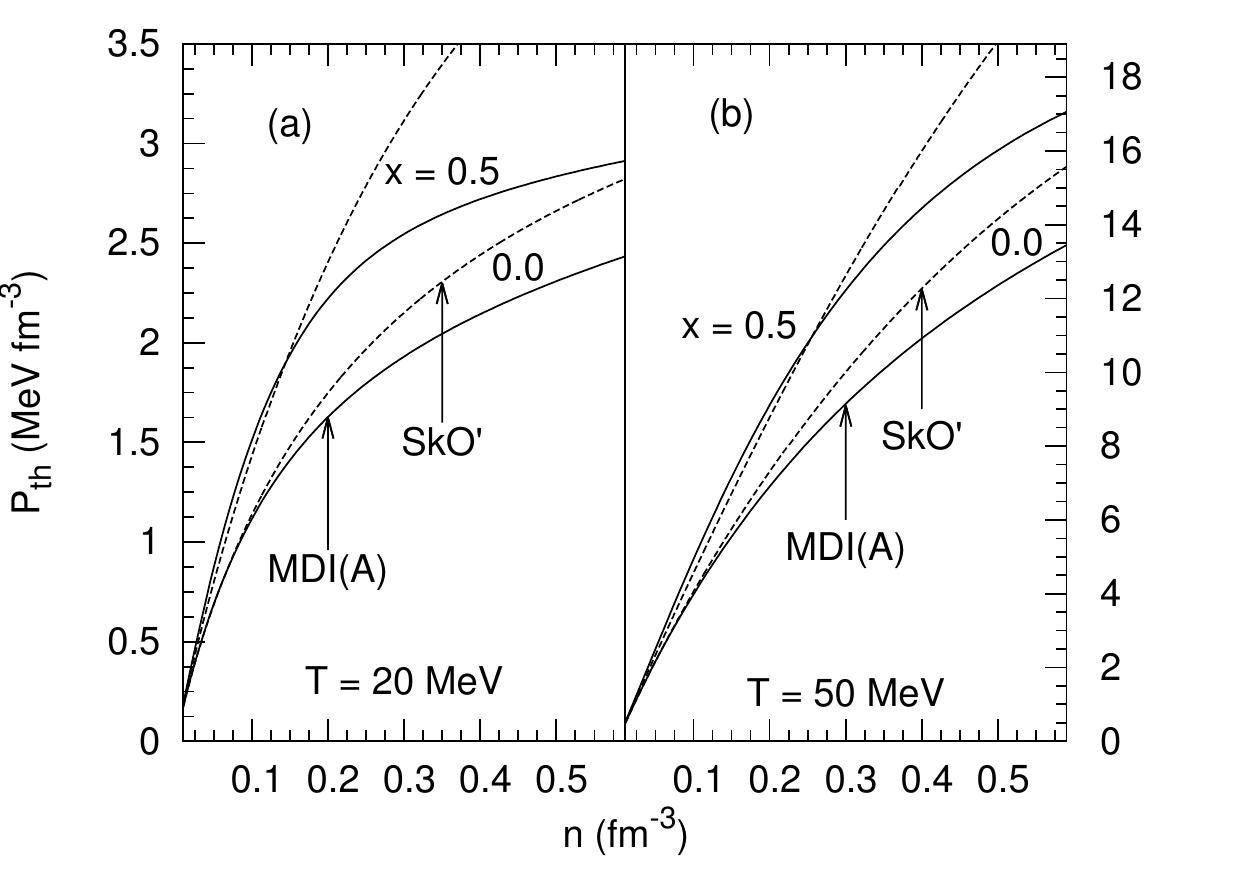}
\caption{Thermal pressure vs $n$ at the indicated proton fractions $x$ and temperatures $T$ for the MDI(A) 
and SkO$^\prime$ models from Eq. (\ref{pres}). }
\label{MDYISk_Pth}
\end{figure}

Figure \ref{MDYISk_S} shows the entropy per baryon for the two models. 
The two models agree at low densities with the best agreement occurring
for pure neutron matter up to about twice the nuclear density.  At larger densities the MDI(A) model predicts that the entropy of symmetric matter converges to that of pure neutron matter.  This feature is also present in SkO$^\prime$ but  occurs at larger 
densities than shown here.

\begin{figure}[!htb]
\includegraphics[width=9.2cm]{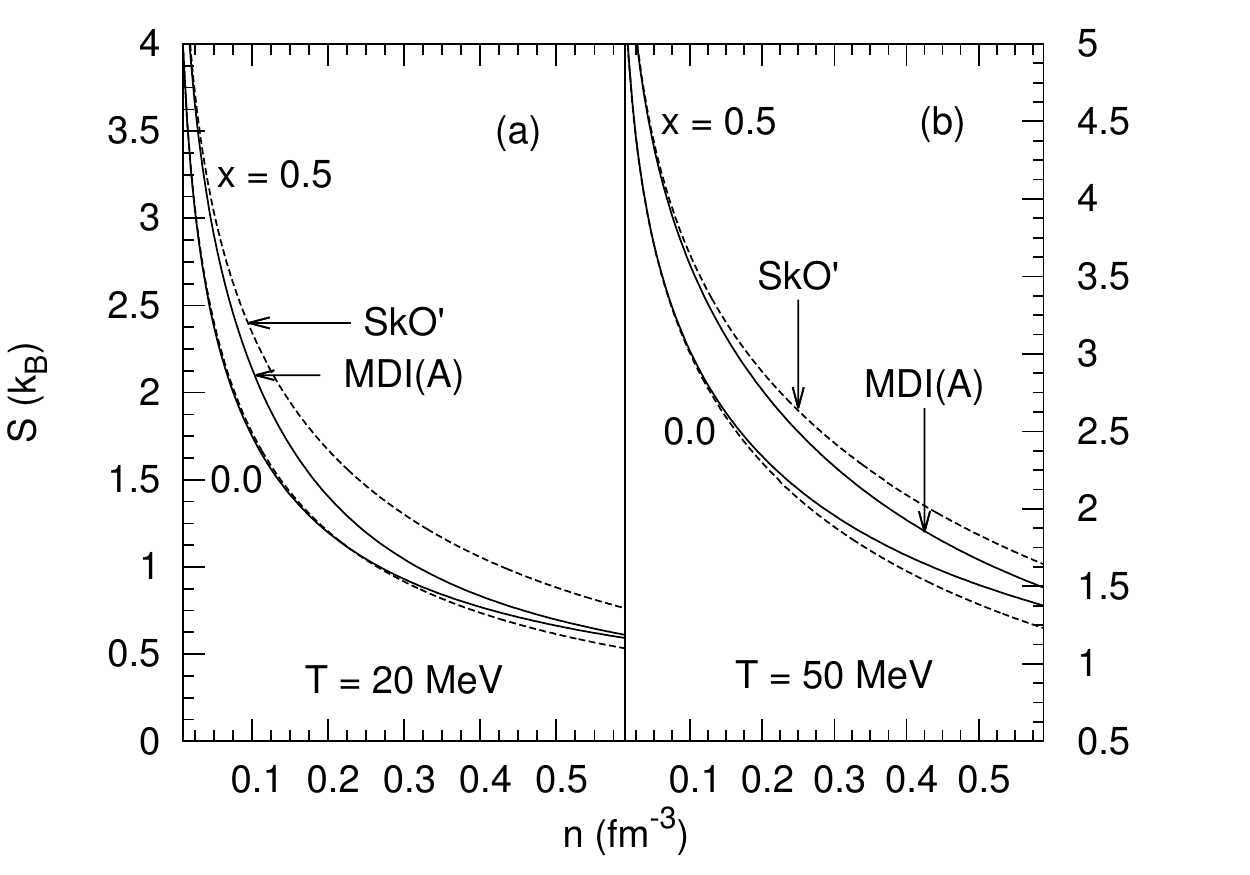}
\caption{Entropy per particle
vs $n$ at the indicated proton fractions $x$ and temperatures $T$ for the MDI(A) and SkO$^\prime$ models from Eq. (\ref{sden}). }
\label{MDYISk_S}
\end{figure}

Isentropic contours in the $n$-$T$ plane are shown in Fig. \ref{MDYISk_EntContours} for a
proton fraction of 0.1 and entropies in the range 0.25-3. Both models show similar trends in that all contours rise quickly until around $n_0/2$, beyond which only a moderate increase in the temperature is observed.
For each entropy contour, the temperature  is systematically larger
for the SkO$^\prime$ model when compared with that of the MDI(A) model. For densities larger than $2n_0$ and
for values of entropy exceeding 1.5, the temperatures predicted by both models are well in 
excess of 50 MeV.

\begin{figure*}[htb]
\centering
\begin{minipage}[b]{0.49\linewidth}
\centering
\includegraphics[width=9.5cm]{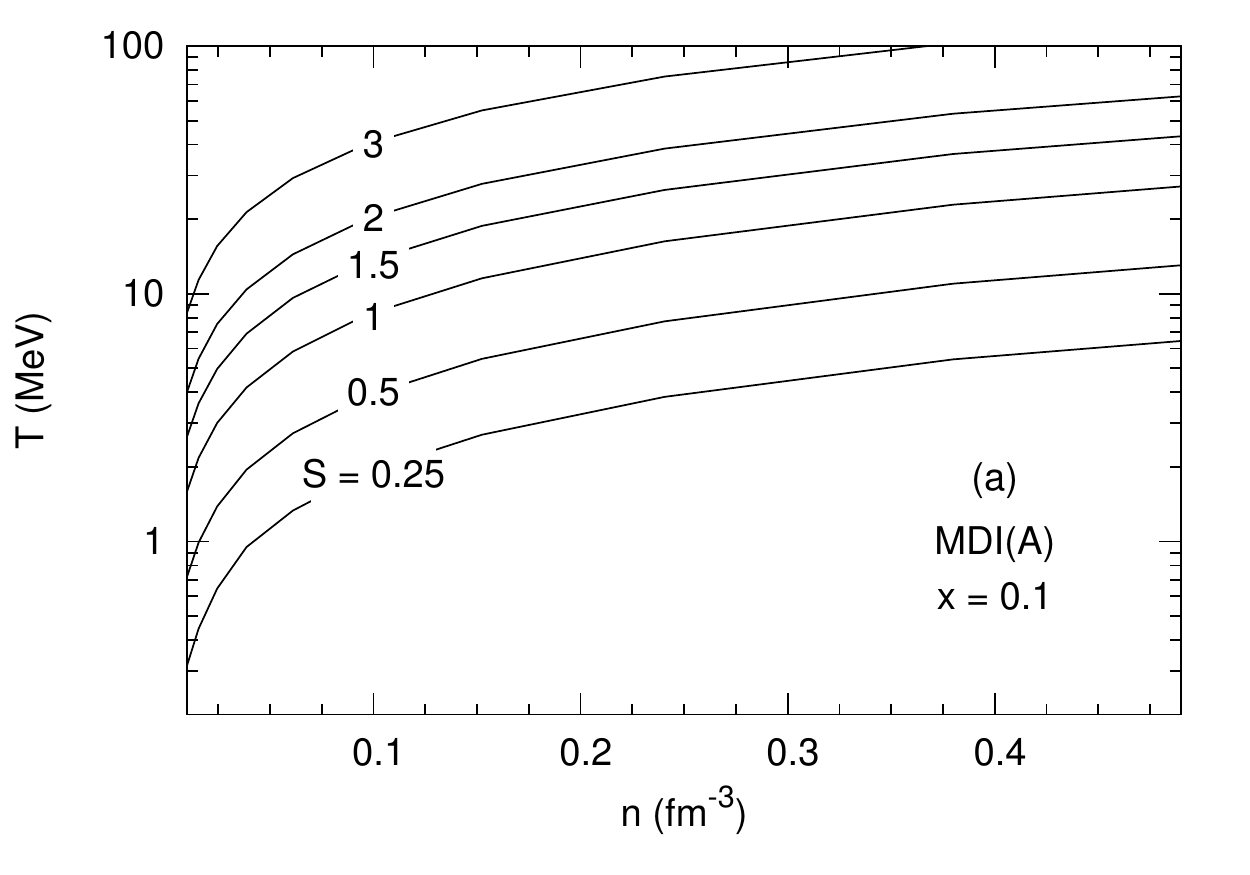}
\end{minipage}
\hspace{0.01cm}
\begin{minipage}[b]{0.49\linewidth}
\centering
\includegraphics[width=9.5cm]{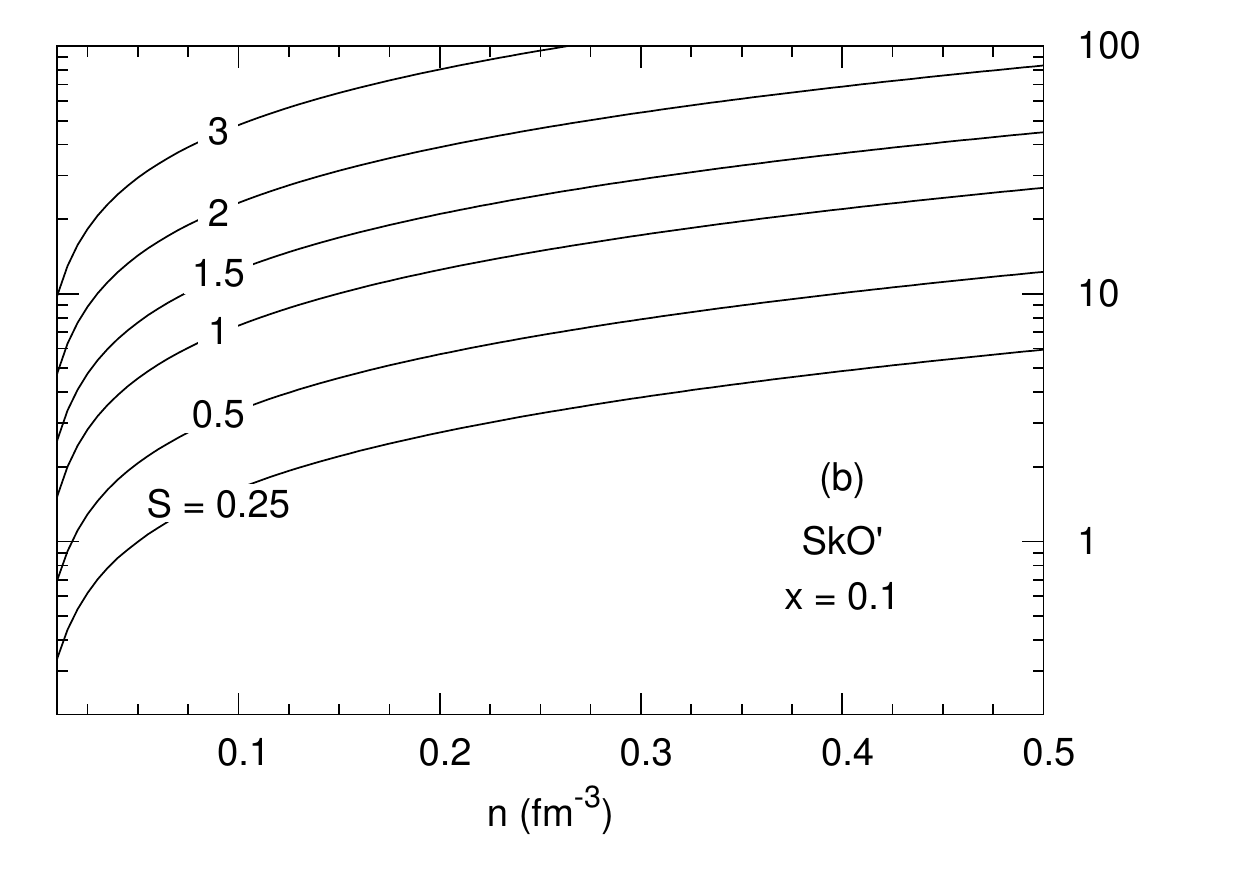}
\end{minipage}
\vskip -0.5cm
\caption{Isentropes in the $n$-$T$ plane for the MDI(A) model in (a)  and those for the  SkO$^\prime$ model in (b).} 
\label{MDYISk_EntContours}
\end{figure*}

The specific heat at constant volume,  $C_V$,  is plotted as a function of baryon 
density in Fig. \ref{MDYISk_Cv} for the two models. 
Noteworthy features at both temperatures are the  peaks occurring at values in excess of 1.5 (the maximum value characteristic of 
free fermi gases at vanishing density, which is also the case for Skyrme models) at finite densities in the MDI(A) model. 
These peaks can be attributed to 
the momentum dependence built into the interaction which produces a
temperature-dependent spectrum via $R(n,p)$. 
This trait, shared with relativistic mean field models (although
there, the $T$-dependence in the spectrum enters through the Dirac
effective mass) \cite{Constantinos:13}, has implications related to the hydrodynamic
evolution of a core-collapse supernova in that $C_V$ controls the density at which
the core rebounds.  
$C_V$ decreases with increasing density and the magnitude of the decrease is larger at the lower 
temperature. As was the case for the entropy per baryon, the $C_V$'s of nuclear and neutron matter
approach each other at large densities. 

\begin{figure}[!htb]
\includegraphics[width=9.2cm]{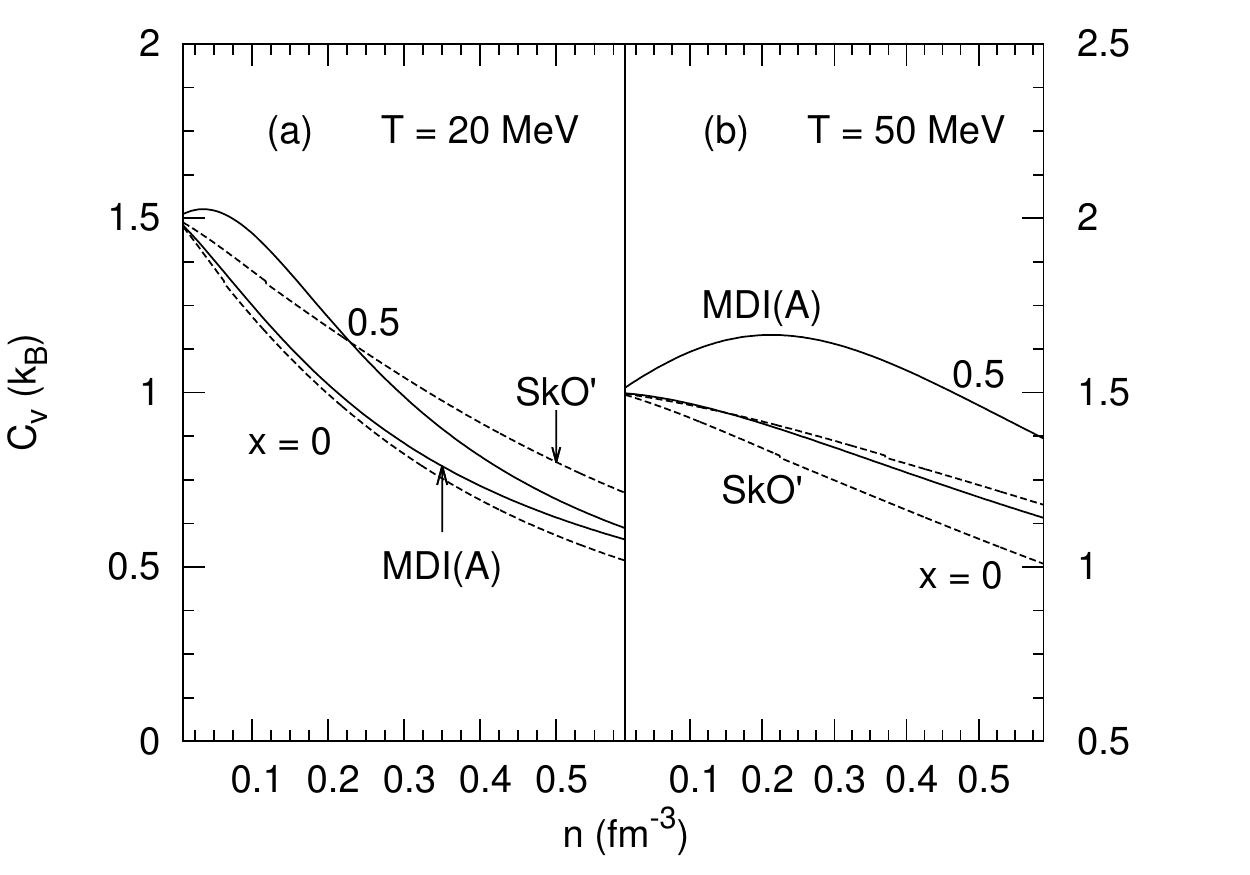}
\caption{[(a) and (b)] Specific heat at constant volume $C_V$ vs baryon density from Eq. (\ref{cv}) for 
the MDI(A) and SkO$^\prime$ models.}
\label{MDYISk_Cv}
\end{figure}

The specific heat at constant pressure, $C_P$, is shown in Fig. \ref{MDYISk_Cp} 
as a function of baryon density. The predominant feature for both models is the sharp
rise in $C_P$ for symmetric nuclear matter at $T = 20$ MeV. This feature arises from the temperature being close to that for  the onset of the liquid-gas phase
transition for which ${dP}/{dn}=0$. At high densities $C_P$ resembles the behaviors seen for 
$C_V$ and the entropy per baryon. 

Figure \ref{MDYIuN2SkOp_Rat_CpCv} shows the ratio of specific heats $C_P/C_V$ vs $n$ for $T=20$ and 50 MeV, respectively, for the two models. In (a), the large variations seen at sub-saturation densities for values of $x$ not too close to that of pure neutron matter are due to the proximity of an incipient liquid-gas phase transition.    
As will be discussed in the next section, 
$C_P/C_V$ is closely related to the adiabatic index, $\Gamma_S$, which provides a measure of the stiffness of the equation of state. In addition, it also determines the speed of 
adiabatic sound wave propagation in hydrodynamic evolution of matter.

\begin{figure}[!htb]
\includegraphics[width=9.2cm]{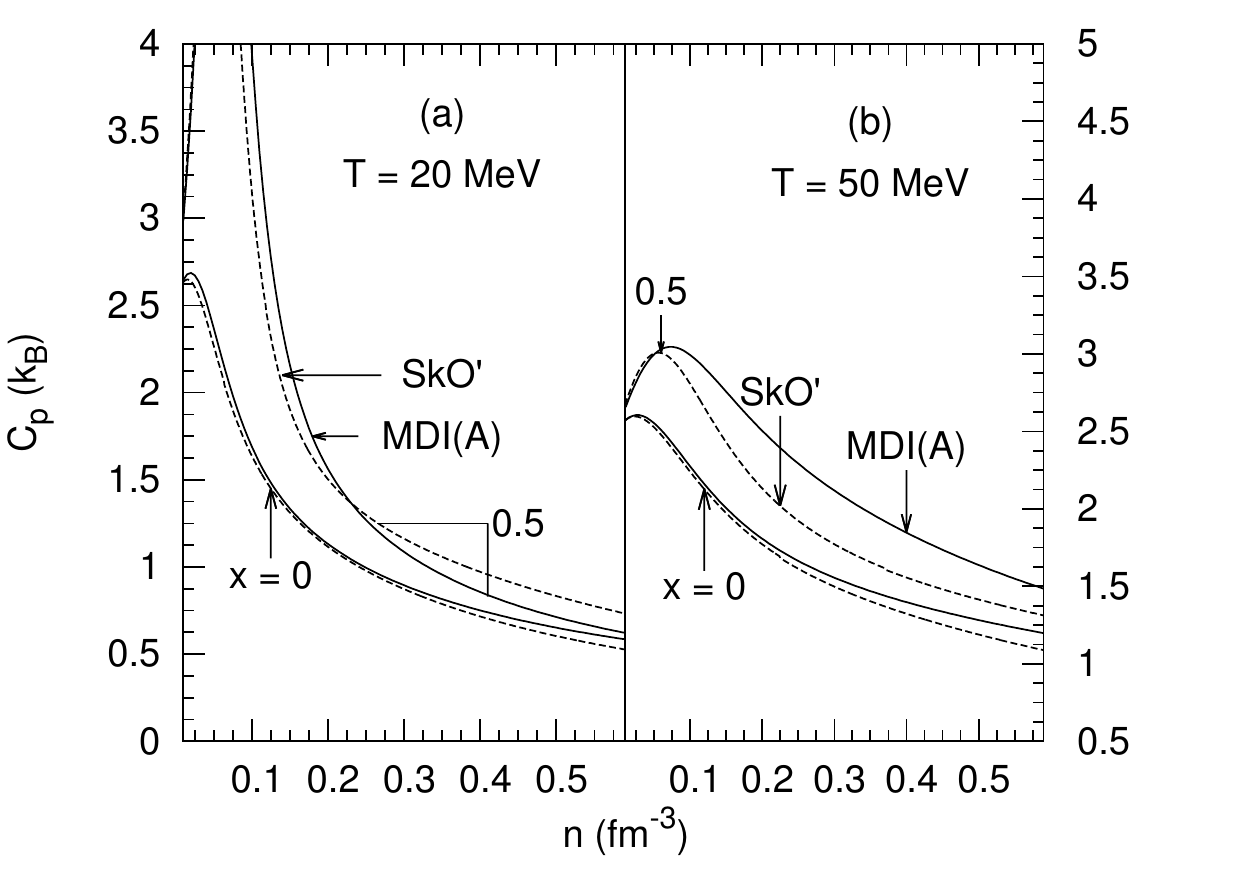}
\caption{[(a) and (b)] Specific heat at constant pressure $C_P$ vs baryon density from Eq. (\ref{cp}).}
\label{MDYISk_Cp}
\end{figure}

\begin{figure}[!h]
\centering
\makebox[0pt][c]{
\hspace{0cm}
\begin{minipage}[b]{\linewidth}
\centering
\includegraphics[width=8.5cm]{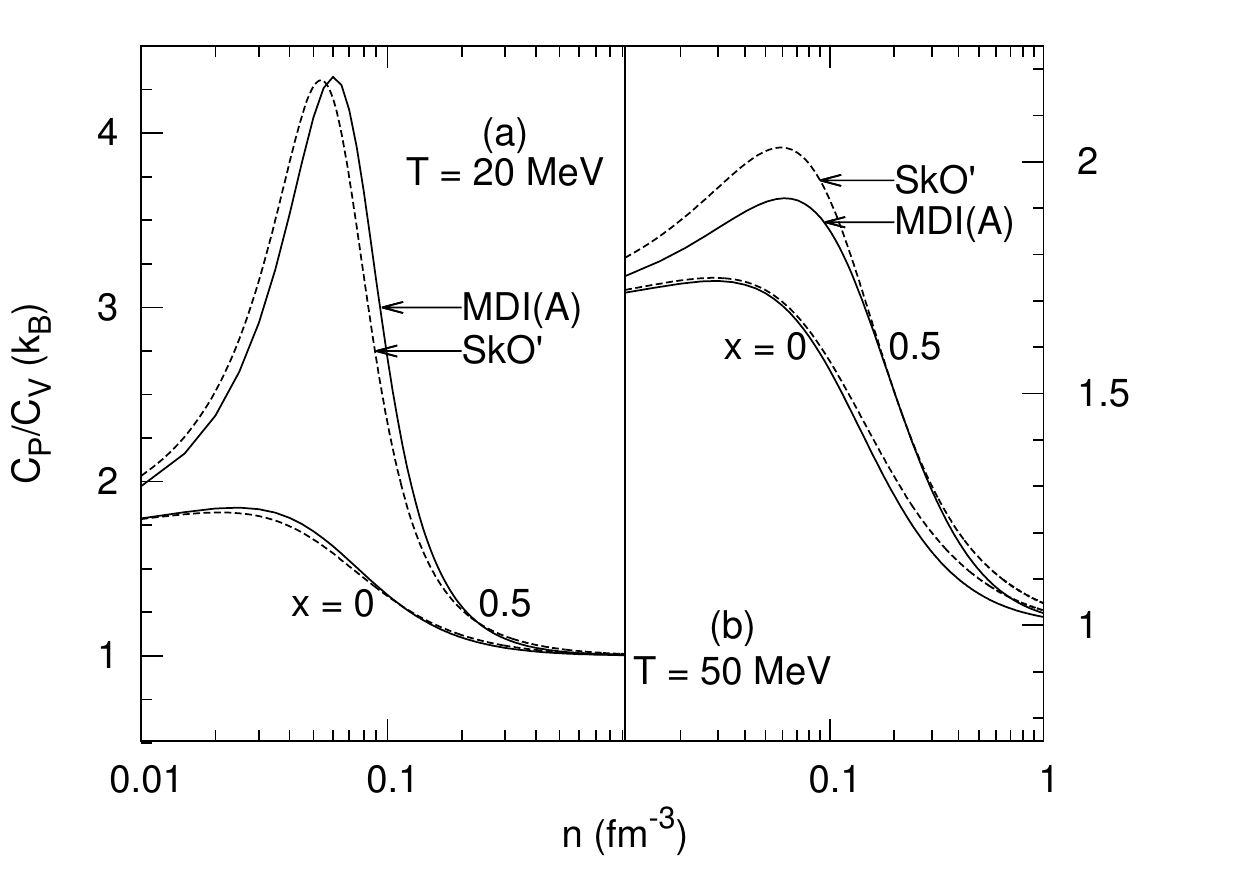}
\end{minipage}
}
\vskip -0.5cm
\caption{Ratio of specific heats at the indicated proton fractions and temperatures.} 
\label{MDYIuN2SkOp_Rat_CpCv}
\end{figure}

\subsection{Analytical results  in limiting cases}
In the cases when degenerate (low $T$, high $n$ such that $T/E_{F_i} \ll 1$) or non-degenerate (high $T$, low $n$ such that $T/E_{F_i} \gg 1$) conditions are met, generally compact analytical 
expressions can be derived. As the densities and temperatures  encountered in the thermal evolution of supernovae, neutron stars, and binary mergers vary over wide ranges, matter could 
be in the degenerate, partially degenerate, or non-degenerate limits depending on the ambient conditions.  A comparison of the exact, but numerical, results with their analytical 
counterparts not only allows for a check of the often involved numerical calculations, but is also helpful in identifying the density and temperature ranges in which matter is one or the 
other limiting case. Because of the varying concentrations of neutrons, 
protons, and leptons, one or the other species may lie in different regimes of degeneracy.

\subsubsection{Degenerate limit}

In this case, we can use Landau's Fermi Liquid Theory (FLT) \cite{ll9,flt} to advantage. The ensuing analytical  
expressions highlight the importance of the effective mass in thermal effects. The explicit forms of the thermal 
energy, thermal pressure, thermal chemical potentials and entropy density in FLT are as follows:
\ba
E_{th} &=& \frac{T^2}{n}\sum_i a_in_i  \label{edeg}\\
P_{th} &=& \frac{2T^2}{3}\sum_i a_in_i\left(1-\frac{3}{2}\frac{n}{m_i^*}\frac{\partial m_i^*}{\partial n}\right) \label{pdeg}\\
\mu_{i,th} &=& -T^2\left(\frac{a_i}{3}+\sum_j \frac{n_ja_j}{m_j^*}\frac{\partial m_j^*}{\partial n_i}\right) \label{mudeg}\\
s &=& 2T\sum_i a_in_i \label{sdeg}
\ea
where $a_i =\frac{\pi^2}{2}\frac{m_i^*}{p_{Fi}^2}$ is the level density parameter. 
In this limit, to lowest order in temperature, $C_V = C_P = s/n$. 
The above relations are quite general in character requiring only the concentrations and effective masses 
(which, in turn, depend on the single particle spectra) of the various constituents in matter. 

\subsubsection{Nondegenerate limit}

Non-degenerate conditions prevail when the fugacities $z_i=e^{\mu_i/T}$ are small. Methods to calculate the state variables in this limit for Skyrme-like models have been amply discussed in the literature (see, {\it e.g.}, Ref. \cite{APRppr} for a recent compilation of the relevant formulas) and will not be repeated here. For the MDI models in which the single particle spectrum receives significant contributions from momentum-dependent interactions, the analysis is somewhat involved. We have developed a method involving next-to-leading order steepest descent calculations that provides an adequate description of the various state variables (see  Appendix \ref{Sec:AppendixB} for details).  The numerical results presented below for the non-degenerate limit are obtained employing the relations in  Appendices \ref{Sec:AppendixB} and  \ref{Sec:AppendixC}.       

\subsection{Numerical vs analytical results}
In this section, the exact numerical results of Sec. \ref{Sec:Teffects} are compared with those using the analytical  
formulas in the degenerate and nondegenerate limits described in the previous section. 
Throughout, results from the MDI(A) model 
are displayed in panels  (a) and (b), whereas panels (c) and (d) contain results
from the SkO$^\prime$ model in Figs. (\ref{MDYISk_Eth_lim})-(\ref{MDYISk_Cp_lim}) below.

Figure \ref{MDYISk_MuNth_lim} contains plots of the exact thermal chemical potential of the neutron
and, its degenenerate and nondegenerate limits. The agreement between the nondegenerate
limit and the exact result is significantly better for SkO$^\prime$ compared with MDI(A). This
is best seen in the $T=50$ MeV results for pure neutron matter for which the nondegenerate limit coincides with
the exact result until about $2n_0$ for SkO$^\prime$ compared with MDI(A) which agrees only to about $n_0$. 
Both models predict convergence between the degenerate and exact results beginning at around $2n_0$ for $T = 20$ MeV and around $4n_0$ for $T = 50$ MeV.

\begin{figure*}[!htb]
\centering
\begin{minipage}[b]{0.49\linewidth}
\centering
\includegraphics[width=9.5cm]{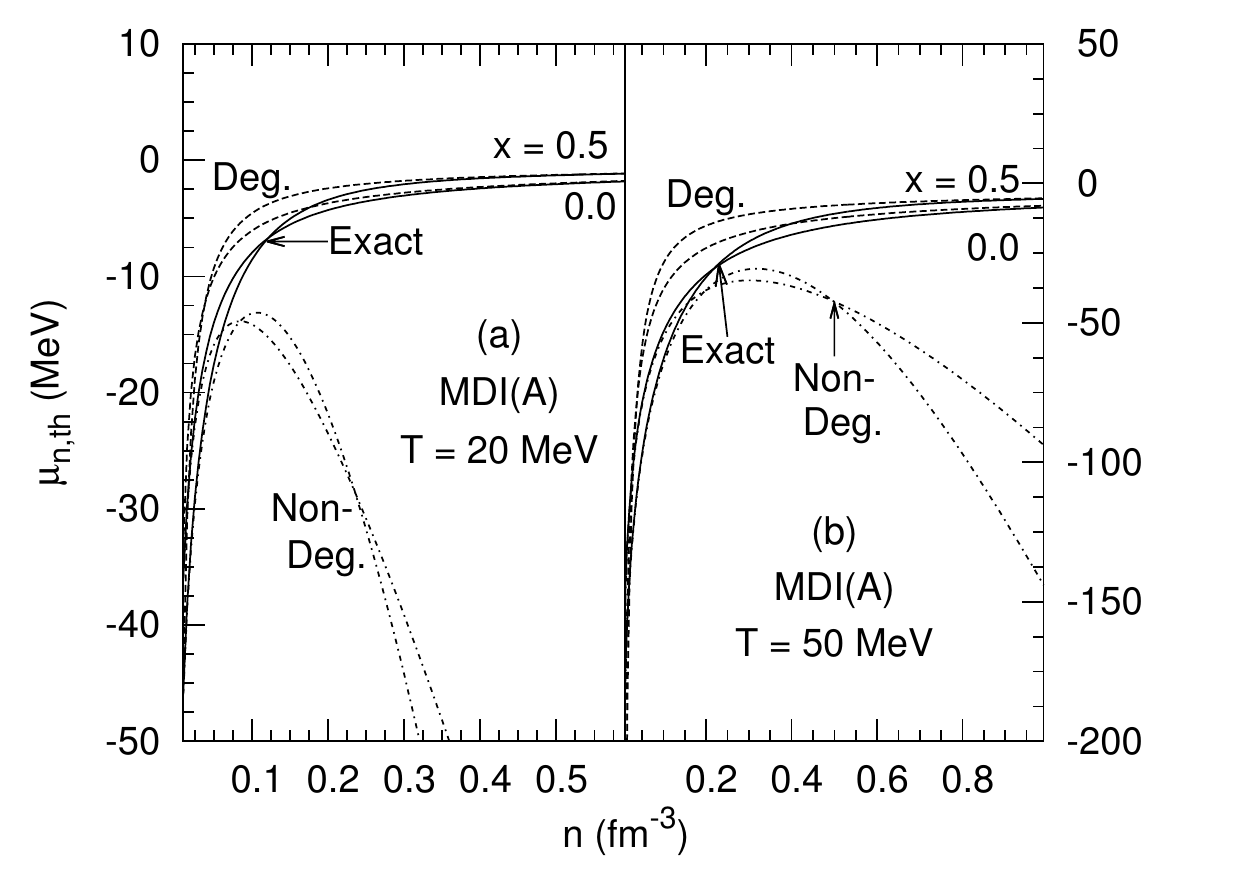}
\end{minipage}
\begin{minipage}[b]{0.49\linewidth}
\centering
\includegraphics[width=9.5cm]{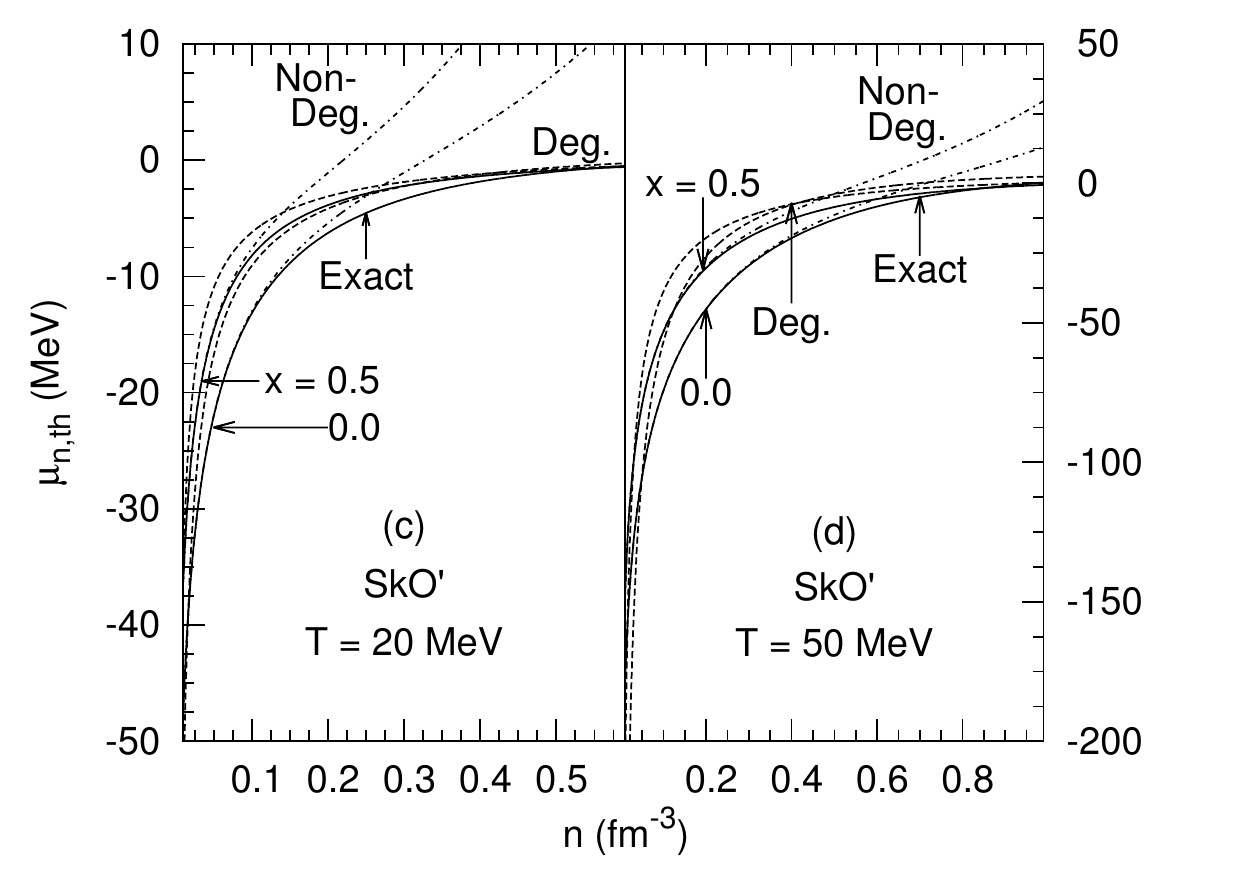}
\end{minipage}
\vskip -0.5cm
\caption{Neutron thermal chemical potentials [Eq. (\ref{ni})] vs $n$ for the MDI(A) model in (a) and (b) and the 
SkO$^\prime$ model in (c) and (d) compared with their limiting cases from Eqs. (\ref{mudeg}) and (\ref{MDYI_Mu_ND}).} 
\label{MDYISk_MuNth_lim}
\end{figure*}
\begin{figure*}[!htb]
\centering
\begin{minipage}[b]{0.49\linewidth}
\centering
\includegraphics[width=9.5cm]{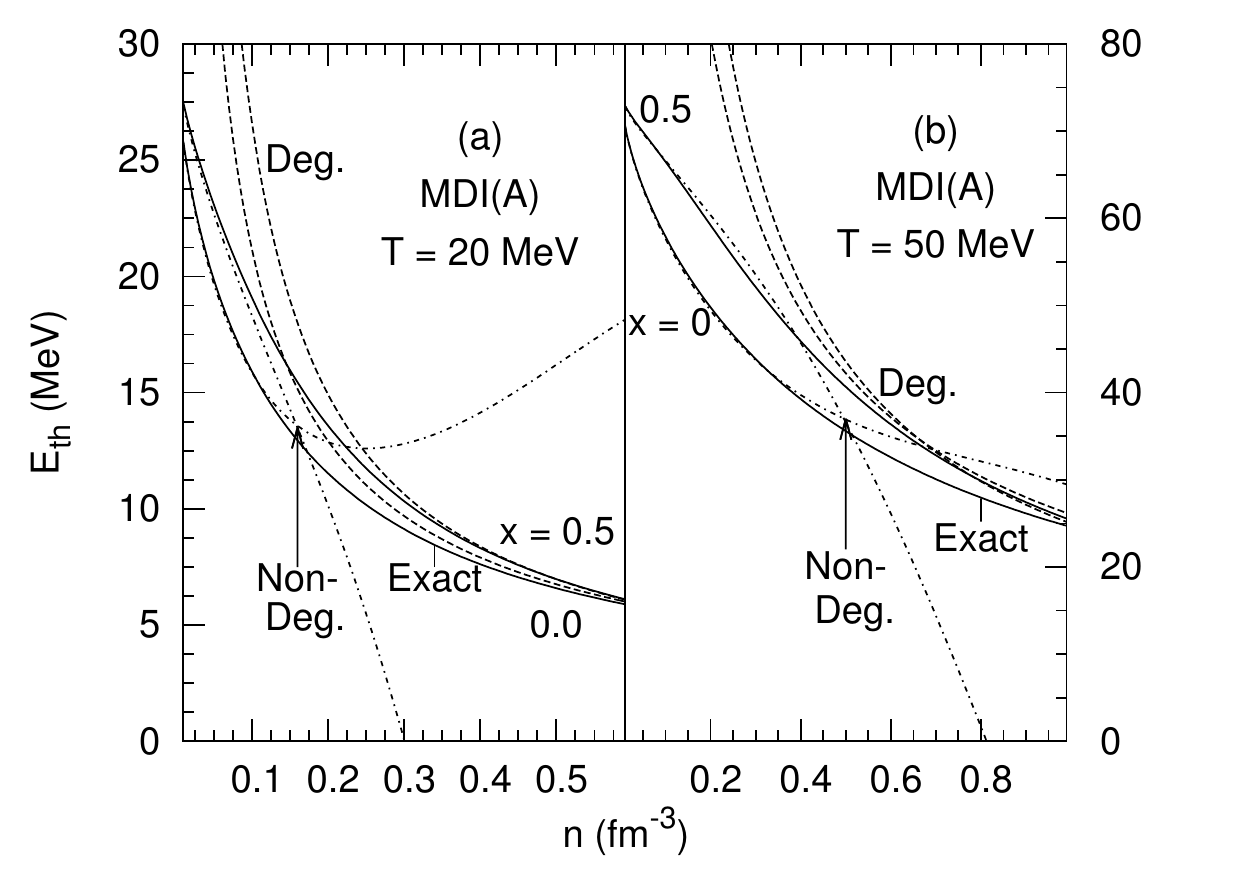}
\end{minipage}
\begin{minipage}[b]{0.49\linewidth}
\centering
\includegraphics[width=9.5cm]{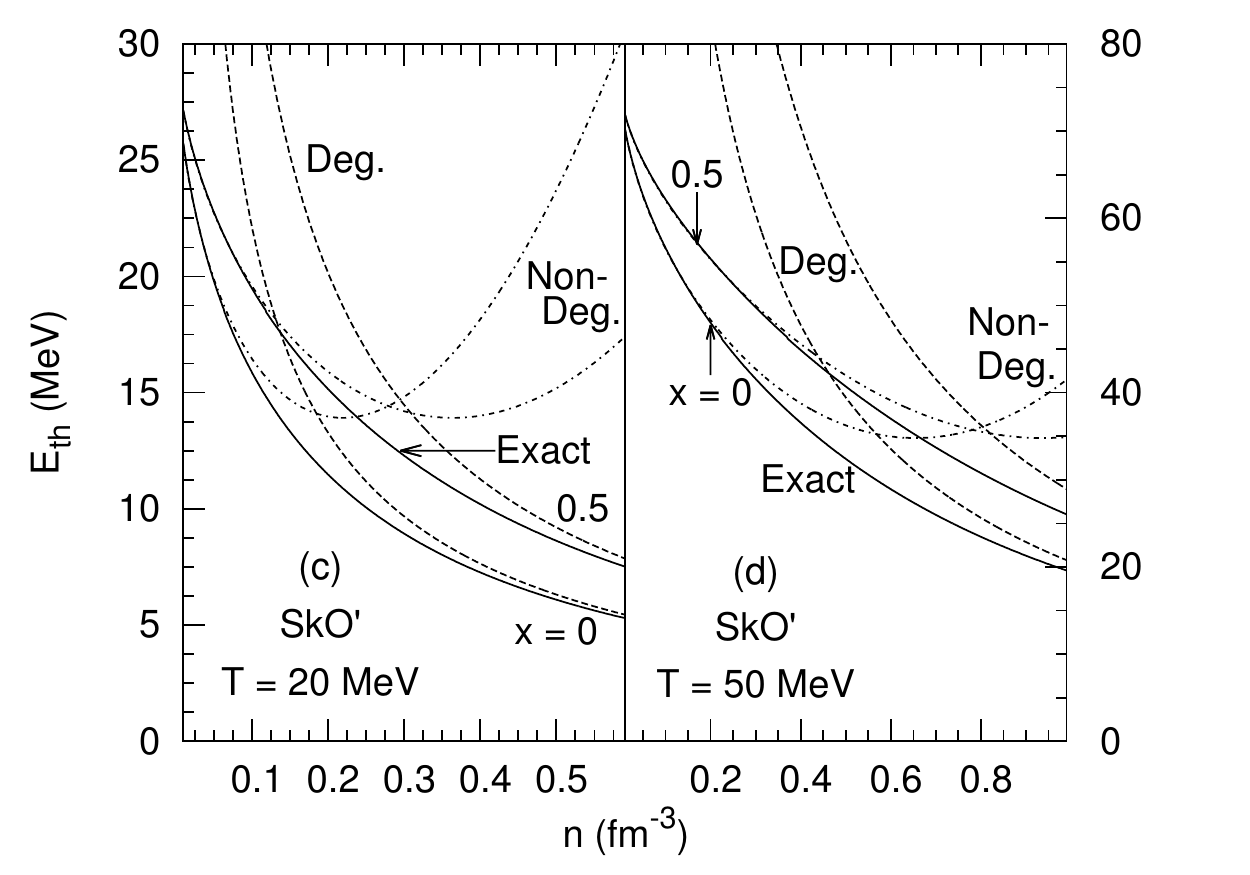}
\end{minipage}
\vskip -0.5cm
\caption{Thermal energy per particle [Eq. (\ref{hmdyi})] and limiting cases [Eqs. (\ref{edeg}) and (\ref{MDYI_E_ND})] vs $n$ for 
the temperatures and proton fractions shown. } 
\label{MDYISk_Eth_lim}
\end{figure*}

The exact thermal energy  and its limits from the two models are shown in 
Fig. \ref{MDYISk_Eth_lim}. The MDI(A) model has a thermal energy that agrees with 
its non-degenerate limit for similar densities compared to that for the SkO$^\prime$
model. For both models the agreement between the exact result and the non-degenerate 
limit is better at high temperatures and for pure neutron matter. Agreement between the degenerate 
limit and the exact solution occurs sooner (lower density) for MDI(A) than for SkO$^\prime$. In both cases,  
the best agreement is for lower temperatures and for symmetric nuclear matter. Note, however, that around nuclear
densities, matter is in the semi-degenerate limit as was the case for APR and Ska models in \cite{APRppr}.

In Fig. \ref{MDYISk_Pth_lim}, we present the thermal pressure and its limiting cases.
For both  models the non-degenerate limit
agrees with the exact result until about $n_0$ at 20 MeV (panels (a) and (c)) and until 
$3n_0$ for $T=50$ MeV (panels (b) and (d). The agreement is better at high temperatures
and for symmetric matter than for  pure neutron matter. At both temperatures and proton fractions, the degenerate
limits come closer to the exact results using the MDI(A) model.

\begin{figure*}[!htb]
\centering
\begin{minipage}[b]{0.49\linewidth}
\centering
\includegraphics[width=9.5cm]{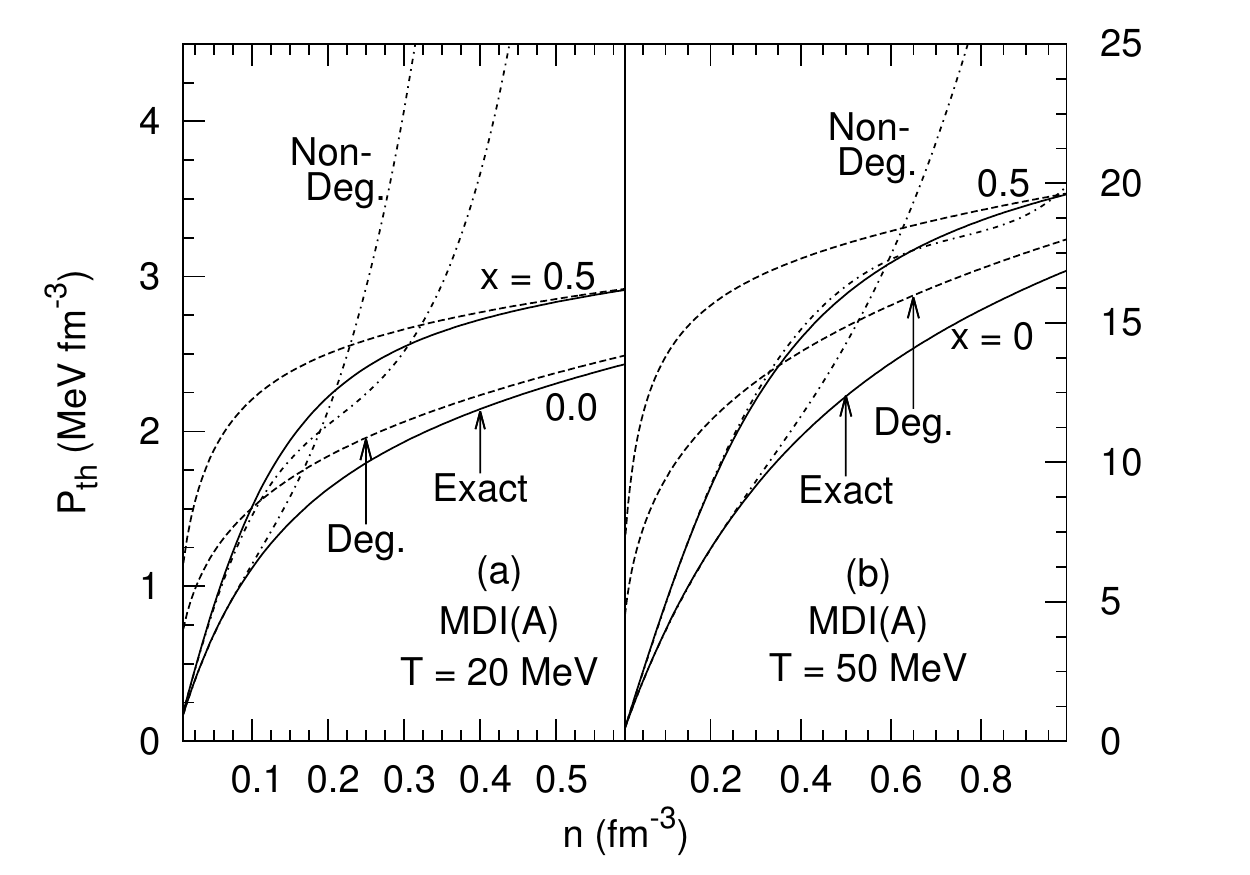}
\end{minipage}
\begin{minipage}[b]{0.49\linewidth}
\centering
\includegraphics[width=9.5cm]{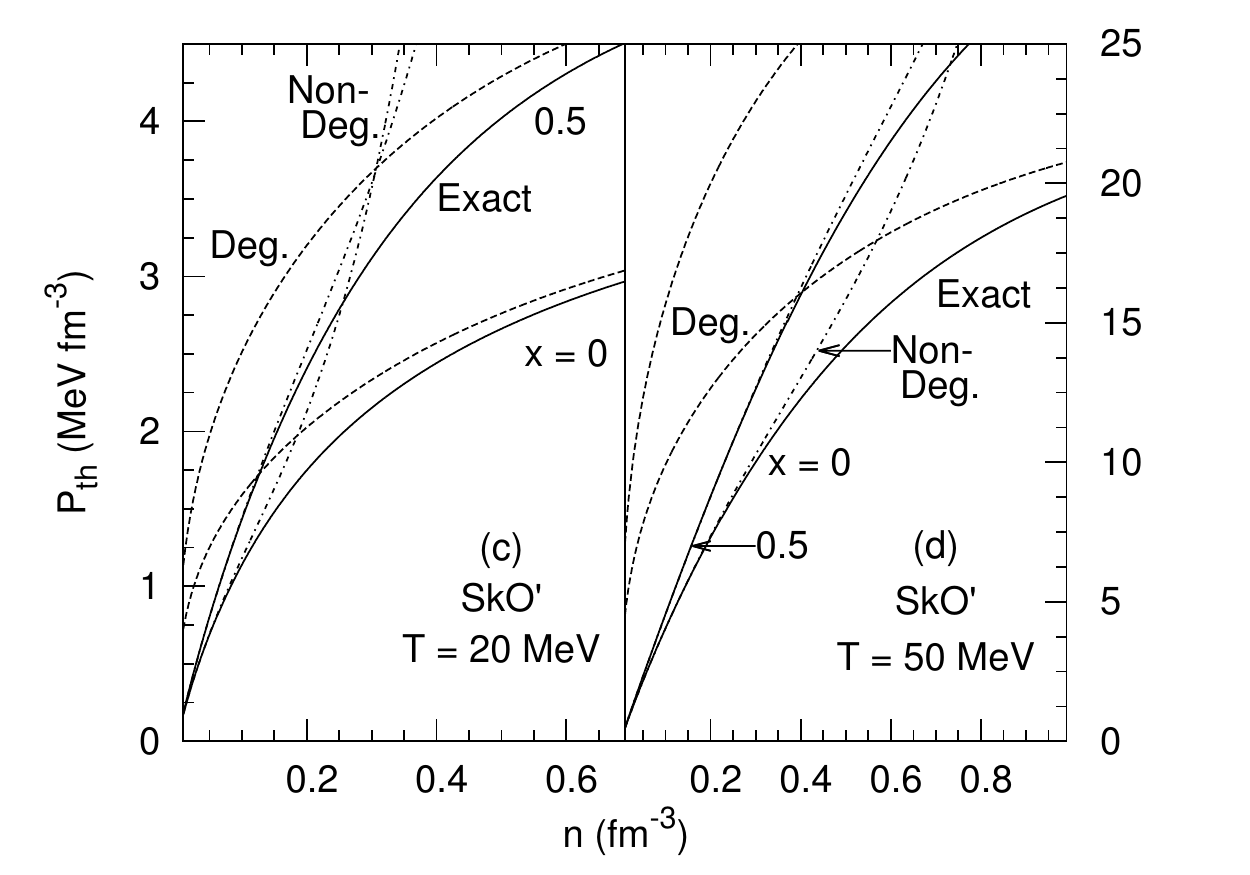}
\end{minipage}
\vskip -0.5cm
\caption{Thermal pressure [Eq. (\ref{pres})] and limiting cases [Eqs. (\ref{pdeg}) and (\ref{MDYI_P_ND})] vs $n$ for the 
temperatures and proton fractions shown. }  
\label{MDYISk_Pth_lim}
\end{figure*}

In Fig. \ref{MDYISk_S_lim}, we present the entropy per baryon and, its degenerate and
non-degenerate limits. The non-degenerate limit has the best agreement using SkO$^\prime$ for symmetric nuclear matter
at $T = 50$ MeV, which extends to about $3n_0$.  The range of densities over which the 
MDI(A) model agrees with the exact result is smaller than that for SkO$^\prime$. For symmetric nuclear matter, for example,  
the agreement does not extend beyond about 1-1.5 $n_0$ even at 50 MeV.
The degenerate limit coincides 
with the exact solution starting approximately around $2n_0$ for the MDI(A) model for symmetric nuclear matter
at $T = 20$ MeV. The SkO$^\prime$ model does notably worse with its best agreement not occurring 
until 3-4 $n_0$ for pure neutron matter at $T = 20$ MeV.

\begin{figure*}[!htb]
\centering
\begin{minipage}[b]{0.49\linewidth}
\centering
\includegraphics[width=9.5cm]{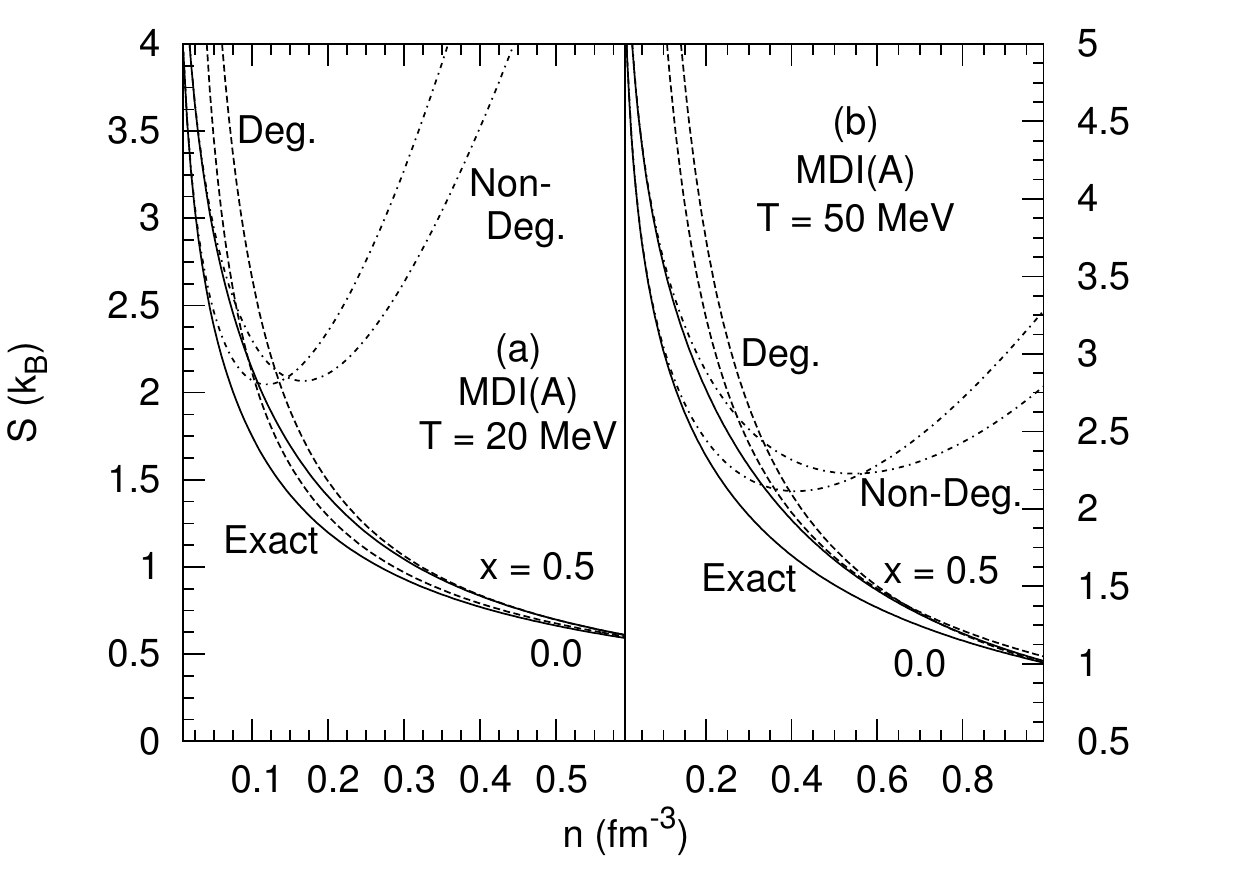}
\end{minipage}
\begin{minipage}[b]{0.49\linewidth}
\centering
\includegraphics[width=9.5cm]{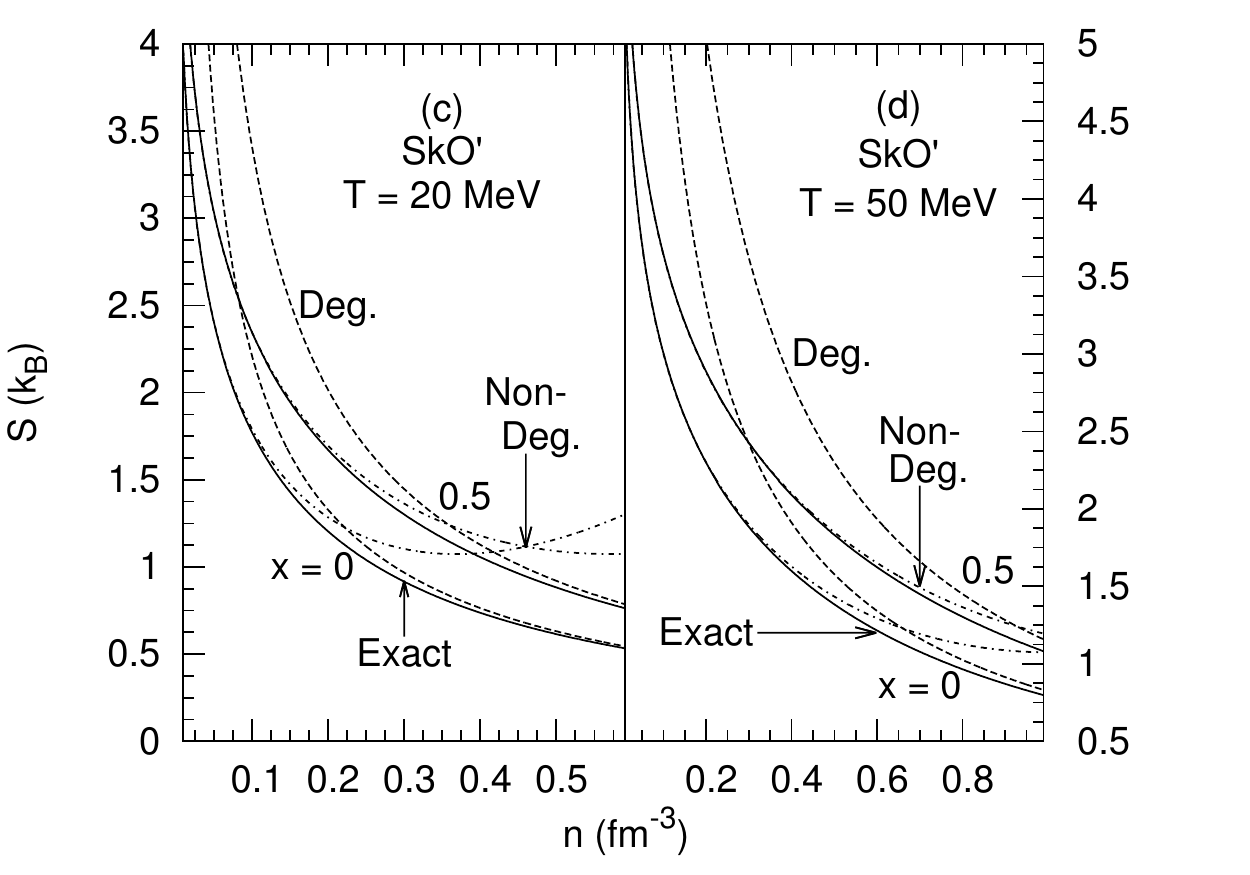}
\end{minipage}
\vskip -0.5cm
\caption{Entropy per particle [Eq. (\ref{sden})] vs $n$ 
for the MDI(A) model in (a) and (b) and the SkO$^\prime$ model in (c) and (d) compared with their limiting cases from 
Eqs. (\ref{sdeg}) and (\ref{MDYI_s_ND}). }
\label{MDYISk_S_lim}
\end{figure*}

Some insight into the behaviors of the thermal variables presented above can be gained from 
the asymptotic behaviors of the single-particle potentials at high
momenta in the two models which lead to two noteworthy effects: (1) The earlier
onset of degeneracy in the MDI model compared to that for SkO$^\prime$ is due to weaker binding at
high densities, and (2) the MDI nucleon effective masses that are nearly
independent of density at high density (while qualitatively the isospin splitting is similar to that of SkO$^\prime$) 
cause the thermal state variables to exhibit less sensitivity to the proton fraction being
changed (cf. FLT equations with $\partial m^*/\partial n \rightarrow 0$).

The specific heat at constant volume vs baryon density and its limiting cases are presented in Figs. \ref{MDYISk_Cv_lim}(a) and (b)
for the MDI(A)  model, while those for the SkO$^\prime$ model are in (c) and (d) of the same figure.   
The best agreement between the results of the exact  and the degenerate limit calculations occurs 
at low temperatures and large densities. Although this is true of both models, the agreement
is better for the MDI(A) model as the degenerate limit comes far closer to the exact solution
than the SkO$^\prime$ model. 
For MDI(A), the degenerate limit has better agreement 
with the exact solution for symmetric nuclear matter as opposed to SkO$^\prime$ which shows 
better agreement for pure neutron matter. The non-degenerate limit coincides with the 
exact solution only for densities much less than the nuclear saturation density. The agreement
between the non-degenerate limit and the exact result is best using the MDI(A) model
for symmetric nuclear matter at high temperatures. The agreement between the non-degenerate
limit and the exact result using the SkO$^\prime$ model is slightly better for pure neutron
matter and at high temperatures.

\begin{figure*}[!htb]
\centering
\begin{minipage}[b]{0.49\linewidth}
\centering
\includegraphics[width=9.5cm]{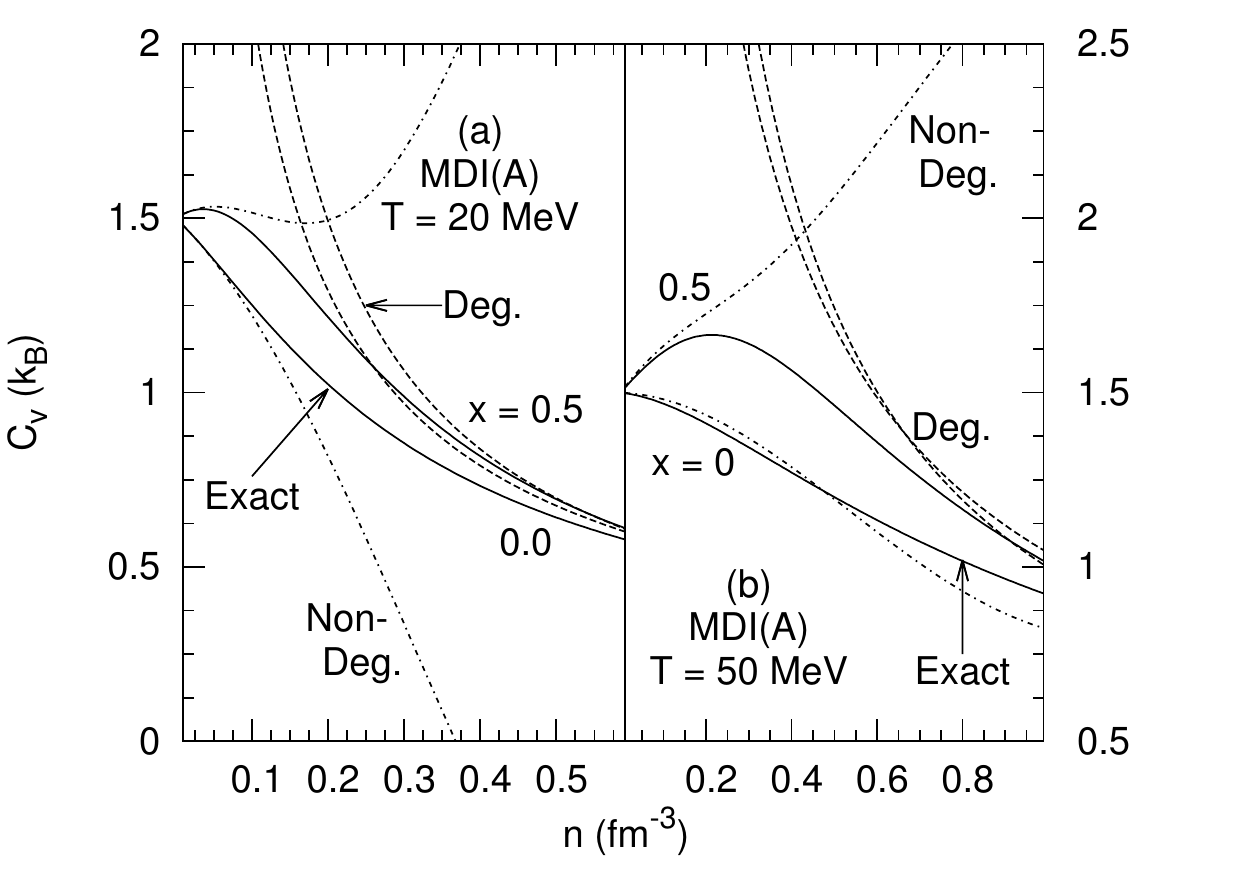}
\end{minipage}
\begin{minipage}[b]{0.49\linewidth}
\centering
\includegraphics[width=9.5cm]{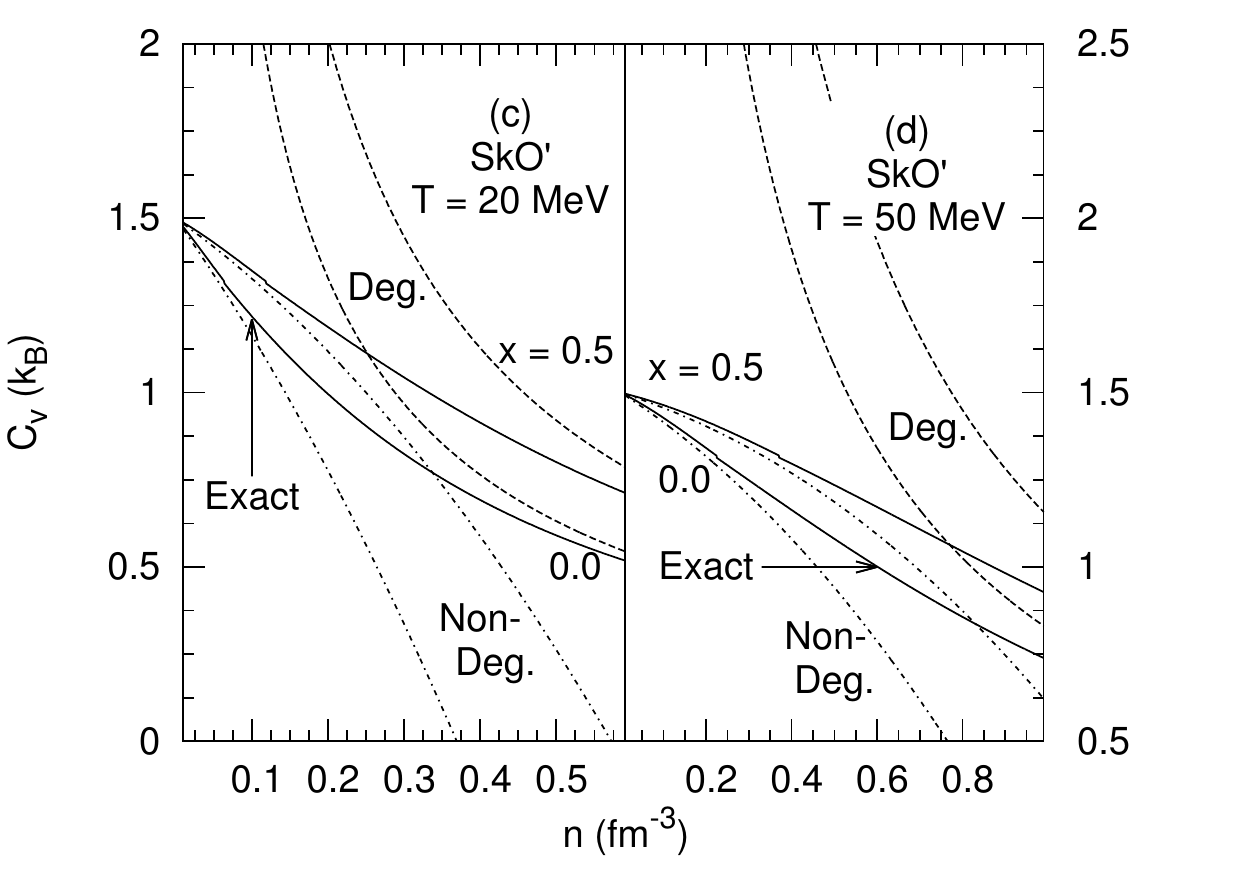}
\end{minipage}
\vskip -0.5cm
\caption{Specific heat at constant volume $C_V$ 
for the MDI(A) model in (a) and (b) and the SkO$^\prime$ model in (c) and (d) compared with their limiting cases from 
Eq. (\ref{sdeg}) as $S=C_V$ in the degenerate limit, and   Eq. (\ref{MDI_Cv}). }
\label{MDYISk_Cv_lim}
\end{figure*}

In Fig. \ref{MDYISk_Cp_lim} we display the specific heat at constant pressure vs baryon density from the 
MDI(A) (panels (a) and (b)) and SkO$^\prime$ (panels (c) and (d)) models.
The agreement between the non-degenerate limit and the exact solution
is remarkably good using the MDI(A) for pure neutron matter at high temperatures 
and low densities. For the MDI(A) model the agreement extends out to about 
$0.4~\rm{fm}^{-3}$, whereas for the SkO$^\prime$ the agreement is up to $0.3~\rm{fm}^{-3}$
for symmetric or pure neutron matter at high temperatures. The 
agreement between the degenerate limit and the exact result for $C_P$ is best for
the MDI(A) model for pure neutron matter at large densities and low temperatures. Using
the SkO$^\prime$ model, the agreement between the degenerate limit and the exact solution
is better for pure neutron matter at large densities and low temperatures.

%
\begin{figure*}[!htb]
\centering
\begin{minipage}[b]{0.49\linewidth}
\centering
\includegraphics[width=9.5cm]{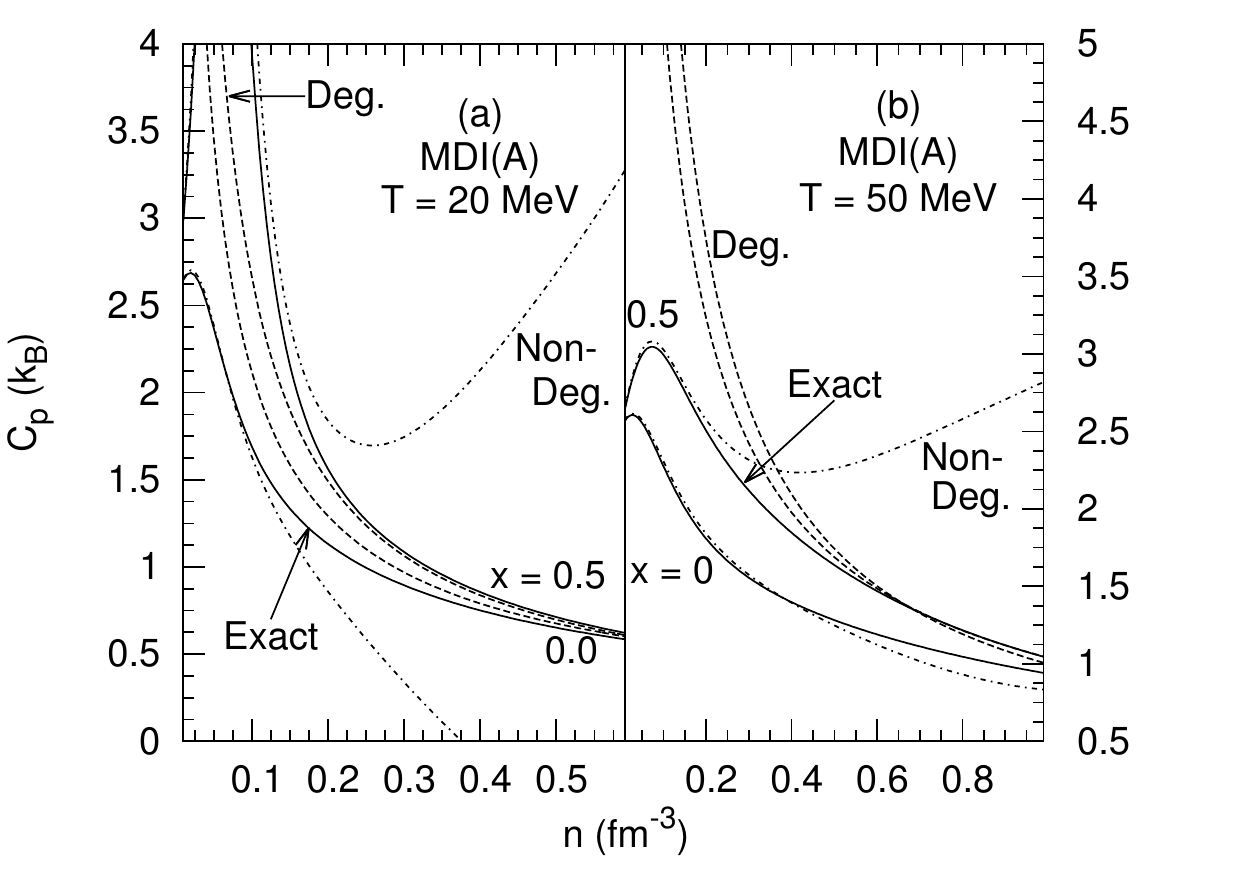}
\end{minipage}
\begin{minipage}[b]{0.49\linewidth}
\centering
\includegraphics[width=9.5cm]{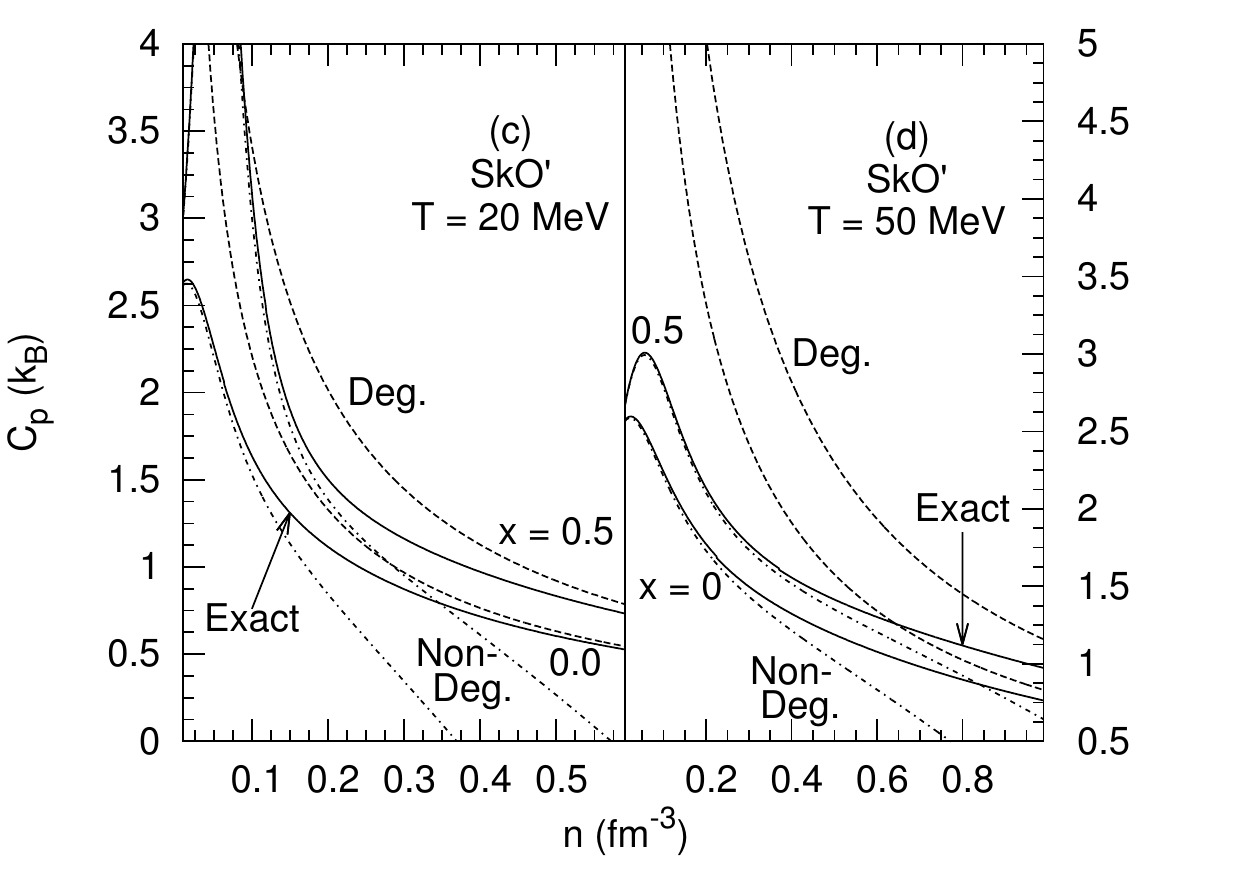}
\end{minipage}
\vskip -0.5cm
\caption{Specific heat at constant pressure $C_P$ 
for the MDI(A) model in (a) and (b) and the SkO$^\prime$ model in (c) and (d) compared with their limiting cases from 
Eq. (\ref{sdeg}) as $S=C_P$ in the degenerate limit, and Eq. (\ref{cp}). }
\label{MDYISk_Cp_lim}
\end{figure*}

\section{Thermal and adiabatic indices} 
\label{Sec:Tindices}
The paradigm for neutron star mergers now seems to be that one begins with two 1.3 -1.4 ${\rm M_G/M_\odot}$ stars, which form  
a hyper-massive remnant stabilized against collapse by rotation, thermal and magnetic effects. It could be differentially
rotating. Loss of differential rotation, and loss of thermal and/or magnetic support leads to an eventual collapse to a
black hole. The timescale is very important, as it will have observable effects on gravitational wave and gamma-ray
burst durations. Thermal effects at supra-nuclear density seem to have little effect, but rotational support means
the average densities of the disc are near saturation density where the thermal effects become substantial (see references below). 
Thus, the thermal support needs to be properly treated.


\subsection{Thermal index}

The inclusion of thermal effects in neutron star merger simulations is often treated using an effective
thermal index $\Gamma_{th}(n)$ defined as
\be
\Gamma_{th}(n) = 1 + \frac {P_{th}}{\varepsilon_{th}} \,.
\label{Gthe}
\ee
Shibata's group commonly uses  $\Gamma_{th}$  to describe finite-temperature effects, favoring the value 1.8 \cite{Hotokesake:13}.
Bauswein et al., \cite{Bauswein:10} prefer the value 2.0; 
see also Janka, et al. \cite{Janka:93}.
In simulations by Foucart et al., \cite{Foucart:14} and Kaplan et al., \cite{Kaplan:14}  
conditions such as $S=5$ at 
$10^{14}$ g cm$^{-3}$ and $S = 10$ at $10^{12}$ g cm$^{-3}$ are reached in the
ejecta. In these works, realistic EOS's with consistent thermal treatments (LS \cite{Lattimer91} or Shen \cite{Shen11}, among others) are used.
An overview has been provided in the work of Cyrol \cite{Cyrol:14} in which the behavior of 
of $\Gamma_{th}$ vs. $n$ has been presented using the tabulated results of non-relativistic and relativistic models.  

Our aim here is to provide a basis for understanding the behavior of $\Gamma_{th}$ from elementary considerations. 
The results of our calculations are relevant only for densities and temperatures for which a bulk homogeneous phase will be present. Inhomogeneous phases present at sub-nuclear densities, and which induce large variations in $\Gamma_{th}$, have not been considered here as they lie beyond the scope of this work.  

In the degenerate limit, the  FLT results in  Eqs. (\ref{edeg}) and (\ref{pdeg}) imply that
\be
\Gamma_{th}(n)  = 1 + \frac 23 \frac {\sum_i a_in_iQ_i } {\sum_i a_in_i} \,, \qquad i=n,p,e
\label{Gth1}
\ee
where the level density parameters are 
\ba
{\displaystyle{
a_i = 
\left\{
\begin{array}{ll} 
\frac {\pi^2m_i^{*}}{2p_{F_i}^2} \,, \quad &  {\rm non-relativistic ~nucleons}\,\\
\frac {\pi^2{\sqrt {p_{F_e}^2 + m_e^2}}}{2p_{F_e}^2} \quad & {\rm relativistic~electrons }\,,\\
\end{array} 
\right\}..
}}
\ea
and 
\be
Q_i = 1 - \frac 32 \frac {n}{m_i^*} \frac {dm_i^*}{dn} \,.
\label{kewi}
\ee
The above equations highlight the role of the effective masses and their behavior with density. 
Note that relativity endows non-interacting electrons with a density-dependent effective mass $m_e^* = E_{F_e} = {\sqrt{p_{F_e}^2 + m_e^2 } }$. Thus, in a pure electron gas, 
\be
Q_e = 1 - \frac 12 \frac {p_{F_e}^2}{p_{F_e}^2 + m_e^2}  \,,
\ee
which has the limit 1/2 for ultra-relativistic electrons ($p_{F_e} \gg m_e$)  and 1 for non-relativistic electrons ($p_{F_e} \ll m_e$). 
These limits help to recover the well known results $\Gamma_{th}=4/3$ in the former case and 5/3 in the latter (also easily obtained  by inspecting the limits of $P_{th}/\varepsilon_{th}$ in the two cases). 

Note that in the degenerate limit, $\Gamma_{th}$ in Eq. (\ref{Gth1})
is independent of temperature. For pure neutron matter (PNM) and symmetric nuclear matter (SNM) without electrons,  Eq. (\ref{Gth1}) reduces to the simple result
\be
\Gamma_{th}(n) = \frac 53 - \frac {n}{m_b^*} \frac {dm_b^*}{dn} \,,
\label{Gth2}
\ee
where the subscript ``$b$'' identifies the appropriate baryons (neutrons in PNM, and neutrons and protons in SNM or isospin asymmetric matter).

For Skyrme models, the above relation is valid for all regions of degeneracy as $P_{th}$ and $\varepsilon_{th}$ can be written in terms of their ideal gas counterparts (calculated with $m^*(n)$ instead of $m$) as  
\ba
P_{th}(n,T) &=& P_{th}^{id}(n,T;m^*) \left( 1 - \frac 32 \frac {n}{m^*} \frac {dm^*}{dn} \right) \nonumber \\
\varepsilon_{th}(n,T) &=& \varepsilon_{th}^{id}(n,T;m^*)\,, \qquad
\frac {P_{th}^{id}}{\varepsilon_{th}^{id}} = \frac 23 \,. 
\ea
These results in conjunction with Eq. (\ref{Gthe}) lead to Eq. (\ref{Gth2}). The simple form of the effective masses in Skyrme models, $m^*=m(1+\beta n)^{-1}$, where the positive constant $\beta$ depends mildly on the proton fraction
(for the SkO$^\prime$ model, $\beta$ lies in the range 0.523-0.724 as $Y_p$ varies from 0-0.5) allows us to obtain 
\be
\Gamma_{th}(n) = \frac 53 + \frac {\beta n}{1+\beta n} = \frac 83 - \frac{m^*}{m}\,.
\ee
which establishes the $T$- independence and very mild dependence on the proton fraction. For Skyrme models therefore, 
$\Gamma_{th}$ of nucleons increases monotonically  from 5/3 to 8/3 as the density increases. \

The analytical expressions for the effective masses and their derivatives with respect to density are more complicated for the MDI models than those for the Skyrme models (see Appendix A). However, they are easily implemented in numerical calculations. 
Results from such computations will be compared with the exact numerical calculations below.  

In the Maxwell-Boltzmann limit [that is, the non-degenerate limit to $\mathcal{O}(z^1)$], the thermal energy 
density and pressure of the MDI model are given, respectively, by
\ba
\varepsilon_{th} &=& \frac{3nT}{2}\exp\left[\frac{R(\sqrt{2mT})-R(\sqrt{4mT})}{T}\right] \nonumber \\
                &+& \frac{n}{2}R(\sqrt{2mT})-\frac{3}{5}\mathcal{T}_Fn  \label{ndeth} \\
P_{th} &=& nT \left[1+\frac{R(\sqrt{2mT})}{2T}\right] -\frac{2}{5}\mathcal{T}_Fn \label{ndpth} \,,
\ea
where $R(p)$ is given by Eq. (\ref{ndR}) with $p_{0R}=\sqrt{2mT}$ and $\mathcal{T}_F = p_F^2/(2m^*)$.
Exploiting the fact that in this regime the interactions are 
weak due to the diluteness of the system, we expand the exponential in Eq. (\ref{ndeth}) in a Taylor series about the zero of its 
argument which leads to 
\be
\varepsilon_{th} = \frac{3nT}{2}\left[1+\frac{4/3 R(\sqrt{2mT})-R(\sqrt{4mT})}{T}\right]
                    -\frac{3}{5}\mathcal{T}_Fn.  \label{ndeth2} 
\ee
Note that in Eqs. (\ref{ndpth}) and (\ref{ndeth2}) the leading terms are proportional to $nT$, the 
interaction terms to $n^2/T^2$ (approximately) and the $T=0$ terms to $n^{5/3}$. Thus in the 
limit of vanishing density, the ratio $P_{th}/\varepsilon_{th}$ goes to $2/3$; consequently,  
$\Gamma_{th}$ approaches $5/3$ as expected for a non-relativistic gas.

%
\begin{figure*}[!htb]
\centering
\begin{minipage}[b]{0.33\linewidth}
\centering
\includegraphics[width=6.3cm,height=11cm]{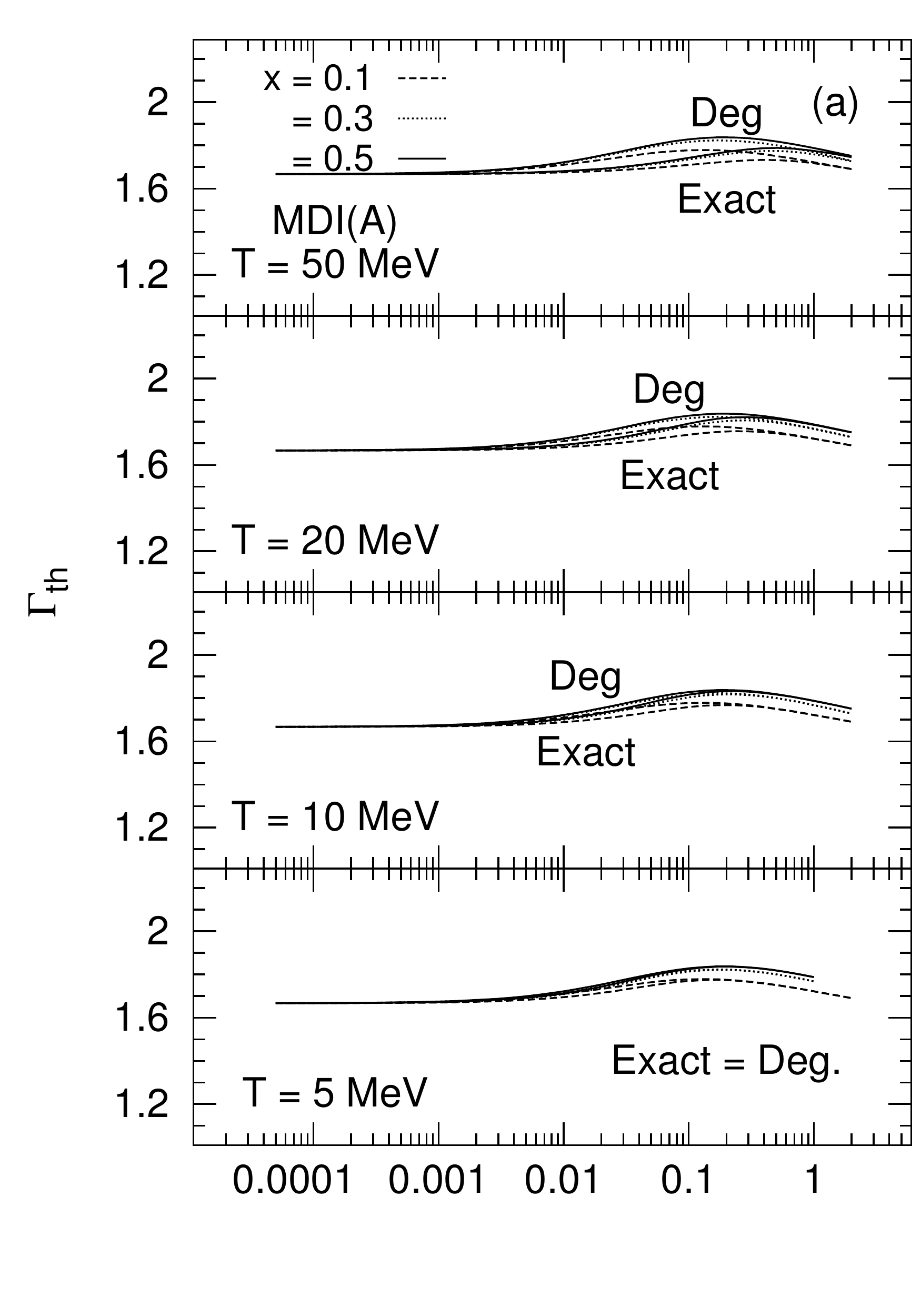}
\end{minipage}
\hspace{-1.04cm}
\begin{minipage}[b]{0.32\linewidth}
\centering
\includegraphics[width=6.4cm,height=11cm]{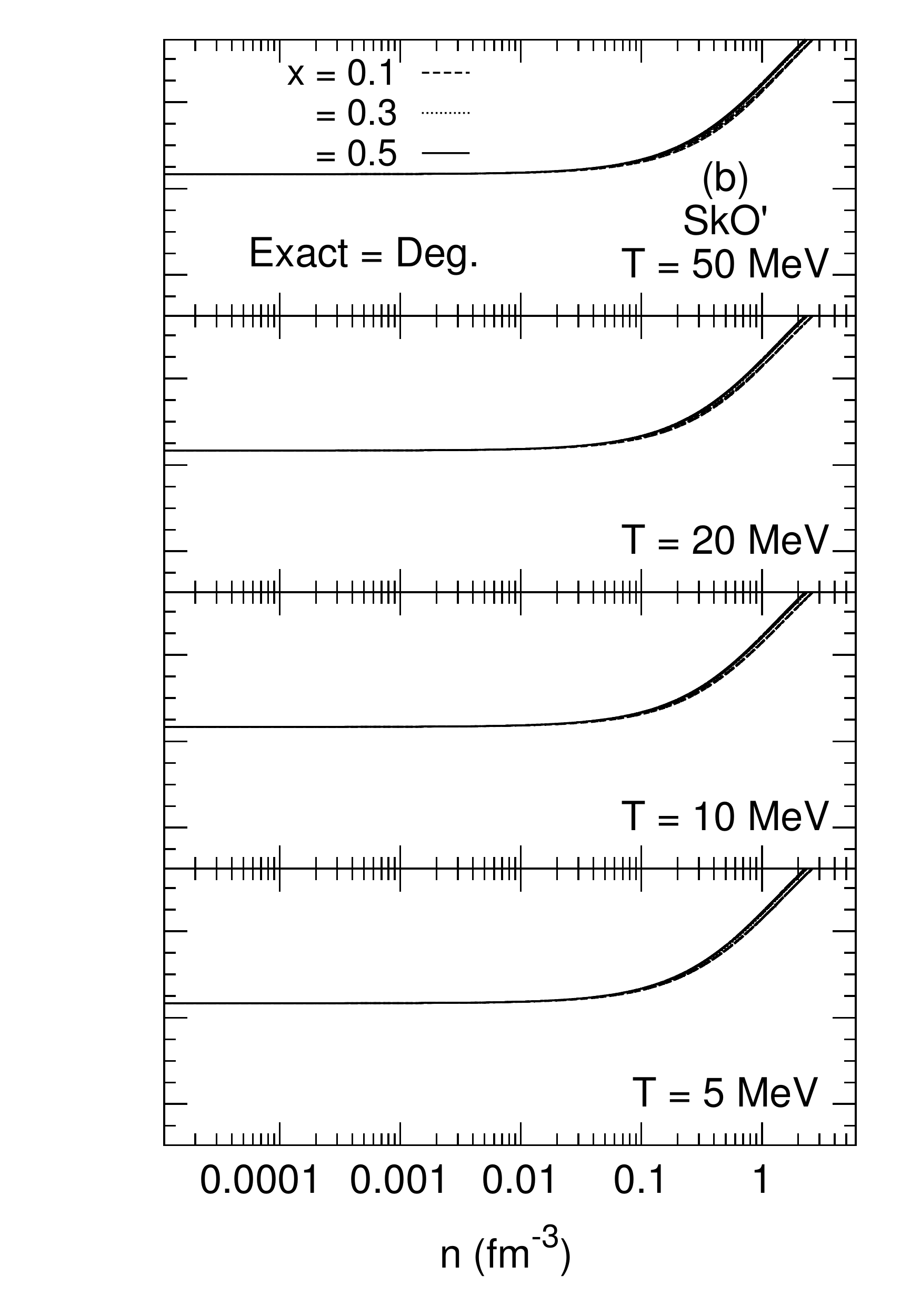}
\end{minipage}
\hspace{-0.032cm}
\begin{minipage}[b]{0.33\linewidth}
\centering
\includegraphics[width=6.3cm,height=11cm]{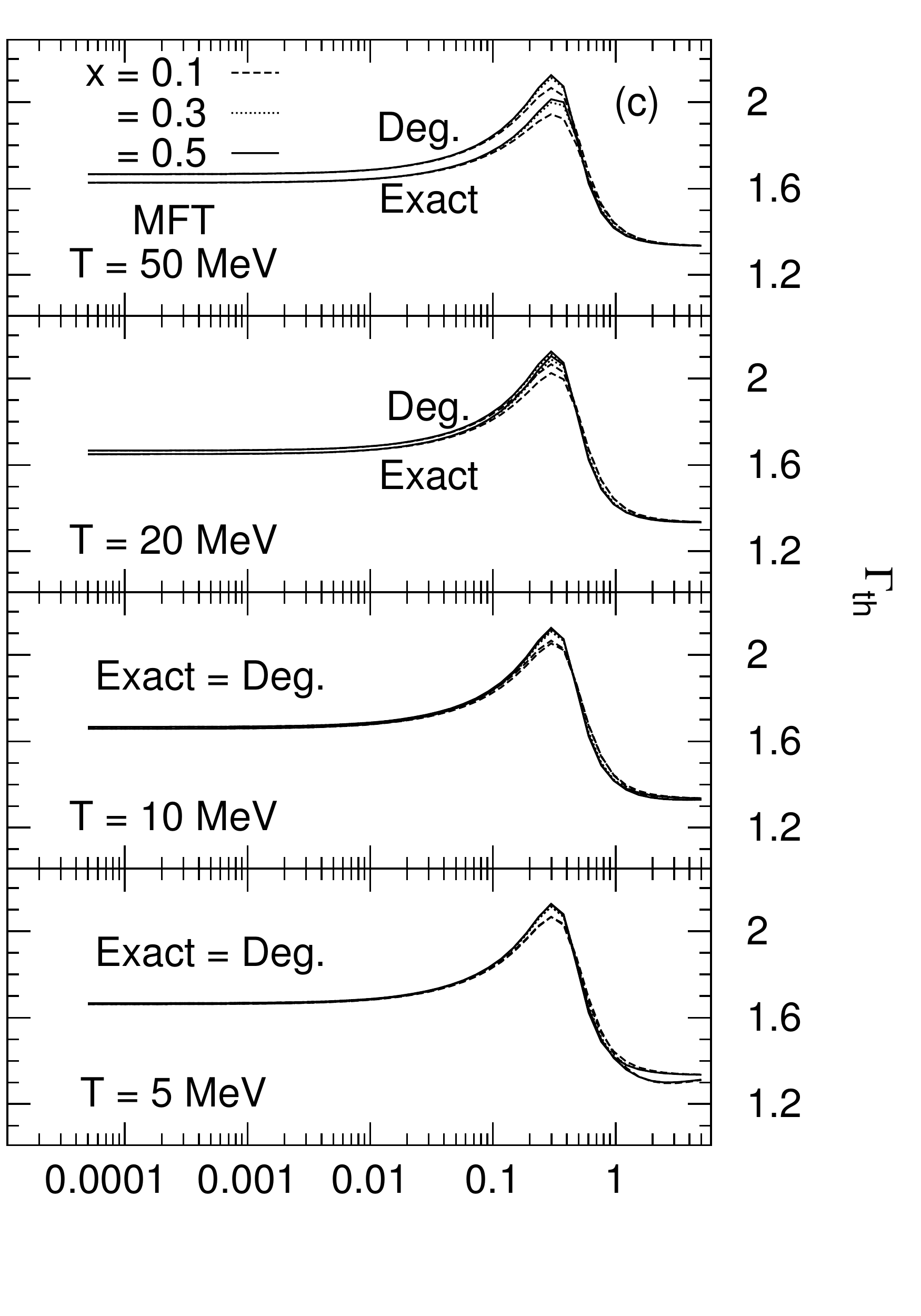}
\end{minipage}
\vskip -0.5cm
\caption{The thermal index $\Gamma_{th}$  vs $n$ for the three models indicated in the figure.  Results for the exact calculations are from Eqs. (\ref{eth}), (\ref{ppth}) and (\ref{Gthe}), whereas those for the degenerate limit are from Eqs. (\ref{Gth1})-(\ref{kewi}). The proton fractions and temperatures are indicated in the figures.  Results shown are for nucleons only.}
\label{Gamma_th_NoLep}
\end{figure*}
  %

In order to appreciate the role of electrons in the behavior of $\Gamma_{th}$ vs $n$, we first show in Fig. \ref{Gamma_th_NoLep} results with nucleons only for three different models. Proton fractions and temperatures are as noted in the figure. Results for the non-relativistic models in this figure are for MDI(A) and SkO$^\prime$ used throughout this work. For contrast, we also show results for a typical 
mean-field theoretical model (labelled MFT in the figure) with up to quartic scalar self-interactions.  The strength parameters of this MFT model yield the zero-temperature  properties of $n_0=0.155~{\rm fm}^{-3},~E_0=16~{\rm MeV},~M^*/M=0.7,~K_0=222~{\rm MeV},~S_2=30~{\rm MeV},~{\rm and}~L=87.0~{\rm MeV}$ \cite{Constantinos:13}.       

In the non-degenerate regime, the exact numerical results in all cases shown tend to 5/3 as expected. 
Also shown in this figure are results from the expression in Eq. (\ref{Gth1}) which agree very well with the exact results in the expected regions of density
(that are very nearly independent of temperature)
for all three models (except the MFT model at $T=50$ MeV, to which we will return below). 
Beginning with the non-relativistic models, we observe a distinct difference between results for the two models in that the MDI(A) model exhibits a pronounced peak for $Y_p=x=0.5$ whereas the SkO$^\prime$ model does not. The origin of these differences can be traced back to the behavior of $m^*$'s of these models with density. The presence or absence of a peak can be ascertained by examining whether or not
\be
\frac {d\Gamma_{th}}{dn} = 0 = \frac {dm^*}{dn} \left(1- \frac {n}{m^*} \frac {dm^*}{dn}  \right) + n \frac {d^2m^*}{dn^2} = 0 
\ee
admits a solution. For the MDI(A) model, the solution of the above equation occurs at $n\simeq 0.15~{\rm fm}^{-3}$ at $T=0$ in good agreement with the exact numerical results. As the temperature increases toward 50 MeV, the estimate from the degenerate limit no longer applies as can be seen from the figure. 

We turn now to analyze results of the MFT model, particularly in the degenerate region where a peak occurs for all values of $Y_p$. In MFT models, the Landau effective masses $m_i^* = E_{F_i}^* = {\sqrt {p_{F_i}^2+M^{*2}}}$, where $M^*$ is the Dirac effective mass obtained from a self-consistent procedure which involves minimizing the energy density (pressure at finite $T$) with respect to density.  

Figure \ref{MFT_Ms_0T} shows $M^*$ and $m_n^*$ of the MFT model employed here for $T=0$. Note that although $M^*$ decreases monotonically with density, $m_n^*$ exhibits a minimum and rises monotonically with density, a characteristic behavior solely due to relativity (at asymptotic densities, mass is overwhelmed by momentum). The density at which the minima occur is easily found from the roots of 
\be
\frac {p_{F_i}}{M^*} + \frac {dM^*}{d{p_{F_i}} }=0 
\ee
for each $Y_p$. The densities at the minima range from 0.52-0.57 fm$^{-3}$ for $Y_p$ in the range 0-0.5. This behavior of $m^*$'s is also at the root of the peaks seen in $\Gamma_{th}$ vs $n$ as the analysis below shows. 

\begin{figure}[!htb]
\includegraphics[width=9.2cm]{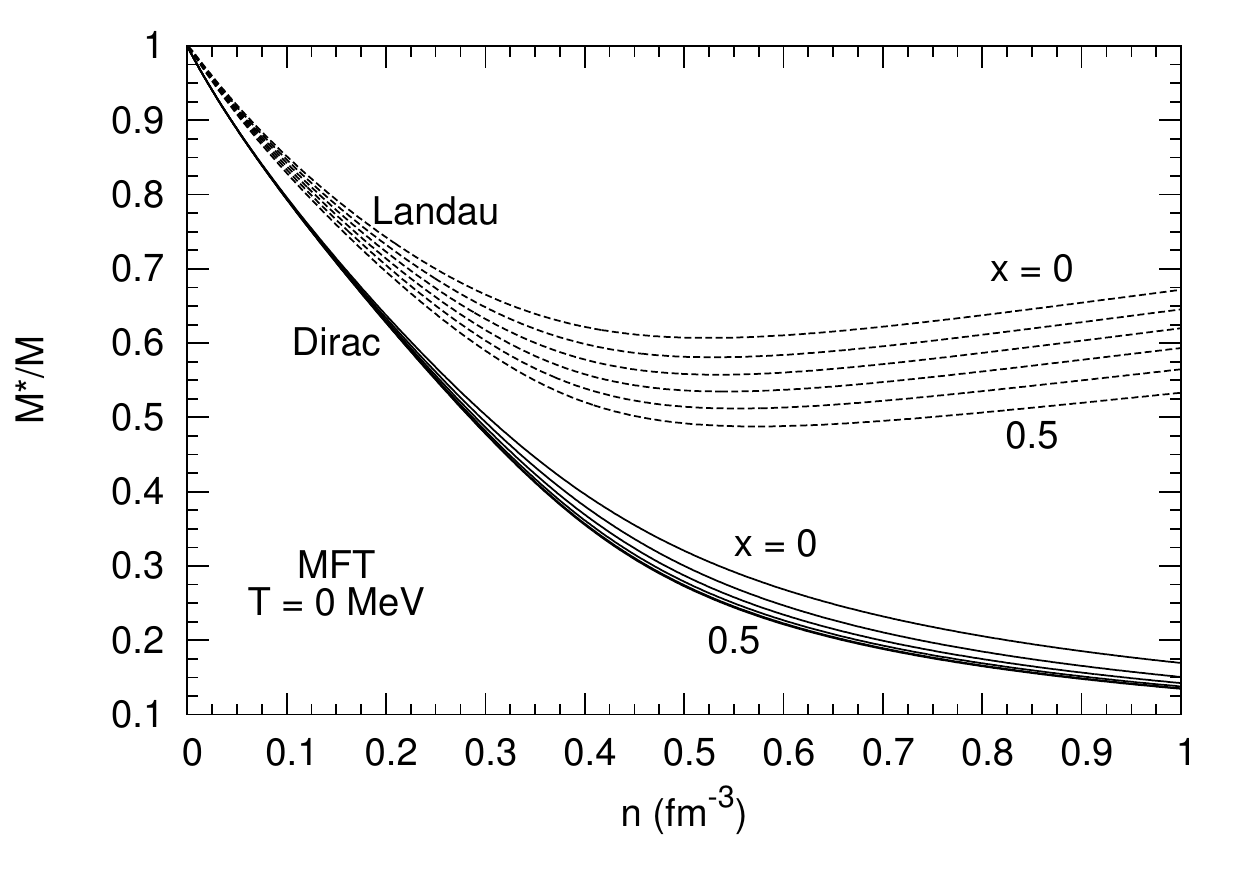}
\caption{The Dirac and Landau effective masses  $M^*$ and $m_n^*$ of  neutrons in a typical mean-field theoretical (MFT) model at zero temperature.}
\label{MFT_Ms_0T}
\end{figure}

We will restrict our analysis to the case of PNM as the location of the peak in $\Gamma_{th}$ is not very sensitive to $Y_p$. 
The degenerate limit expressions for the thermal pressure and energy density are \cite{Prakash87}:
\ba
P_{th} &=& \frac 13 naT^2 \left[1 + \left( \frac {M^*}{E_F^*}\right)^2 \left (1- 3 \frac {n}{M^*} \frac {dM^*}{dn}\right) \right]  \\
\varepsilon_{th} &=& naT^2 \,,
\ea
where the level density parameter $a=\pi^2E_F^*/(2p_F^2)$.  The thermal index then becomes
\be
\Gamma_{th} = \frac 43 + \frac 13 \left[ \left( \frac {M^*}{E_F^*}\right)^2 \left (1- 3 \frac {n}{M^*} \frac {dM^*}{dn}\right) \right]  \,.
\ee
In the non-relativisitic limit, $M^*/E_F^* \rightarrow 1$ and the logarithmic derivative of $M^*$ with respect to $n$ tends to zero leading to $\Gamma_{th} =5/3$. In the ultra-relativistic limit, $M^* \rightarrow 0$ so that  $\Gamma_{th} =4/3$.  The density at which 
the maximum occurs in $\Gamma_{th}$ can determined from
\be
\frac {d\Gamma_{th}}{dp_F} = 0 = \frac {d}{dp_F} \left( 
\frac {p_F^2}{E_F^{*2}} + \frac {p_FM^*}{E_F^{*2}} \frac {dM^*}{dp_F} \right) \,.
\ee
The result is $n\simeq 0.27~{\rm fm^{-3}}$ in good agreement with the exact results. Performing the appropriate calculations for a mixture of neutrons and protons for $Y_p$ in the range 0.1-0.5 yields densities in the range 0.28-0.31 $\rm fm^{-3}$ also in good agreement with the exact numerical results shown Fig. \ref{Gamma_th_NoLep}.

The $T=50$ MeV results for the MFT model requires some explanation. Note that for densities below the peak, the exact and degenerate results are somewhat different although the qualitative trend is maintained. This is due to the fact that in MFT models, $M^*$ acquires a temperature dependence with progressively increasing $T$, whereas the degenerate limit results are calculated with the zero temperature $M^*$.  

\begin{figure}[!htb]
\includegraphics[width=9.2cm]{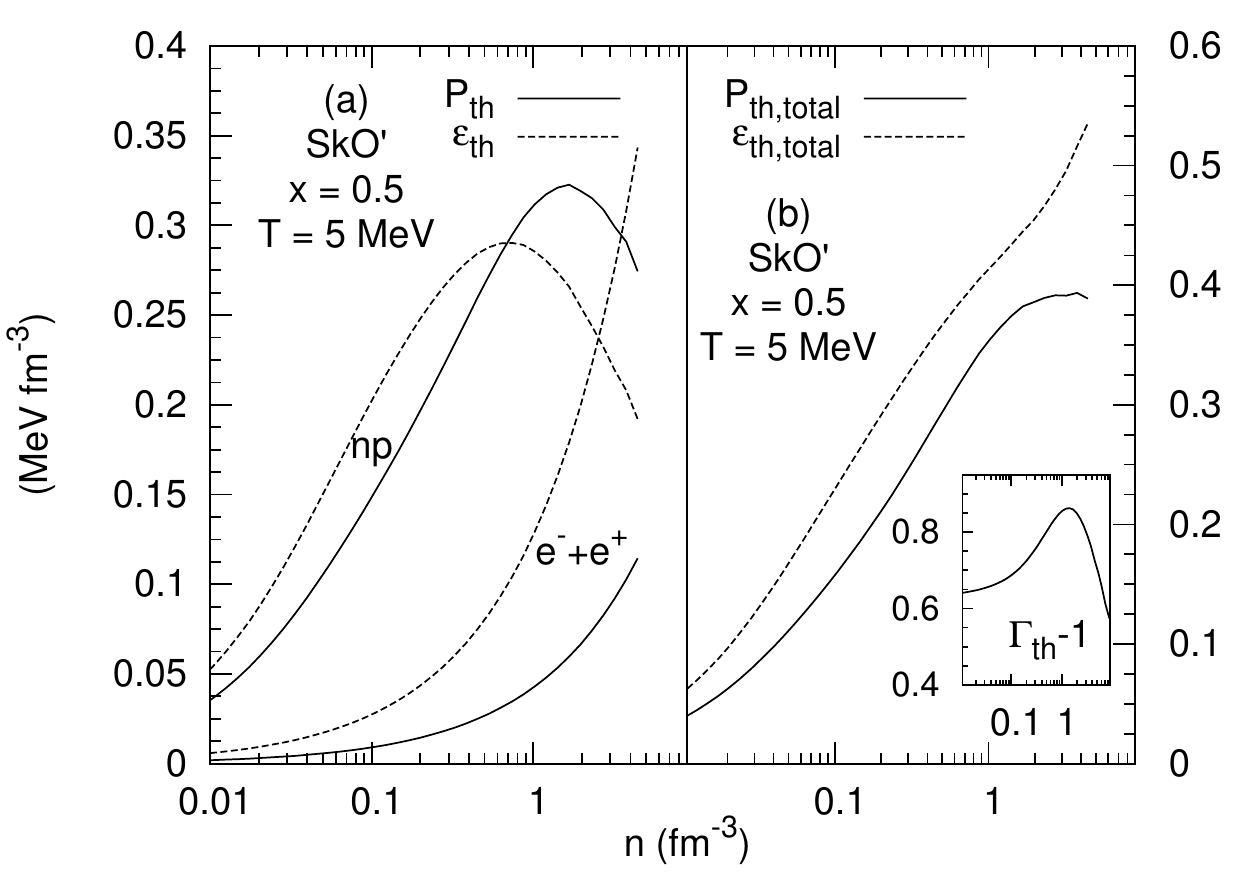}
\caption{(a) Individual contributions from nucleons and leptons to thermal pressure, $P_{th}$, and thermal energy density, 
$\varepsilon_{th}$, for the SkO$^\prime$ model at the indicated proton fraction and temperature.  (b) The total $P_{th}$ and 
$\varepsilon_{th}$; the inset shows their ratio.}
\label{SkOp_NPL_EP_T5_x5}
\end{figure}

The preceding analysis sets the stage to assess the role leptons ($e^-$ and $e^+$) play in determining $\Gamma_{th}$. 
As Skyrme models lend themselves to a straightforward 
analysis (results for other models are similar although quantitative differences exist), we show in Fig. \ref{SkOp_NPL_EP_T5_x5}(a) the contributions from nucleons and leptons to $P_{th}$ and $\varepsilon_{th}$ at $T=5$ MeV and $Y_p=0.5$. The ratio $P_{th}/\varepsilon_{th}$ for leptons remains close to 1/3 for all densities with negligible contributions from positrons. With increasing density, the ideal gas value of $P_{th}/\varepsilon_{th}=2/3$ for nucleons changes significantly due to corrections from the density dependence of $m^*$.  From the degenerate result for $\varepsilon_{th}$, it is easy to show that its maximum value is reached at $n=1/(2\beta)\cong 0.69~{\rm fm}^{-3}$ using 
$\beta=0.724~{\rm fm}^3$ for $x=0.5$ for this model.  Use of this density yields the peak value of $\varepsilon_{th} \simeq 0.29~{\rm MeV~fm}^{-3}$ in good agreement with the exact result.  Likewise, the degenerate expression for $P_{th}$ yields the density at which its peak occurs as $n=0.5(1+{\sqrt {1.8}})/\beta \cong 1.62~{\rm fm}^{-3}$ and a peak value of $P_{th} \simeq 0.32~{\rm MeV~fm}^{-3}$, again in good agreement with the exact result.

Figure \ref{SkOp_NPL_EP_T5_x5}(b) shows the total thermal pressure and energy density  along with its ratio in the inset. It is evident that the contributions from the leptons  remain subdominant except at very low and very high densities at this temperature.  
With the inclusion of leptons the ratio of $P_{th}/\varepsilon_{th}$ is less than one at all densities. At large densities, the total 
thermal pressure approaches a constant value whereas $\varepsilon_{th}$ continues to increase. Thus, 
the ratio $P_{th}/\varepsilon_{th}$, and thereby $\Gamma_{th}$, decreases at large densities
resulting in the maximum observed in the inset. The location of this
maximum can be calculated by solving for the density at which 
\be
\frac {d}{dn} \left( \frac {2a_nQ_n + a_eQ_e}{2a_p+a_e}\right) = 0 
\ee
appropriate for SNM with $Q_n=Q_p$, $a_n=a_p$, and $n_n=n_p=n_e$.  Using results from FLT,  the above equation can be cast in the form
\be
\frac{1}{\gamma}\frac{d\gamma}{dn}+\frac{1}{Q_n}\frac{dQ_n}{dn} = 0
\ee
with
\be
\gamma = \left(1+\frac{Q_e {m_e}^*}{2 Q_n {m_n}^*}\right)\left(1+\frac{{m_e}^*}{2{m_n}^*}\right)^{-1}
\ee
a relation that involves density-dependent effective masses of the neutron ($Q_n= 5/3 - m_n^*/m$) and the electron.
A straightforward numerical evaluation yields the result $\simeq 1.37~{\rm fm^{-3}}$ for the density at which the peak in 
$\Gamma_{th}$ occurs in agreement with the exact result. 

\begin{figure}[!htb]
\includegraphics[width=9.2cm]{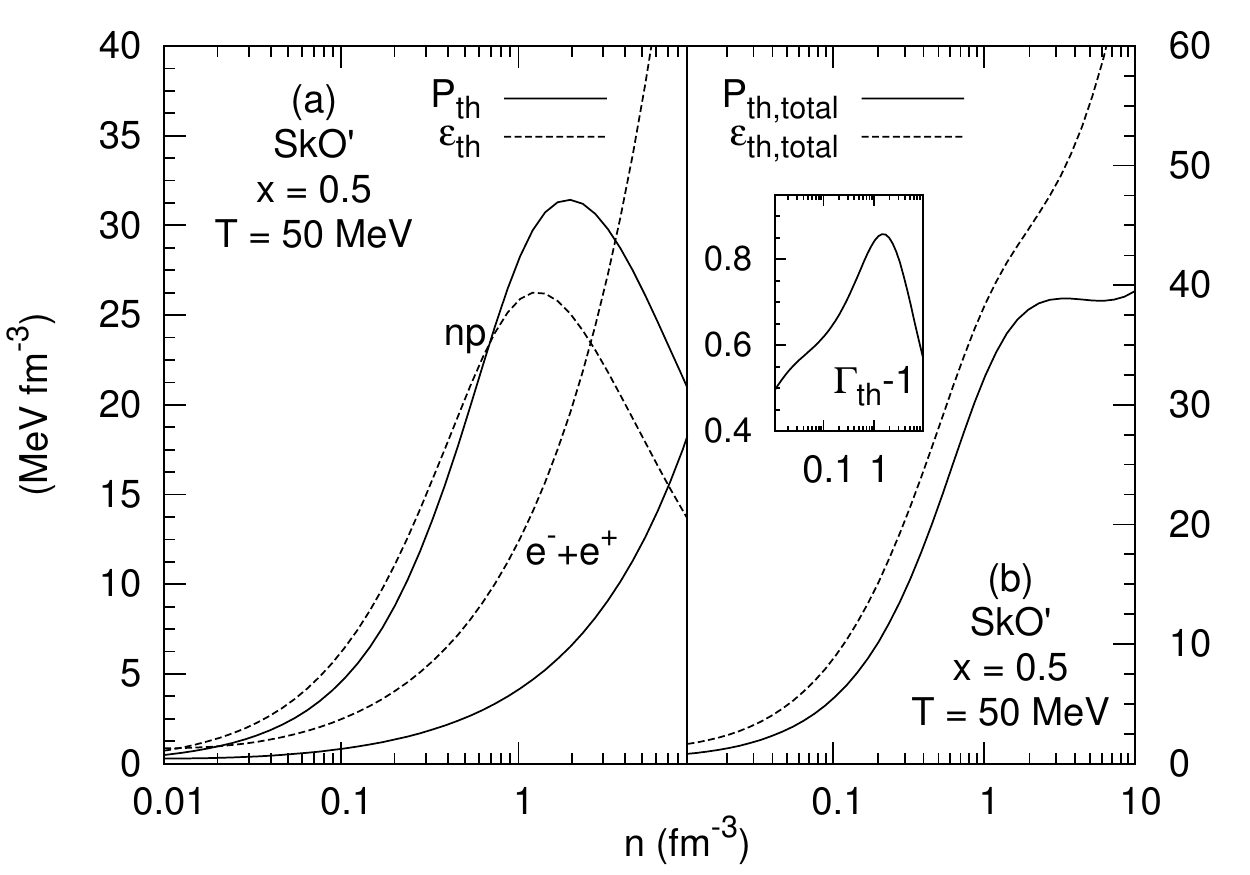}
\caption{Same as Fig. \ref{SkOp_NPL_EP_T5_x5}, but at $T=50$ MeV. }
\label{SkOp_NPL_EP_T50_x5}
\end{figure}

For contributions from the leptons ($e^-$ and $e^+$), we adopt the JEL scheme \cite{jel} in all regions of degeneracy in our 
exact numerical calculations.  Contributions from photons (significant at high values of $T$ and $n$) are easily incorporated using 
\be
\varepsilon_\gamma = \frac {\pi^2}{15} \frac {T^4}{(\hbar c)^3}\,, \quad P_\gamma = \frac {\varepsilon_\gamma}{3}\,, \quad {\rm and} \quad 
s_\gamma = \frac {4}{3} \frac {\varepsilon_\gamma}{T} \,,
\ee
respectively.  

In Figs. \ref{SkOp_NPL_EP_T50_x5}(a) and (b), results similar to those of Fig. \ref{SkOp_NPL_EP_T5_x5} are shown, but at $T=50$ MeV. Contributions from the leptons are such that  $P_{th}/\varepsilon_{th}$ remains at 1/3 as for $T=5$ MeV, with minimal contributions from positrons. 
At very low densities the nucleonic contributions are those of nearly free and non-degenerate fermions. With increasing density, however, nucleons enter the semi-degenerate region for which a transparent 
analysis is not possible. 
While the qualitative features are similar to those of  Fig. \ref{SkOp_NPL_EP_T5_x5}, quantitative differences are due to the higher temperature in this case.

Results of $\Gamma_{th}$ for the three models considered in this section including the contributions from leptons and photons are shown in Fig. \ref{Gamma_th} at the temperatures and proton fractions shown in the figure. Regions of density and temperature for which the degenerate or non-degenerate approximation is valid are apparent in this figure. Note that the behavior of 
$\Gamma_{th}$ at supra-nuclear nuclear densities for the non-relativistic models is significantly altered from the results with nucleons only 
(see Fig. \ref{Gamma_th_NoLep} for comparison) primarily because of the contributions from leptons.


%
\begin{figure*}[!htb]
\centering
\begin{minipage}[b]{0.33\linewidth}
\centering
\includegraphics[width=6.3cm,height=11cm]{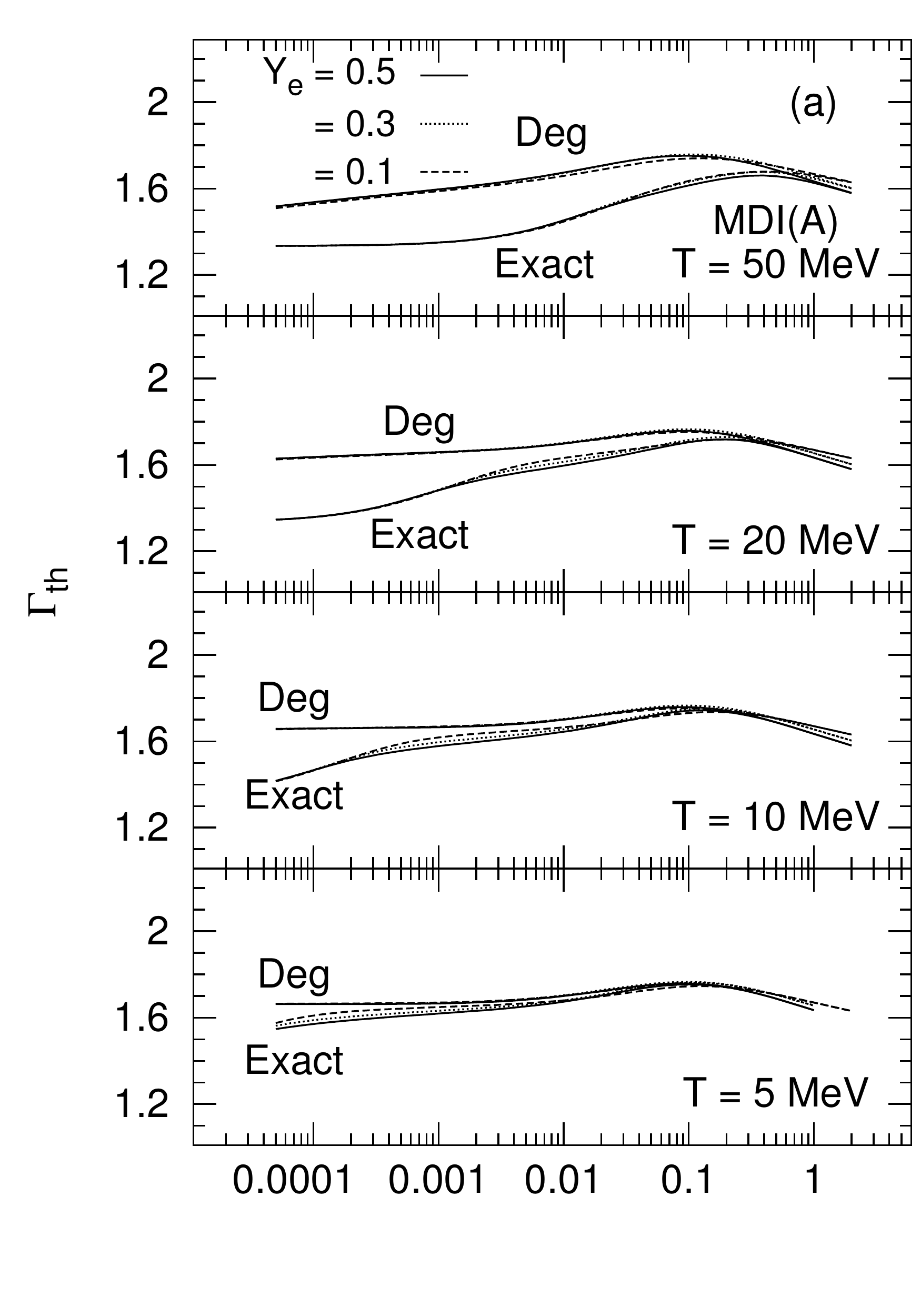}
\end{minipage}
\hspace{-1.04cm}
\begin{minipage}[b]{0.32\linewidth}
\centering
\includegraphics[width=6.4cm,height=11cm]{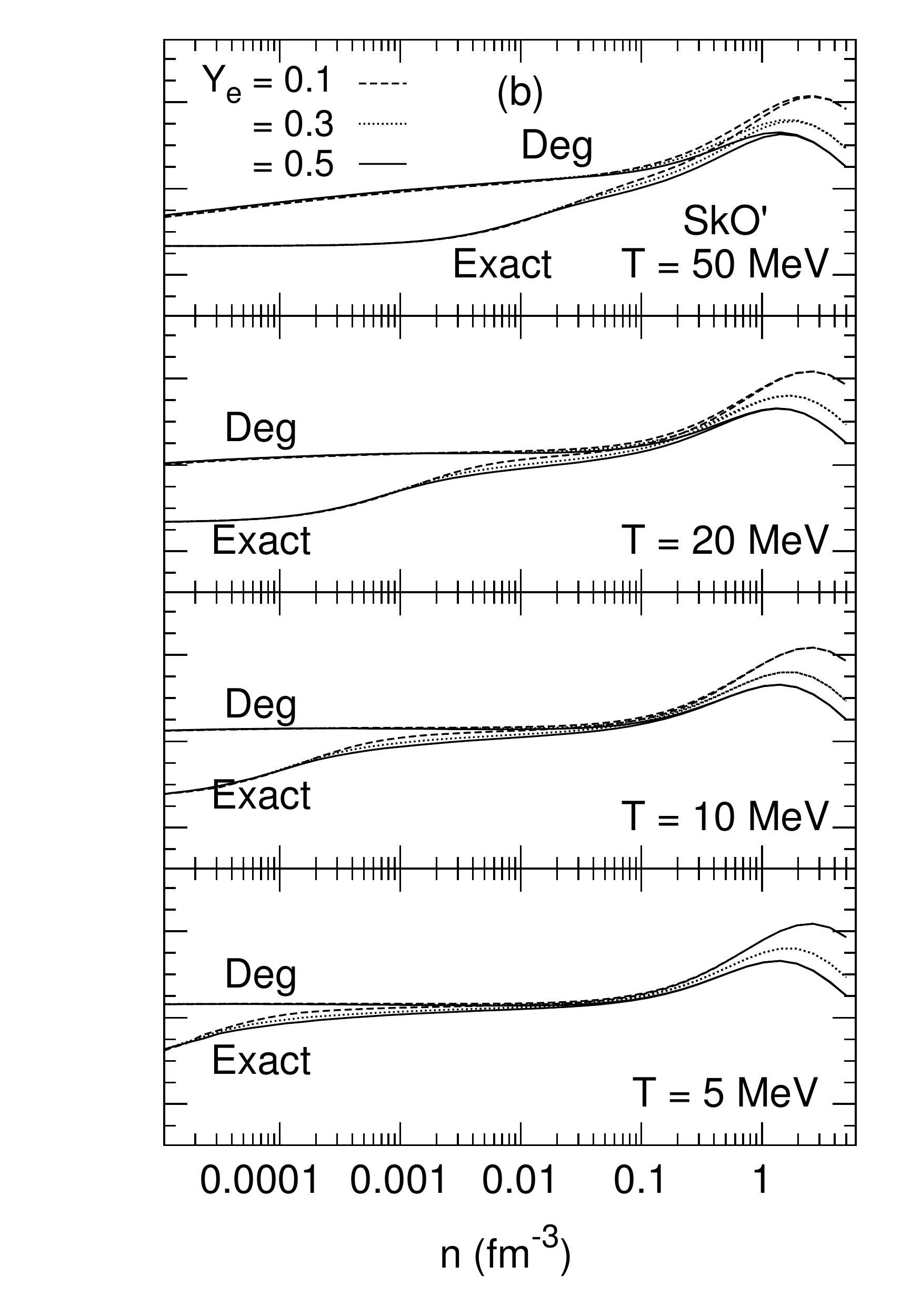}
\end{minipage}
\hspace{-0.032cm}
\begin{minipage}[b]{0.33\linewidth}
\centering
\includegraphics[width=6.3cm,height=11cm]{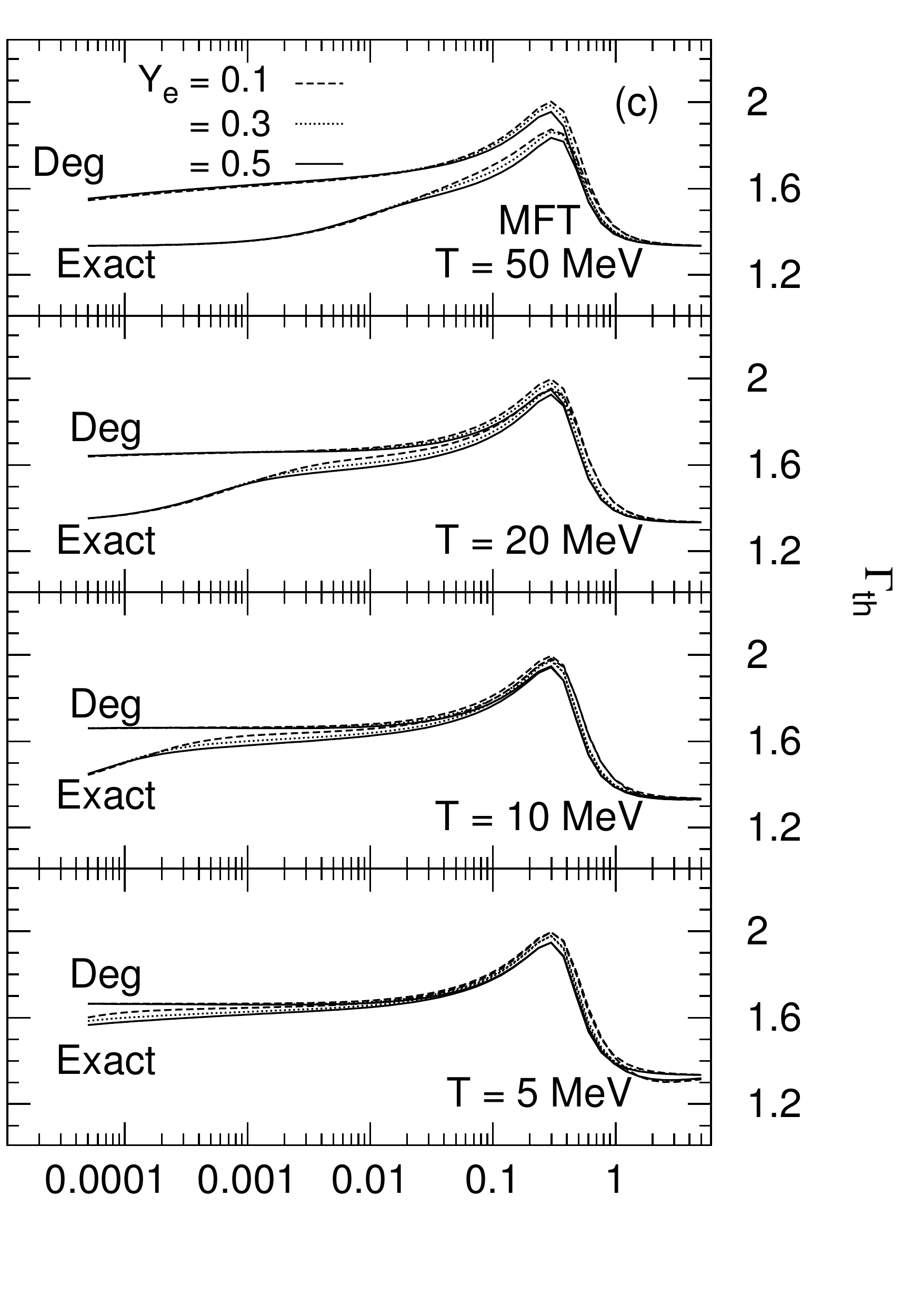}
\end{minipage}
\vskip -0.5cm
\caption{Same as Fig. \ref{Gamma_th_NoLep}, but with the inclusion of leptons and photons.}
\label{Gamma_th}
\end{figure*}

\subsection{Adiabatic index}

In hydrodynamics, compressions and rarefactions are adiabatic (that is, isentropic) rather than isothermal. The 
adiabatic index defined by
\begin{equation}
\Gamma_S = \left.\frac{\partial \ln P}{\partial \ln n}\right|_S
       = \frac{n}{P}\left.\frac{\partial P}{\partial n}\right|_S
       \label{gamS1}
\end{equation}
is  a gauge of the fractional variation of local pressure with density and hence the stiffness of the EOS. 
For our purposes it is more convenient to transform to the variables $(n, T)$ which is
achieved by the use of Jacobians \cite{LLI}:
\begin{eqnarray}
\Gamma_S &=& \frac{n}{P}\frac{\frac{\partial(P,S)}{\partial(P,T)}\partial(P,T)}
                           {\frac{\partial(n,S)}{\partial(n,T)}\partial(n,T)}  \nonumber \\
       &=& \frac{n}{P}\frac{\left.\frac{\partial S}{\partial T}\right|_P}
                           {\left.\frac{\partial S}{\partial T}\right|_n}
                            \frac{\partial(P,T)}{\partial(n,T)}                \               
       =  \frac{C_P}{C_V} \frac{n}{P}  \left.\frac{\partial P}{\partial n}\right|_T .
       \label{gamS2}
\end{eqnarray}
The calculation of $\Gamma_S$ is facilitated by the isentropes 
 in the $n$-$T$ plane shown in Fig. \ref{Scontours}. In contrast to the results shown in 
Fig. \ref{MDYISk_EntContours},  contributions from leptons and photons  in addition to those from nucleons are included in the results of Fig. \ref{Scontours}. 
In order to calculate $\Gamma_S$ from Eq. (\ref{gamS2}) one first selects values of $(n,T)$ 
for which $S(n,T)$ is a prescribed constant. These $n$ and $T$  are in turn used to calculate  $P$, its derivative $(\partial P/\partial n)|_T$, and the specific heats $C_P$ and $C_V$.  Alternatively, $P$ can be expressed analytically as a function of $(S,n)$ (as for example, in the degenerate and non-degenerate limits). We have verified that these two approaches yield identical results.

\begin{figure*}[!htb]
\centering
\begin{minipage}[b]{0.5\linewidth}
\centering
\includegraphics[width=8.8cm]{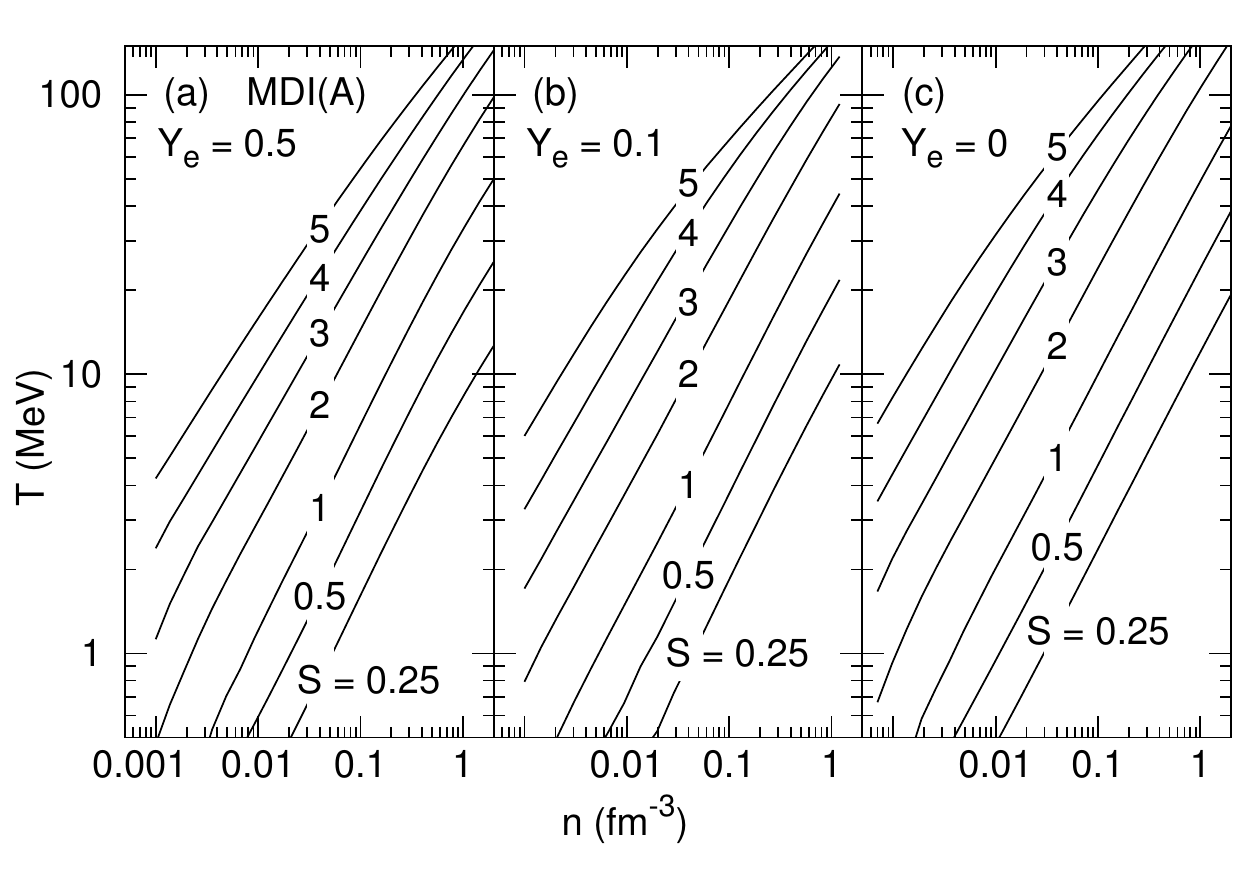}
\end{minipage}
\hspace{-0.5cm}
\begin{minipage}[b]{0.5\linewidth}
\centering
\includegraphics[width=8.8cm]{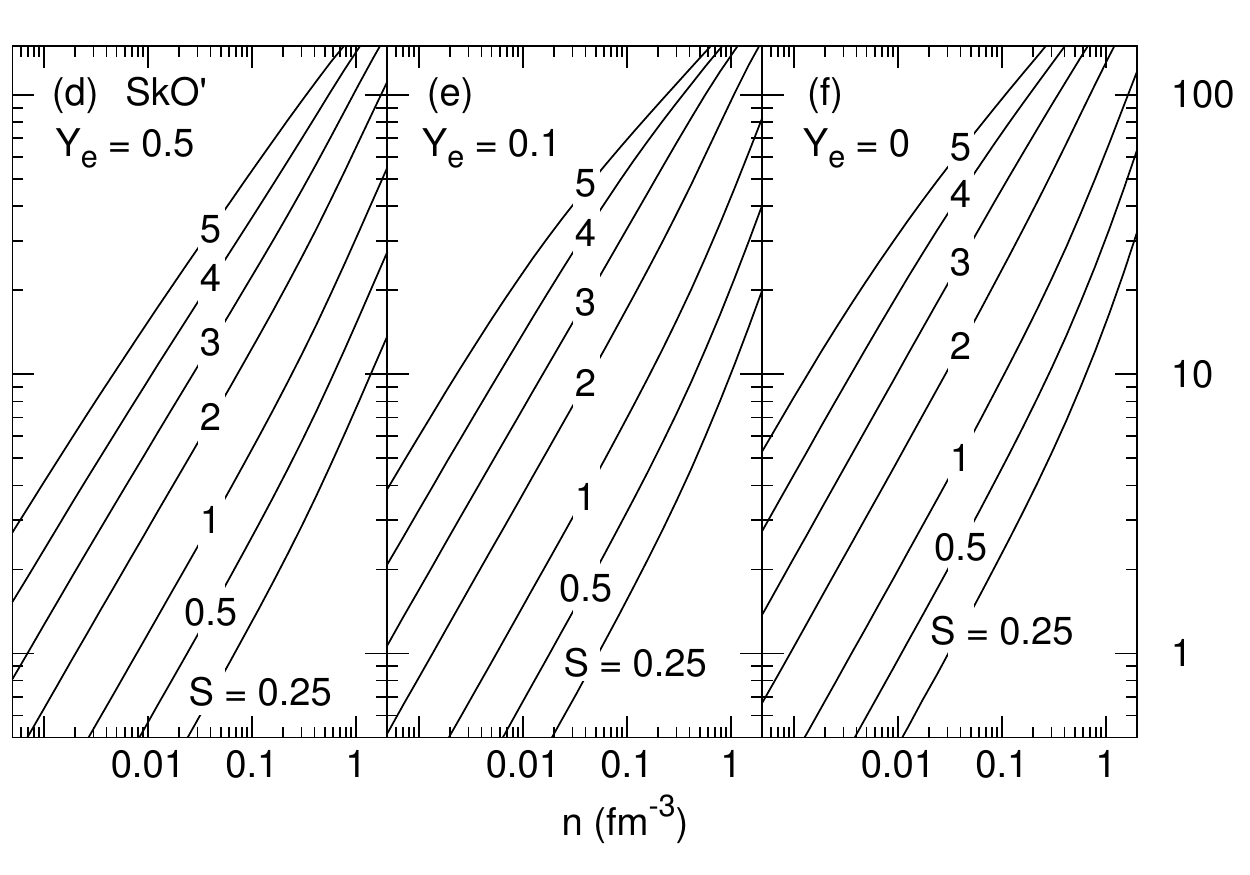}
\end{minipage}
\vskip -0.5cm
\caption{Isentropes for the MDI(A) and SkO$^\prime$ models at the indicated lepton fractions for matter with
nucleons, leptons and photons.}
\label{Scontours}
\end{figure*}

In the derivation of a degenerate limit expression for $\Gamma_S$ we must be 
mindful of the fact that even though $C_P\simeq C_V$ in FLT, the ratio $C_P/C_V$ 
cannot be set equal to unity. This is because $C_P$ and $C_V$ are polynomials 
in odd powers of $T$ in this limit, and thus
\begin{equation}
\frac{C_P}{C_V} = 1 + C(n)T^2 + \ldots ~~~\,,
\end{equation}
where $C(n)$ is a density dependent factor.
The $T^2$ term cannot be ignored at this level of approximation as it is 
of the same order in $T$ as the thermal pressure, $P_{th}$. We circumvent this 
problem by turning to Eq. (\ref{cp}), and write the adiabatic index as 

\begin{eqnarray}
\Gamma_S &=& \frac{n}{P_0 + P_{th}}\left\{\left.\frac{\partial(P_0+P_{th})}{\partial n}\right|_T
           + \frac{T}{n^2 C_V}\left[\left.\frac{\partial(P_0+P_{th})}{\partial T}\right|_n\right]^2\right\} \nonumber \\
       &=& \frac{n}{P_0 + P_{th}}\left[\frac{K}{9}+\left.\frac{\partial P_{th}}{\partial n}\right|_T
           + \frac{T}{n^2 C_V}\left(\left.\frac{\partial P_{th}}{\partial T}\right|_n\right)^2\right]    
           \label{gamS3}
\end{eqnarray}
where $P_0$ and $K$ are the cold pressure and incompressibility, respectively,
the baryonic components of which are listed in Appendix \ref{Sec:AppendixA}.

Under degenerate conditions (when contributions from photons are not significant), $P_{th}$ can also be expressed in terms of $S$ (to leading order) as \cite{APRppr,Prakash:97}
\be
P_{th} = \frac {nS^2}{6}  \frac {\sum_i a_i Y_i Q_i}{\left( \sum_i a_i Y_i\right)^2 }, \quad i=n,p,e
\label{gamS4}
\ee
which allows $\Gamma_S$ to be calculated in a relatively simple manner and provides a check on results from Eq. (\ref{gamS3}).  

When non-degenerate conditions prevail, analytical expressions for $P_{th}$, $S$, $C_V$, $(\partial P/\partial T)|_n$, and $(\partial P/\partial n)|_T$ in Eqs. (160), (161), (168), (170) and (171) of Ref. \cite{APRppr} are useful, and  are employed here. In the semi-degenerate region, numerical calculations are unavoidable. 

\subsubsection*{Results at zero temperature}

As $\Gamma_S$ receives significant contributions from the zero temperature pressure and its variation with density, we begin our analysis by examining $\Gamma_{S=0}$ which highlights the role of contributions from electrons. Figure \ref{MDYISkOp_Gam_0T}
shows $\Gamma_{S=0}$ at select values of $Y_e$ for the two models employed here. The dashed curves in this figure are for nucleons only, whereas the solid curves include the contributions from leptons.   Noteworthy features of the results for nucleons are: 
(i) they diverge at values of $n$ for those $Y_p=Y_e$ for which the pressure vanishes, (ii) they are negative in some regions of sub-saturation densities indicating mechanical instability, and (iii) they approach constant values for asymptotically low and high densities. The inclusion of charge-balancing electrons renders $\Gamma_{S=0}$ to  (i) be positive for all $n$ thereby restoring mechanical stability, (ii) exhibit a gradual variation from a low to high value with the largest variation occurring  at sub-saturation densities, and (iii) approach constant values for asymptotically low and high densities. These features can be quantitatively understood by examining  the nucleon- and lepton- pressures as functions of density 
(see Fig. \ref {MDYISkOp_nuc_lep_pres}) as the following analysis shows.


%
\begin{figure*}[!htb]
\centering
\begin{minipage}[b]{0.5\linewidth}
\centering
\includegraphics[width=9.2cm]{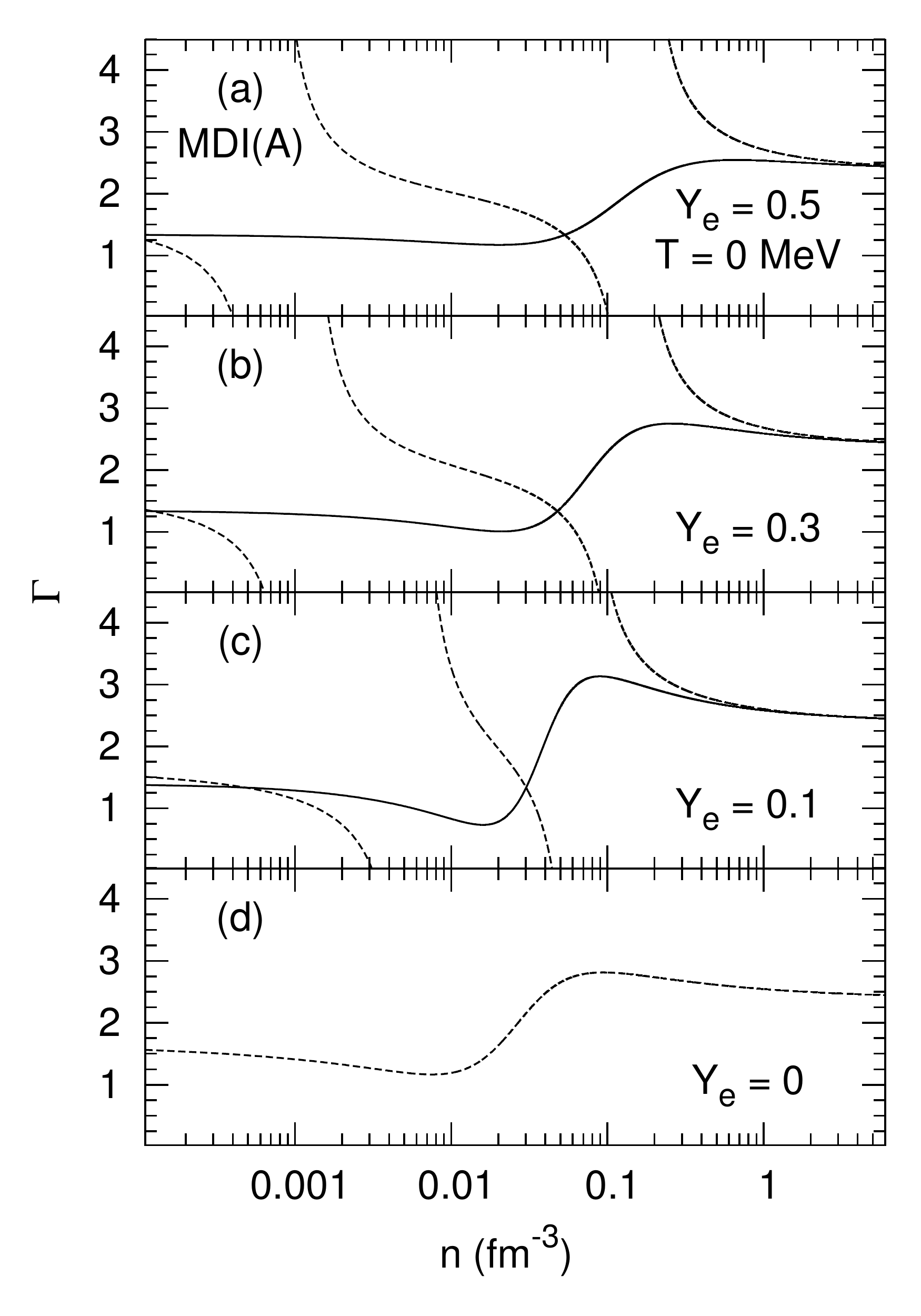}
\end{minipage}
\hspace{-1cm}
\begin{minipage}[b]{0.5\linewidth}
\centering
\includegraphics[width=9.2cm]{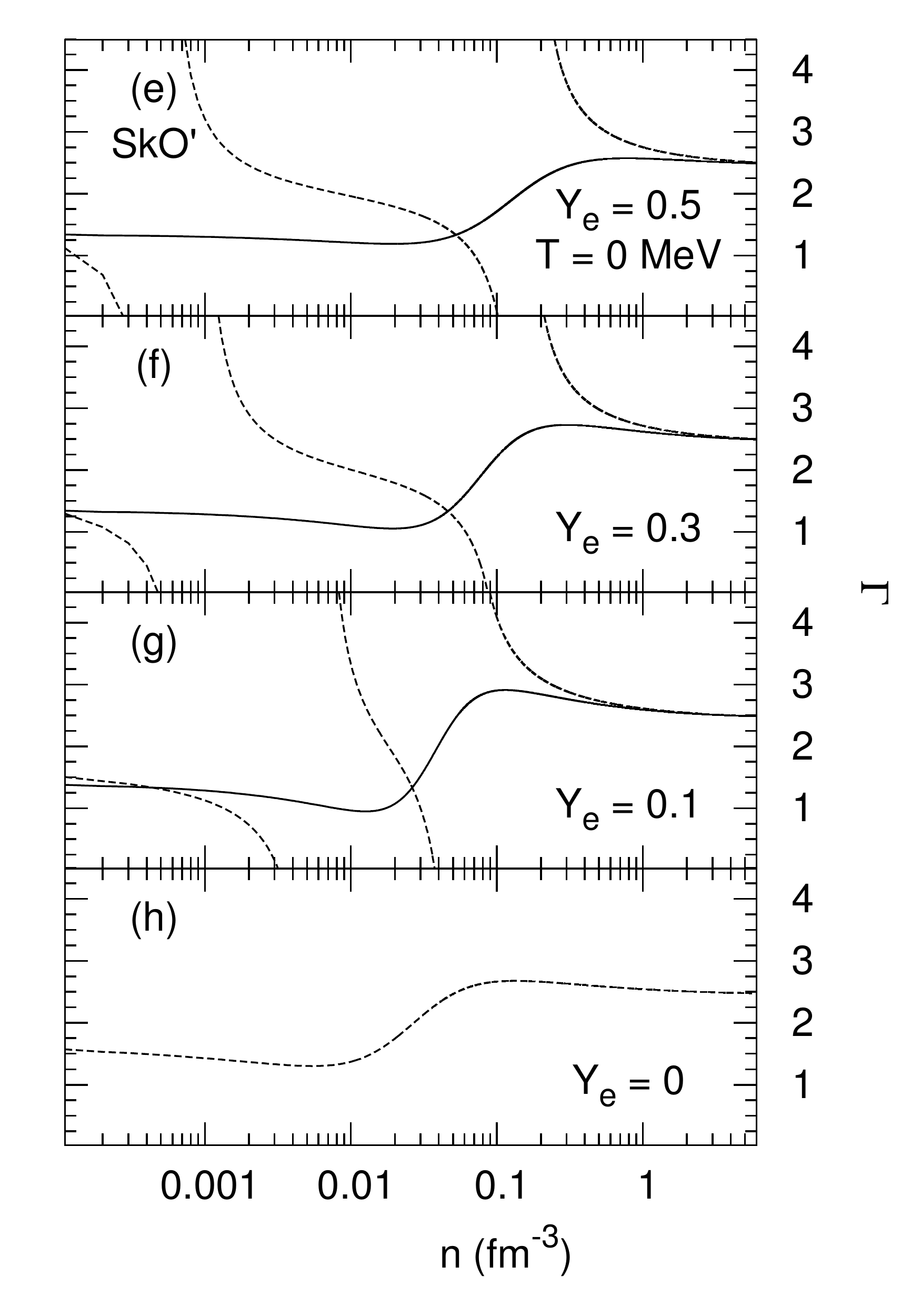}
\end{minipage}
\vskip -0.5cm
\caption{The zero temperature adiabatic index $\Gamma_{S=0}$ for the MDI(A) and SkO$^{\prime}$ models at 
at the indicated proton fractions. Dashed curves include only nucleons while solid lines include
nucleons and leptons.}
\label{MDYISkOp_Gam_0T}
\end{figure*}

\begin{figure}[!htb]
\includegraphics[width=9.2cm]{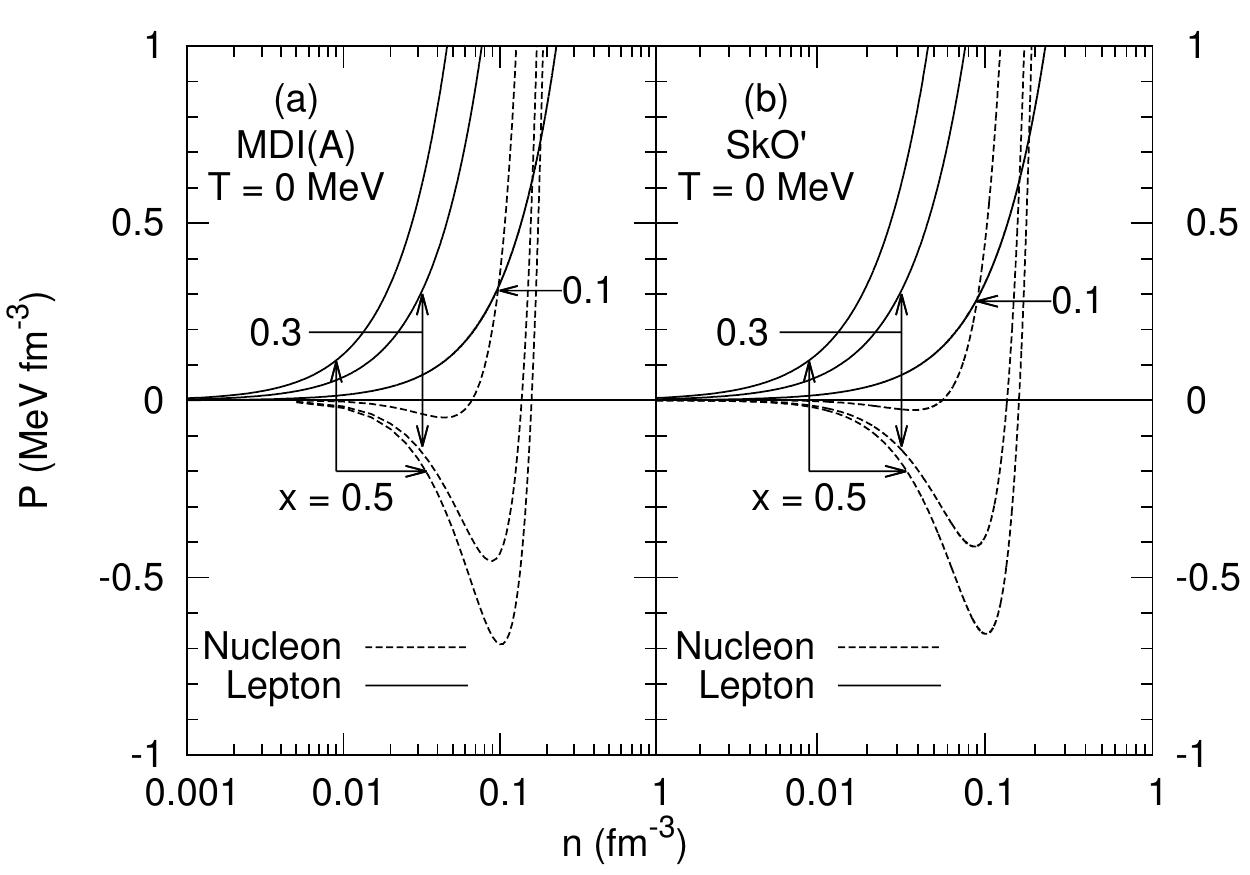}
\caption{Zero temperature nucleon (dashed curves) and lepton (solid curves) pressures at the 
indicated proton fractions.}
\label{MDYISkOp_nuc_lep_pres}
\end{figure}

At zero temperature (entropy), the pressure of relativistic leptons (electrons) can be written as 
\be
P_l = \frac {\hbar  c}{4} n_0 (3\pi^2n_0)^{1/3} Y_e^{4/3} \left( \frac {n}{n_0}\right)^{4/3} \,.
\label{plep}
\ee
Thus, the adiabatic index of matter with baryons and leptons takes the form
\be
\Gamma_{S=0} = \left( \frac 43 + \frac {n}{P_l} \frac {dP_b}{dn} \right) \left( 1 + \frac {P_b}{P_l} \right)^{-1}\,,
\label{GamS0}
\ee
where use of $dP_l/dn = (4/3)P_l$ has been made. For $n$ not too far from $n_0$, the pressure of baryons is well approximated by 
\ba
P_b \simeq P_b(n,x=1/2) &+& n_0(1-2x)^2 u^2 \frac {dS_2}{du} \nonumber \\
P_b(n,x=1/2) &\simeq& \frac {K_0n_0}{9}~u^2(u-1) \,,
\label{pbar}
\ea
where $u=n/n_0$. As demonstrated  below, the above relations allow us to gain an analytical understanding of the magnitude of $\Gamma_{S=0}$ in terms of quantities accessible to nuclear experiments. 

For SNM, $P_b=0$ at $u=1$ so that $\Gamma_{S=0}$ takes the simple form 
\be
\Gamma_{S=0} (u=1)= \frac 43 + \frac {(K_0/9)}{(P_l/n_0)} \,.
\ee
With $P_l=5.26~{\rm MeV~fm}^{-3}$ and values of $K_0$ and $n_0$ from Table. \ref{propsMDYINSkO'} at $u=1$, we obtain $\Gamma_{S=0}=2.12(2.09)$ for the MDI(A)[SkO$^\prime$] model, in very good agreement with the exact results in 
Fig. \ref{MDYISkOp_Gam_0T}.

The density at which mechanical instability of cold baryons-only matter (often referred to as the spinodal instability) sets in is determined by $dP_b/dn = 0$. In this case, 
\be
\Gamma_{S=0}^{(sp)} = \frac 43 \left( 1 + \frac {P_b}{P_l} \right)^{-1} \,,
\ee 
where the superscript $(sp)$ stands for spinodal. Figure \ref{MDYISkOp_nuc_lep_pres} shows the baryon and lepton pressures vs density from which the density regions in which mechanical or spinodal instability occurs for baryons-only matter for the two models can be ascertained. For SNM, Eq. (\ref{pbar}) implies that $u_{(sp)} = 2/3$, in good agreement with the exact results for both models. From Eqs. (\ref{plep}) and 
(\ref{pbar}), the leptonic and baryonic pressures at $u_{(sp)}$ are
\ba
P_l(u_{(sp)}) &=& 3.06~{\rm MeV~fm^{-3}} \nonumber \\
P_b(u_{(sp)}) &=& -  (4/243)~K_0n_0 \,,
\ea
which yield $\Gamma_{S=0}^{(sp)} \simeq 1.67(1.65) $ for the MDI(A)[SkO$^\prime$] model in good accord with the exact results.

Equation (\ref{pbar}) further implies that the density at which the derivative of the baryonic pressure, $dP_b/du$ is maximum occurs at $u=1/3$.  At this density, 
\ba
P_b(u=1/3) &=& - (2/243)~K_0n_0 \nonumber \\
P_l (u=1/3) &=& 1.22~{\rm MeV~fm^{-3}} \,. 
\ea
The corresponding expression for $\Gamma_{S=0}$ is
\be
\Gamma_{S=0}(u=1/3) = \left( \frac 43 - \frac {1}{81} \frac {K_0n_0}{P_l}\right)
\left(1- \frac{2}{243}  \frac {K_0n_0}{P_l}\right)^{-1} \,,  
\ee
which yields a numerical value of $\simeq 1.28$ for both the MDI(A) and SkO$^\prime$ models. Note that $u=1/3$ marks the density at which $\Gamma_{S=0}$ begins to rise for both models. 

For matter with $Y_e = 0$ (PNM), 
\be
\frac {dP_b}{dn} = \frac {K(n)}{9} + 2u S_2^\prime + u^2 S_2^{\prime\prime} \,,
\ee 
using which we can write 
\ba
\Gamma_{S=0} &=& \left[\frac {K_0}{9} (u-1) + S_2^\prime\right]^{-1} \nonumber \\
&\times & \left[\frac {K_0}{9} (3u-2) + 2S_2^\prime + uS_2^{\prime\prime}\right] \,,
\ea
where the primes above denote derivatives with respect to $u$. The relation above highlights the roles of the first and second derivatives of the symmetry energy $S_2(n)$ at sub-saturation densities of SNM. 
For the special case of PNM at $u=1$, the adiabatic index therefore becomes
\be
\Gamma_{S=0}(u=1) = 2 + \frac 13 \frac {K_0+K_v}{L_v} \,.
\ee
Utilizing the values of $K_0$, $K_v$ and $L_v$ in Table \ref{propsMDYINSkO'}, we obtain $ \Gamma_{S=0}(u=1) = 2.82 (2.69)$ for the MDI(A)[SkO$^\prime$] model in excellent agreement 
with the exact results of 2.77 and 2.67, respectively.   

For PNM at $u=2/3$, $\Gamma_{S=0}$ simplifies to 
\be
\Gamma_{S=0}(u=2/3) = 2 \left( 1- \frac {1}{27} \frac {K_0}{S_2^\prime}  \right)^{-1}
\left(1+ \frac 13 \frac {S_2^{\prime\prime} }{S_2^\prime} \right) \,,
\ee
where the derivatives above are evaluated at $u=2/3$.  The values of $S_2^\prime$ and $S_2^{\prime\prime}$ at $u=2/3$ are 24.93(-12.15) and 26.48(-12.9) for the 
MDI(A) [SkO$^\prime$], respectively. The corresponding values of
$\Gamma_{S=0}(u=2/3)$ are 2.56 and 2.43 for the two models, which match closely with
results of exact numerical calculations that yield 2.81 and 2.67, respectively. The differences from the exact results can be attributed to terms involving $S_4$ and its derivatives.

For values $0 < Y_e < 0.5$, an analysis similar to that presented above can be performed using Eq. (\ref{pbar}). 
The density at which the spinodal instability sets in, determined by solving 
\be
(3u-2) + (1-2x)^2 \left(\frac {18S_2^\prime}{K_0} + \frac {9uS_2^{\prime\prime}}{K_0}  \right) =0 \,, 
\ee
steadily shifts to lower values of $u$ as $x$ decreases toward 0.  Corrections due to to the skewness in $E(n,x)$ and contributions from $S_4(n,x)$ {\it etc.,} are small and do not affect the qualitative behaviors.     
As for $Y_e=0.5$, the contribution from electrons to the total pressure entirely 
removes  the mechanical (spinodal) instability present in baryons-only matter for all $Y_e$
as also confirmed by the exact numerical results. 

At asymptotically low densities and for $Y_e\neq 0$,  $\Gamma_{S=0}$ approaches 4/3 - the value characteristic for relativistic electrons. For PNM, the corresponding value is 5/3, as expected for non-relativistic neutrons.  For asymptotically high densities and for all $Y_e$, $\Gamma_{S=0}$ is controlled by a combination of the highest powers of density in the expression for the pressure of baryons; both models yield the value of $\simeq 2.5$.  

\subsubsection*{Results for finite entropies}

In Fig. \ref{SkOp_InterpS1_Gamma}, $\Gamma_{S=0}$ and  $\Gamma_{S=1}$ for matter with baryons, leptons and photons are contrasted for several values of $Y_e$.  A characteristic feature to note is that $\Gamma_{S=1} < \Gamma_{S=0}$ for densities beyond a certain $Y_e$-dependent density, whereas the opposite is the case below that density for each $Y_e$. The densities at which  $\Gamma_{S=1} = \Gamma_{S=0}$ occurs at sub-saturation densities and ranges from $\sim 0.02$-0.1 fm$^{-3}$ for $Y_e$ in the range 0-0.5.  The approach to the corresponding low- and high- density asymptotic values also depends on $Y_e$. The largest differences between the results of $S=0$ and $S=1$ are for PNM, the least differences being those for SNM. 

As $S=1$ represents matter that is degenerate for all but the very low densities, the ingredients that cause the behaviors in 
Fig. \ref{SkOp_InterpS1_Gamma} can be identified by utilizing the results of FLT. Here we provide relations applicable to the cases of SNM and PNM. Those for intermediate $Y_e$'s are straightforward to develop, albeit  lengthy. We start from
\be
\Gamma_S = \left.\left[ \frac {n}{P_0+P_{th}}   \frac {d}{dn}(P_0+P_{th})\right|_S \right]
\label{GamFinS1}
\ee
with $P_0=P_{b0}+P_{l0}$. The leptonic pressure and its derivative at zero temperature (entropy), $P_{l0}$  and $dP_{l0}/dn$, are obtained from Eq. (\ref{plep}).  The cold baryonic pressure $P_{b0}$ is that from an appropriate model for baryons (MDI(A) or SkO$^\prime$ in this paper), and its derivative $dP_{b0}/dn = K(n)/9$. In the degenerate limit of SNM in which $p_{F_{b}} = p_{F_{e}} = 
(3\pi^2n/2)^{1/3}$, the thermal pressure and its derivative are
\ba
P_{th} &=& \frac {S^2}{6} n~\left(a_bQ_b + \frac {a_e}{4} \right) \left(a_b+\frac {a_e}{2} \right)^{-2} \nonumber \\
\frac {dP_{th}}{dn} &=& \frac {P_{th}}{n} \left[ 1 + \frac 43 \left(a_bQ_b+\frac {a_e}{4}\right) \left(a_b+\frac {a_e}{2}\right)^{-1} \right.\nonumber \\
&+&\left. \left(a_bn\frac {dQ_b}{dn} - \frac {2}{3} a_bQ_b^2 - \frac {a_e}{12} \right) \left(a_bQ_b+\frac {a_e}{4}   \right)^{-1} \right] \,.\nonumber \\
\label{pthsdpths}
\ea
The quantities $a_b=\pi^2m_b^*/(2p^2_{F_{b}})$ and $a_e\simeq\pi^2/(2p_{F_{e}})$ are the level density parameters of the baryons and relativistic electrons, respectively.  The quantity $Q_b$ is as in Eq. (\ref{kewi}), and for the SkO$^\prime$ model,
\be
Q_b = 1 + \frac 32 \left( 1 -\frac {m_b^*}{m} \right);  \quad n \frac {dQ_b}{dn} = \frac 32 \frac {m_b^*}{m} \left (1 - \frac  {m_b^*}{m} \right)\,,
\ee
which exemplifies the role of the effective masses in determining $P_{th}$. Also, use of 
$Q_e\simeq 1/2$ has been made in obtaining Eq. (\ref{pthsdpths}) using which $\Gamma_S$ is readily calculated for SNM. 

Relations appropriate for PNM are obtained simply by setting $a_e=0$ in Eq. (\ref{pthsdpths})  with the result
\ba
P_{th} &=& \frac {S^2}{6} \left( \frac {nQ_b}{a_b} \right) \nonumber \\
\frac {dP_{th}}{dn} &=& \frac {P_{th}}{n} \left(1 + \frac 23 Q + \frac {n}{Q_b} \frac {dQ_b}{dn} \right) \,.
\ea
Consequently, the adiabatic index for PNM becomes
\be
\Gamma_s = \frac {n}{P_0+P_{th}} \left[ \frac {K}{9} + \frac {S^2}{6a} \cdot 5 \left(\frac 43 - \frac {m_b^*}{m}  \right) \right] \,
\ee
for the SkO$^\prime$ model, which exhibits the role of the effective mass in determining the thermal component for finite $S$. The result for the MDI(A) model is straightforward to obtain, but is somewhat more lengthy due to its lengthy expression for the baryon effective mass. We have verified that the degenerate limit result for $\Gamma_s$ accurately reproduces the exact results for all values of $Y_e$  shown in Fig. \ref{SkOp_InterpS1_Gamma}. 

Based on a mean-field theoretical model, Shen {\em et al}., \cite{shen2011new}
have presented results for $\Gamma_{S=1}$  for various $Y_e$'s including inhomogeneous phases at sub-saturation densities, and find that it rises abruptly at the transition from non-uniform to uniform matter. In the region of densities for which an appropriate comparison can be made ({\em i.e}., the homogeneous phase), our results agree semi-quantitatively with those of Ref. \cite{shen2011new} differences being due to the different model parameters used.

\begin{figure}[!htb]
\includegraphics[width=7.5cm]{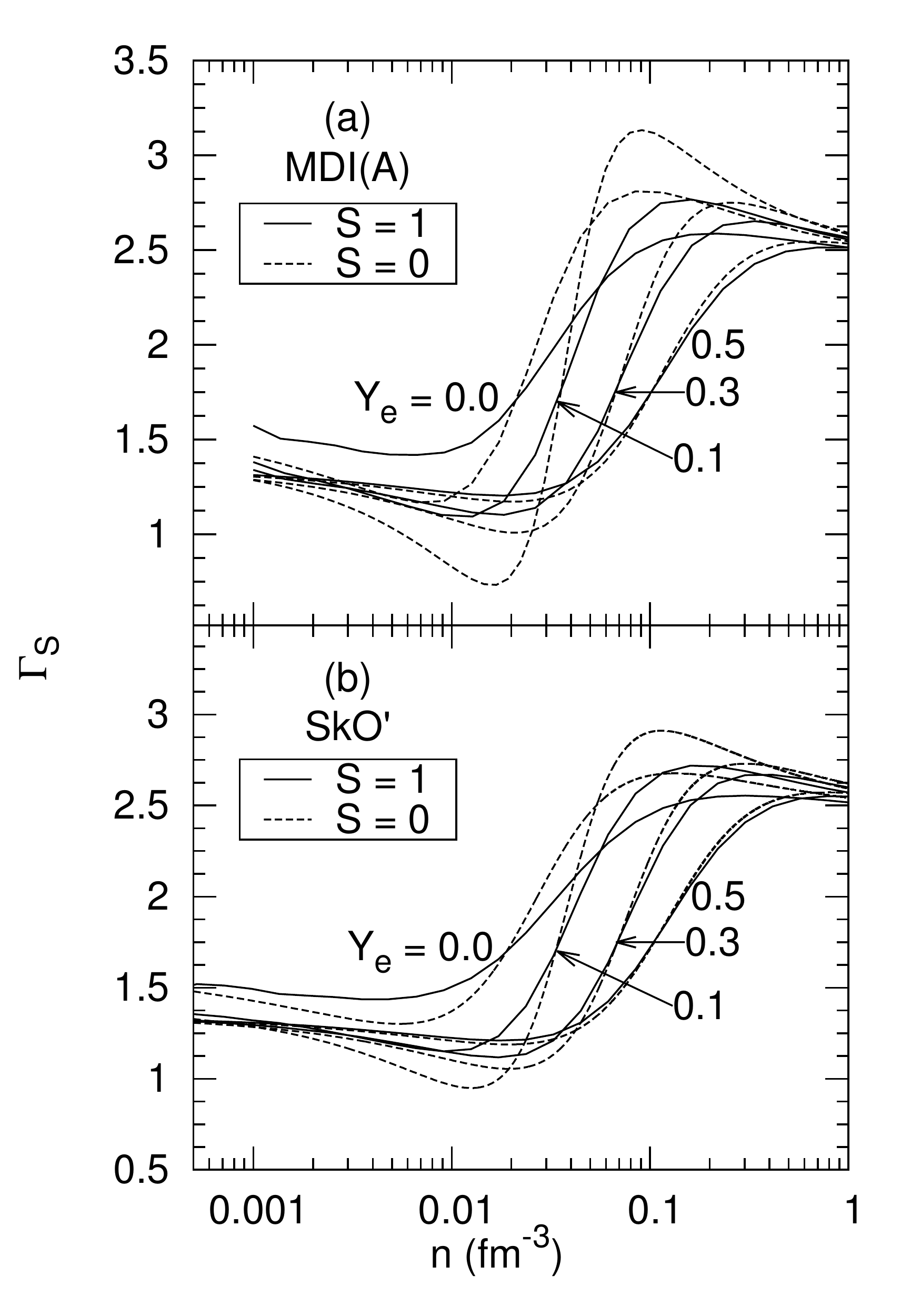}
\caption{Adiabatic index at fixed entropy at the indicated proton fractions for the 
SkO$^\prime$ and MDI(A) models. Results at zero and finite entropies are calculated
using Eqs. (\ref{GamS0}) and (\ref{GamFinS1}), respectively.}
\label{SkOp_InterpS1_Gamma}
\end{figure}

Figure \ref{SkOp_InterpS234_x5_Gamma} shows the effect of increasing S for $Y_e=0.5$. Also shown are results from the degenerate and non-degenerate limit expressions discussed above. The remarkable feature of the results shown is that 
the finite and zero entropy adiabatic indices are equal at nearly the same density, $n_\times \simeq 0.08$-0.09 fm$^{-3}$, 
for a wide range of entropies.  
For a fixed $Y_e$, the intersection density $n_\times$ can be determined by setting $\Gamma_{S=0} = \Gamma_S$ which can be rearranged to
\be
\frac {1}{P_0} \frac {dP_0}{dn} = \left. \left( \frac {1}{P_{th}} \frac {dP_{th}}{dn} \right)\right|_S \,.
\ee
Note that in the degenerate limit, the right hand side becomes independent of $S$ (see the relations in Eq. 
(\ref{pthsdpths})), which explains the near independence with $S$ of $n_\times$ which is mainly determined by the density dependent terms on the left hand side and $a_b, a_e$ and $Q_b$ on the right hand side. 
Insofar as the pressure can be expressed as of function of $Y_i$'s and to higher orders in $S$, a mild dependence of $n_\times$ on $S$ may be expected.   Not surprisingly, the non-degenerate approximation reproduces the exact results for $\Gamma_S$ vs $n$ better than the degenerate approximation with increasing values of $S$.

As noted before,  
the logarithmic derivative $d\ln P_0/d\ln n$ exhibits its largest variation at sub-saturation densities (see Fig. \ref{MDYISkOp_nuc_lep_pres}), whereas 
$(d\ln P_{th}/d\ln n)|_S$ remains relatively  flat (with values that decrease slightly from $\sim 5/3$ with increasing $S$) around the 
intersection density $n_\times$.  In addition, $d\ln P_0/d\ln n > (d\ln P_{th}/d\ln n)|_S$ for $n>n_\times$ but the reverse trend prevails for $n<n_\times$. These general features for all $Y_e$'s make  $\Gamma_S < \Gamma_{S=0}$ for densities larger than $n_\times$ and  $\Gamma_S \geq \Gamma_{S=0}$ for densities lower.

\begin{figure}[!htb]
\includegraphics[width=7.5cm]{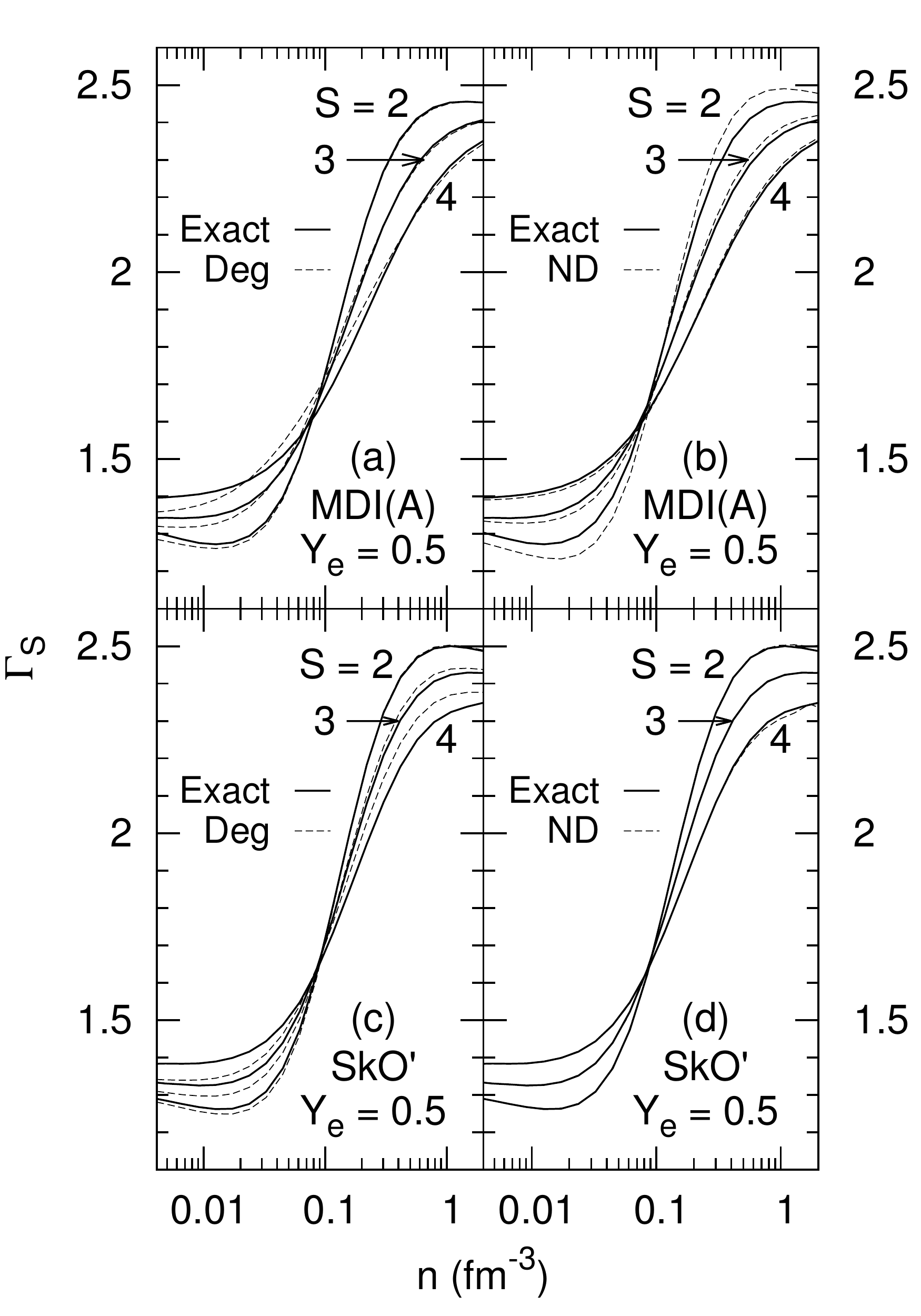}
\caption{Adiabatic index at fixed entropy from the SkO$^\prime$ and MDI(A) models 
for SNM using Eq. (\ref{GamFinS1}). Results for the degenerate 
(left) and non-degenerate (right) limits are from Eqs. (\ref{GamFinS1}) and (\ref{pthsdpths}), respectively.}
\label{SkOp_InterpS234_x5_Gamma}
\end{figure}

The degenerate and non-degenerate approximations to $\Gamma_S$ are compared with the exact numerical results in Fig. \ref{SkOp_GamLims} for $S=2,3,$ and 4. For both models, and for both $Y_e$'s shown, the degenerate approximation reproduces the exact results only for $S=2$ at supra-nuclear densities. As high temperatures are required at these densities for 
$S$ exceeding 2, the non-degenerate approximation fares better being indistinguishable from the exact results.

%
\begin{figure*}[!htb]
\centering
\hspace{-1.25cm}
\begin{minipage}[b]{0.25\linewidth}
\centering
\includegraphics[width=5.75cm]{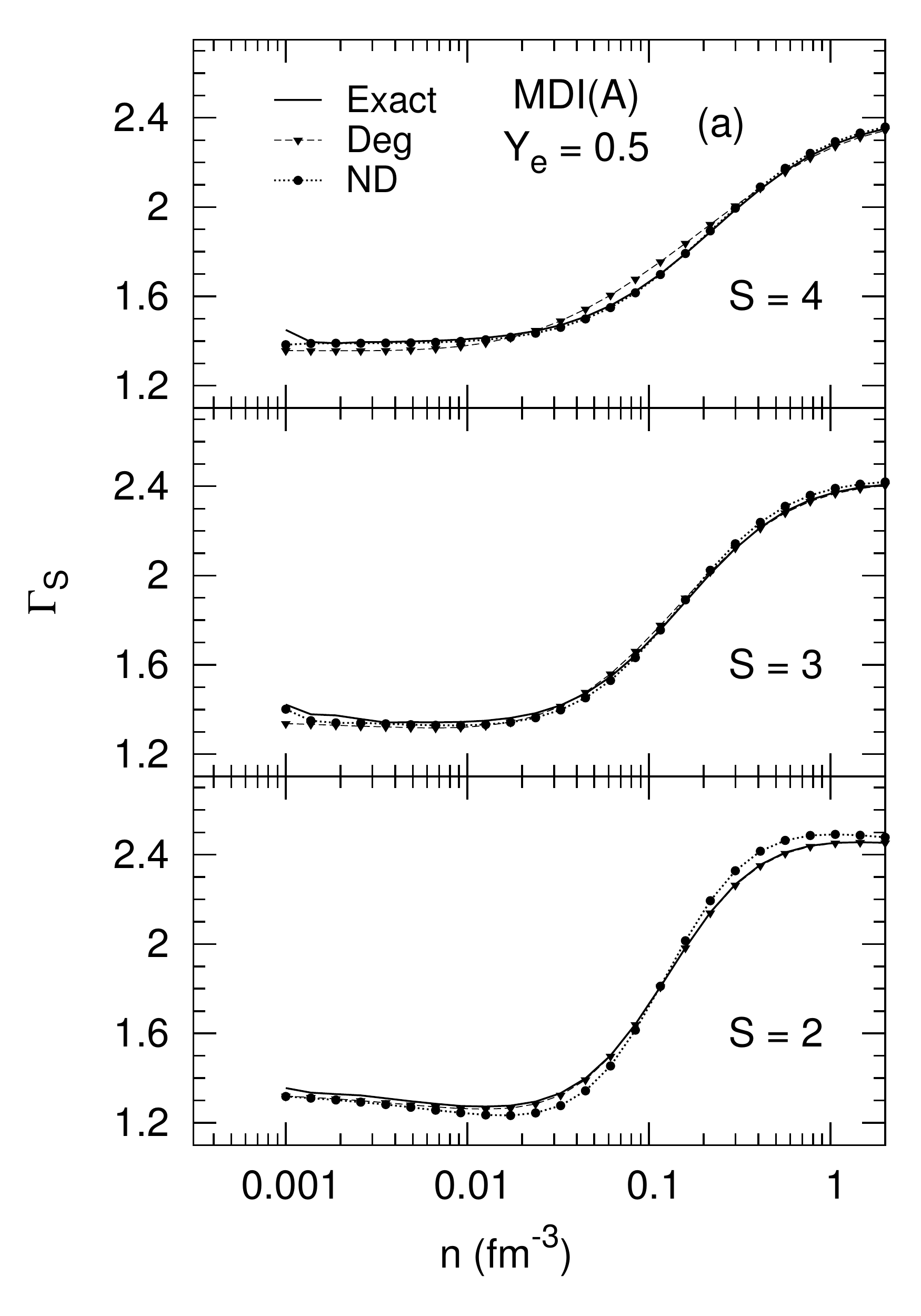}
\end{minipage}
\hspace{-0.05cm}
\centering
\begin{minipage}[b]{0.25\linewidth}
\includegraphics[width=5.75cm]{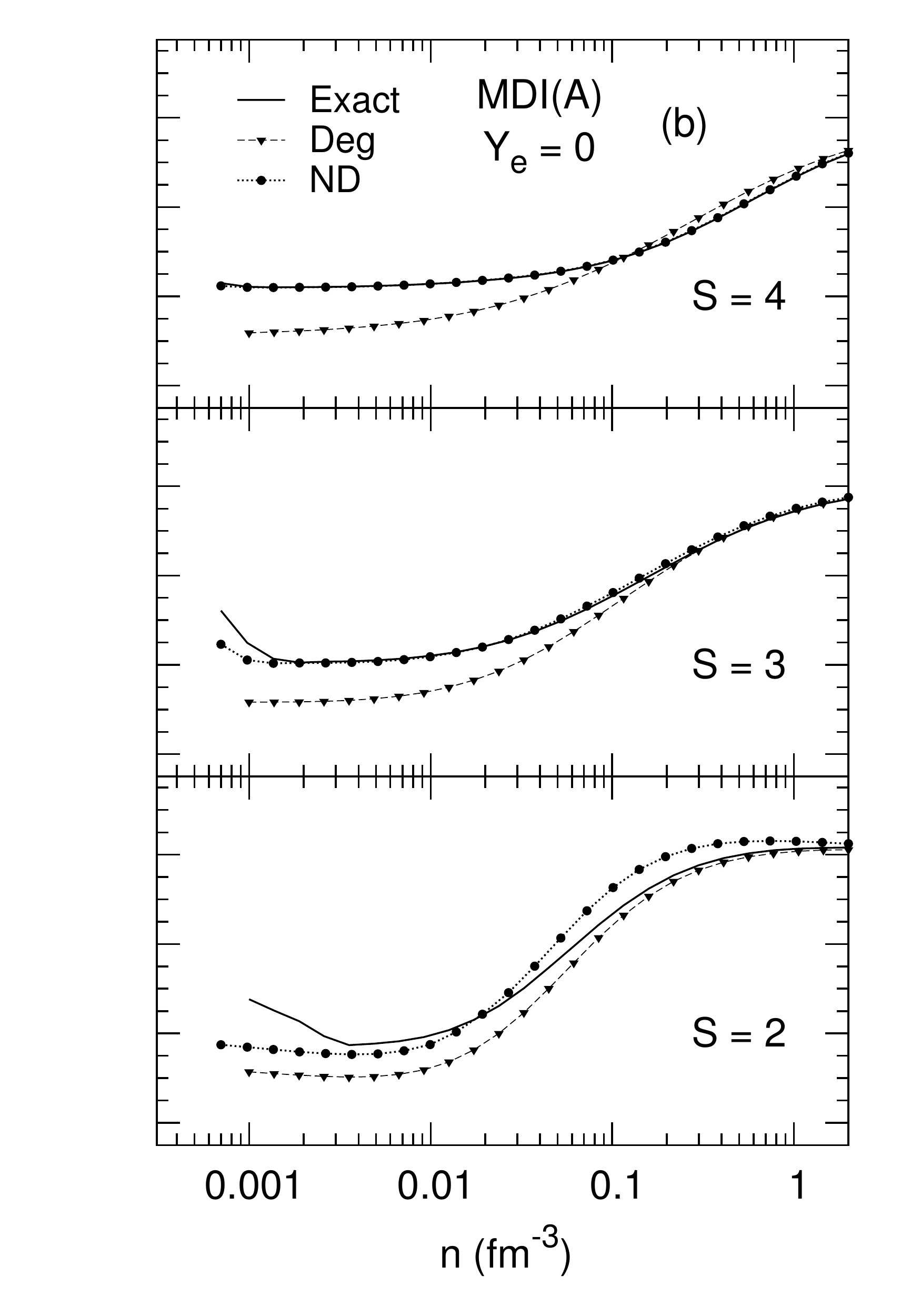}
\end{minipage}
\centering
\hspace{-0.3cm}
\begin{minipage}[b]{0.25\linewidth}
\includegraphics[width=5.75cm]{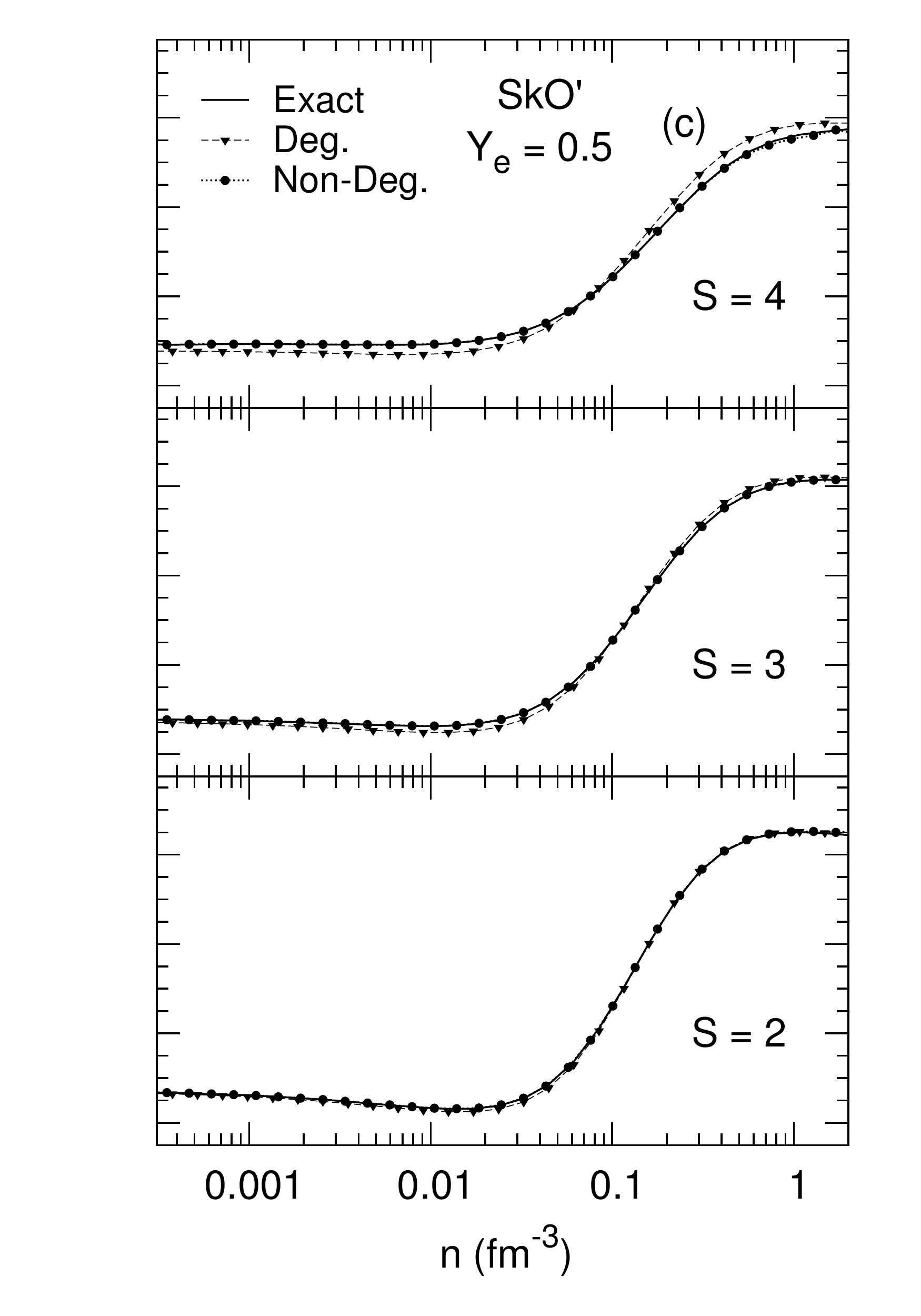}
\end{minipage}
\centering
\hspace{0.3cm}
\begin{minipage}[b]{0.25\linewidth}
\includegraphics[width=5.75cm]{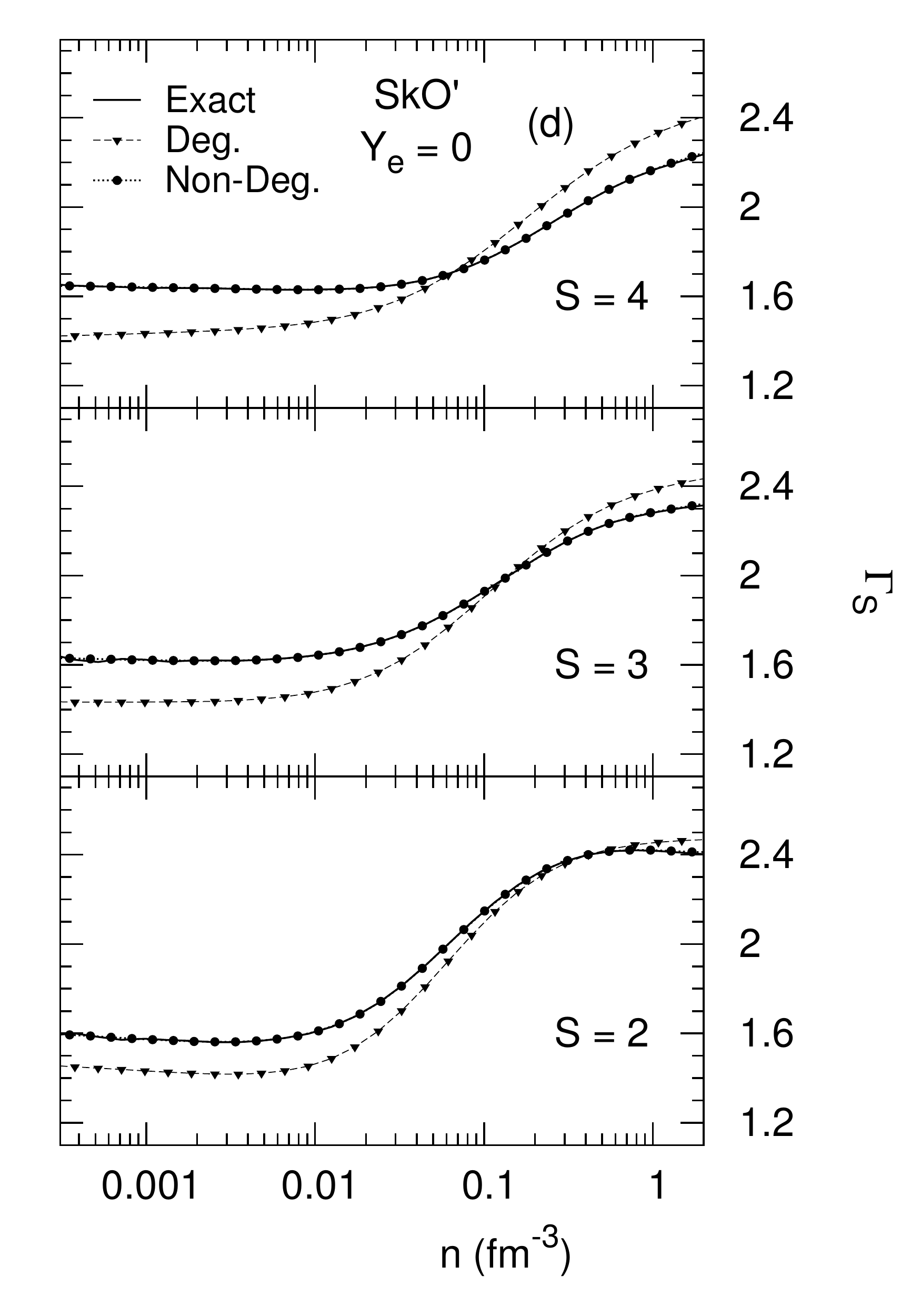}
\end{minipage}
\vskip -0.5cm
\caption{Adiabatic index at constant entropy for the SkO$^\prime$ and MDI(A) models using
Eqs. (\ref{GamFinS1}) and (\ref{pthsdpths}). Results for SNM [(a) and (c)] and 
for PNM [(b) and (d)]  are shown along with their limiting cases.}
\label{SkOp_GamLims}
\end{figure*}

In Fig. \ref{Scontribution}, we show the contributions of nucleons, leptons and photons for fixed entropies $S=1$ and 4. 
For both proton fractions shown, the dominant contributions are from nucleons except at very high densities when contributions from leptons begin to become equally important. 
The density-independent contributions from photons are significant only for large values of $S$.

\begin{figure*}[!htb]
\centering
\begin{minipage}[b]{0.5\linewidth}
\centering
\includegraphics[width=8.75cm]{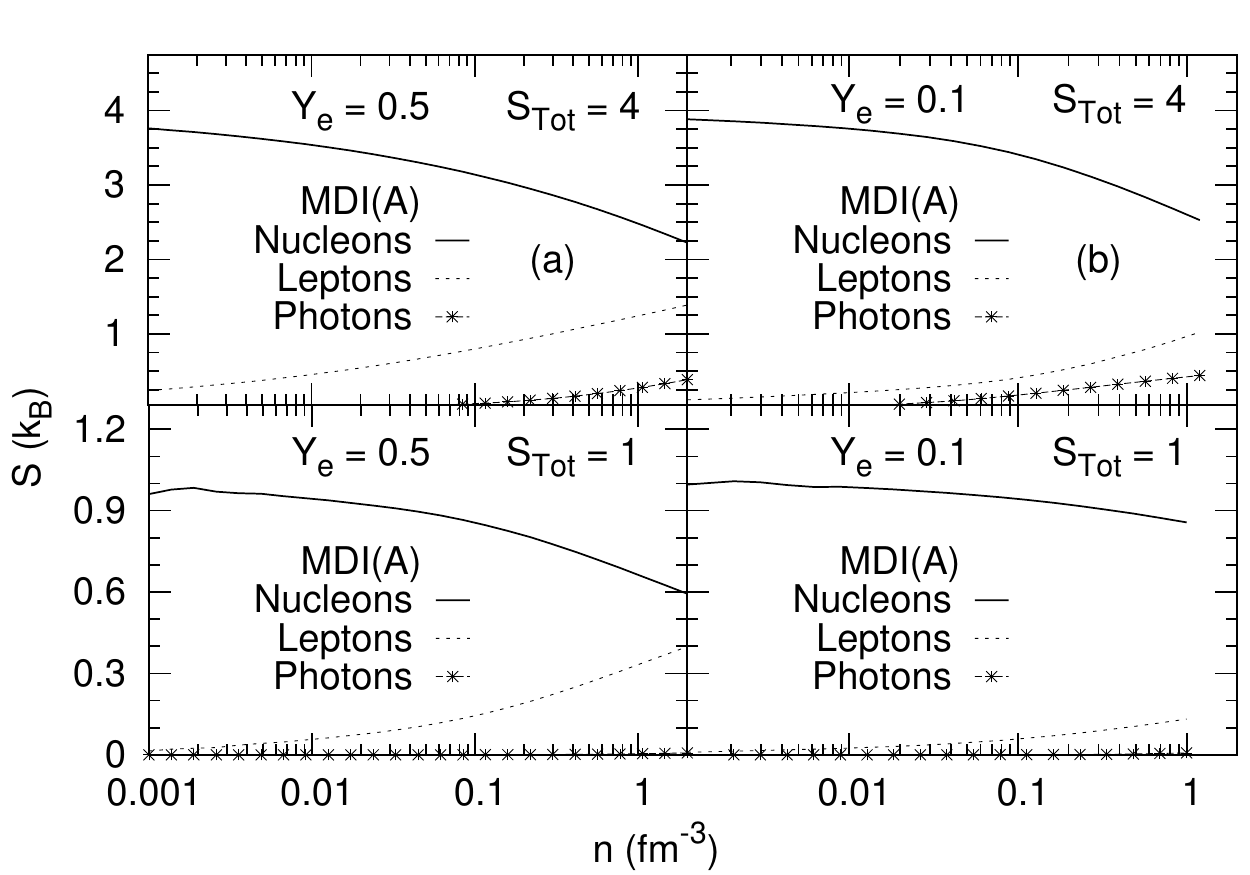}
\end{minipage}
\hspace{-0.5cm}
\begin{minipage}[b]{0.5\linewidth}
\centering
\includegraphics[width=8.75cm]{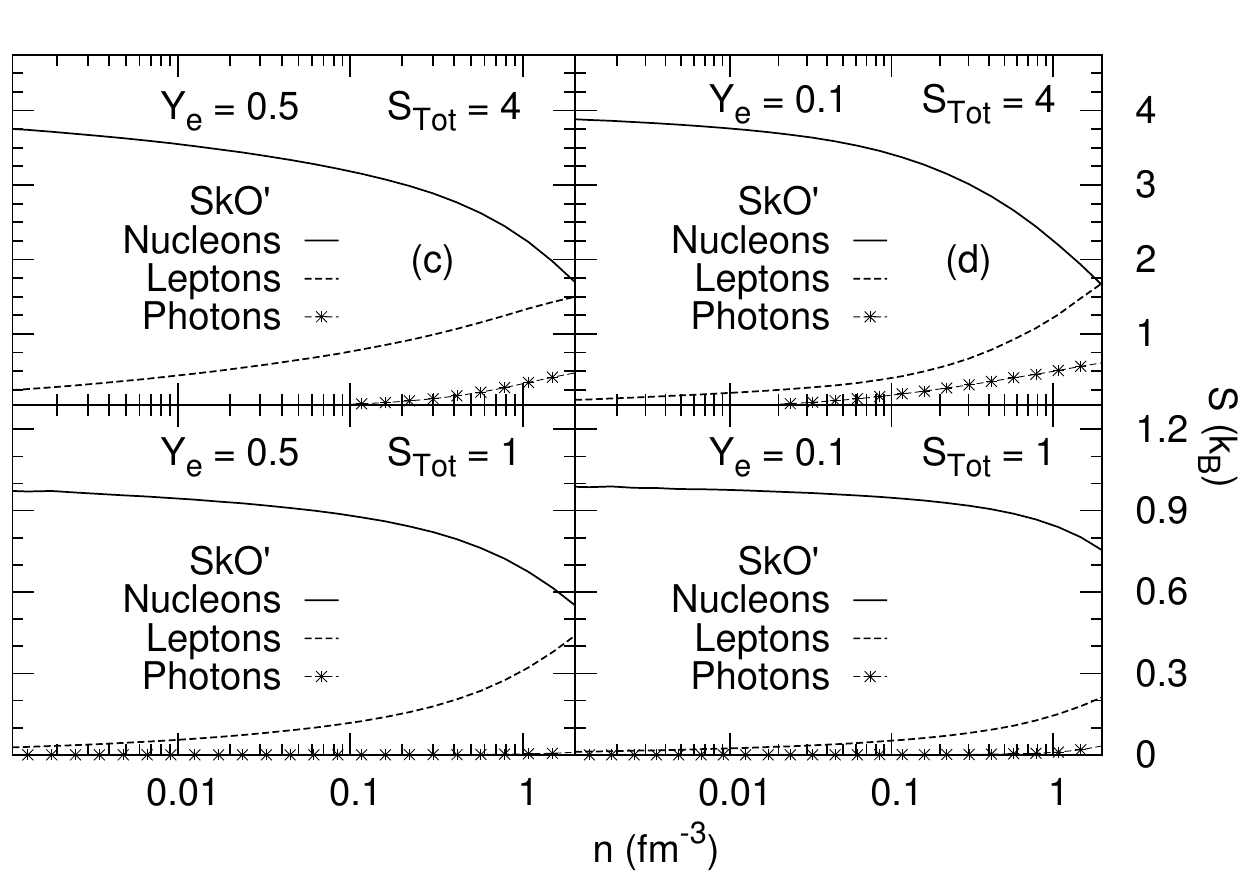}
\end{minipage}
\caption{Contributions from nucleons, leptons and photons to the total entropy at the indicated
proton fractions for the MDI(A) and SkO$^\prime$ models.}
\label{Scontribution}
\end{figure*}

\subsection{Speed of sound}

We conclude this section with a brief discussion of a quantity closely related to $\Gamma_S$, namely the adiabatic speed of sound $c_s$ given by
\be
\left(\frac{c_s}{c}\right)^2 = \left.\frac{dP}{d\epsilon}\right|_{S,Y_e}  \label{cs1} \,,
\ee
where $c$ is the speed of light.  The energy density $\epsilon$ above is inclusive of the rest-mass density; in relativistic 
approaches this is so by default but in nonrelativistic ones, such as in our treatment of nucleons, one must write 
$\epsilon=\varepsilon + mn$, where the first term here is the internal energy density.

Equation (\ref{cs1}) can be expressed in terms of derivatives of the density: 
\ba
\left(\frac{c_s}{c}\right)^2 &=& \left.\frac{dP}{dn}\right|_{S,Y_e}\left(\left.\frac{d\epsilon}{dn}\right|_{S,Y_e}\right)^{-1} \\
                            &=& \frac{\left.\frac{dP}{dn}\right|_{S,Y_e}}{E + m + n\left.\frac{dE}{dn}\right|_{S,Y_e}}.       \label{cs2}
\ea
Then by taking advantage of the total differential of the energy per particle
\be
dE = TdS + \frac{P}{n^2}dn + (\mu_p+\mu_e-\mu_n)dY_e
\ee
we transform Eq. (\ref{cs2}) to
\ba
\left(\frac{c_s}{c}\right)^2 &=& \left.\frac{dP}{dn}\right|_{S,Y_e}\frac{1}{E + m+ \frac{P}{n}}  \label{cs3} \\
                            &=& \Gamma_S\frac{P}{h + mn}  \label{cs4}
\ea
thereby making the connection between $c_s$ and $\Gamma_S$ explicit. 
Here, $h = \varepsilon +P$ is the enthalpy density. Equation (\ref{cs4}) generalizes the definition of the squared speed of sound to finite entropy. This result can also be derived from a time-dependent analysis of traveling sound waves in a medium (see, {\it e.g}, 
\cite{Weinberg1971,Guichelaar74}).  
Note that Eqs. (\ref{cs3}) and(\ref{cs4}) define $c_s$ in 
terms of previously-calculated quantities and, furthermore, allow one to write $c_s$ in the degenerate limit as an explicit 
function of the entropy. 

An interesting feature of the degenerate limit is that the cold and thermal contributions to $c_s$ can be separated. We 
begin from Eq. (\ref{cs3}) which we convert to
\be
\left(\frac{c_s}{c}\right)^2 = \frac{\frac{K}{9}+\left.\frac{dP_{th}}{dn}\right|_{S,Y_e}}{(H_0+m)\left(1+\frac{H_{th}}{H_0+m}\right)} \,,
\ee
where $H = E+ P/n$ is the enthalpy per particle. In the degenerate region $H_{th}\ll H_0$, so that
\ba
\left(\frac{c_s}{c}\right)^2 &\simeq& \frac{\left(\frac{K}{9}+\left.\frac{dP_{th}}{dn}\right|_{S,Y_e}\right)\left(1-\frac{H_{th}}{H_0+m}\right)}
                                          {H_0+m}  \label{cs5}  \\
          &\simeq& \frac{\frac{K}{9} - \frac{K}{9}\frac{H_{th}}{H_0+m} + \left.\frac{dP_{th}}{dn}\right|_{S,Y_e}}{H_0+m} 
                   \label{cs6}    \\
          &=& \left(\frac{c_{s0}}{c}\right)^2 + \frac{1}{H_0+m}\left[\left.\frac{dP_{th}}{dn}\right|_{S,Y_e} 
                                                                      - \left(\frac{c_{s0}}{c}\right)^2H_{th}\right]\,. \nonumber \\
\ea
In the transition from Eq. (\ref{cs5}) to Eq. (\ref{cs6}), we have discarded the term proportional to 
$dP_{th}/dn \times H_{th}$ as it represents higher-order corrections. 

Results for the squared sound speed vs density for the two models are shown in Fig. \ref{MDYIuN2SkOp_InterpS_Cs2_lin} for values of $S$ up to 4. Contributions from photons are straightforwardly included in $\Gamma_S$, $P$ and $h$ of Eq. (\ref{cs4}). 
Finite entropy contributions to $c_s^2$ depend on the density, and are positive up to some high density beyond which the  trend is reversed as implied in Eq. (\ref{cs6}). We note that non-relativistic models tend to go acausal at some high density. For a  prescription to render them causal at finite $S$,  see the erratum \cite {APRerr} to the procedure outlined in Ref. \cite{APRppr}.

\begin{figure}[!htb]
\includegraphics[width=9cm]{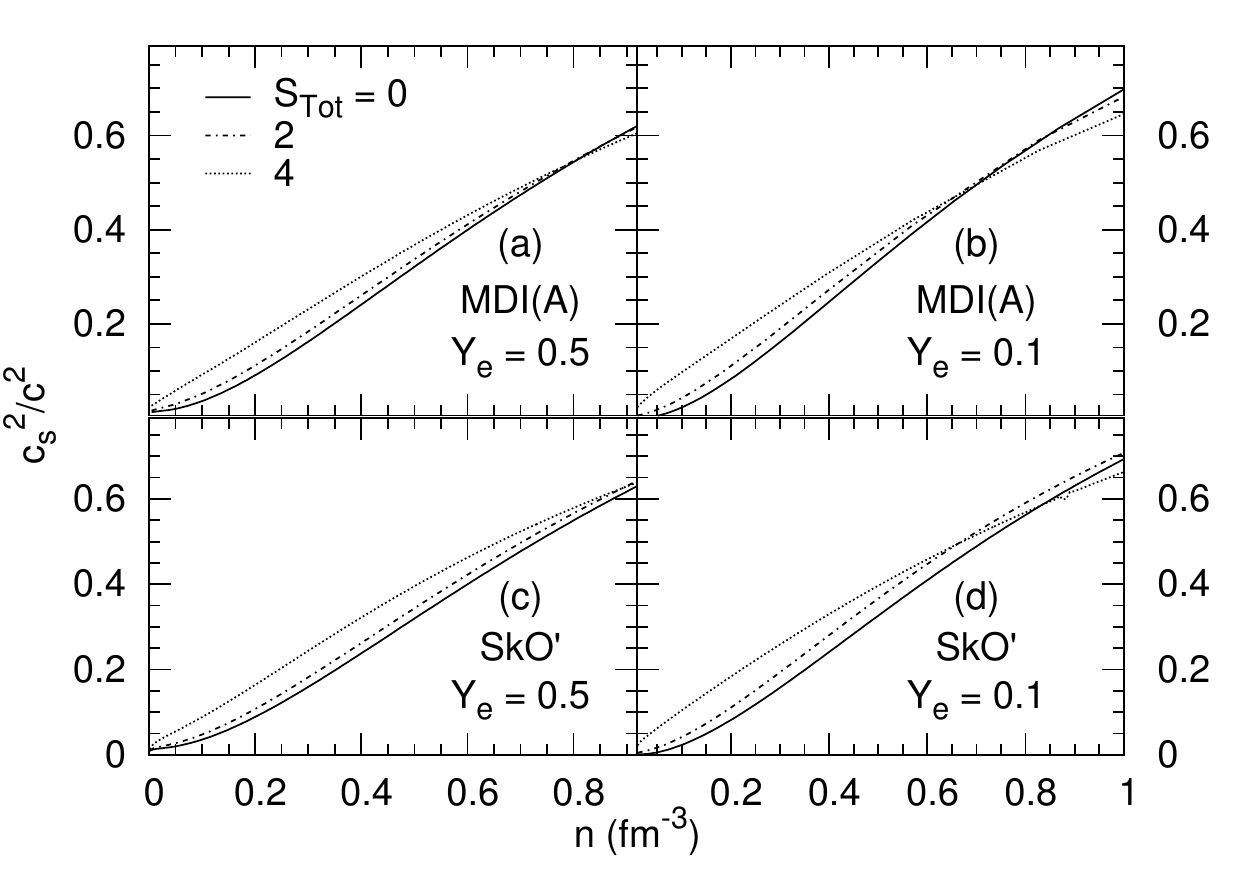}
\caption{Squared sound speed vs density for various entropy per baryon from Eq. (\ref{cs4}).}
\label{MDYIuN2SkOp_InterpS_Cs2_lin}
\end{figure}

\section{ SUMMARY AND CONCLUSION}
\label{Sec:Conclusion}

In this work, we have sought answers to the following two questions, often raised in the community. 

\noindent  (1)   Are equations of state that describe the observed collective flow in heavy-ion collisions in the range 0.5-2 GeV/A able to account for the largest of the recently well-measured neutron star masses  $\simeq 2~{\rm M}_\odot$? 

\noindent (2) How do the thermal properties of such equations of state compare with those currently used in simulations 
of core-collapse supernovae, proto-neutron star evolution, and binary mergers of compact stars?

In order to account for the measured mean transverse momentum per nucleon vs rapidity, flow angles, azimuthal distributions, and radial flow in medium-energy heavy-ion collisions, density and momentum-dependent mean fields that matched the real part of the optical model potential fits to nucleon-nucleus scattering data are required \cite{Gale87,Prakash88b,Welke88,Gale90,Danielewicz:00,Danielewicz:02}.  Such mean fields are also the result of microscopic many-body calculations of dense nucleonic matter \cite{w88,Zuo14}.   A characteristic feature of these mean fields is that for large momenta, they saturate unlike those given by zero-range Skyrme models which rise quadratically with momentum (use of Skyrme model mean fields yields  flow observables that are too large compared with data).   For rapid use in large-scale transport model  calculations of heavy-ion collisions, features of microscopic calculations have been successfully parametrized in schematic models using {\it e.g.}, finite-range Yukawa-type forces \cite{Welke88,Gale90,Das03}. In this work, we have adapted the MDI(0) model of Das {\em et al}. \cite{Das03} in which the mean field is both isospin and momentum dependent in order to study both the structural aspects of neutron stars and the thermal properties of isospin asymmetric matter.  Our principal findings are summarized below along with conclusions. 

The MDI(0) model of Das {\em et al}. yields a neutron star maximum mass of 1.88 ${\rm M}_\odot$, which falls slightly below the central value of the largest well-measured mass of $2.01\pm 0.04~{\rm M}_\odot$.  
The radius  of the 1.4 ${\rm M}_\odot$ star is 11.27 km to be compared with $11.5\pm 0.7$ km recommended in Refs. \cite{Steiner10,Steiner13,Lat12}.

To bring the maximum mass closer to the observed value, we devised a revised parametrization labeled MDI(A) that gave ${\rm M}_{max} = 1.97~{\rm M}_\odot$ and a radius of 11.58 km for the  1.4 ${\rm M}_\odot$ star.  
At the edges of the 1-$\sigma$ errors of the empirical properties $K_0,~L_v$ and $K_v$ 
at the nuclear equilibrium density [with attendant small changes in the strength and range parameters of MDI(A)], 
results obtained were 
${\rm M}_{max} = 2.15~{\rm M}_\odot$ and a radius of 12.13 km for the  1.4 ${\rm M}_\odot$ star.  
The conclusion from these results is that  a momentum-dependent EOS that reproduces heavy-ion flow data is well able to yield 
neutron stars in excess of 2  ${\rm M}_\odot$.  

The pure neutron matter to symmetric nuclear matter energy difference of the finite range MDI models receives finite contributions from terms involving 
potential interactions beyond the quadratic term in the neutron-proton asymmetry.  The magnitudes of such terms are, however, relatively small compared 
to the leading order quadratic term.  Such beyond-the-leading-order contributions may also be expected to be present in 
microscopic calculations of isospin asymmetric matter such as BHF, DBHF, GFMC, EFT, {\em etc.} that employ finite range interactions. 
The general practice has been to parametrize results of these microscopic calculations with a quadratic term only.  
Such potential contributions to $S_l$ for $l>2$ are absent in zero-range Skyrme models by construction.

Our investigations of thermal effects were performed for the finite-range MDI(A) model and the zero-range Skyrme model called SkO$^\prime$ to provide contrasts. The two models give nearly identical results for the energy per particle, pressure, chemical potentials, {\it etc.}, at zero temperature  for both nuclear matter and pure neutron matter at all densities. Consequently, bulk properties accessible in experiments involving nuclei near the  equilibrium density of nuclear matter are similar for the two models as are their neutron star attributes. However, the momentum dependences of their mean field potentials are distinctly different, particularly at large momenta. The MDI(A) potential saturates logarithmically at high momenta whereas the potential of the SkO$^\prime$ model rises
quadratically with momentum as in all Skyrme models. These differences are in turn reflected in their effective masses, both in their magnitudes and density dependences. The influence of  their differing effective masses on the thermal components of energies, pressures, chemical potentials, and the specific heats at constant volume and pressure were explored in detail for both symmetric nuclear matter and pure neutron matter. Formulas to calculate the thermal components for all proton fractions from 0-0.5 were derived
and are presented in the text and appendices. 

The isospin splitting of the effective masses in the MDI(A) and SkO$^\prime$ models is similar in that $m_n^* > m_p^*$ in neutron-rich matter for all densities. Such is not the case  in many Skyrme models (in which $m_n^* < m_p^*$ for neutron-rich matter) which have been successful in predicting the ground state properties and collective excitations of laboratory nuclei. 
This reversal in behavior of the expected isospin splitting of the neutron and proton effective masses are exemplified in the finite-range  MDI(B) model and the zero-range SLy4 model.   
This curious feature led us to establish relations involving the strength and range parameters of Skyrme and MDI models in which 
 $m_n^* > m_p^*$, a feature exhibited in microscopic calculations of dense matter  \cite{Ma04,Dalen05,Sammarruca05}.  
In astrophysical settings where wide ranges of densities, compositions, and temperatures are encountered with varying degrees of degeneracy,  the proper disposition of the neutron and proton effective masses with respect to isospin content is important insofar they govern all of the thermal variables and their response functions.  

The conclusion that emerges from comparing the exact numerical results of the finite-range MDI(A) and zero-range SkO$^\prime$ models is that the agreement or disagreement depends on the specific thermal variable in question as well as the values of all three independent variables ($n,x,T$).  For example, at $T=20$ MeV the thermal energies of symmetric nuclear matter begin to 
differ for $n>n_0$, whereas they do so for $n<2n_0$ at T=50 MeV. For pure neutron matter, there is good agreement for all densities at $T=20$ MeV, but not so for $n>n_0$ at $T=50$ MeV.  The thermal pressures on the other hand begin to deviate from each other for $n>n_0$ for both $x=0$ and $x=0.5$ at $T=20$ MeV, whereas they do so for $n>2n_0$ at $T=50$ MeV. Similar dependences on $(n,x,T)$, albeit in different regimes of ($n,x$),  are seen for the thermal chemical  potentials and the entropy per baryon. The influence of finite range interactions is particularly evident when one compares the specific heats at constant volume, $C_V$,  and at constant pressure, $C_P$, with those of the zero-range interaction model.   To varying degrees, differences are manifest in $C_V$ for wide ranges of $n$. Likewise, departures  in $C_P$ also depend on the values of ($n,x,T$).  

Analytical formulas that were derived in the limiting situations of degenerate and non degenerate matter shed some light on the origin of the differences described above.  In the degenerate limit, the differences between the magnitudes and density dependences of the nucleon effective masses between the two models control the behavior of the thermal state variables and their specific heats. Due the saturating decline of effective masses with density in the MDI models,  thermal variables are generally different in magnitudes than those of the Skyrme models.  
In the non degenerate regime, the influence of the finite-range interactions appears to be distinctly different from those of zero-range interactions. 

Some points are worthy of mention. Given the vast number of Skyrme parametrizations in the literature, many of which are able to successfully predict ground state and exitation properties of laboratory nuclei, it is quite possible that many of their thermal properties (exceptions being response functions such as specific heats, susceptibilities, {\it etc.})  can be made to resemble those of models with finite range interactions. However, their inability to describe heavy-ion data satisfactorily would remain due to the  quadratically rising momentum dependence of their mean field potentials.     
It must be also be noted that at sub-saturation densities, inhomogeneous phases of matter must be considered in astrophysical situations. Whether models with finite-range interactions would yield thermal properties of  such phases that differ significantly from those that result from the use of zero-range interactions is  unexplored and is a subject for future study.  

Astrophysical simulations of core-collapse supernovae, proto-neutron star evolution and mergers of compact binary stars are also sensitive to thermal effects in widely varying regions of density and temperature.  We have therefore conducted an in-depth analysis of the thermal and adiabatic indices,  $\Gamma_{th}$ and $\Gamma_S$ (which are measures of the stiffness of the equation of state), and the speed of sound $c_s$ (a guide for the rapidity with which hydrodynamic evolution proceeds) of hot and dense matter containing nucleons, leptons and photons. 
Our focus here was on the homogeneous phase of matter results of which provide a benchmark against which effects of inhomogeneous phases  known to exist at sub-saturation densities and low entropies can be assessed.  Studies in this regard are in progress and will  be reported separately.
Our study here has highlighted the influence of nucleon effective masses, particularly their iso-spin and density dependences, on the thermal properties of baryons. 
Equally important is the role of leptons which serve to remove the mechanical (or spinodal) instability of baryons-only matter at sub-saturation densities. 
We find that substantial variations in $\Gamma_{th}$, $\Gamma_S$ and $c_s$ begin to occur at sub-saturation densities   before asymptotic values at supra-nuclear densities are reached. Consequently,  laboratory experiments and theoretical studies involving neutron-rich matter can pin down these quantities.

\section*{ACKNOWLEDGEMENTS}
This work was supported by
by the U.S. DOE under grants No. DE-FG02-93ER-40756 (for B.M. and M.P) and
No. DE-FG02-87ER-40317 (for J.M.L).  

\appendix

\section{MDI STATE VARIABLES AT $T=0$}
\label{Sec:AppendixA}

In this appendix, analytical expressions for pressure, chemical potentials, the nuclear matter incompressibility, the symmetry energy and its density derivatives, single particle potentials, and the density derivative of the nucleon effective masses at zero temperature are collected. We will use the decomposition of the MDI Hamiltonian density as $\mathcal{H} = \mathcal{H}_k + \mathcal{H}_d + \mathcal{H}_m$ to isolate the contributions arising from kinetic sources, density-dependent  and momentum-dependent interactions to the various quantities of interest.

\subsection{Pressure P}
\label{Sec:AppendixA_P}
\ba
P &=& n\frac{\partial \mh}{\partial n}-\mh  \label{Pin}\\
\frac{\partial \mh}{\partial n} &=& \frac{\partial \mh_k}{\partial n}+\frac{\partial \mh_d}{\partial n}
                                 +\frac{\partial \mh_m}{\partial n}    \\
\frac{\partial \mh_k}{\partial n} &=& \frac{5}{3n}\mh_k  \\
\frac{\partial \mh_d}{\partial n} &=& A_1\frac{n}{n_0}+A_2\frac{n}{n_0}(1-2x)^2 \nonumber \\
                                &+& B\left(\frac{n}{n_0}\right)^{\sigma}\left[1-y(1-2x)^2\right]  \\
\frac{\partial \mh_m}{\partial n} &=& \frac{C_l}{n_0}\left(\frac{p_{Fn}}{3n}\frac{\partial I_{nn}}{\partial p_{Fn}} 
                                +\frac{p_{Fp}}{3n}\frac{\partial I_{pp}}{\partial p_{Fp}}\right)     \nonumber  \\
                                &+& \frac{2C_u}{n_0} \left(\frac{p_{Fn}}{3n}\frac{\partial I_{np}}{\partial p_{Fn}} 
                                +\frac{p_{Fp}}{3n}\frac{\partial I_{np}}{\partial p_{Fp}}\right)  \\
\frac{\partial I_{np}}{\partial p_{Fi}} &=& \frac{8\pi^2\Lambda^2}{(2\pi\hbar)^6}\left\{4p_{Fi}^2p_{Fj}+ 4\Lambda p_{Fi}^2\right.  \nonumber \\
       &\times& \left[\arctan\left(\frac{p_{Fi}-p_{Fj}}{\Lambda}\right)
                                            -\arctan\left(\frac{p_{Fi}+p_{Fj}}{\Lambda}\right)\right] \nonumber \\
      &+& \left. p_{Fi}(\Lambda^2+p_{Fj}^2-p_{Fi}^2)
           \ln\left[\frac{(p_{Fi}+p_{Fj})^2+\Lambda^2}{(p_{Fi}-p_{Fj})^2+\Lambda^2}\right]\right\}   \label{Pfin}\nonumber \\ 
\mbox{Here,} && i\ne j.
\ea

\subsection{Chemical potentials $\mu_i$}
\label{Sec:AppendixA_Mu}
\ba
\mu_i &=&  \frac{\partial \mh}{\partial n_i} = \frac{\partial \mh_k}{\partial n_i}+\frac{\partial \mh_d}{\partial n_i}
                                 +\frac{\partial \mh_m}{\partial n_i}  \label{muin}  \\
\frac{\partial \mh_k}{\partial n_i} &=& \frac{1}{2m}p_{Fi}^2  \\
\frac{\partial \mh_d}{\partial n_i} &=& A_1\frac{n}{n_0}\pm A_2\frac{n}{n_0}(1-2x)
                  +B\left(\frac{n}{n_0}\right)^{\sigma}               \\
               &\times& \left\{1-\frac{y(\sigma-1)}{\sigma+1}(1-2x)^2
                   \left[1\pm \frac{2}{(\sigma-1)(1-2x)}\right]\right\}  \nonumber \\
     && + ~\mbox{for neutrons},~~ - ~\mbox{for protons}  \nonumber \\
\frac{\partial \mh_m}{\partial n_i} &=&  \frac{C_l}{n_0}\frac{p_{Fi}}{3n_i}\frac{\partial I_{ii}}{\partial p_{Fi}} 
                                +\frac{2C_u}{n_0}\frac{p_{Fi}}{3n_i}\frac{\partial I_{np}}{\partial p_{Fi}} \label{mufin}
\ea

\subsection{Nuclear matter incompressibility $K$}
\label{Sec:AppendixA_K}
\ba
K &=& 9n\frac{\partial^2 \mh}{\partial n^2} = 9n \left(\frac{\partial^2 \mh_k}{\partial n^2}
                                 +\frac{\partial^2 \mh_d}{\partial n^2}
                                 +\frac{\partial^2 \mh_m}{\partial n^2}\right) \nonumber \\ \\
\frac{\partial^2 \mh_k}{\partial n^2} &=& \frac{10}{9n^2}\mh_k  \\
\frac{\partial^2 \mh_d}{\partial n^2} &=& \frac{A_1}{n_0}+\frac{A_2}{n_0}(1-2x)^2 \nonumber \\
                &+& \sigma B\frac{n^{\sigma-1}}{n_0^{\sigma}}\left[1-y(1-2x)^2\right]  \\
\frac{\partial^2 \mh_m}{\partial n^2} &=& \frac{C_l}{n_0}\sum_i\left[\left(\frac{p_{Fi}}{3n}\right)^2
                                  \frac{\partial^2 I_{ii}}{\partial p_{Fi}^2}  
                                -\frac{2p_{Fi}}{9n^2}\frac{\partial I_{ii}}{\partial p_{Fi}}\right] \nonumber \\
                           &+&  \frac{2C_u}{n_0} \left\{\sum_i\left[\left(\frac{p_{Fi}}{3n}\right)^2
                                 \frac{\partial I_{np}^2}{\partial p_{Fi}^2} 
                                -\frac{2p_{Fi}}{9n^2}\frac{\partial I_{np}}{\partial p_{Fi}}\right]\right. \nonumber \\  
                            &+& \left.\frac{2p_{Fn}p_{Fp}}{9n^2}\frac{\partial^2 I_{np}}{\partial p_{Fn}\partial p_{Fp}}\right\} \\
\frac{\partial^2 I_{np}}{\partial p_{Fi}^2} &=& \frac{8\pi^2\Lambda^2}{(2\pi\hbar)^6}\left\{12p_{Fi}p_{Fj}+8\Lambda p_{Fi}\right. \nonumber \\
                         &\times& \left[\arctan\left(\frac{p_{Fi}-p_{Fj}}{\Lambda}\right)
                                            -\arctan\left(\frac{p_{Fi}+p_{Fj}}{\Lambda}\right)\right]  \nonumber  \\
      &+& \left. (\Lambda^2+p_{Fj}^2-3p_{Fi}^2)
           \ln\left[\frac{(p_{Fi}+p_{Fj})^2+\Lambda^2}{(p_{Fi}-p_{Fj})^2+\Lambda^2}\right]\right\}  \nonumber \\
\mbox{Here,}       && i\ne j.     \\
\frac{\partial^2 I_{np}}{\partial p_{Fn}\partial p_{Fp}} &=& \frac{8\pi^2\Lambda^2}{(2\pi\hbar)^6}
                   2p_{Fn}p_{Fp} \ln\left[\frac{(p_{Fn}+p_{Fp})^2+\Lambda^2}{(p_{Fn}-p_{Fp})^2+\Lambda^2}\right]
\ea

\subsection{Symmetry energy $S_2$}

\label{Sec:AppendixA_S2}
\ba
S_2 &=& \frac{1}{8n}\left.\frac{\partial^2 \mh}{\partial x^2}\right|_{x=1/2} 
         = \frac{1}{8n} \left(\left.\frac{\partial^2 \mh_k}{\partial x^2}\right|_{x=1/2}\right. \nonumber \\
                                 &+& \left.\left.\frac{\partial^2 \mh_d}{\partial x^2}\right|_{x=1/2}
                                 +\left.\frac{\partial^2 \mh_m}{\partial x^2}\right|_{x=1/2}\right) \\                     
\left.\frac{\partial^2 \mh_k}{\partial x^2}\right|_{\frac{1}{2}} &=& \frac{40}{9}\left.\mh_k\right|_{x=1/2}  \\
\left.\frac{\partial^2 \mh_d}{\partial x^2}\right|_{\frac{1}{2}} &=& \frac{4A_2}{n_0}n^2 
                                   -\frac{8By}{\sigma+1}\frac{n^{\sigma+1}}{n_0^{\sigma}}  \\
\left.\frac{\partial^2 \mh_m}{\partial x^2}\right|_{\frac{1}{2}} &=& \frac{8C_l}{9n_0}D_l + \frac{16C_u}{9n_0}D_u  \\
D_l &=& \frac{8\pi^2\Lambda^2}{(2\pi\hbar)^6}\left[8p_F^4-2\Lambda^2p_F^2\ln\left(1+\frac{4p_F^2}{\Lambda^2}\right)\right] \nonumber \\ \\
D_u &=& \frac{8\pi^2\Lambda^2}{(2\pi\hbar)^6}\left[4p_F^4\right. \nonumber \\
                   &-& \left. (4p_F^4+\Lambda^2p_F^2)\ln\left(1+\frac{4p_F^2}{\Lambda^2}\right)\right]  \\
p_F &=& \left(\frac{3\pi^2n\hbar^3}{2}\right)^{1/3}
\ea

\subsection{Derivative of the symmetry energy $L_2$}
\label{Sec:AppendixA_L2}
\ba  
L_2 &=& 3n\frac{dS_2}{dn}
       = \frac{3}{8} \left(\left.\frac{d}{dn}\frac{\partial^2 \mh_k}{\partial x^2}\right|_{x=1/2}\right. \nonumber \\
                                 &+& \left.\left.\frac{d}{dn}\frac{\partial^2 \mh_d}{\partial x^2}\right|_{x=1/2}
                  +\left.\frac{d}{dn}\frac{\partial^2 \mh_m}{\partial x^2}\right|_{x=1/2}\right) -3S_2 \nonumber \\ \\ 
\left.\frac{d}{dn}\frac{\partial^2 \mh_k}{\partial x^2}\right|_{\frac{1}{2}} &=& \frac{200}{27n}\left.\mh_k\right|_{x=1/2}  \\
\left.\frac{d}{dn}\frac{\partial^2 \mh_d}{\partial x^2}\right|_{\frac{1}{2}} &=& \frac{8A_2}{n_0}n 
                                   -8By\left(\frac{n}{n_0}\right)^{\sigma}  \\
\left.\frac{d}{dn}\frac{\partial^2 \mh_m}{\partial x^2}\right|_{\frac{1}{2}} &=& \frac{8C_l}{9n_0}\frac{p_F}{3n}\frac{dD_l}{dp_F} 
                                       + \frac{16C_u}{9n_0}\frac{p_F}{3n}\frac{dD_u}{dp_F} \\
\frac{dD_l}{dp_F} &=& \frac{8\pi^2\Lambda^2}{(2\pi\hbar)^6}\left[16p_F^3+\frac{64p_F^5}{4p_F^2+\Lambda^2}\right. \nonumber \\
                            &-& \left. 4\Lambda^2p_F\ln\left(1+\frac{4p_F^2}{\Lambda^2}\right)\right]     \\
\frac{dD_u}{dp_F} &=& \frac{8\pi^2\Lambda^2}{(2\pi\hbar)^6}\left[8p_F^3 \right. \nonumber \\
                   &-& \left. 2(8p_F^3+\Lambda^2p_F)\ln\left(1+\frac{4p_F^2}{\Lambda^2}\right)\right]
\ea

\subsection{Curvature of the symmetry energy $K_2$}
\label{Sec:AppendixA_K2}
\ba
K_2 &=& 9n^2\frac{d^2S_2}{dn^2}
     =  \frac{9n}{8} \left(\left.\frac{d^2}{dn^2}\frac{\partial^2 \mh_k}{\partial x^2}\right|_{x=1/2}\right. \nonumber \\
                                 &+& \left.\left.\frac{d^2}{dn^2}\frac{\partial^2 \mh_d}{\partial x^2}\right|_{x=1/2}
                  +\left.\frac{d^2}{dn^2}\frac{\partial^2 \mh_m}{\partial x^2}\right|_{x=1/2}\right) -6L_2 \nonumber \\ \\ 
\left.\frac{d^2}{dn^2}\frac{\partial^2 \mh_k}{\partial x^2}\right|_{\frac{1}{2}} &=& \frac{400}{81n^2}\left.\mh_k\right|_{x=1/2}  \\
\left.\frac{d^2}{dn^2}\frac{\partial^2 \mh_d}{\partial x^2}\right|_{\frac{1}{2}} &=& \frac{8A_2}{n_0} 
                                   -\frac{8By\sigma}{n_0^{\sigma}}n^{\sigma-1}  \\
\left.\frac{d^2}{dn^2}\frac{\partial^2 \mh_m}{\partial x^2}\right|_{\frac{1}{2}} &=& 
                  \frac{8C_l}{9n_0}\left[\left(\frac{p_F}{3n}\right)^2\frac{d^2D_l}{dp_F^2}- \frac{2p_F}{9n^2}\frac{dD_l}{dp_F}\right] \nonumber \\
                &+& \frac{16C_u}{9n_0}\left[\left(\frac{p_F}{3n}\right)^2\frac{d^2D_u}{dp_F^2}- \frac{2p_F}{9n^2}\frac{dD_u}{dp_F}\right]  \\
\frac{d^2D_l}{dp_F^2} &=& \frac{8\pi^2\Lambda^2}{(2\pi\hbar)^6}
         \left[\frac{16p_F^2(96p_F^4+36p_F^2\Lambda^2+\Lambda^4)}{(4p_F^2+\Lambda^2)^2}\right. \nonumber \\
                            &-& \left. 4\Lambda^2\ln\left(1+\frac{4p_F^2}{\Lambda^2}\right)\right]     \\
\frac{d^2D_u}{dp_F^2} &=& \frac{8\pi^2\Lambda^2}{(2\pi\hbar)^6}\left[8p_F^2-\frac{64p_F^4}{4p_F^2+\Lambda^2} \right. \nonumber \\
                   &-& \left. 2(24p_F^2+\Lambda^2)\ln\left(1+\frac{4p_F^2}{\Lambda^2}\right)\right]
\ea

\subsection{Single-particle potentials $U_i$}
\label{Sec:AppendixA_SPP}
\ba
U_i(n_i,n_j,p_i) &=& A_1\frac{n}{n_0}\pm A_2\frac{n}{n_0}(1-2x)
                  +B\left(\frac{n}{n_0}\right)^{\sigma}  \nonumber \\
                &\times& \left\{1-\frac{y(\sigma-1)}{\sigma+1}(1-2x)^2 \right. \nonumber \\
                &\times& \left.   \left[1\pm \frac{2}{(\sigma-1)(1-2x)}\right]\right\}  \nonumber \\
                &+& \frac{2C_l}{n_0}R_{ii}(n_i,p_i) + \frac{2C_u}{n_0}R_{ij}(n_j,p_i) ~~;~~ i\ne j \nonumber \\
                && + ~\mbox{for neutrons}, ~~- ~\mbox{for protons}   \\
R_{ii}(n_i,p_i) &=& \frac{\Lambda^3}{4\pi^2\hbar^3}
      \left\{\frac{2p_{Fi}}{\Lambda}  \right.  \nonumber \\
    &-&  2\left[\arctan\left(\frac{p_i+p_{Fi}}{\Lambda}\right)
                                            -\arctan\left(\frac{p_i-p_{Fi}}{\Lambda}\right)\right] \nonumber \\
      &+& \left. \frac{(\Lambda^2+p_{Fi}^2-p_i^2)}{2\Lambda p_i}
           \ln\left[\frac{(p_i+p_{Fi})^2+\Lambda^2}{(p_i-p_{Fi})^2+\Lambda^2}\right]\right\}  \\
R_{ij}(n_j,p_i) &=& \frac{\Lambda^3}{4\pi^2\hbar^3}
      \left\{\frac{2p_{Fj}}{\Lambda} \right.   \nonumber  \\ 
     &-& 2\left[\arctan\left(\frac{p_i+p_{Fj}}{\Lambda}\right)
                                            -\arctan\left(\frac{p_i-p_{Fj}}{\Lambda}\right)\right]  \nonumber \\
      &+& \left. \frac{(\Lambda^2+p_{Fj}^2-p_i^2)}{2\Lambda p_i}
           \ln\left[\frac{(p_i+p_{Fj})^2+\Lambda^2}{(p_i-p_{Fj})^2+\Lambda^2}\right]\right\}  ~;~i\ne j \nonumber \\
\ea

\subsection{Derivatives of the effective masses $dm_i^*/dn_i$}
\label{Sec:AppendixA_dMs}
\ba
\frac{\partial m_i^*}{\partial n_i} &=& \frac{p_{Fi}}{3n_i}(-m_i^{*2})\left(\frac{2C_l}{n_0}G_{ii}
                                           +\frac{2C_u}{n_0}G_{1,ij}\right) ~~;~~i\ne j \nonumber \\ \\
\frac{\partial m_i^*}{\partial n_j} &=& \frac{p_{Fj}}{3n_j}(-m_i^{*2})\frac{2C_u}{n_0}G_{2,ij} ~~;~~i\ne j \\
\frac{\partial m_i^*}{\partial n} &=& (1-x)\frac{\partial m_i^*}{\partial n_n}+x\frac{\partial m_i^*}{\partial n_p} \\
G_{ii} &=& \frac{\Lambda^2}{2\pi^2\hbar^3}\left[\frac{-3}{p_{Fi}^2}+\frac{4}{4p_{Fi}^2+\Lambda^2}  \right. \nonumber  \\
         &+& \left. \frac{2p_{Fi}^2+3\Lambda^2}{4p_{Fi}^4}\ln\left(1+\frac{4p_{Fi}^2}{\Lambda^2}\right)\right]  \\
G_{1,ij} &=& \frac{\Lambda^2}{2\pi^2\hbar^3}  \nonumber \\
&\times& \left\{\frac{-p_{Fj}[p_{Fi}^4+4p_{Fi}^2(\Lambda^2-p_{Fj}^2)+3(\Lambda^2+p_{Fj}^2)^2]}
        {p_{Fi}^3[(p_{Fi}+p_{Fj})^2+\Lambda^2][(p_{Fi}-p_{Fj})^2+\Lambda^2]}\right. \nonumber \\  
   &+& \left.\frac{3(\Lambda^2+p_{Fj}^2)+p_{Fi}^2}{4p_{Fi}^4}
           \ln\left[\frac{(p_{Fi}+p_{Fj})^2+\Lambda^2}{(p_{Fi}-p_{Fj})^2+\Lambda^2}\right]\right\} \nonumber \\  \\
G_{2,ij} &=& \frac{\Lambda^2}{2\pi^2\hbar^3}  \nonumber \\
&\times& \left\{\frac{2p_{Fj}^2(-p_{Fi}^2+\Lambda^2+p_{Fj}^2)}
        {p_{Fi}^2[(p_{Fi}+p_{Fj})^2+\Lambda^2][(p_{Fi}-p_{Fj})^2+\Lambda^2]}\right. \nonumber \\
   &-& \left.\frac{p_{Fj}}{2p_{Fi}^3}
           \ln\left[\frac{(p_{Fi}+p_{Fj})^2+\Lambda^2}{(p_{Fi}-p_{Fj})^2+\Lambda^2}\right]\right\} 
\ea
In the last two expressions $i\ne j$.


\section{MDI MODELS IN THE NON-DEGENERATE LIMIT}
\label{Sec:AppendixB}
In this appendix we 
describe the method by which analytical expressions for the state variables valid in the limit 
of small degeneracy $(z \ll 1)$ were obtained.

This method is based primarily on two elements the first of which is the assumption that in the nondegenerate (ND) 
limit the system is sufficiently dilute such that the term $R(p)$ in the single-particle energy spectrum depends 
weakly on momentum $p$ and therefore the location $p_0$ of the peak of a given thermodynamic integrand is the same as 
that of a free gas:

\be
\frac{d}{dp}\left.\left(\frac{p^{\alpha}}{1+z^{-1}e^{\frac{p^2}{2mT}}}\right)\right|_{p=p_0} = 0 
\ee
\be
\Rightarrow p_0 \stackrel{z\ll1}{\longrightarrow} (\alpha mT)^{1/2}\left[1+\frac{1}{2e^{\alpha/2}}z
                      -\frac{(1+2\alpha)}{8e^{\alpha}}z^2 + \ldots\right]. \label{p0}
\ee

The second 
element of the method is the happenstance
that the Taylor series expansion about $z=0$ of the state 
functions results in integrals of the form 
\be
I = \int dp~g(p)~e^{-\beta f(p)}
\ee
which are amenable to estimation
via the saddle point approximation \cite{Butler}:
\ba
I &\simeq& \frac{\sqrt{2\pi}g_0e^{-\beta f_0}}{\sqrt{\beta f_0''}}\left[1+\beta^{-1}\left(\frac{5}{24}\frac{f_0'''^2}{f_0''^3}
                                       -\frac{1}{8}\frac{f_0^{iv~2}}{f_0''^2}\right.\right.  \nonumber \\
      &+&  \left.\left.\frac{g_0''}{2g_0f_0''}-\frac{g_0'f_0'''}{2g_0f_0''^2}\right)+O(\beta^{-2})\right] \label{sd}
\nonumber \\
\ea
where the primes denote differentiation with respect to $p$ and the subscripts 0 evaluation at $p=p_0$. The dummy 
variable $\beta$ - to be set equal to 1 at the end of the calculation - keeps track of the order of the asymptotic expansion.

From Eq. (\ref{sd}) it is clear that 
\ba
\int dp~e^{-\beta f(p)} &\simeq& \frac{\sqrt{2\pi}e^{-\beta f_0}}{\sqrt{\beta f_0''}}  \nonumber \\
               &\times& \left[1+\beta^{-1}\left(\frac{5}{24}\frac{f_0'''^2}{f_0''^3}
                                       -\frac{1}{8}\frac{f_0^{iv~2}}{f_0''^2}\right)+O(\beta^{-2})\right].
\nonumber \\
\ea
Thus, to $O(\beta^{-1})$ in asymptotics, we can write
\be
I \simeq g_0\left[1+\beta^{-1}\left(\frac{g_0''}{2g_0f_0''}-\frac{g_0'f_0'''}{2g_0f_0''^2}\right)\right]
                    \int dp~e^{-\beta f(p)}  \label{sd2}
\ee
which is
convenient for our purposes because, as will be shown subsequently, the integral in Eq. (\ref{sd2}) 
can be evaluated exactly. Henceforth, we shall make use of the definition
\be
G_0 \equiv g_0\left(1+\frac{g_0''}{2g_0f_0''}-\frac{g_0'f_0'''}{2g_0f_0''^2}\right). \label{G0}
\ee

At this point we must note a caveat with regards to the use of the saddle point approximation in the ND limit: 
High temperatures tend to spread the integrands over a wider range and thus incur more error in the approximation. 
This is not a problem for the temperatures relevant to supernovae and compact objects $(T\le 50$ MeV) but nevertheless, 
the method cannot be expected to work for arbitrarily high temperatures. Such broadening can often be eliminated 
by a suitable variable change. However, no systematic way exists for finding the appropriate transformation.

In what follows, we demonstrate the method for a single-component gas for the sake of simplicity. Generalization 
to multi-component gases is straightforward. We begin with $R(p)$ given by
\be
R(p) = \frac{2C}{n_0}\frac{2}{(2\pi\hbar)^3}\int d^3p'~\frac{1}{1+\left(\frac{\vec p - \vec p'}{\Lambda}\right)^2}f_p'
\label{rapp}
\ee
where $C=C_l$ for pure neutron matter and $C=C_l+C_u$ for symmetric nuclear matter.
Performing the angular integrals leads to 
\be
R(p) =  \frac{C}{2n_0}\frac{\Lambda^2}{\pi^2\hbar^3}\frac{1}{p}
       \int dp'~p'^2 \ln\left[\frac{\Lambda^2+(p+p')^2}{\Lambda^2+(p-p')^2}\right]^{1/p'}f_p'
\ee
The assumption of weak dependence on $p'$ allows us to treat $\ln[\ldots]^{1/p'}$ as constant evaluated at 
$p_{0R} = (2mT)^{1/2}$ (where only the leading term of Eq. (\ref{p0}) is kept) and take it out of the integral:
\be
R(p)=\frac{C}{2n_0}\frac{\Lambda^2}{p}\ln\left[\frac{\Lambda^2+(p+p_{0R})^2}{\Lambda^2+(p-p_{0R})^2}\right]^{\frac{1}{p_{0R}}}
         \frac{1}{\pi^2\hbar^3}\int dp'~p'^2 f_p'  .
\ee
Thus, here, we effectively keep only the first term in the saddle point approximation. This, as well as the 
truncated $p_{0R}$, induce little error because in the ND limit $R(p)$ is itself a small correction to the 
spectrum.

Noting that 
\be
n = \frac{g}{2\pi^2\hbar^3}\int dp~p^2 f_p \label{napp}
\ee
where $g$ is the degeneracy factor, we write $R(p)$ as
\be
R(p)=\frac{C}{n_0}\frac{\Lambda^2}{pp_{0R}}\ln\left[\frac{\Lambda^2+(p+p_{0R})^2}{\Lambda^2+(p-p_{0R})^2}\right]\frac{n}{g}.
\label{ndR}
\ee
In the next step, we expand Eq. (\ref{napp}) in a Taylor series about $z=0$:
\ba
n &\simeq& \frac{g}{2\pi^2\hbar^3}\left[z\int dp~p^2~e^{-\frac{p^2}{2mT}-\frac{R(p)}{T}} \right. \nonumber \\
    &-& \left. z^2\int dp~p^2~e^{-\frac{p^2}{mT}-\frac{2R(p)}{T}} + \ldots \right]  \\
 &=&  \frac{g}{2\pi^2\hbar^3}\left[z\int dp~e^{-\frac{R(p)}{T}}~e^{-\frac{p^2}{2mT}+2\ln p} \right. \nonumber \\
    &-& \left. z^2\int dp~e^{-\frac{2R(p)}{T}}~e^{-\frac{p^2}{mT}+2\ln p} + \ldots \right]  
\ea
where 
terms in the single-particle energy spectrum that depend only on the density have been absorbed in $z$.
For the first integral we identify
\be
g_{n1}(p) = e^{-R/T},~~f_{n1}(p) = \frac{p^2}{2mT}-2\ln p,~~p_{n1} = (2mT)^{1/2}
\ee
and for the second
\be
g_{n2}(p) = e^{-2R/T},~~f_{n2}(p) = \frac{p^2}{mT}-2\ln p,~~p_{n2} = (mT)^{1/2}.
\ee
Therefore,
\ba
n &\simeq& \frac{g}{2\pi^2\hbar^3}\left(z G_{n1}\int dp~e^{-\frac{p^2}{2mT}+2\ln p} \right. \nonumber \\
    &-& \left. z^2 G_{n2}\int dp~e^{-\frac{p^2}{mT}+2\ln p}\right)  \\
  &=& zgG_{n1}n_Q - \frac{z^2}{2^{3/2}}gG_{n2}n_Q  \label{nz}
\ea
where $G_{n1},~G_{n2}$ are given by Eq. (\ref{G0}) and $n_Q = (mT/2\pi\hbar^2)^{3/2}$ is the quantum concentration.
         
By perturbative inversion of Eq. (\ref{nz}) to second order in $n/n_Q$ we obtain
\be
z = \frac{n}{G_{n1}gn_Q} + \frac{1}{2^{3/2}}\left(\frac{n}{G_{n1}gn_Q}\right)^2\frac{G_{n2}}{G_{n1}}.
\ee
The chemical potential $\mu$ is related to $z$ according to
\be
\mu = T\ln z +u(n) \label{MDYI_Mu_ND}
\ee
where $u(n)$ are the terms in the single-particle potential which depend only on the density.

We now  
turn our attention to the kinetic energy density:
\ba
\tau &=& \frac{g}{2\pi^2\hbar^3}\int dp~p^4 f_p  \\
 &&\stackrel{z\ll1}{\longrightarrow}  \frac{g}{2\pi^2\hbar^3}\left[z\int dp~e^{-\frac{R(p)}{T}}~e^{-\frac{p^2}{2mT}+4\ln p}
                 \right. \nonumber \\
    &-& \left. z^2\int dp~e^{-\frac{2R(p)}{T}}~e^{-\frac{p^2}{mT}+4\ln p} + \ldots \right]  
\ea
%
With the identification
\ba
g_{\tau1}(p) &=& e^{-R/T},~~f_{\tau1}(p) = \frac{p^2}{2mT}-4\ln p,~~p_{\tau1} = (4mT)^{1/2}  \nonumber \\ \\
g_{\tau2}(p) &=& e^{-2R/T},~~f_{\tau2}(p) = \frac{p^2}{mT}-4\ln p,~~p_{\tau2} = (2mT)^{1/2} \,, \nonumber \\
\ea
%
we obtain
\ba
\tau &=& 3mTG_{\tau1}gn_Q\left(z-\frac{z^2}{2^{5/2}}\frac{G_{\tau2}}{G_{\tau1}}\right)  \\
     &=& 3mTn\frac{G_{\tau1}}{G_{n1}}\left[1+\frac{1}{2^{3/2}}\left(\frac{n}{G_{n1}gn_Q}\right)
                               \left(\frac{G_{n2}}{G_{n1}}-\frac{G_{\tau2}}{2G_{\tau1}}\right)\right]. \nonumber \\
\ea

The starting point for the calculation of the potential energy density is
\be
V = \frac{C}{n_0}\frac{2g}{(2\pi\hbar)^6}\int d^3p~d^3p' \frac{f_pf_p'}
                             {1+\left(\frac{\vec p - \vec p'}{\Lambda}\right)^2} \,,
\label{vapp}
\ee
where the factor $2g$ would have been $g^2$ if $C_l$ was equal to $C_u$.
Invoking the definition of $R(p)$ [Eq. (\ref{rapp})], we recast Eq. (\ref{vapp}) as
\ba
V &=& \frac{g}{4\pi^2\hbar^3}\int dp~p^2f_pR(p)  \\
  &=& G_V \frac{n}{2}.
\ea
The functions pertaining to the saddle point calculation of $G_V$ are
\be
g_{V}(p) = R(p),~~f_{V}(p) = \frac{p^2}{2mT}-2\ln p,~~p_{V} = (2mT)^{1/2}
\ee
Strictly speaking, one should use $f_V =-\ln (p^2f_p)$ and $p_V$ as in Eq.(\ref{p0}) although at the expense of 
simplicity.

With complete expessions for $\tau$ and $V$, the total energy density is acquired from
\be
\varepsilon = \frac{\tau}{2m} + \mathcal{H}_d + V\label{MDYI_E_ND}
\ee
where  $\mathcal{H}_d$ is given by Eq. ({\ref{MDYI_H_0T_d}).

The entropy density in this scheme is obtained from Eq. (\ref{sden}) which, upon Taylor expansion about $z=0$, 
yields
\ba
s &=& \frac{-g}{2\pi^2\hbar^3}\int dp~p^2\left[z~e^{-\frac{p^2}{2mT}-\frac{R}{T}}
                                  \left(-1-\frac{p^2}{2mT}-\frac{R}{T}+\ln z\right)\right. \nonumber \\
  &-& \left. z^2e^{-\frac{p^2}{mT}-\frac{2R}{T}}
                                  \left(-\frac{1}{2}-\frac{p^2}{2mT}-\frac{R}{T}+\ln z\right)\right]  \\
 &=& \frac{\tau}{2mT} + \frac{2V}{T} - n\ln z  \nonumber \\
 &+& \frac{g}{2\pi^2\hbar^3}\left(z \int dp~e^{-\frac{R}{T}}e^{-\frac{p^2}{2mT}+2\ln p}\right. \nonumber \\
 &-& \left. \frac{z^2}{2} \int dp~e^{-\frac{2R}{T}}e^{-\frac{p^2}{mT}+2\ln p}\right)  \\
 &=& \frac{\tau}{2mT} + \frac{2V}{T} - n\ln z 
     + zgG_{n1}n_Q - \frac{z^2}{2^{5/2}}gG_{n2}n_Q \\
 &=& \frac{\tau}{2mT} + \frac{2V}{T} - n\ln z 
     + n\left[1+\frac{n}{2^{5/2}}\left(\frac{n}{G_{n1}gn_Q}\right)\frac{G_{n2}}{G_{n1}}\right] \nonumber\label{MDYI_s_ND} \\
\ea
Finally, the pressure is given by the thermodynamic identity
\be
P = -\varepsilon +Ts +\mu n\label{MDYI_P_ND}.
\ee
and the specific heats by
\ba
C_V &=& \frac{1}{n}\left.\frac{\partial \varepsilon}{\partial T}\right|_n  \\
C_P &=& C_V + \frac {T}{n^2} \frac{\left(\left.\frac{\partial P}{\partial T}\right|_n \right)^2 }
               {\left. \frac{\partial P}{\partial n}\right|_T} \,. 
\ea
We point out that, besides the usual practice of going to higher powers of $z$ in $n$ and $\tau$, the accuracy 
of this procedure can be improved by calculating $R(p)$ to $O(\beta^{-1})$ in the saddle point approximation and 
by using a beyond-leading-order $p_V$ in the determination of $V$.


\section{SPECIFIC HEATS}
\label{Sec:AppendixC}

\subsection*{Finite-range models}

The main difficulty in calculating $C_V$ and $C_P$ in the MDI model arises from 
the quantity $R(p)$, which is both density- and momentum-dependent.  Hence, it cannot 
be absorbed in the chemical potential. To make this statement explicit, 
consider the spectrum of a single-species in the MDI model:
\be
\epsilon_p = \frac{p^2}{2m} + U(n,p) = \frac{p^2}{2m} + R(p) + u(n) \,,
\label{spes}
\ee
where
\ba
R(p) &=& \frac{C}{n_0}\frac{\Lambda^2}{2\pi^2\hbar^3}\frac{1}{p}
         \int dp'~p'\ln\left[\frac{\Lambda^2+(p+p')^2}{\Lambda^2+(p-p')^2}\right]f'  
         \label{Rp} \\
C &=& \left\{\begin{array}{ll}
              C_l & \mbox{for~PNM}  \\
              C_l+C_u & \mbox{for~SNM}
             \end{array}\right.  \\
u(n) &=&  \left\{\begin{array}{ll}
              \frac{(A_1+A_2)}{n_0}n + \frac{B(1-y)}{n_0^{\sigma}}n^{\sigma} & \mbox{for~PNM}  \\
              \frac{A_1}{n_0}n + \frac{B}{n_0^{\sigma}}n^{\sigma} & \mbox{for~SNM}
             \end{array}\right. \label{un}.
\ea
The acronyms PNM and SNM above stand for pure neutron mater and symmetric nuclear matter, respectively. 
The last term in Eq. (\ref{spes}) has no momentum dependence and can thus 
be subsumed in the chemical potential for purposes of integrating over 
momentum. However, $R(p)$ cannot be treated the same way and so the "reduced" 
spectrum 
\be
\varepsilon_p = \frac{p^2}{2m} + R(p)
\ee
is implicitly density-dependent causing complications when one attempts to take 
derivatives with respect to $\mu$ and $T$ because now the Fermi-Dirac 
distribution has the form
\be
f = \frac{1}{1+\exp\left[\frac{\frac{p^2}{2m}+R(p,\mu,T)-\mu}{T}\right]} \,,
\ee
and therefore
\ba
\left.\frac{\partial f}{\partial \mu}\right|_{T} &=&      
        \left.\frac{\partial f}{\partial \mu}\right|_{R,T} 
        + \left.\frac{\partial f}{\partial R}\right|_{\mu,T}
            \left.\frac{\partial R}{\partial \mu}\right|_{T}  \\
 &=&\frac{1}{T}f(1-f)\left(1-\left.\frac{\partial R}{\partial \mu}\right|_{T}\right) 
  \label{dfdmu} \\
\left.\frac{\partial f}{\partial T}\right|_{\mu} &=&      
        \left.\frac{\partial f}{\partial T}\right|_{R,\mu} 
        + \left.\frac{\partial f}{\partial R}\right|_{\mu,T}
            \left.\frac{\partial R}{\partial T}\right|_{\mu}  \\
  &=& \frac{1}{T}f(1-f)
  \left[\ln\left(\frac{1-f}{f}\right)-\left.\frac{\partial R}{\partial T}\right|_{\mu}\right].
  \label{dfdt}   
\ea
We obtain $\left.\partial R/\partial \mu\right|_{T}$ and 
$\left.\partial R/\partial T\right|_{\mu}$ by taking derivatives of Eq. (\ref{Rp}) 
with respect to the appropriate variables:
\ba
\left.\frac{\partial R}{\partial \mu}\right|_{T} &=&
\frac{C}{n_0}\frac{\Lambda^2}{2\pi^2\hbar^3}\frac{1}{Tp}
         \int dp'~p'\ln\left[\frac{\Lambda^2+(p+p')^2}{\Lambda^2+(p-p')^2}\right] \nonumber \\
      &\times&   f'(1-f')\left(1-\left.\frac{\partial R'}{\partial \mu}\right|_{T}\right)  
         \label{drdmu}\\
\left.\frac{\partial R}{\partial T}\right|_{\mu} &=&
\frac{C}{n_0}\frac{\Lambda^2}{2\pi^2\hbar^3}\frac{1}{Tp}
         \int dp'~p'\ln\left[\frac{\Lambda^2+(p+p')^2}{\Lambda^2+(p-p')^2}\right] \nonumber \\
    &\times&   f'(1-f')\left[\ln\left(\frac{1-f'}{f'}\right)
         -\left.\frac{\partial R'}{\partial T}\right|_{\mu}\right] 
         \label{drdt}  \,,
\ea
where Eqs. (\ref{dfdmu})-(\ref{dfdt}) have been used for 
$\left.\partial f/\partial \mu\right|_{T}$ 
and $\left.\partial f/\partial T\right|_{\mu}$.     

As $\left.\partial R'/\partial \mu\right|_{T}$ and 
$\left.\partial R'/\partial T\right|_{\mu}$ are momentum-dependent, they 
cannot be taken out of the integrals. We must therefore solve 
Eqs. (\ref{drdmu})-(\ref{drdt}) self-consistently in a manner similar to 
the one used to compute $R$ itself.

Equations (\ref{dfdmu})-(\ref{drdt}) are the necessary ingredients for calculating 
the derivatives (with respect to $\mu$ and $T$) of the kinetic energy 
density $\tau$ and the number density $n$ :
\ba
\left.\frac{\partial n}{\partial \mu}\right|_{T} &=&
  \frac{g}{2\pi^2\hbar^3}\int dp~p^2 \left.\frac{\partial f}{\partial \mu}\right|_{T} 
  \label{dndmu}  \\
\left.\frac{\partial n}{\partial T}\right|_{\mu} &=&
  \frac{g}{2\pi^2\hbar^3}\int dp~p^2 \left.\frac{\partial f}{\partial T}\right|_{\mu} 
  \label{dndt}  \\
\left.\frac{\partial \tau}{\partial \mu}\right|_{T} &=&
  \frac{g}{2\pi^2\hbar^3}\int dp~p^4 \left.\frac{\partial f}{\partial \mu}\right|_{T} 
  \label{dtaudmu}  \\
\left.\frac{\partial \tau}{\partial T}\right|_{\mu} &=&
  \frac{g}{2\pi^2\hbar^3}\int dp~p^4 \left.\frac{\partial f}{\partial T}\right|_{\mu}. 
  \label{dtaudt}  
\ea
Here, $\left.\partial f/\partial \mu\right|_{T}$ 
and $\left.\partial f/\partial T\right|_{\mu}$ are given by 
Eqs. (\ref{dfdmu})-(\ref{dfdt}) in which $\left.\partial R/\partial \mu\right|_{T}$ 
and $\left.\partial R/\partial T\right|_{\mu}$ are the self-consistent 
solutions of Eqs. (\ref{drdmu})-(\ref{drdt}). We also define the double 
integrals (pertaining to the finite-range terms of MDI) 
\ba
I_1 &\equiv& \frac{g}{8\pi^4\hbar^6}\int\int dp~dp'~p~p'  \nonumber \\
   &\times&  \ln\left[\frac{\Lambda^2+(p+p')^2}{\Lambda^2+(p-p')^2}\right]f~f'  
   \label{i1} \\
I_2 &\equiv& \frac{g}{\pi^4\hbar^6}\int\int dp~dp'~p^2~p'^2 \nonumber \\
   &\times& \frac{(\Lambda^2-p^2+p'^2)}{[\Lambda^2+(p+p')^2][\Lambda^2+(p-p')^2]}f~f'  
   \label{i2}
\ea
for the derivatives of which similar considerations hold:
\ba
\left.\frac{\partial I_1}{\partial \mu}\right|_{T} &=&
     \frac{g}{4\pi^4\hbar^6}\int\int dp~dp'~p~p'  \nonumber \\
  &\times& \ln\left[\frac{\Lambda^2+(p+p')^2}{\Lambda^2+(p-p')^2}\right]f'
      \left.\frac{\partial f}{\partial \mu}\right|_{T} 
      \label{di1dmu} \\
\left.\frac{\partial I_1}{\partial T}\right|_{\mu} &=&
     \frac{g}{4\pi^4\hbar^6}\int\int dp~dp'~p~p'  \nonumber \\
  &\times& \ln\left[\frac{\Lambda^2+(p+p')^2}{\Lambda^2+(p-p')^2}\right]f'
      \left.\frac{\partial f}{\partial T}\right|_{\mu} 
      \label{di1dt} \\
\left.\frac{\partial I_2}{\partial \mu}\right|_{T} &=& 
    \frac{2g}{\pi^4\hbar^6}\int\int dp~dp'~p^2~p'^2  \nonumber  \\
  &\times& \frac{(\Lambda^2-p^2+p'^2)}{[\Lambda^2+(p+p')^2][\Lambda^2+(p-p')^2]}f'
      \left.\frac{\partial f}{\partial \mu}\right|_{T} 
      \label{di2dmu}  \\
\left.\frac{\partial I_2}{\partial T}\right|_{\mu} &=& 
    \frac{2g}{\pi^4\hbar^6}\int\int dp~dp'~p^2~p'^2   \nonumber \\
  &\times& \frac{(\Lambda^2-p^2+p'^2)}{[\Lambda^2+(p+p')^2][\Lambda^2+(p-p')^2]}f'
      \left.\frac{\partial f}{\partial T}\right|_{\mu} . \label{di2dt} 
      \nonumber \\
\ea
Note that a factor of 2 is gained in Eqs. (\ref{di1dmu})-(\ref{di2dt}) 
due to the interchangeability of $f$ and $f'$.             
                        
For the specific heat at constant volume, we use 
\ba
C_V &=& \left.\frac{\partial E}{\partial T}\right|_{n}
    = \frac{1}{n}\left.\frac{\partial \varepsilon}{\partial T}\right|_{\mu} \label{MDI_Cv}\\
    &=& \frac{1}{n} \left(\left.\frac{\partial \varepsilon}{\partial T}\right|_{\mu} 
    -\frac{\left.\frac{\partial \varepsilon}{\partial \mu}\right|_{T}
              \left.\frac{\partial n}{\partial T}\right|_{\mu}}
           {\left.\frac{\partial n}{\partial \mu}\right|_{T}}\right). 
           \label{MDI_CV1}
\ea
With the aid of Eqs. (\ref{un}) and (\ref{i1}), the energy density of MDI 
can be expressed as 
\be
\varepsilon = \frac{1}{2m}\tau + \int u(n)dn + \frac{C \Lambda^2}{n_0}I_1
\ee
and therefore, its derivatives with respect to $\mu$ and $T$ are given by 
\ba
\left.\frac{\partial \varepsilon}{\partial \mu}\right|_{T} &=& 
  \frac{1}{2m}\left.\frac{\partial \tau}{\partial \mu}\right|_{T}
  +u(n)\left.\frac{\partial n}{\partial \mu}\right|_{T}
  +\frac{C \Lambda^2}{n_0}\left.\frac{\partial I_1}{\partial \mu}\right|_{T}  \\
\left.\frac{\partial \varepsilon}{\partial T}\right|_{\mu} &=& 
  \frac{1}{2m}\left.\frac{\partial \tau}{\partial T}\right|_{\mu}
  +u(n)\left.\frac{\partial n}{\partial T}\right|_{\mu}
  +\frac{C \Lambda^2}{n_0}\left.\frac{\partial I_1}{\partial T}\right|_{\mu}    
\ea
where Eqs. (\ref{dndmu})-(\ref{dtaudt}) and (\ref{di1dmu})-(\ref{di1dt}) 
are to be utilized. These relations provide a cross-check of evaluating $C_V$ in Eq. (\ref{MDI_CV1})  directly through appropriate tabulations of the non degenerate expression for $\epsilon$ (see Appendix \ref{Sec:AppendixB})  as  a function of $\mu$ and $T$.

The specific heat at constant pressure is obtained from 
\be
C_P = C_V + \frac{T}{n^2}\frac{\left(\left.\frac{\partial P}{\partial T}\right|_{n}\right)^2}
                      {\left.\frac{\partial P}{\partial n}\right|_{T}} \,,\label{MDI_Cp}
\ee
where [by means of a Jacobi transformation from $(n,T)$ to $(\mu,T)$]
\ba
\left.\frac{\partial P}{\partial T}\right|_{n} &=&
\left.\frac{\partial P}{\partial T}\right|_{\mu} 
    -\frac{\left.\frac{\partial P}{\partial \mu}\right|_{T}
              \left.\frac{\partial n}{\partial T}\right|_{\mu}}
           {\left.\frac{\partial n}{\partial \mu}\right|_{T}}  \\
\left.\frac{\partial P}{\partial n}\right|_{T} &=&
     \frac{\left.\frac{\partial P}{\partial \mu}\right|_{T}}
           {\left.\frac{\partial n}{\partial \mu}\right|_{T}}.
\ea

\medskip

Starting from Eq. (\ref{pres2}), and observing that 
\ba
\frac{\partial U(n,p)}{\partial p} &=& \frac{\partial R}{\partial p} \\
 &=&  -\frac{R}{p}
   +\frac{2C}{n_0}\frac{\Lambda^2}{\pi^2\hbar^3}\frac{1}{p}\int dp'~p'^2  \nonumber \\
   &\times& \frac{(\Lambda^2-p^2+p'^2)}{[\Lambda^2+(p+p')^2][\Lambda^2+(p-p')^2]}f'
\ea
we can write the pressure as 
\be
P = \frac{1}{3m}\tau + \int u(n)dn + \frac{C \Lambda^2}{3n_0}I_1   
          + \frac{C \Lambda^2}{3n_0}I_2.
\ee
Hence
\ba
\left.\frac{\partial P}{\partial \mu}\right|_{T} &=& 
  \frac{1}{3m}\left.\frac{\partial \tau}{\partial \mu}\right|_{T}
  +u(n)\left.\frac{\partial n}{\partial \mu}\right|_{T}  \nonumber \\
  &+& \frac{C \Lambda^2}{3n_0}\left.\frac{\partial I_1}{\partial \mu}\right|_{T} 
  + \frac{C \Lambda^2}{3n_0}\left.\frac{\partial I_2}{\partial \mu}\right|_{T} \\
\left.\frac{\partial P}{\partial T}\right|_{\mu} &=& 
  \frac{1}{3m}\left.\frac{\partial \tau}{\partial T}\right|_{\mu}
  +u(n)\left.\frac{\partial n}{\partial T}\right|_{\mu}  \nonumber \\
  &+& \frac{C \Lambda^2}{3n_0}\left.\frac{\partial I_1}{\partial T}\right|_{\mu}
  + \frac{C \Lambda^2}{3n_0}\left.\frac{\partial I_2}{\partial T}\right|_{\mu} . 
\ea        

As with $C_V$, the derivatives above provide an alternate means to check the 
evaluation of $C_P$ in Eq. (\ref{MDI_Cp})  directly through tabulations 
of the relevant quantities as functions of $\mu$ and $T$.

\subsection*{Zero-range models}                      
                      
For a generic single-species Skyrme model with the spectrum
\be
\epsilon_p = \frac{p^2}{2m} + Anp^2 + \frac{\partial \mathcal{H}_d}{\partial n} \,,
\ee
we have
\ba
\left.\frac{\partial f}{\partial \mu}\right|_{T} &=&      
  \frac{1}{T}f(1-f)\left(1-Ap^2\left.\frac{\partial n}{\partial \mu}\right|_{T}\right)  \\
\left.\frac{\partial f}{\partial T}\right|_{\mu} &=&      
  \frac{1}{T}f(1-f)  \nonumber \\
  &\times& \left[\ln\left(\frac{1-f}{f}\right)-Ap^2\left.\frac{\partial n}{\partial T}\right|_{\mu}\right] \,,
\ea
and thus 
\ba
\left.\frac{\partial n}{\partial \mu}\right|_{T} &=&  
  \frac{g}{2\pi^2\hbar^3}\int dp~p^2\frac{1}{T}f(1-f)  \nonumber \\
  &\times& \left(1-Ap^2\left.\frac{\partial n}{\partial \mu}\right|_{T}\right)  \\
\left.\frac{\partial n}{\partial T}\right|_{\mu} &=& 
 \frac{g}{2\pi^2\hbar^3}\int dp~p^2     
  \frac{1}{T}f(1-f)   \nonumber  \\
  &\times& \left[\ln\left(\frac{1-f}{f}\right)-Ap^2\left.\frac{\partial n}{\partial T}\right|_{\mu}\right] .
\ea
Collecting $\left.\partial n/\partial \mu \right|_{T}$ and
$\left.\partial n/\partial T \right|_{\mu}$ to the left hand side, we get
\ba
\left.\frac{\partial n}{\partial \mu}\right|_{T} &=& 
\frac{\frac{g}{2\pi^2\hbar^3}\frac{1}{T}\int dp~p^2f(1-f)}
     {1+\frac{gA}{2\pi^2\hbar^3}\frac{1}{T}\int dp~p^4f(1-f)}  \\
\left.\frac{\partial n}{\partial T}\right|_{\mu} &=& 
\frac{\frac{g}{2\pi^2\hbar^3}\frac{1}{T}\int dp~p^2f(1-f)\ln\left(\frac{1-f}{f}\right)}
     {1+\frac{gA}{2\pi^2\hbar^3}\frac{1}{T}\int dp~p^4f(1-f)}.  
\ea
With the $\mu$ and $T$ derivatives of $f$ and $n$ completely determined, the 
same can be done for $\tau$ and consequently for $\varepsilon$, $P$, and the 
specific heats. The results are identical to those obtained using the method we used in Ref. \cite{APRppr}.  

\medskip


\bibliographystyle{h-physrev3}
\bibliography{mdyi}

\end{document}